\documentclass[12pt]{article}
\usepackage{graphics,amssymb,float}
\usepackage{graphicx} 
\usepackage{rotating} 
\usepackage{dcolumn}
\usepackage{bm}
\usepackage{cite}
\usepackage{amssymb,amsmath,amsbsy}
\usepackage{mathrsfs}
\usepackage{dsfont}
\usepackage{caption}
\usepackage{indentfirst} 

\usepackage[colorlinks=true
,urlcolor=blue 
,anchorcolor=blue
,citecolor=blue
,filecolor=blue
,linkcolor=blue
,menucolor=blue
,pagecolor=blue
,linktocpage=true
]{hyperref}
\usepackage{epstopdf}
\usepackage{slashed,xcolor}

\DeclareMathOperator\sgn{sgn}
\newenvironment{Eqnarray}%
         {\arraycolsep 0.14em\begin{eqnarray}}{\end{eqnarray}}
\def\br{{\rm BR}}

\def\sina{\sin\alpha}

\def\cosa{\cos\alpha}
\def\tanb{\tan\beta}
\def\sinb{\sin\beta}
\def\cosb{\cos\beta}
\def\logZ{\log_{10}|Z_6|}
\def\cotb{\cot\beta}
\def\sbma{s_{\beta-\alpha}}
\def\cbma{c_{\beta-\alpha}}
\def\sbmaii{s^2_{\beta-\alpha}}
\def\cbmaii{c^2_{\beta-\alpha}}

\def\cbmaiii{c^3_{\beta-\alpha}}

\def\gam{\gamma}
\def\ie{{\it i.e.}}
\def\eg{{\it e.g.}}

\def\gev{~{\rm GeV}}

\def\bit{\begin{itemize}}
\def\eit{\end{itemize}}
\def\ben{\begin{enumerate}}
\def\een{\end{enumerate}}
\def\beq{\begin{equation}}
\def\eeq{\end{equation}}

\def\Eq#1{Eq.~(\ref{#1})}

\def\phm{\phantom{-}}

\def\bea{\begin{Eqnarray}}
\def\eea{\end{Eqnarray}}

\def\cbma{\cos(\beta-\alpha)}
\def\pb{~{\rm pb}}
\def\fb{~{\rm fb}}

\def\beq{\begin{equation}}
\def\eeq{\end{equation}}
\def\beqa{\begin{Eqnarray}}
\def\eeqa{\end{Eqnarray}}

\def\cbma{c_{\beta-\alpha}}

\def\sbma{s_{\beta-\alpha}}

\def\phm{\phantom{-}}

\def\beq{\begin{equation}}
\def\eeq{\end{equation}}
\def\ifmath#1{\relax\ifmmode #1\else $#1$\fi}

\def\tb{t_{\beta}}

\def\sb  {s_{\beta}}
\def\cb  {c_{\beta}}

\def\sba  {s_{\beta-\alpha}}
\def\cba  {c_{\beta-\alpha}}
\def\tanb{\tan\beta}
\def\cotb{\cot\beta}
\def\sinb{\sin\beta}
\def\cosb{\cos\beta}
\def\sina{\sin\alpha}
\def\cosa{\cos\alpha}

\def\sbmaii{s^2_{\beta-\alpha}}
\def\cbmaii{c^2_{\beta-\alpha}}

\def\hl{h}
\def\ha{A}
\def\hh{H}
\def\hpm{{H^\pm}}

\def\hp{{H^+}}
\def\hm{{H^-}}

\def\mha{m_{\ha}}
\def\mhl{m_{\hl}}
\def\mhh{m_{\hh}}
\def\mhpm{m_{\hpm}}
\def\mhat{\overline{m}}

\def\ls#1{\ifmath{_{\lower1.5pt\hbox{$\scriptstyle #1$}}}}
\def\lss#1{\ifmath{^{\,\lower2.5pt\hbox{$\scriptstyle #1$}}}}

\def\half{\tfrac{1}{2}}

\def\ie{{\it i.e.}}
\def\eg{{\it e.g.}}

\def\typei{Type~I}
\def\typeii{Type~II}

\def\lsim{\mathrel{\raise.3ex\hbox{$<$\kern-.75em\lower1ex\hbox{$\sim$}}}}
\def\gsim{\mathrel{\raise.3ex\hbox{$>$\kern-.75em\lower1ex\hbox{$\sim$}}}}
\def\ifmath#1{\relax\ifmmode #1\else $#1$\fi}

\def\eq#1{Eq.~(\ref{#1})}
\def\Eq#1{Eq.~(\ref{#1})}

\def\eqs#1#2{Eqs.~(\ref{#1}) and (\ref{#2})}
\def\eqss#1#2#3{Eqs.~(\ref{#1}), (\ref{#2}) and (\ref{#3})}
\def\eqst#1#2{Eqs.~(\ref{#1})--(\ref{#2})}

\begin{document}
\begin{titlepage}
\begin{center}

\vspace*{-1.5cm}
\begin{flushright}
LPSC15274,\\
UCD-2015-002,\\  
SCIPP-15/13\\
\end{flushright}

\vspace*{1.5cm}
{\Large\bf Scrutinizing the Alignment Limit\\[2mm] in Two-Higgs-Doublet Models} 

\vspace*{0.8cm}

{\large\bf\boldmath Part~2: $m_H=125$~GeV} 

\vspace*{0.8cm}

\renewcommand{\thefootnote}{\fnsymbol{footnote}}

{\large
J\'er\'emy~Bernon$^{1}$\footnote[1]{Email: bernon@lpsc.in2p3.fr},
John~F.~Gunion$^{2}$\footnote[2]{Email: jfgunion@ucdavis.edu},
Howard~E.~Haber$^{3}$\footnote[3]{Email: haber@scipp.ucsc.edu},
Yun~Jiang$^{2,4}$\footnote[4]{Email: yunjiang@nbi.ku.dk},
Sabine~Kraml$^{1}$\footnote[5]{Email: sabine.kraml@lpsc.in2p3.fr}  
}

\renewcommand{\thefootnote}{\arabic{footnote}}

\vspace*{0.8cm}
{\normalsize \it
$^1\,$Laboratoire de Physique Subatomique et de Cosmologie, Universit\'e Grenoble-Alpes,
CNRS/IN2P3, 53 Avenue des Martyrs, F-38026 Grenoble, France\\[2mm]
$^2\,$Department of Physics, University of California, Davis, CA 95616, USA \\[2mm]
$^3\,$Santa Cruz Institute for Particle Physics, Santa Cruz, CA 95064, USA\\[2mm]
$^4\,$ NBIA and Discovery Center, Niels Bohr Institute, University of Copenhagen, \\
Blegdamsvej 17, DK-2100, Copenhagen, Denmark
}
\vspace{0.8cm}

\begin{abstract}
In the alignment limit of a multi-doublet Higgs sector, one of the Higgs mass eigenstates aligns in field space with the direction of the scalar field vacuum expectation values, and its couplings approach those of the Standard Model (SM) Higgs boson.  We consider CP-conserving Two-Higgs-Doublet Models (2HDMs) of Type~I and Type~II near the alignment limit in which the heavier of the two CP-even Higgs bosons,
$H$, is the SM-like state observed with a mass of $125\gev$, and
the couplings of $H$ to gauge bosons approach those of the SM.
We review the theoretical structure and analyze the phenomenological implications of 
this particular realization of the alignment limit, where decoupling of the extra states cannot occur given that the lighter CP-even state $h$ must, by definition, have a mass below $125\gev$. 
For the numerical analysis, we perform scans of the 2HDM parameter space employing the software packages \texttt{2HDMC} and \texttt{Lilith}, taking into account all relevant pre-LHC constraints, constraints from the measurements of the $125\gev$ Higgs signal at the LHC,  as well as the most recent limits coming from searches for other Higgs-like states. Implications for Run~2 at the LHC, including expectations for observing the other scalar states, are also discussed.

\end{abstract}

\end{center}
\end{titlepage}

\section{Introduction}\label{sec:intro}

While the Higgs boson measurements at Run~1 of the LHC \cite{Aad:2015gba,Khachatryan:2014jba,ATLAS-CONF-2015-044} (see also \cite{Bernon:2014vta,Corbett:2015ksa}) show no deviations from Standard Model (SM) expectations, conceptually there is no reason why the Higgs sector should be minimal. Indeed a non-minimal Higgs sector is theoretically very attractive and, if confirmed, would shine a new light on the dynamics of electroweak symmetry breaking. The challenge for Run~2 of the LHC, and other future collider programs, is to determine whether the observed state with mass 125~GeV is \textit{the} SM Higgs boson, or whether it is part of a non-minimal Higgs sector of a more fundamental theory. 

In models with a multi-doublet Higgs sector such as the Two-Higgs-Doublet Model~\cite{Gunion:1989we,Branco:2011iw}
 (2HDM), which is the focus of our study, a special situation arises when one of the Higgs mass eigenstates is approximately aligned in field space with the direction of the scalar field vacuum expectation values (vevs).  This motivates the introduction of the Higgs basis, in which the scalar vev resides entirely in one linear combination of scalar fields.  In the approach to the so- called \textit{alignment limit},~\!\footnote{Aspects of the alignment limit were first emphasized in
\cite{Gunion:2002zf} and investigated further in \cite{Delgado:2013zfa, Craig:2013hca, Asner:2013psa, Carena:2013ooa,
Haber:2013mia,Dev:2014yca}.}
the $W^\pm$ and $Z$ gauge bosons dominantly acquire their masses from the Higgs doublet of the Higgs basis with the non-zero vev, and the coupling of that Higgs boson to $W^+W^-$ (and $ZZ$) tends toward its SM value, $C_V\to 1$.\footnote{We use the notation of coupling scale factors, or {\em reduced couplings},  employed in \cite{Bernon:2014vta}: $C_V$ ($V=W,Z$) for the coupling to gauge bosons, $C_{U,D}$ for the couplings to up-type and down-type fermions and $C_{\gamma,g}$ for the loop-induced couplings to photons and gluons.}  
Of course, for consistency with the LHC measurements, this Higgs boson must then be identified with the observed SM-like state at 125~GeV.  

In a recent paper \cite{Bernon:2015qea}, we provided a comprehensive study of the alignment limit in the context of CP-conserving 2HDMs of Type~I and Type~II, assuming that the observed $125\gev$ state is the lighter of the two CP-even Higgs bosons, $h$, in these models. 
Whereas alignment is automatically attained in the decoupling limit in the case of $m_h=125\gev$ when the additional Higgs states $H,\, A,\, H^\pm$ are very heavy, it can also occur when the additional Higgs states  are light, i.e.~alignment without decoupling~\cite{Gunion:2002zf}. The purpose of \cite{Bernon:2015qea} was to investigate the phenomenological consequences of alignment without decoupling in the $m_h=125\gev$ scenario and contrast the resulting phenomenology to the case of decoupling.
In this paper, we now focus on the equally interesting but much less studied possibility that the observed $125\gev$ state is the heavier $H$ of the two CP-even Higgs bosons of the 2HDM. In this case, alignment is always attained without decoupling, since by definition $h$ is lighter than $H$. Furthermore, the masses of the CP-odd Higgs boson $A$ and the charged Higgs boson $H^\pm$ are limited by the requirements of stability, perturbativity and electroweak precision measurements.  
The alignment limit in the $m_H=125\gev$ scenario therefore offers a very specific phenomenology that is worth contrasting to that of the  $m_h=125\gev$ scenario. 

The paper is organized as follows. In Section~\ref{modelreview} we elaborate on theoretical considerations that are specific to the alignment limit in the case of $m_H=125\gev$. 
The numerical results of our study are presented in Section~\ref{results-lighth}.  
Two aspects are considered in detail: the precision measurements of the couplings and signal strengths of the SM-like Higgs boson at 125~GeV, and the ways to discover the additional Higgs states of the 2HDM when they are light. 
Section~\ref{conclusions} presents our conclusions. 
Throughout the paper we follow the notation and conventions used in \cite{Bernon:2015qea}. 
The setup of the numerical analysis and the constraints applied also follow \cite{Bernon:2015qea}. 
In addition, we consider the new CMS limit~\cite{Khachatryan:2015baw} for light neutral Higgs bosons with masses between 25 GeV and 80 GeV, produced in association with a pair of $b$ quarks and decaying into $\tau\tau$. 
Moreover, we take into account the CMS limits \cite{CMS-PAS-HIG-15-001} 
on $gg\to A\to Zh$ with $Z\to \ell\ell$ and $h\to b\bar b$ or $\tau\tau$, which significantly constrain  the 
scenario studied in this paper (but are much less relevant for the analysis of \cite{Bernon:2015qea}). 
Details on the CMS $gg\to A\to Zh$ limits and their impact on the 2HDM parameter space are given in the Appendix.

\section{Theoretical considerations}
\label{modelreview}

In this section, we expand on the theoretical discussion in \cite{Bernon:2015qea} (see also \cite{Haber:2015pua}), treating questions that are relevant specifically for a SM-like $H$ at 125~GeV. 
It is convenient to work in the Higgs basis~\cite{Branco:1999fs,Davidson:2005cw}, where the vev, $v=2m_W/g\simeq 246$~GeV, resides entirely in one of the two Higgs doublet fields,
\beq \label{vevs}
\langle H_1^0\rangle=v/\sqrt{2}\quad  {\rm and}\quad  \langle H_2^0\rangle=0\,.
\eeq
The scalar potential in the Higgs basis is
\beqa
\mathcal{V}&=& Y_1 H_1^\dagger H_1+Y_2 H_2^\dagger H_2
+Y_3[H_1^\dagger H_2+{\rm h.c.}]
+\half Z_1(H_1^\dagger H_1)^2
+\half Z_2(H_2^\dagger H_2)^2
+Z_3(H_1^\dagger H_1)(H_2^\dagger H_2)
\nonumber\\[8pt]
&&\quad\quad 
+Z_4(H_1^\dagger H_2)(H_2^\dagger H_1)
+\left\{\half Z_5(H_1^\dagger H_2)^2
+\big[Z_6(H_1^\dagger H_1)
+Z_7(H_2^\dagger H_2)\big]
H_1^\dagger H_2+{\rm h.c.}\right\}, \label{potZ}
\eeqa
where $Y_1=-\half Z_1 v^2$ and $Y_3=-\half Z_6 v^2$ at the scalar potential minimum.
For simplicity, we assume that the field $H_2$ can be rephased such that the potentially complex parameters $Z_5$, $Z_6$ and $Z_7$ are real, in which case the scalar potential and Higgs vacuum are CP-conserving.  Henceforth, we will always adopt such a ``real basis''.\footnote{No rephasing of $H_1$ is permitted since by assumption the vev $v$ is real and positive.}  In order to preserve perturbativity and tree-level unitarity~\cite{Huffel:1980sk,Maalampi:1991fb,Kanemura:1993hm,Akeroyd:2000wc,Ginzburg:2005dt,Kanemura:2015ska}, the dimensionless couplings $Z_i$ cannot be taken arbitrary large.  Generically, the $Z_i$ are $\mathcal{O}(1)$ constants, although it is possible for some of the $Z_i$ to be as large as $\sim 10$ without violating any low-energy constraints.\footnote{Taking the $Z_i$ significantly larger than $\mathcal{O}(1)$ will lead to Landau poles at an energy scale below the Planck scale~\cite{Chakrabarty:2014aya,Das:2015mwa,Ferreira:2015rha,Chowdhury:2015yja}.   However, we shall take an agnostic view in our scans by treating the 2HDM as an effective low-energy theory with no assumptions on its behavior at higher energies.}  

Under the assumption of a CP-conserving Higgs sector, the Higgs mass spectrum is easily determined.  
The squared-masses of the charged Higgs and CP-odd Higgs bosons are given by
\beqa
m_{H^\pm}^2&=&Y_2+\half Z_3 v^2 \label{chhiggs}\,,\\
\mha^2&=& m_{H^\pm}^2+\half(Z_4-Z_5)v^2\label{cpodd}\,,
\eeqa
and the two CP-even squared masses are obtained by diagonalizing the CP-even Higgs squared-mass matrix,
 \beq \label{Hmm}
\mathcal{M}_H^2=\begin{pmatrix} Z_1 v^2 & \quad Z_6 v^2 \\  Z_6 v^2 & \quad m_A^2+Z_5 v^2\end{pmatrix}\,.
\eeq
The physical mass eigenstates are
\beqa
\hh &=&(\sqrt{2}\,{\rm Re\,}H_1^0-v)\cbma-
\sqrt{2}\,{\rm Re\,}H_2^0\,\sbma\,,\label{Hscalareigenstates}\\
\hl &=&(\sqrt{2}\,{\rm Re\,}H_1^0-v)\,\sbma+
\sqrt{2}\,{\rm Re\,}H_2^0\,\cbma\,,
\label{hscalareigenstates}
\eeqa
where $m_h\leq m_H$.  In \eqs{Hscalareigenstates}{hscalareigenstates}, the CP-even Higgs mixing angle in the Higgs basis is denoted by $\alpha-\beta$ and the notation $c_{\beta-\alpha}\equiv \cos(\beta-\alpha)$ and $s_{\beta-\alpha}\equiv\sin(\beta-\alpha)$ is employed.  The resulting CP-even Higgs squared-masses are given by
\beq \label{cpevenmasses}
m^2_{H,h}=\half\biggl[m_A^2+(Z_5+Z_1)v^2 \pm \sqrt{[m_A^2+(Z_5-Z_1)v^2]^2+4Z_6^2 v^4}\,\biggr]\,.
\eeq

In light of \eq{vevs},
if $\sqrt{2}\,{\rm Re}~H_1^0-v$ were a mass eigenstate, then its tree-level couplings to SM particles and its self-couplings would be precisely those of the SM Higgs boson.  That is,  if one of the neutral CP-even Higgs mass eigenstates is approximately aligned in field space with the direction of the vev (the so-called alignment limit), then the couplings of this Higgs boson are SM-like. 
In  \cite{Bernon:2015qea}, we examined the case of a SM-like $h$, 
where alignment can be achieved in two ways. First, in the decoupling limit of the 2HDM with $m_{A}\gg v$, the mixing of states in \eq{Hmm} is automatically negligible. Second, alignment without decoupling occurs if $|Z_6| v^2\ll Z_1 v^2<m_A^2+Z_5 v^2$. In both cases, $h\simeq\sqrt{2}\,{\rm Re}~H_1^0-v$, corresponding to $|\cbma|\ll 1$.

In this paper we consider the case in which the heavier of the two CP-even neutral scalars~$H$ is identified as the SM-like Higgs boson with a mass of 125 GeV.  
Equations~\eqref{Hscalareigenstates} and \eqref{hscalareigenstates} then imply that 
$H\simeq\sqrt{2}\,{\rm Re}~H_1^0-v$, corresponding to $|\sbma|\ll 1$.  A SM-like $H$ can only be achieved~if 
\beqa
&& m_A^2+Z_5 v^2<Z_1 v^2\,,\label{ineq1}\\
&& |Z_6|v^2\ll |m_A^2+(Z_5-Z_1)v^2|\,,\label{ineq2}
\eeqa
in which case the CP-even Higgs squared masses are given by
\beqa
\mhh^2&=&Z_1 v^2 +\frac{Z_6^2 v^4}{|m_A^2+(Z_5-Z_1)v^2|}+\mathcal{O}(Z_6^4)\,,\label{mhhapp}\\
\mhl^2&=&m_A^2+Z_5 v^2-\frac{Z_6^2 v^4}{|m_A^2+(Z_5-Z_1)v^2|}+\mathcal{O}(Z_6^4)\label{mhlapp}\,,
\eeqa
in the approach to the alignment limit.
In contrast to the case of a SM-like $h$, there is no analog of the decoupling limit in which $H$ is SM-like since 
$h$ is necessarily lighter than $H$, and the masses of $H^\pm$ and $A$ are typically of $\mathcal{O}(v)$ in light of \eqs{cpodd}{ineq1}.

The physical masses of the neutral scalars are related by the following exact expressions,
\beqa
Z_1 v^2&=&\mhl^2 s^2_{\beta-\alpha}+\mhh^2 c^2_{\beta-\alpha}\,,\label{z1v}\\
Z_6 v^2&=&(\mhl^2-\mhh^2)\sbma\cbma\,,\label{z6v} \\
Z_5 v^2&=&\mhh^2 s^2_{\beta-\alpha}+\mhl^2 c^2_{\beta-\alpha}-m_A^2\,.\label{z5v}
\eeqa
Using \eqs{cpodd}{z5v}, it also follows that
\beq \label{z4v}
Z_4 v^2=\mhh^2 s^2_{\beta-\alpha}+\mhl^2 c^2_{\beta-\alpha}+\mha^2-2\mhpm^2\,.
\eeq
These equations exhibit all the features discussed above.  In the exact alignment limit where $H$ is identified as the SM Higgs boson and $\sbma=0$, we see that $Z_6=0$, $m_h^2=m_A^2+Z_5 v^2$, $m_H^2=Z_1 v^2$ and the inequalities of \eqs{ineq1}{ineq2} are satisfied.  Indeed, all Higgs boson masses are of $\mathcal{O}(v)$ in this limit.  In contrast, if $Z_6=0$ is satisfied by taking $\cbma=0$, it is evident that it is $h$ that is SM-like.  

The consequences of \eqst{z1v}{z5v} obtained in \cite{Bernon:2015qea} were written in a form that was convenient for the case in which $h$ is the SM-like Higgs boson, i.e.~where $|\cbma|\ll 1$.  In the case where $H$ is the SM-like Higgs boson, i.e.~$|\sbma|\ll 1$, it is more useful rewrite the expressions obtained from \eqst{z1v}{z5v} as follows,
\beqa
\mhh^2&=&\left(Z_1 -Z_6 \frac{\sbma}{\cbma}\right)v^2\,,\label{hh2mass} \\
\mhl^2&=&m_A^2+\left(Z_5+Z_6 \frac{\sbma}{\cbma}\right)v^2\,.\label{hl2mass} 
\eeqa
Note that \eqs{z1v}{z6v} imply that 
\beqa 
&& Z_6\sbma\cbma\leq 0\,,\label{sgnz6} \\[8pt] 
&&|Z_6| v^2=\sqrt{\bigl(\mhh^2-Z_1 v^2)(Z_1 v^2-\mhl^2\bigr)}\,.\label{Hrelation}
\eeqa

One can also derive expressions for $\cbma$ and $\sbma$ directly from \eqs{z1v}{z6v}. In light of \eq{sgnz6}, the sign of the product $\sbma\cbma$ is fixed by
the sign of $Z_6$.  However, since $\beta-\alpha$ is defined only modulo $\pi$, we are free to choose
a convention where either $\cbma$ or $\sbma$ is always non-negative.
A convenient choice of convention is dictated by the form of the couplings of the neutral CP-even Higgs bosons to $VV$ (where $VV=W^+ W^-$ or $ZZ$).   Denoting the $\phi VV$ couplings 
($\phi=h,H$) normalized to the corresponding coupling of $VV$ to the SM Higgs boson by $C^\phi_V$, 
it follows that~\cite{Gunion:1989we,Branco:2011iw}
\beq
C_V^{h}=\sbma\,,\qquad\quad C_V^H=\cbma\,,
\eeq
as shown in Table~\ref{tab:2hdm-couplings}.

Since $|\sbma|\ll 1$ for a SM-like $H$, the value of $|\cbma|$ must be close to 1.
In order to be consistent with the standard presentation of the SM Higgs Lagrangian, we shall choose a convention where $\cbma$ is non-negative (so that $C_V^H\to + 1$ in the alignment limit).  In this convention,
\beq \label{sbmaeq}
\sbma=-\sgn(Z_6)\sqrt{\frac{\mhh^2-Z_1 v^2}{\mhh^2-\mhl^2}}=\frac{-Z_6 v^2}{\sqrt{(\mhh^2-\mhl^2)(Z_1 v^2-\mhl^2)}}\,.
\eeq
In the approach to the alignment limit, we may use \eqs{mhhapp}{mhlapp} to write
\beq
\sbma=\frac{-Z_6 v^2}{|m_A^2+(Z_5-Z_1)v^2|}+\mathcal{O}(Z_6^4)\,,
\label{sbmaeq-approx}
\eeq
subject to the inequality given in \eq{ineq2}.

Having adopted the real Higgs basis in which all the potentially complex parameters of the scalar potential are real,  one can still perform a field redefinition $H_2\to -H_2$, which would flip the signs of $Y_3$, $Z_6$ and $Z_7$.  Such a field redefinition has no physical consequence in the most general CP-conserving 2HDM.  In particular, the sign of $Z_6$ is unphysical. Indeed, as previously noted below \eq{Hrelation}, only the sign of the product $Z_6\cbma\sbma$ is meaningful.

We have emphasized above that if $H$ is SM-like, then we expect all Higgs boson masses to be of $\mathcal{O}(v)$.
Nevertheless, a parameter regime exists in which $A$ and/or $H^\pm$ can be considerably heavier than $H$.
To see how this can arise, we rewrite \eqs{z5v}{z4v} as follows,
\beqa
\mha^2&=&\mhh^2 s_{\beta-\alpha}^2+\mhl^2 c^2_{\beta-\alpha}-Z_5 v^2\,,\label{amasssq}\\
\mhpm^2&=&\mhh^2 s_{\beta-\alpha}^2+\mhl^2 c^2_{\beta-\alpha}-\half
(Z_4+Z_5)v^2\,.\label{chiggsmass}
\eeqa
Consequently,
\beqa
&& m_A\gg m_{H}\,,\,m_{H^\pm}\,,\quad\text{if $Z_5$ is large and negative and {$|Z_4+Z_5|\lsim\mathcal{O}(1)$}\,,}\label{c1}\\
&& m_{H^\pm}\gg m_H\,,\,m_A\,,\quad\text{if $Z_4+Z_5$ is large and negative and {$|Z_5|\lsim\mathcal{O}(1)$}\,,}\label{c2}\\
&& m_A\,,\, m_{H^\pm}\gg m_H\,,\quad \text{if $Z_5$ is large and negative and {$|Z_4|\lsim\mathcal{O}(1)$}\,,}\label{c3}
\eeqa
under the condition that the magnitudes of $Z_4$ and $Z_5$ are consistent with tree-level unitarity bounds and that the inequality given in \eq{ineq1} is satisfied.  
However, none of these three cases above corresponds to a decoupling limit, since in each case the low-energy effective Higgs theory contains at least one additional scalar state (namely~$h$) beyond the SM-like Higgs boson.  
Note that in the parameter regime where Eqs.~(\ref{c1}) or (\ref{c2}) is satisfied, a second scalar state beyond $h$ may be present whose mass lies below $m_H=125$~GeV.  
If the conditions of \eq{c1} are satisfied, then $m_{H^\pm}<m_H$ if $(Z_4+Z_5)v^2> -2(m_H^2-m_h^2)c^2_{\beta-\alpha}$ [cf.~\eq{chiggsmass}].  Similarly, if  the conditions of \eq{c2} are satisfied, then $m_A<m_H$ if $Z_5 v^2>-(m_H^2-m_h^2)c^2_{\beta-\alpha}$ [cf.~\eq{amasssq}].

So far, the above discussion is applicable to the scalar sector of the most general CP-conserving 2HDM.   
If we now add the most general Higgs-fermion Yukawa couplings, we encounter tree-level Higgs-mediated flavor-changing neutral currents (FCNCs) that are too large and thus in conflict with experimental data~\cite{Cheng:1987rs,Atwood:1996vj}.  It is well known that these FCNCs can be eliminated by introducing a new basis of scalar fields $\{\Phi_1,\Phi_2\}$, and a $\mathbb{Z}_2$ discrete symmetry, $\Phi_1\to +\Phi_1$ and $\Phi_2\to -\Phi_2$ under which the dimension-four terms of the Higgs scalar potential are invariant.  This new basis of scalar fields (in which the 
$\mathbb{Z}_2$ symmetry is manifest) is designated as the $\mathbb{Z}_2$-basis, and is  
given in terms of the Higgs basis fields by\footnote{The translation between the Higgs and $\mathbb{Z}_2$-bases is discussed in detail in \cite{Gunion:2002zf,Davidson:2005cw,Bernon:2015qea,Haber:2015pua}.}
\beq \label{z2basis}
\Phi_1\equiv H_1\cb-H_2\sb\,,\qquad \Phi_2\equiv \sb H_1+\cb H_2\,,
\eeq 
where
\beq \label{cbsb}
\cb\equiv\cos\beta=v_1/v\,,\qquad\quad \sb\equiv\sin\beta=v_2/v\,,
\eeq
and $\langle\Phi_i^0\rangle=v_i$ ($i=1,2$).  
The $\mathbb{Z}_2$ symmetry is then extended to the Higgs-fermion Yukawa couplings such that two of the four Higgs-quark Yukawa interaction terms (and their hermitian conjugates) vanish (and similarly for the Higgs couplings to leptons).  
There are a number of ways to accomplish this~\cite{Glashow:1976nt,Paschos:1976ay}.  The Type~I model is defined by taking all right-handed fermion fields to be odd under the $\mathbb{Z}_2$ symmetry in the Higgs-fermion interactions, whereas in Type~II Higgs-fermion interactions, only the up-type right-handed fermion field is odd under the $\mathbb{Z}_2$ symmetry~\cite{Hall:1981bc}.\footnote{All left-handed fermion fields are even under the $\mathbb{Z}_2$ symmetry in both Type~I and Type~II models.  Note that the down-type quark and charged lepton fields transform in the same way under the $\mathbb{Z}_2$.  In principle, two more models can be constructed in which 
the Higgs--quark Yukawa couplings are of Type I and the Higgs--lepton Yukawa couplings are of Type II, or vice 
versa~\cite{Barger:1989fj,Akeroyd:1996he,Aoki:2009ha,Barger:2009me,Su:2009fz}.
We do not consider these model types in this paper.}
The $\mathbb{Z}_2$ symmetry guarantees that the neutral Higgs-fermion couplings are diagonal in flavor space when expressed in the fermion mass-eigenstate basis.  
The tree-level Type~I and Type~II couplings of the neutral Higgs bosons to the fermions, normalized to the corresponding couplings of the SM Higgs boson, are given in Table~\ref{tab:2hdm-couplings}.  
%
\begin{table}[t!]
\begin{center}
\caption{Tree-level vector boson couplings $C_V$ ($V=W,Z$) and fermionic couplings $C_U$ and $C_D$
normalized to their SM values for the two CP-even scalars $h,H$ and the CP-odd scalar $A$
in Type I and Type II 2HDMs~\cite{Hall:1981bc}.}
\label{tab:2hdm-couplings}
\begin{tabular}{|c|c|c|c|c|c|}
\hline
\ & Types I and II  & \multicolumn{2}{c|}  {Type I} & \multicolumn{2}{c|}{Type II} \\
\hline
Higgs & $VV$ & up quarks & down quarks  & up quarks & down quarks  \\
&  &  &  and leptons &  &  and leptons \cr
\hline
 $h$ & $\sin(\beta-\alpha)$ & $\cosa/ \sinb$ & $\cosa/ \sinb$  &  $\cosa/\sinb$ & $-{\sina/\cosb}$   \\
\hline
 $H$ & $\cos(\beta-\alpha)$ & $\sina/ \sinb$ &  $\sina/ \sinb$ &  $\sina/ \sinb$ & $\phm\cosa/\cosb$ \\
\hline
 $A$ & 0 & $\cotb$ & $-\cotb$ & $\cotb$  & $\tanb$ \\
\hline
\end{tabular}
\end{center}
\end{table}

In particular, note that the normalized couplings of the SM-like $H$,
\beqa
\frac{\sin\alpha}{\sin\beta}&=&\cbma-\sbma\cot\beta\,,\label{coup1}\\
\frac{\cos\alpha}{\cos\beta}&=&\cbma+\sbma\tan\beta\,,\label{coup2}
\eeqa
approach unity in the limit of $\sbma\to 0$ in the convention where $\cbma$ is non-negative.
That is, in Type~I models for small $|\sbma|$, we have $C_U^H=C_D^H\equiv C_F^H \simeq C_V^H\simeq 1$ unless $\tan\beta$ is very small (which is excluded by constraints related to stability, unitarity and perturbativity, in particular the requirement of perturbativity of the top Yukawa coupling). Thus,  $C_F^H\to 1$ in the alignment limit.   In Type~II models we arrive at a similar conclusion for the up-type fermion--Higgs Yukawa coupling, namely $C^H_U\to 1$ in the alignment limit. However,  $C_D^H\simeq 1$ only when $|\sbma\tan\beta|\ll 1$, which means that if $\tan\beta\gg 1$ then the approach to the alignment limit, $C^H_D\to 1$, is delayed.

In the $\mathbb{Z}_2$-basis of scalar fields defined in \eq{z2basis}, $\alpha$ is the CP-even Higgs mixing angle, which is defined modulo~$\pi$, and $\tan\beta=v_2/v_1$ is the ratio of neutral Higgs vevs [cf.~\eq{cbsb}].   In a convention in which $\cbma$ is non-negative, the sign of $\sbma$ is determined by \eq{sgnz6}.  
However, in introducing the $\mathbb{Z}_2$-basis, there is freedom to impose an additional sign
convention such that the vevs $v_1$ and $v_2$ are non-negative.  That is, we assume henceforth that $\tan\beta$ is non-negative, or equivalently,
\beq \label{brange}
0\leq\beta\leq\half\pi\,.
\eeq
For non-negative $\tan\beta$ and $\cbma$, it is clear from \eqs{coup1}{coup2} that the sign of $\sbma$ is physically relevant.  In light of \eq{sbmaeq}, it also follows that 
the sign of $Z_6$ is now meaningful.  These observations can be understood in another way as follows.  Since $\tan\beta$ is assumed to be non-negative, one is no longer permitted to perform field redefinitions that change the relative sign of $\Phi_1$ and $\Phi_2$.  Using \eq{z2basis}, it then follows that one is no longer permitted to perform a redefinition of the Higgs basis field $H_2\to -H_2$.   But, we previously used such a field redefinition to conclude that the signs of $Z_6$ and $\sbma$ are unphysical [see the text following \eq{sbmaeq-approx}].  This is no longer possible once we fix a convention where $\tan\beta$ is non-negative. 

Another common choice in the literature is to take $-\half\pi\leq\alpha\leq\half\pi$ along with \eq{brange}, in which case parameter regimes exist in which $\cbma$ and/or $\sbma$ can take on either sign.  
It is a simple matter to translate among the various conventions.
For example, given any 2HDM parameter point $(\alpha, \beta)$ with $-\half\pi\leq\alpha\leq\half\pi$ and 
$0\leq\beta\leq\half\pi$, one can compute the values of $\sbma$ and $\cbma$.  Then, to convert to the convention of non-negative $\cbma$, one would simply replace 
\beq
   (\sbma,\cbma)\to (-\sbma,-\cbma) 
  \label{changeconvention}
\eeq   
if $\cbma$ is initially negative.  Only the \textit{relative} sign of $\sbma$ and $\cbma$ 
is physical for a given sign choice of $Z_6$ in light of \eq{sgnz6}.

Let us now examine the number of parameters that govern the CP-conserving 2HDM of Types I and II.
In order to ensure the absence of tree-level Higgs-mediated FCNCs, it is sufficient to impose
the $\mathbb{Z}_2$ discrete symmetry introduced above on all dimension-four terms of the Higgs Lagrangian.  Thus, we are free to include terms in the Higgs Lagrangian that softly break the symmetry.
There exists precisely one term of this type, namely the following dimension-two term that can be added to the scalar potential in the $\mathbb{Z}_2$-basis,
\beq
V_{\rm soft}=-m_{12}^2\Phi_1^\dagger\Phi_2+{\rm h.c.}
\label{vsoft}
\eeq
The squared-mass parameter $m_{12}^2$ is related to the Higgs basis parameters and the angle $\beta$ via
\beq \label{monetwo}
m_{12}^2=\half\bigl(Y_2+\half Z_1 v^2\bigr)\sin 2\beta+\half Z_6 v^2\cos 2\beta\,.
\eeq
Note that if $\sin 2\beta=0$, then the $\mathbb{Z}_2$ basis and the Higgs basis coincide\footnote{Strictly speaking, if $\beta=\half\pi$ then we need to interchange the Higgs basis fields $H_1$ and $H_2$ to be consistent with the definition of the Higgs basis specified in \eq{vevs}.}  (cf.~Eqs.~(24) and (25) of \cite{Bernon:2015qea}) and $Z_6=Z_7=0$. The latter implies that $m_{12}^2=0$ in light of \eq{monetwo}. 
  That is, if $\beta=0$ or $\half\pi$ then the $\mathbb{Z}_2$ discrete symmetry, under which the 
 Higgs basis fields $H_1\to +H_1$ and $H_2\to -H_2$, is exact (and cannot be softly-broken\footnote{Due to the potential minimum condition $Y_3=-\half Z_6 v^2$, it follows that $Z_6=Z_7=0$ in the Higgs basis is sufficient to guarantee that the entire scalar potential respects the $\mathbb{Z}_2$ symmetry.}).
 
As an aside, we note that the so-called Inert Doublet Model (IDM)~\cite{Deshpande:1977rw,Barbieri:2006dq,LopezHonorez:2006gr}  can be defined as a Type~I 2HDM with $Z_6=Z_7=0$~\cite{Asner:2013psa}, in which all fields (excepting $H_2$) are even under the $\mathbb{Z}_2$ discrete symmetry.
In light of the above discussion, we see that the alignment limit is exact in the IDM and 
$\sqrt{2} H_1^0-v$ is identified as the SM Higgs boson, which can either be the lighter or the heavier of the two CP-even Higgs bosons, depending on whether the inequality given in \eq{ineq1} is satisfied.  The phenomenology of the IDM has been treated in detail in \cite{Goudelis:2013uca,Belanger:2015kga,Ilnicka:2015jba}, so we do not pursue this case further in this paper.

Henceforth, we assume that $\sin 2\beta\neq 0$.
The eight free parameters of the CP-conserving, softly-broken $\mathbb{Z}_2$-symmetric 2HDM are $v$, $\alpha$, $\beta$, $\mhl$, $\mhh$, $\mha$, $\mhpm$ and $m_{12}^2$, where $0<\beta<\half\pi$ and $|\beta-\alpha|\leq\half\pi$.
The same counting can be performed in the Higgs basis.  Note that the imposition of the discrete $\mathbb{Z}_2$ symmetry on the quartic terms of the scalar potential yields two relations among the $Z_i$ as shown in \cite{Bernon:2015qea,Haber:2015pua}
\beqa 
Z_2&=& Z_1+2(Z_6+Z_7)\cot 2\beta\,,\label{ztwo}\\
Z_{3}&=&Z_1-Z_4-Z_5+2Z_6\cot 2\beta-(Z_6-Z_7)\tan 2\beta\,,\label{z345}
\eeqa
under the assumption\footnote{For $\beta=\tfrac{1}{4}\pi$, \eq{z345} implies that $Z_6=Z_7$, in which case $Z_3$ must be considered as an independent quantity.}  
that
$\beta\neq\tfrac{1}{4}\pi$.  
Thus, the eight independent parameters in the Higgs basis can be chosen as $v$, $Y_2$, $Z_1$, $Z_3$, $Z_4$, $Z_5$, $Z_6$, and $\beta$ (with $0<\beta<\half\pi$) since $Y_1$ and $Y_3$ are determined by the scalar potential minimum conditions, and $Z_2$ and $Z_7$ can be determined from \eqs{ztwo}{z345}.  
The parameter $\beta$ serves to fix the $\mathbb{Z}_2$-basis relative to the Higgs basis.  
The translation between $\{Y_2, Z_1, Z_4, Z_5, Z_6\}$
and $\{\mhl, \mhh, \mha, \mhpm, \beta-\alpha\}$ is governed by Eqs.~(\ref{chhiggs}), (\ref{cpodd}) and 
(\ref{z1v})--(\ref{z5v}).
Specifying the parameter $m_{12}^2$ then allows one to determine $Z_2$, $Z_3$ and $Z_7$.   In particular, it is convenient to introduce [cf.~\eq{monetwo}],
\beq \label{mbardef}
\overline{m}^2\equiv \frac{2m^2_{12}}{\sin 2\beta}=Y_2+\half Z_1 v^2+Z_6 v^2\cot 2\beta\,.
\eeq
Using \eqss{chhiggs}{cpodd}{z345}, it follows that
\beq \label{useful}
\overline{m}^2=\mha^2+Z_5 v^2+\half(Z_6-Z_7)v^2\tan 2\beta\,.
\eeq
Combining the results of \eqst{z1v}{z4v} with Eq.~(\ref{ztwo}), (\ref{z345}) and (\ref{useful}), we obtain,\footnote{The Higgs basis parameter $Z_2$ only appears in the quartic Higgs couplings, which we do not address in this paper.  We only provide \eq{z2v} for the sake of completeness.}
\beqa
Z_2 v^2&=&\mhl^2(\sbma+2\cbma \cot 2\beta)^2+\mhh^2(\cbma-2\sbma\cot 2\beta)^2-4\overline{m}^2\cot^2 2\beta\,,\label{z2v}\\
Z_3 v^2&=& \mhl^2 s_{\beta-\alpha}^2+\mhh^2 c_{\beta-\alpha}^2+2(\mhl^2-\mhh^2)\cbma\sbma\cot 2\beta+2(\mhpm^2-\overline{m}^2)\,,\label{z3v}\\
Z_7 v^2&=&2\bigl(\mhl^2 c_{\beta-\alpha}^2+\mhh^2 s_{\beta-\alpha}^2-\overline{m}^2\bigr)\cot 2\beta+(\mhl^2-\mhh^2)\sbma\cbma\,.\label{z7v}
\eeqa
Note that the following conditions are necessary (although not sufficient) to guarantee that the scalar potential in the Higgs basis is bounded from below~\cite{Deshpande:1977rw,Gunion:2002zf};
\beq \label{z123cond}
Z_1>0\,,\qquad\quad Z_2>0\,,\,\qquad\quad Z_3>-\sqrt{Z_1 Z_2}\,.
\eeq
The condition $Z_1>0$ is automatically satisfied in light of \eq{z1v}.  The constraints on $Z_2$ and $Z_3$ 
imposed by \eq{z123cond} place mild constraints on the Higgs parameters employed in our numerical scans.

Finally, we examine the trilinear Higgs self-couplings, focusing on those involving the $H$.  Explicit expressions for these couplings in terms of the $Z_i$ and $\beta-\alpha$ have been given in \cite{Bernon:2015qea}.  
The corresponding three-Higgs vertex Feynman rules
(including the corresponding symmetry factor for identical particles but excluding an overall factor of $i$) are given by
\beqa
\!\!\!\!\!\! g\ls{\hh\hh\hh} &=& {-3v}\bigl[
    Z_1\cbmaiii+Z_{345}\cbma\sbmaii-3Z_6\sbma\cbmaii
    -Z_7 s^3_{\beta-\alpha}\bigr]\,,\label{hhhhhh} \\[4pt]
\!\!\!\!\!\! g\ls{\hh\hl\hl} &=& -{3v}\bigl[
   Z_1\cbma\sbmaii
   +Z_{345}\cbma\left(\tfrac{1}{3}-\sbmaii\right)-Z_6\sbma(1-3\cbmaii)
    -Z_7\cbmaii\sbma\bigr]\,, \label{hhhlhl} \\[4pt]
\!\!\!\!\!\! g\ls{\hh\ha\ha} &=&
   {-v}\bigl[(Z_3+Z_4-Z_5)\cbma-Z_7\sbma\bigr]\,, \label{hhhaha}\\[4pt]
\!\!\!\!\!\! g\ls{\hh\hp\hm} &=&
   {-v}\bigl[Z_3\cbma-Z_7\sbma\bigr]\,, \label{hhhphm}
   \eeqa
where we have introduced the notation,
\beq \label{z345def}
Z_{345}\equiv Z_3+Z_4+Z_5\,.
\eeq

In the alignment limit $|\sbma|\to 0$, \eqs{hh2mass}{hhhhhh} yield:
\beq \label{HHHalign3}
g_{HHH}= g^{\rm SM}_{HHH}\biggl[1-\frac{2Z_6}{Z_1}\sbma+\left(\frac{Z_{345}}{Z_1}-\frac{2Z_6^2}{Z_1^2}-\frac{3}{2}\right)
s_{\beta-\alpha}^2+\mathcal{O}(s_{\beta-\alpha}^3)\biggr]\,,
\eeq
where the self-coupling of the SM Higgs boson is given by
\beq
 g^{\rm SM}_{HHH}=-\frac{3m_H^2}{v}=-3v\left(Z_1-Z_6\frac{\sbma}{\cbma}\right)\,.
 \eeq
 It is convenient to make use of \eq{sbmaeq} [in a convention where $\cbma\geq 0$] to write
\beq \label{sz}
\sbma=-\eta Z_6\,,
\eeq
where
\beq
\label{etastuff}
\eta\equiv \frac{v^2}{\sqrt{(\mhh^2-\mhl^2)(Z_1 v^2-\mhl^2)}}
\eeq
is a positive $\mathcal{O}(1)$ parameter.  In the approach to the alignment limit, \eqs{mhhapp}{mhlapp} yield
\beq
\eta=\frac{v^2}{|m_A^2+(Z_5-Z_1)v^2|}+\mathcal{O}(Z_6^2)\,.
\eeq
Inserting \eq{sz} in \eq{HHHalign3} yields
\beq \label{previous}
g_{HHH}= g^{\rm SM}_{HHH}\biggl\{1+\biggl[\bigl(Z_{345}-\tfrac{3}{2}Z_1\bigr)\eta^2+2\eta\biggr]\frac{Z_6^2}{Z_1}
+\mathcal{O}(Z_6^3)\biggr\}\,.
\eeq

\noindent In light of \eqs{z345}{z345def}, the parameter $Z_{345}$ depends on
$Z_6$.  We can therefore rewrite \eq{previous} as
\beq \label{gHHH}
g_{HHH}= g^{\rm SM}_{HHH}\biggl\{1+\biggl[\bigl(Z_7\tan 2\beta-\tfrac{1}{2}Z_1\bigr)\eta^2+2\eta\biggr]\frac{Z_6^2}{Z_1}
+(2\cot 2\beta-\tan 2\beta)\eta^2\frac{Z_6^3}{Z_1}+\mathcal{O}(Z_6^3)\biggr\}\,,
\eeq
where the term designated by $\mathcal{O}(Z_6^3)$ contains no
potential enhancements in the limit of $s_{2\beta}\to 0$ or
$c_{2\beta}\to 0$.    The $HHH$ coupling can thus be either suppressed or enhanced with respect to the SM.  
For example $g_{HHH}>g_{HHH}^{\rm SM}$ is possible in two cases.  If
$\tan\beta\sim 1$, then one must satisfy $(Z_7-Z_6)\eta\tan
2\beta\gsim \half Z_1\eta-2$.
Alternatively, if $\tan\beta\gg 1$, then one must satisfy
$Z_6\eta\cot 2\beta\gsim \frac{1}{4}Z_1\eta-1$.
In both cases, the $HHH$ coupling is enhanced even when
$|Z_6|$ is significantly smaller than 1.

Next, consider the $Hhh$ and $HAA$ couplings given in
\eqs{hhhlhl}{hhhaha}.  Both these couplings approach nonzero values in
the alignment limit ($\sbma\to 0$),
\beqa
g_{Hhh}&=& -vZ_{345}+\mathcal{O}(\sbma)\,, \\
g_{HAA}&=& -v(Z_{345}-2Z_5)+\mathcal{O}(\sbma)\,.
\eeqa
One can now eliminate $Z_{345}$ in favor of $Z_6$ as before.  After
employing \eq{hh2mass}, we end up with
\beq \label{Hhhalign}
g_{Hhh}=-\frac{1}{v}\biggl\{\mhh^2-(Z_6-Z_7)v^2\tan 2\beta+2Z_6 v^2\cot 2\beta+\mathcal{O}(Z_6)\biggr\}\,,
\eeq
where the term designated by $\mathcal{O}(Z_6)$ contains no
potential enhancements in the limit of $s_{2\beta}\to 0$ or
$c_{2\beta}\to 0$. 

Last but not least, it is noteworthy that 
\beq \label{HHHlim}
g_{HH^+ H^-}=-v Z_3+\mathcal{O}(\sbma)\,,
\eeq
approaches a finite nonzero value in the alignment limit.  This is
relevant for the analysis of the loop-induced process $H\to\gamma\gamma$,
which has a contribution that is mediated by a $H^\pm$ loop.
Recall that the charged Higgs mass is given by \eq{chiggsmass},
which cannot be much heavier than $\mathcal{O}(v)$.  Hence,
the charged Higgs loop is parametrically of the same order
as the corresponding SM loop contributions, thereby leading to a shift
of the effective $H\gamma\gamma$ coupling from its SM value.  
This is in stark contrast to the behavior of tree-level Higgs couplings, which
approach their SM values in the alignment limit.

Although we expect $\mhpm\lsim\mathcal{O}(v)$ over most of the 2HDM parameter space when $H$ is a SM-like Higgs boson, 
there exists a parameter regime [cf.~\eqs{c2}{c3}] in which $\mhpm\gg \mhh$.   Indeed, a heavy charged Higgs mass  is required in Type~II to avoid conflict with the observed rate for $b\to s\gamma$~\cite{Misiak:2015xwa}.
In light of \eqs{chhiggs}{cpodd},
suppose that $Y_2\ll Z_3 v^2$ where $Z_3$ is large 
[say, of $\mathcal{O}(10)$] but still consistent with the unitarity bounds.
In order to satisfy  the inequality given in \eq{ineq1},
$Z_4+Z_5$ must be negative and its magnitude must be large (but not too large in order
to satisfy the unitarity bounds).
It then follows that $\mhpm^2\simeq\half Z_3 v^2$, in which case \eq{hhhphm} yields
\beq \label{H3}
g_{HH^+ H^-}\simeq -\frac{2m_{H^\pm}^2}{v}+\mathcal{O}(\sbma)\,,
\eeq
in the approach to the alignment limit.

One can also obtain \eq{H3} by expressing $g_{\hh\hp\hm}$ in terms of the Higgs masses and the squared-mass parameter $\overline{m}^2$ defined in \eq{mbardef}.
Inserting \eqs{z3v}{z7v} into \eq{hhhphm} yields~\cite{Ferreira:2014naa,Bernon:2015qea}
\beq
 g\ls{\hh\hp\hm} = -\frac{1}{v}\biggr\{\bigl[\mhh^2+2(\mhpm^2-\overline{m}^{\,2})\bigr]\cbma-2\cot 2\beta(\mhh^2-\overline{m}^{\,2})\sbma\biggr\}\,.
\eeq
In the alignment limit where $\cbma\to 1$ (or equivalently, $Z_6\to 0$),
\beq
g_{\hh\hp\hm}=-{1\over v}(\mhh^2+2\mhpm^2-2\overline{m}^2)+\mathcal{O}(\sbma)\,,
\label{ghhhphmform}
\eeq
where $\mhh\simeq 125\gev$.  In the parameter regime where  $\mhpm$ is large [such that $Y_2\ll Z_3 v^2$ as discussed above \eq{H3}], it follows from \eq{mbardef} that $\overline{m}^2\sim\mathcal{O}(v^2)$.\footnote{If $|\cot 2\beta|\gg 1$, it is more convenient to invoke \eqs{ineq1}{useful} to conclude that $\overline{m}^2\sim\mathcal{O}(v^2)$.}
Thus, in the alignment limit with $\mhpm$ large compared to $v$, one again obtains the asymptotic result of \Eq{H3}.

We denote the one-loop $H\to\gamma\gamma$ amplitude normalized to the corresponding SM value by $C^H_\gamma$.
The coupling given in \eq{H3} matches precisely the $HH^+H^-$ interaction term of
\beq
\mathscr{L}_{\rm int}=-\frac{gm_t}{2m_W}\overline{t}tH+gm_W W_\mu^+ W^{\mu-}H-\frac{gm_{H^\pm}^2}{m_W} H^+ H^- H\,,
\eeq
given in Eq.~(2.15) of \cite{Gunion:1989we}.  Hence, we can immediately obtain an 
estimate for $C^H_\gamma$ in the alignment limit by employing the asymptotic forms for the contributions to the $H\to\gamma\gamma$ amplitude, $F_i$ (corresponding to a particle in the loop with spin $i=0,\half,1$) given in Eq.~(2.21) of \cite{Gunion:1989we},
\beq \label{chgamma}
C^H_\gamma =\frac{F_0+F_1+3e_t^2F_{1/2}}{F_1+3e_t^2F_{1/2}}\simeq 0.94\,,
\eeq
where $e_t=\tfrac{2}{3}$ is the charge of the top quark in units of $e$,
$F_0=-\tfrac{1}{3}$, $F_{1/2}=-\tfrac{4}{3}$ and $F_1=7$.\footnote{These asymptotic forms are valid when $4m_i^2/m_H^2\gg 1$, where $m_i$ is the mass of the particle in the loop.  Nevertheless, these approximations work quite well even for the $t$-quark and the $W$ boson.}
A more complete calculation taking into account finite-mass effects yields a very similar result, $C^H_\gamma\simeq 0.95$. 
That is, the contribution of the charged Higgs loop asymptotically yields a 5\% reduction in $C_\gamma^H$.

In contrast, in the case of lighter charged Higgs boson masses (which are allowed in Type~I), the 
approximate form for the $HH^+H^-$ coupling given in \eq{H3} and the asymptotic form for $F_0$ employed in 
\eq{chgamma} are no longer valid.   In particular, in the approach to the alignment limit, $g_{HH^+H^-}\simeq -vZ_3$.
When $Z_3>0$ [as in \eq{H3} where $Z_3\sim 2\mhpm^2/v^2$], the charged Higgs loop interferes destructively with the $W$ boson loop. 
However, for small values of $\mhpm$ 
there exist regions of the 2HDM parameter space where $Z_3<0$ [consistent with the bounds given in \eq{z123cond}], which
then yields an $HH^+ H^-$ coupling of the opposite sign.  In this case, the charged Higgs boson loop interferes constructively with the $W$ boson loop, thereby generating a value of $C_\gamma^H>1$.  Using
\Eq{ghhhphmform}, it follows that the sign flip of $g_{\hh\hp\hm}$ occurs roughly when $2\overline{m}_{12}^2 > 2m_{H^\pm}^2 + m_H^2$.  In practice, as we shall see in Section~\ref{results-lighth}, positive values of $\overline{m}^2$ do not exceed about $(150\gev)^2$, which implies that a light charged Higgs boson with a mass of about $\mhpm< 160\gev$ is required for $C_\gamma^H>1$.  The non-decoupling of the charged Higgs contribution and the possible sign flip in $g_{\hh\hp\hm}$ was also addressed in Appendix B of \cite{Dumont:2014wha}.

\section{\boldmath Numerical results}\label{results-lighth}

Let us now turn to the numerical scan of the 2HDM parameter space.  
The free parameters in our analysis are the four physical Higgs masses $m_h, m_H, m_{H^\pm}, m_A$, the squared-mass parameter $m_{12}^2$, the ratio of the two Higgs vacuum expectation values $\tan\beta$ and the mixing angle $\alpha$ of the CP-even Higgs squared-mass matrix. Setting $m_H\equiv 125.5\gev$,\footnote{Having performed 
the parameter scans before the publication of \cite{Aad:2015zhl} which reports a central value of the Higgs mass of 125.09 GeV, we use $125.5\gev$ as the observed Higgs mass in this analysis.} we allow the 2HDM parameters to vary in the following ranges
\beqa
&\alpha\in[-\pi/2,\pi/2],\ \ \  \tan\beta\in[0.5, 60], \ \ \ m_{12}^2\in[-(2000 \text{ GeV})^2,(2000\text{ GeV})^2], \nonumber\\[8pt]
&m_h\in[10\text{ GeV}, 121.5\text{ GeV}], \ \ \ m_{H^\pm}\in[m^{*\!},2000\text{ GeV}], \ \ \ m_A\in[5\text{ GeV}, 2000\text{ GeV}], 
\label{eq:parameterscan}
\eeqa
where $m^*$ is the lower bound on the charged Higgs mass in Type~I or Type~II. 
Note that, as in \cite{Bernon:2015qea}, the degenerate case $m_h\simeq m_H$ is not considered in this study.
Instead, we require a 4~GeV mass splitting between $h$ and $H$ in order to avoid $h$ contamination of the $H$ signal.
Since we are primarily interested in the near-alignment case, we allow at most a 1\% deviation from $|C^H_V|=1$, 
which translates into $|\sba|\lesssim 0.14$. 
We also note that although the scan was performed in terms of $\alpha$ and $\tan\beta$ as given in Eq.~\eqref{eq:parameterscan}, we will present our results in the convention of $\cbma\geq 0$. That means, given a point $(\alpha,\tanb)$ in our scan with $\cbma<0$, we convert to the positive $\cbma$ convention by making the replacement $\alpha\to\alpha+\pi$ [see also the discussion around Eq.~\eqref{changeconvention}].  Thus our alignment condition translates into $\cba\ge 0.99$ whereas $\sba$ can have either sign.

As in \cite{Bernon:2015qea}, the public tools used include 
\texttt{2HDMC}~\cite{Eriksson:2009ws} for computing couplings and decay widths and for testing theoretical constraints in the 2HDM, \texttt{Lilith 1.1.2}~\cite{Bernon:2015hsa} (with database version 15.04) for evaluating the Higgs signal strength constraints, and \texttt{SusHi-1.3.0}~\cite{Harlander:2012pb} and \texttt{VBFNLO-2.6.3}~\cite{Arnold:2008rz} for computing production cross sections at the LHC. The setup of the analysis and the experimental constraints imposed are exactly the same as in \cite{Bernon:2015qea}, with three additions. 
First, we include the recent update of the bound on the charged Higgs mass in Type~II, $\mhpm>480$~GeV at 95\%~CL~\cite{Misiak:2015xwa}, based on the observed rates for 
the weak radiative $B$-meson decay, $\overline{B}\to X_s\gamma$. 
Second, we take into account the new CMS result~\cite{CMS-PAS-HIG-15-001} on the search for a new heavy resonance decaying to a $Z$ boson and a light resonance, followed by $Z\to \ell^+\ell^-$ and the light resonance decaying to $b\bar b$ or $\tau\tau$. In particular, the cross section upper limit for the $\ell\ell b\bar b$ final state in the plane of the masses of the two resonances, Fig.~5b of \cite{CMS-PAS-HIG-15-001}, puts a very severe constraint on $gg\to A\to Zh$ with $Z\to \ell\ell$ and $h\to b\bar b$ in our study.\footnote{The corresponding ATLAS search for $A\to Zh$ with $Z\to\ell\ell$ and $h\to b\bar b$ (or $\tau\tau$)~\cite{Aad:2015wra} assumes a SM-like $h$ with 125~GeV mass and thus does not apply here. (It would apply to $A\to ZH$ in our study, but does not give any relevant constraint for this case.)}
This is in contrast to the $\mhl\simeq 125\gev$ case studied in \cite{Bernon:2015qea}, where this limit has almost no effect. There are two reasons for the larger impact in the $m_H\simeq 125\gev$, $\cba\geq 0.99$ scenarios studied here: first, the $Z\!Ah$ coupling is proportional to $\cbma$ and therefore $\br(A\to Zh)$ is typically large; second, $\br(h\to b\bar b)\approx 1$ since the $h$ is always light.
Finally, the CMS constraint~\cite{Khachatryan:2015baw} on neutral Higgs bosons with masses between 25 GeV and 80 GeV, produced in association with a pair of $b$ quarks, followed by the decay into $\tau\tau$, is also applied in our analysis. 
We find that this constraint eliminates a substantial part of the Type~II parameter space at large $\tanb$.
Unless otherwise stated, all parameter space points shown in the following satisfy all of the latest constraints. For more details on the numerical procedure, we refer the reader to \cite{Bernon:2015qea}.

\subsection{Parameters}
\label{sec:parameters}

\begin{figure}[t!]\centering
\includegraphics[width=0.5\textwidth]{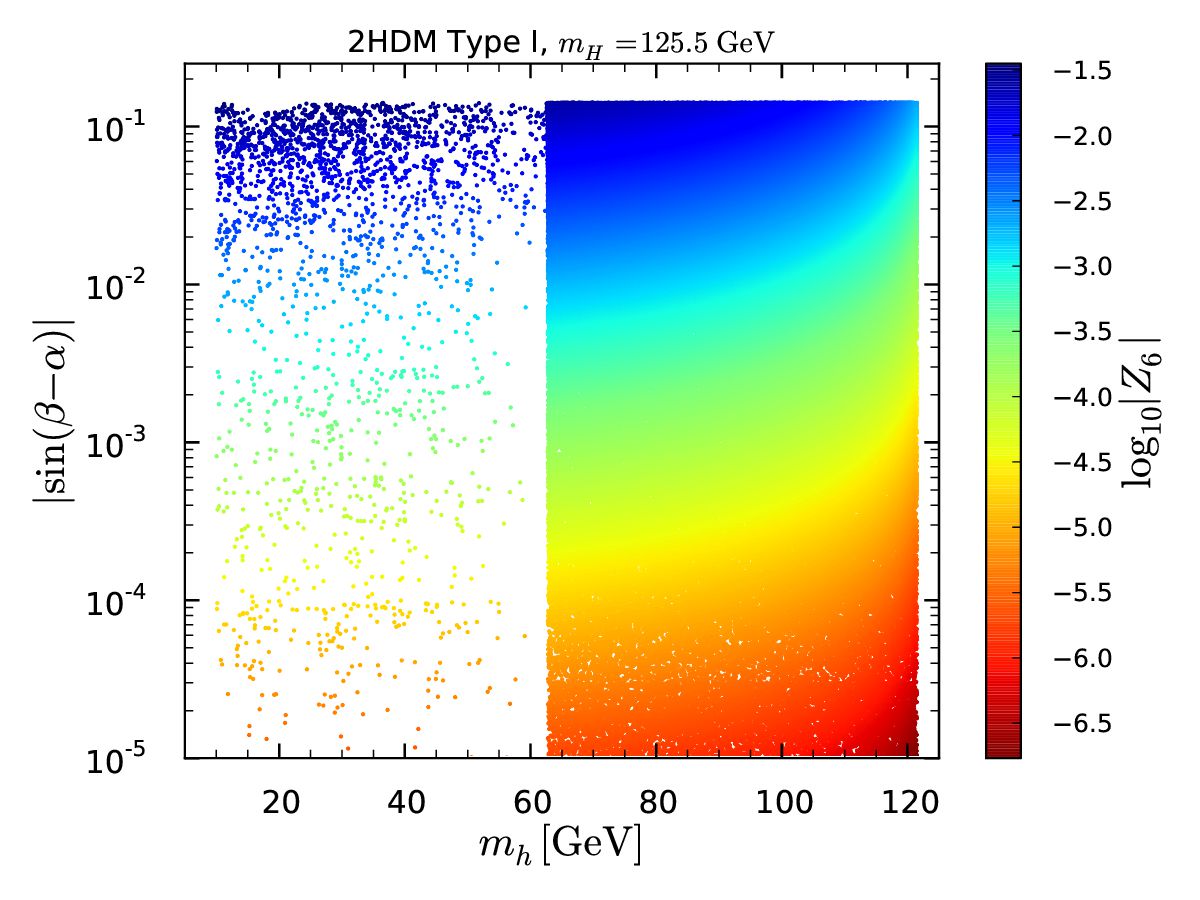}\includegraphics[width=0.5\textwidth]{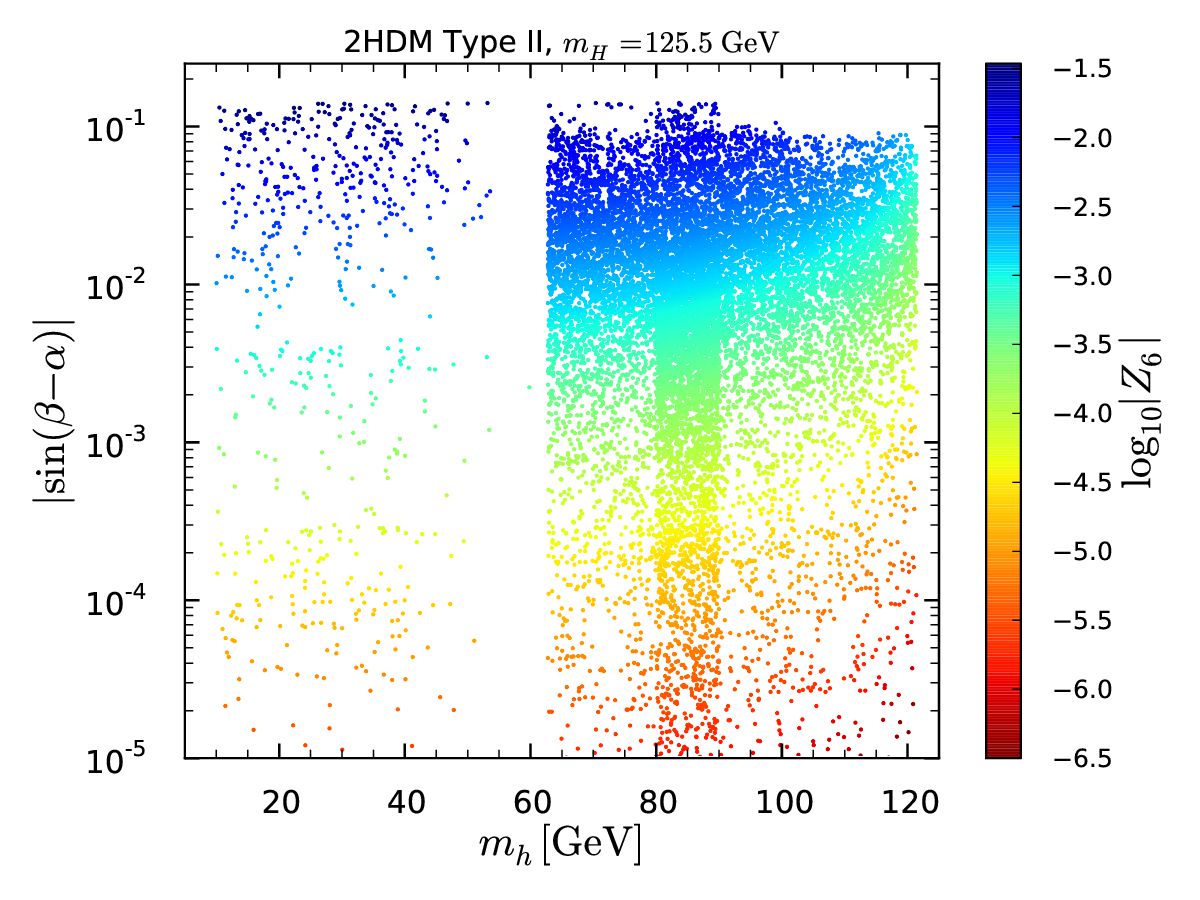}
  \caption{$|\sba|$ versus $m_h$ in Type~I (left) and Type~II (right) with $\log_{10}|Z_6|$ color code. Points are plotted in the order of high to low $\log_{10}|Z_6|$ values.}
\label{mh_sba_logZ6}
\end{figure}

We start by illustrating the parameter space of the analysis. 
Figure~\ref{mh_sba_logZ6} shows the relation between $m_h$, $|\sba|$ and $\logZ$.\footnote{In this and subsequent figures, we give 3d information on a 2d plot by means of a color code in the third dimension. To this end, we must chose a definite plotting order. Ordering the points from high to low values in the third dimension, as done for $\log_{10}|Z_6|$ in Fig.~\ref{mh_sba_logZ6}, means that the highest values are plotted first and lower and lower values are plotted on top of them. As a consequence, regions with low values may (partly) cover regions with high values. The opposite is of course true for the ordering from low to high values. To avoid a proliferation of plots, in each figure we show only one ordering, trying to choose the one that gives most information. The figures with inverted plotting order are available upon request.} 
The expected correlation between the three parameters is clearly observed. In particular, larger values of $m_h$ imply smaller $|Z_6|$ for the same value of $|\sbma|$, and for each $m_h$, $\logZ$ can be as small as desired if $|\sba|$ is allowed to be correspondingly small. Here, we show results down to $|\sba|=10^{-5}$; we  have checked that this catches all the phenomenology of the scenarios under consideration. Because of the absence of a decoupling limit, $|Z_6|$ does not exceed $\sim 10^{-1.5}$ in our scan. This illustrates that in the scenario under consideration, alignment is solely controlled by the smallness of $|Z_6|$. Note also that the region of $m_h\leq \half m_H$ requires subtle correlations among the 2HDM parameters 
to ensure that  $\br(H\to hh)$ is sufficiently small to be in agreement with the experimental constraints~\cite{Bernon:2014nxa}.
This explains the relatively low density of points in this region. 
On the other hand, the higher density of points seen in Type~II for $m_h\in[80\gev,90\gev]$ arises because light neutral states $X=h,A$ with masses below 80 GeV are severely constrained by the CMS $b\bar{b}X$ with $X\to\tau\tau$ search~\cite{Khachatryan:2015baw},\footnote{The CMS analysis given in \cite{Khachatryan:2015baw} considers only $pp\to b\bar bA$ production with $A\to\tau\tau$. However, the same limit should also apply to $pp\to b\bar bh$ with $h\to\tau\tau$; see \eg~\cite{Gunion:1996xu}.} while masses above 90 GeV are also constrained by the ATLAS~\cite{Aad:2014vgg} and CMS~\cite{Khachatryan:2014wca} searches for $X\to\tau\tau$ decays in both the $gg\to X$ and $b\bar{b}X$ production modes. 

\begin{figure}[t!]\centering
\includegraphics[width=0.495\textwidth]{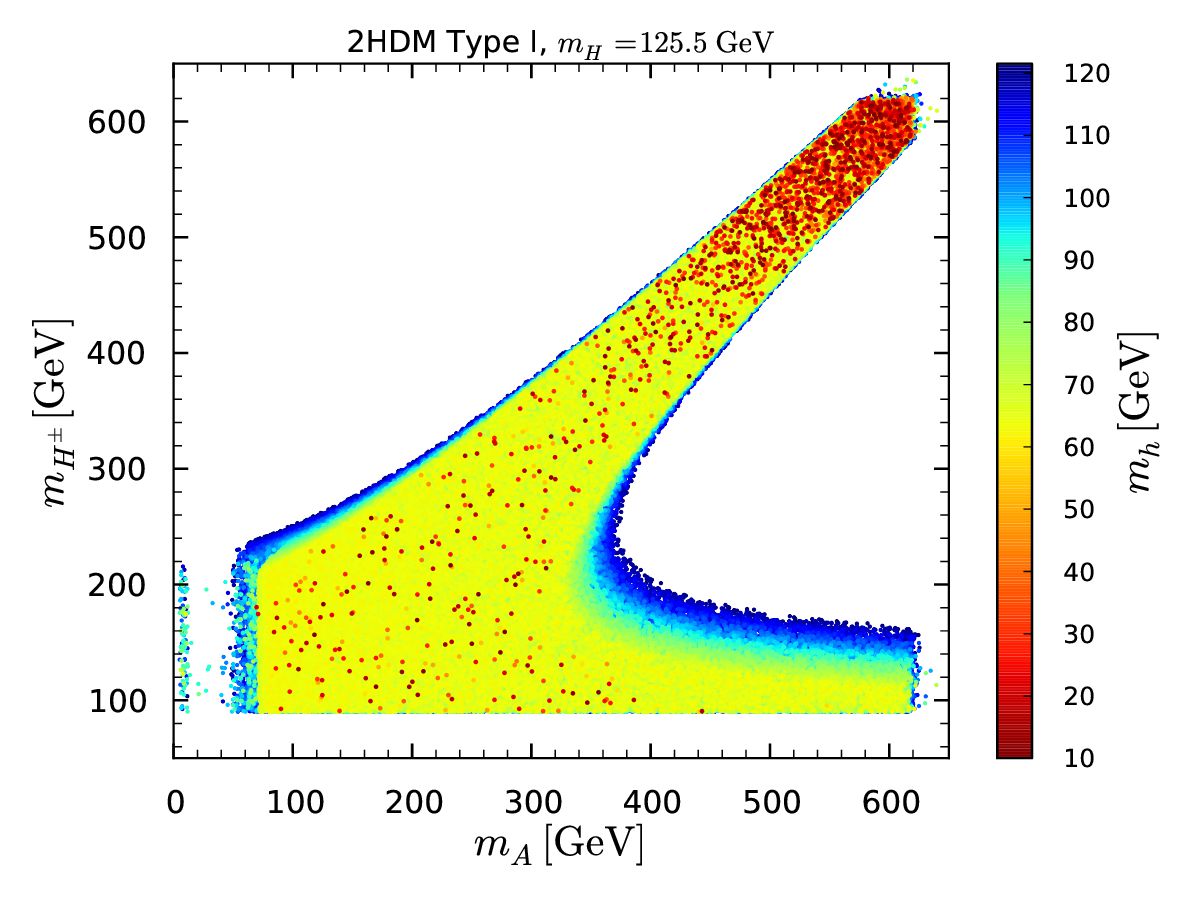} \includegraphics[width=0.495\textwidth]{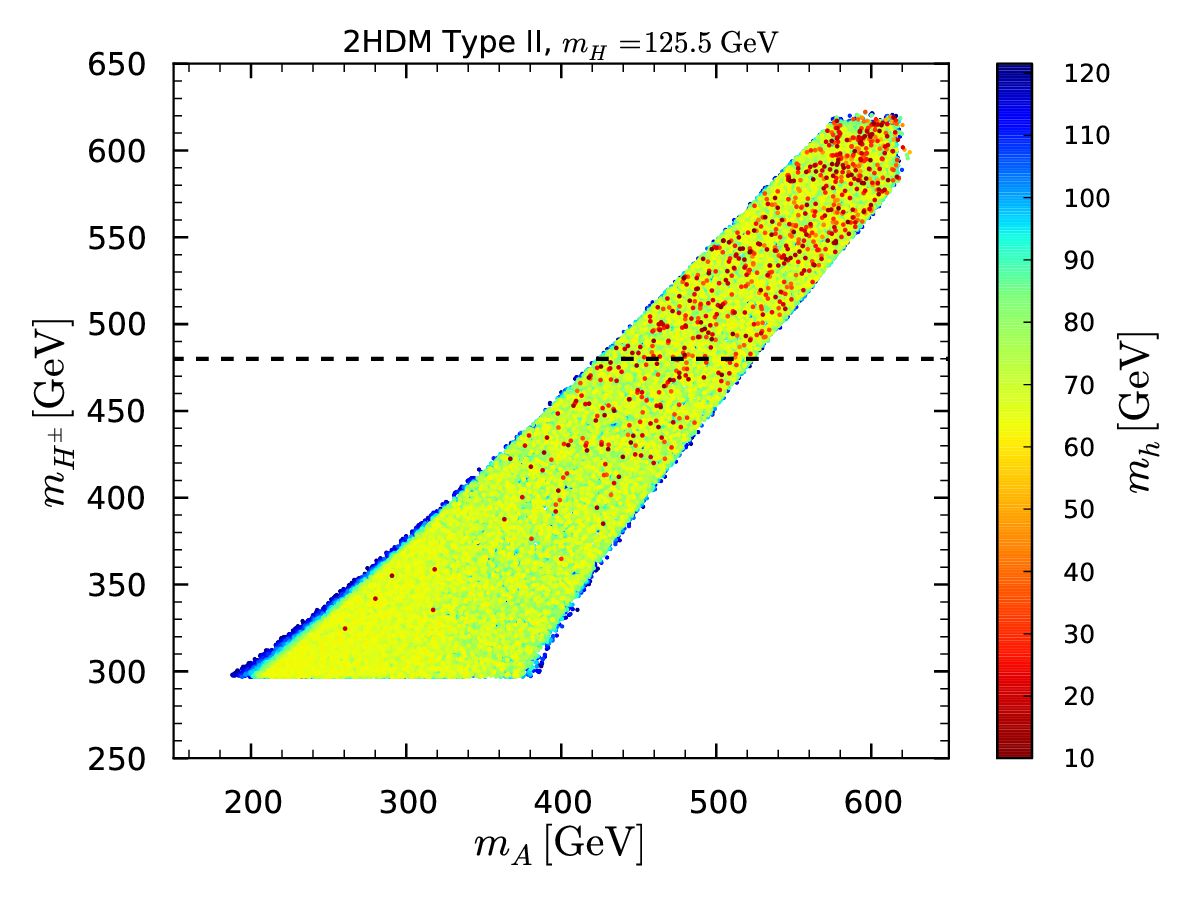}
  \caption{$m_{H^\pm}$ versus $m_A$ in Type~I (left) and Type~II (right) with $m_h$ color code. Points are ordered from high to low $m_h$ values. The right plot shows the whole parameter space scanned over for Type~II, with the horizontal line indicating the updated limit on the charged Higgs mass, $\mhpm>480$~GeV.}
  \label{mA_mC_mh}
\end{figure}

The relation between the three free Higgs masses, $m_A$, $\mhpm$ and $m_h$ is shown in Fig.~\ref{mA_mC_mh}. The absence of a decoupling limit results in an upper bound on the CP-odd and charged Higgs masses, $m_A, \mhpm \lesssim 630$~GeV, which depends on the allowed values of $\cba$. Indeed, without the $\cba\ge 0.99$ constraint that we imposed to focus on the alignment scenario, one would find instead $m_A, \mhpm\lesssim 800$~GeV, where the bound is saturated for $\sba\simeq0.7$~\cite{Dumont:2014wha}. The characteristic correlation between $m_A$ and $\mhpm$ is a consequence of the precision electroweak measurements, primarily the $T$ parameter~\cite{Peskin:1991sw}. In Type~II, a large part of the parameter space is excluded by weak radiative $B$ meson decays for which agreement with observations sets a strong lower bound on the charged Higgs mass, $\mhpm>480$~GeV at 95\% CL, which is practically independent of $\tan\beta$ for $\tan\beta>2$, and is even stronger for $\tan\beta<2$~\cite{Misiak:2015xwa}. 
This constraint, in association with the distinctive $m_A$--$\mhpm$ correlation, sets a bound on the CP-odd Higgs mass. 
We find that $m_A\gtrsim 420$~GeV, which rules out the region of $m_A\leq \half m_H$ in Type~II. We also note that this forces the CP-odd and charged Higgs states to be relatively 
close in mass.  

In contrast to Type~II, the charged Higgs mass is much less impacted from flavor physics constraints~\cite{Mahmoudi:2009zx,Branco:2011iw} in Type~I.
For $\mhpm\lesssim 160$~GeV, the CP-odd state can have any mass below 630~GeV in Type I, as shown in the left panel in Fig.~\ref{mA_mC_mh}.
Moreover, whereas $m_h\leq \half m_H$ can only be found for $\mha, \mhpm\gtrsim 400\gev$ in Type~II, 
such a light $h$ is possible for most of the allowed combinations of $m_A$ and $\mhpm$ in Type~I---with the notable exception of the light $m_A\leq\half m_H$ region, since LEP constraints imply that $A$ and $h$ cannot both have a mass 
below $\half m_H$ simultaneously~\cite{Bernon:2014nxa}. However, there are narrow bands at the border of the allowed $m_A$ vs.\ $\mhpm$ region that unambiguously lead to values of $m_h\gtrsim 100$ GeV.  One such region is the blue band in the left panel of  Fig.~\ref{mA_mC_mh} with $m_A\gtrsim 350\gev$ and $\mhpm\lsim 200\gev$.
Such mass correlations may be used to predict or cross-check the validity of the scenario in the case that two or three extra Higgs states are discovered in the future. 
Finally, as discussed in \cite{Bernon:2014nxa}, $\mhl$ values below about $60\gev$ are only possible for $\tanb\lsim 2$ in 
\typeii.  Hence,  if such a low mass $\hl$ is observed and its properties require a high value of $\tanb$, 
then the \typeii\ model would be eliminated.

\subsection{Couplings}

%
%

\begin{figure}[t!]\centering
\includegraphics[width=0.5\textwidth]{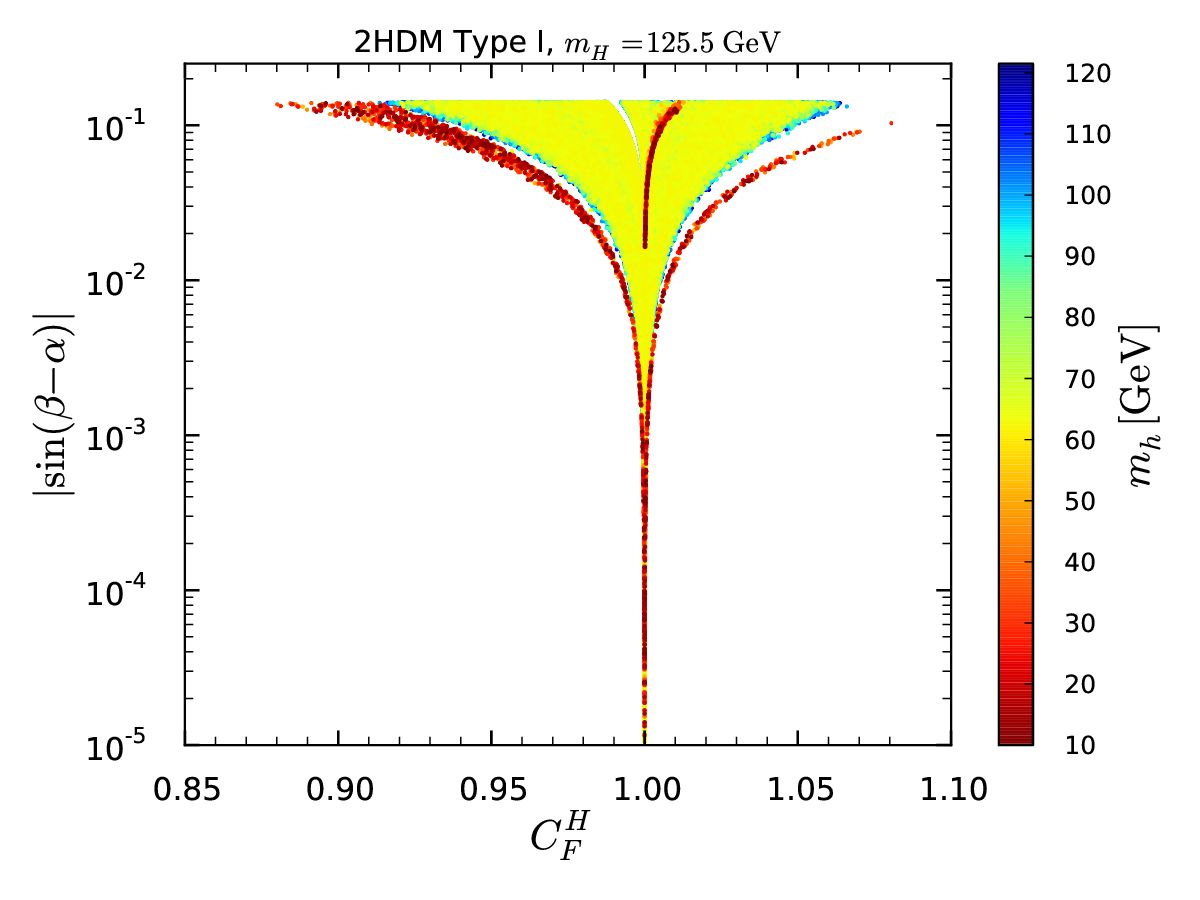}\includegraphics[width=0.5\textwidth]{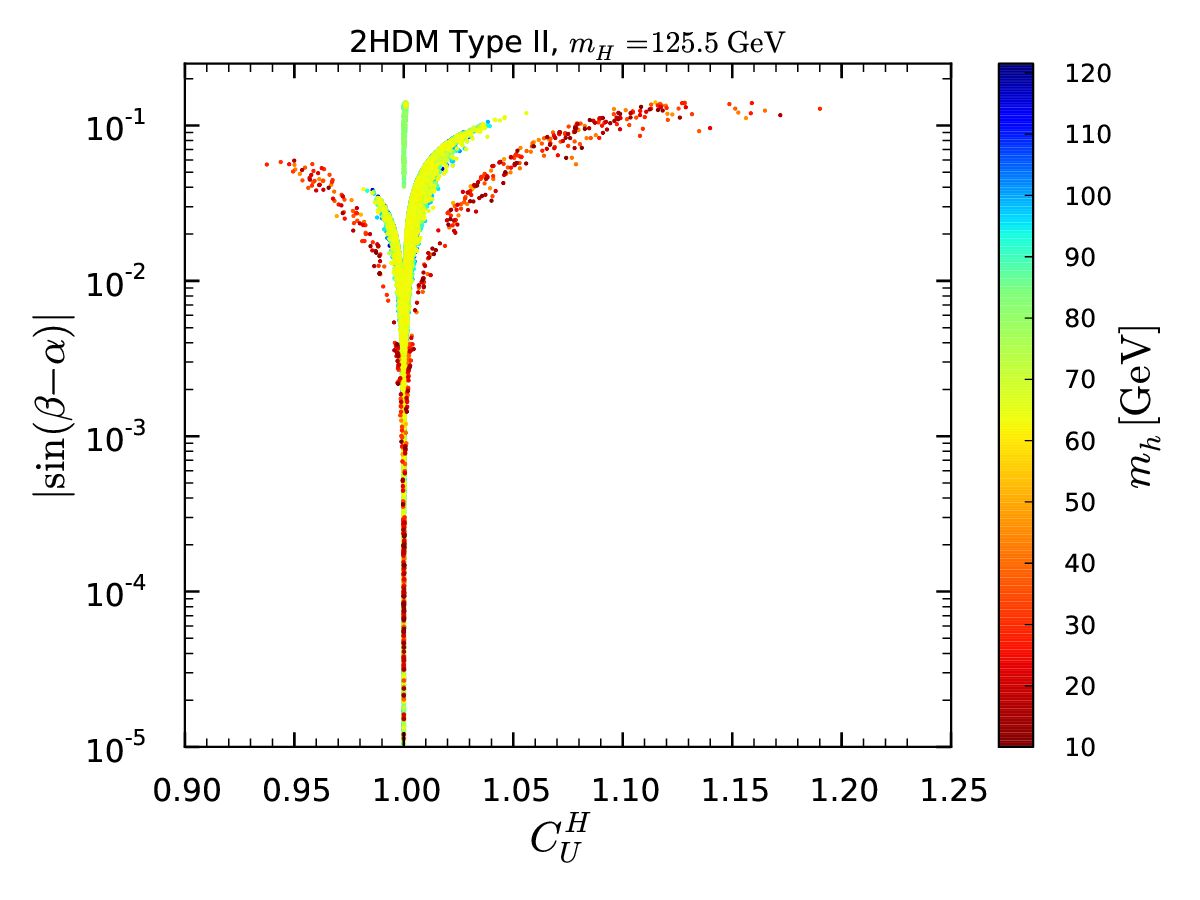}\\
\includegraphics[width=0.5\textwidth]{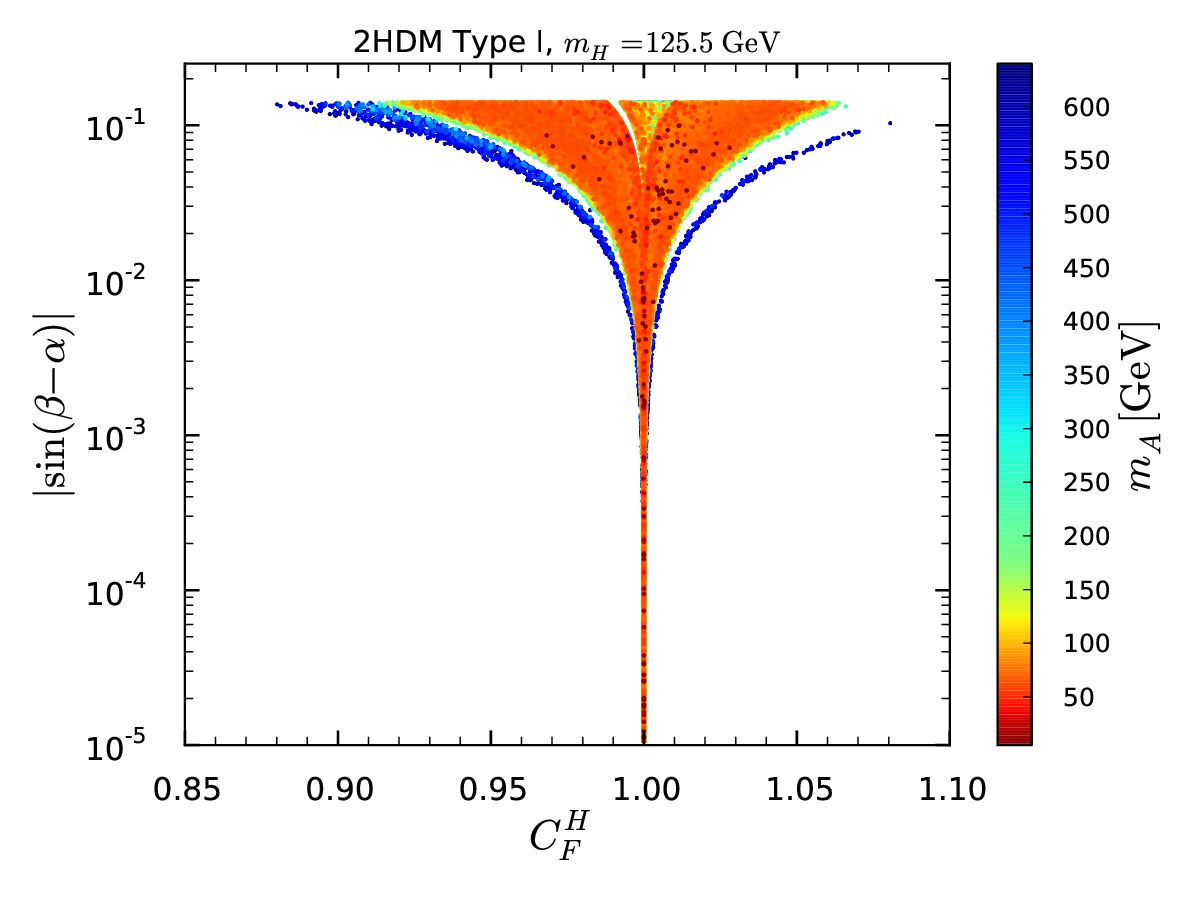}\includegraphics[width=0.5\textwidth]{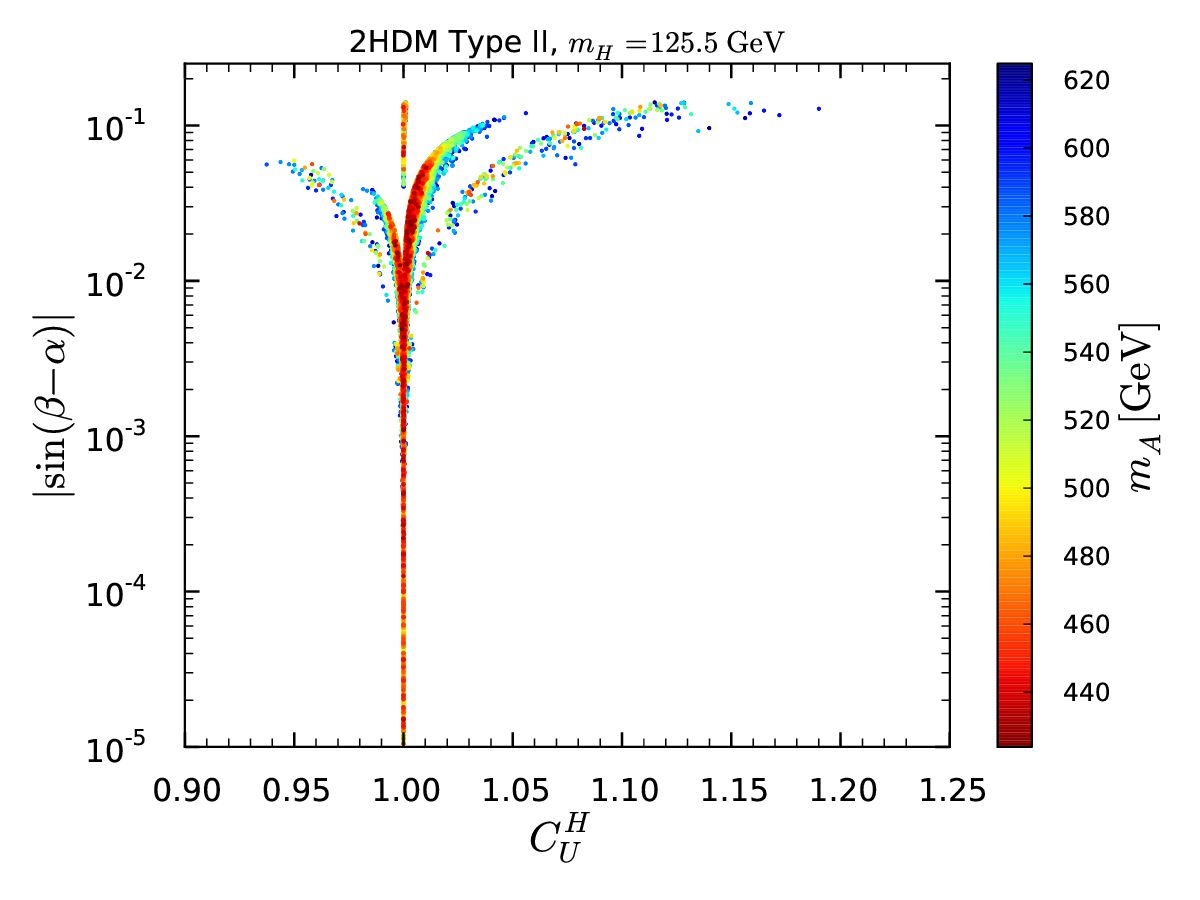}
\caption{$|\sba|$ versus reduced fermionic coupling $C_F^H$ in Type~I (left) and $C_U^H$ in Type~II (right). In the upper two panels, the color code shows the dependence on $m_h$; in the lower two panels the dependence on $m_A$, with points ordered from high to low $m_h$ and $m_A$ values, respectively. Note that the color scales for $m_A$ are different for Type~I and Type~II because of the very different allowed ranges of $m_A$.} 
  \label{CV_CF_H125}
\end{figure}

Let us now turn to the properties of the $H$ in the (near-)alignment regime. 
Figure~\ref{CV_CF_H125} shows the possible variation of the coupling of $H$ to up-type fermions,  
$C_U^H=C_D^H\equiv C_F^H$ in Type~I and $C_U^H$ in Type~II. 
Deviations from unity of around $\pm 10\%$ are possible in Type~I for $|\sba| \sim 0.14$, 
while in Type~II the deviations range from $-7$\% to $+20$\%.  As expected, in both types $C_U^H$ quickly approaches unity as $|\sba|$ decreases. It is interesting to note that, while $C_U^H=\sin\alpha/\sin\beta$ in both Type~I and Type~II, the actual values that can be reached are different in the two models because of constraints involving the down-type coupling. 
The largest deviations occur for large $h$--$A$ mass splitting, when $m_h$ is below 60~GeV, while $m_A$ is close to its upper bound. As discussed in \cite{Bernon:2014nxa}, $\tanb$ is very close to 1 in this case.  In Type~I, there is also another region with $m_h<\half m_H$ at larger values of $\tanb$, although this is only achieved when $|\sba|\gtrsim 10^{-2}$. This is seen as the narrow banana-shaped red strip with $C_F\approx 1$--$1.01$ in the upper left panel of Fig.~\ref{CV_CF_H125}.
Also noteworthy are the white gaps between the regions filled with valid scan points: these are caused by the CMS limits~\cite{CMS-PAS-HIG-15-001} on $A\to Zh\to \ell\ell b\bar b$ (and $\ell\ell \tau\tau$) and will appear in many of the subsequent figures. 
The impact of this limit is discussed in more detail in the Appendix. 

\begin{figure}[t!]\centering
\includegraphics[width=0.5\textwidth]{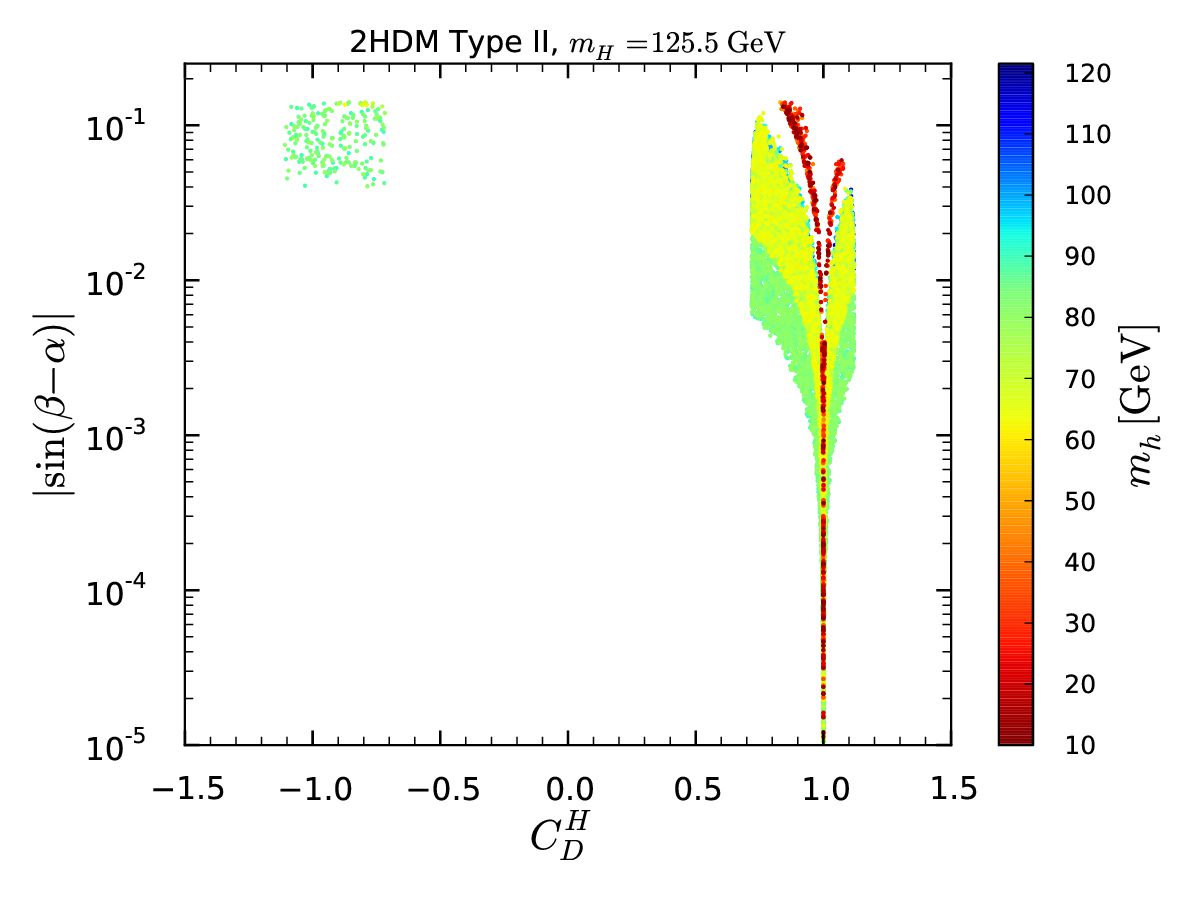}\includegraphics[width=0.5\textwidth]{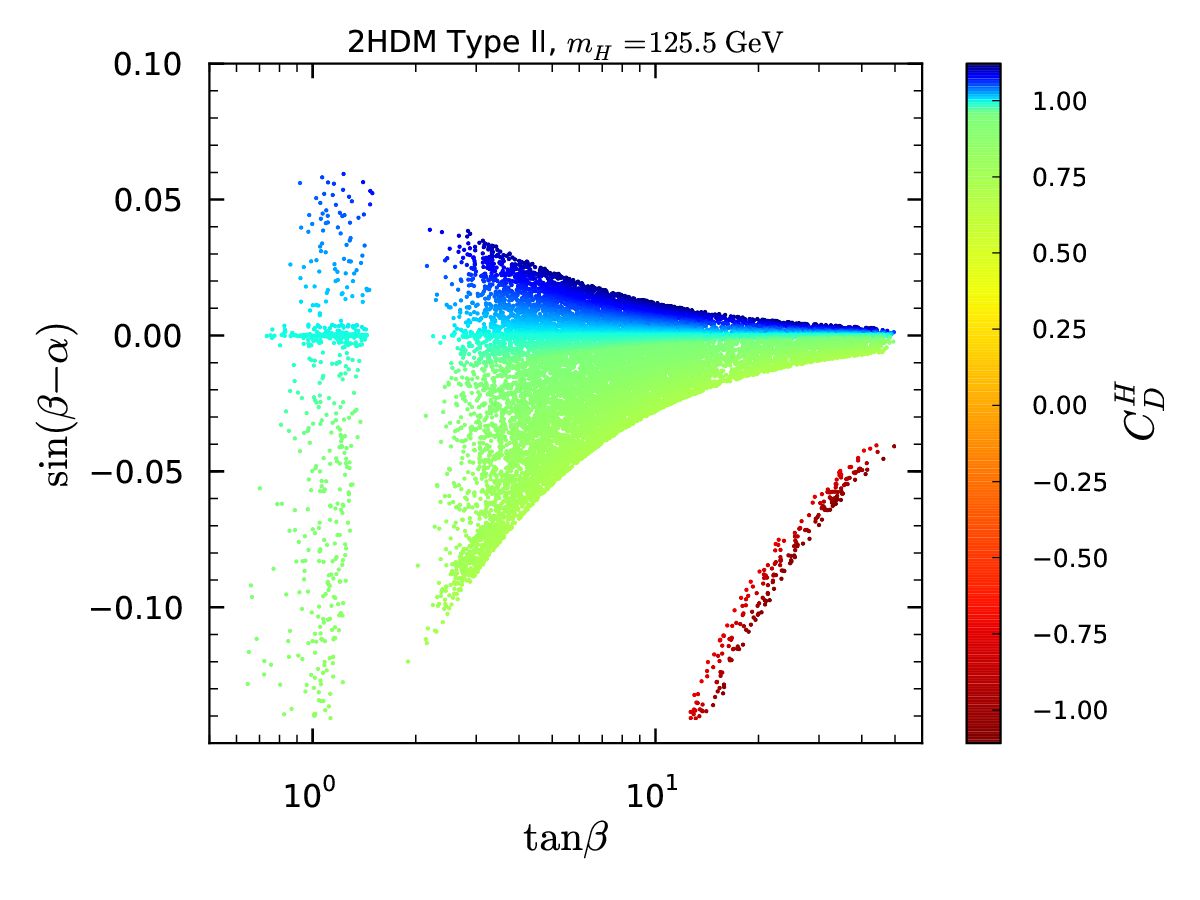}
\caption{In the left panel, we exhibit $|\sba|$ versus $C_D^H$ in Type~II. The color code shows the dependence on $m_h$, with points ordered from high to low $m_h$ values. 
In the right panel, we exhibit $\sba$ vs.~$\tanb$ with the color code showing $C_D^H$ ordered from high to low values.} 
  \label{CV_CD_H125}
\end{figure}

The possible variation of the coupling to down-type fermions, $C_D^H$, in Type~II is shown in Fig.~\ref{CV_CD_H125}. 
Let us first consider the left panel. 
As in the $m_h=125\gev$ case, there are two solutions: one where $C_V^H$, $C_U^H$ and $C_D^H$ all have the same sign (as is the case in the SM), and one where $C_D^H$ has opposite sign relative to $C_U^H$ and $C_V^H$~\cite{Ferreira:2014naa}. 
In the normal (same) sign region, deviations from the predicted SM coupling in the range of roughly $-30\%$ to $+12\%$ are possible even for rather low $|\sba|\sim 5\times 10^{-3}$, as long as the $H\to hh$ decay mode is closed. 
If the $H\to hh$ decay  (which is constrained to $\br(H\to hh)\lesssim 0.27$ at 95\%~CL by the fit to the 125~GeV signal strength measurements) contributes to the total width, then $C_D^H$ is confined to the range $[0.83,\,1.08]$ and quickly converges to unity as $|\sba|$ decreases.
Note however that $C_D^H$ is never exactly 1 unless $|\sba|$ is at the level of few times $10^{-3}$ or smaller. 
The gap between the red and the yellow/green/blue points is again caused by the CMS limits on $A\to Zh$. 
On the other hand, the opposite-sign region, $C_D^H\in [-1.1, -0.7]$, requires $\sba\lesssim -0.04$ due to the fact that  $\sba$ and $\tan\beta$ are correlated in Type II as illustrated in the right panel of Fig.~\ref{CV_CD_H125}.  We see that $C_D^H=\cba+\sba\tb\gtrsim 1$ for $\sba\gtrsim 0$ but decreases below 1 when $\sba$ turns negative. Consequently, for moderately negative $\sbma$ and large enough $\tanb$, $C_D^H$ flips sign. Values of $|C_D^H|\lesssim 0.7$ are excluded by the fit of the signal strengths, but the opposite-sign solution with $C_D^H\approx -1$ is still phenomenologically viable.
The region of $\tanb\gtrsim 50$ is excluded because of the strong constraints on $A\to\tau\tau$ decays from ATLAS~\cite{Aad:2014vgg} and CMS~\cite{Khachatryan:2014wca}; see also~\cite{Dumont:2014wha}. The CP-odd scalar mass $m_A$ does not have much influence, since it can only vary over the very limited range 420--630~GeV in Type~II. Large deviations of $C_D^H$ may imply large excursions of the $H$ signal strengths away from 1, the details of which will be studied in Section~\ref{sec:signalstrengths}.

%
%

\begin{figure}[t!]\centering
\includegraphics[width=0.5\textwidth]{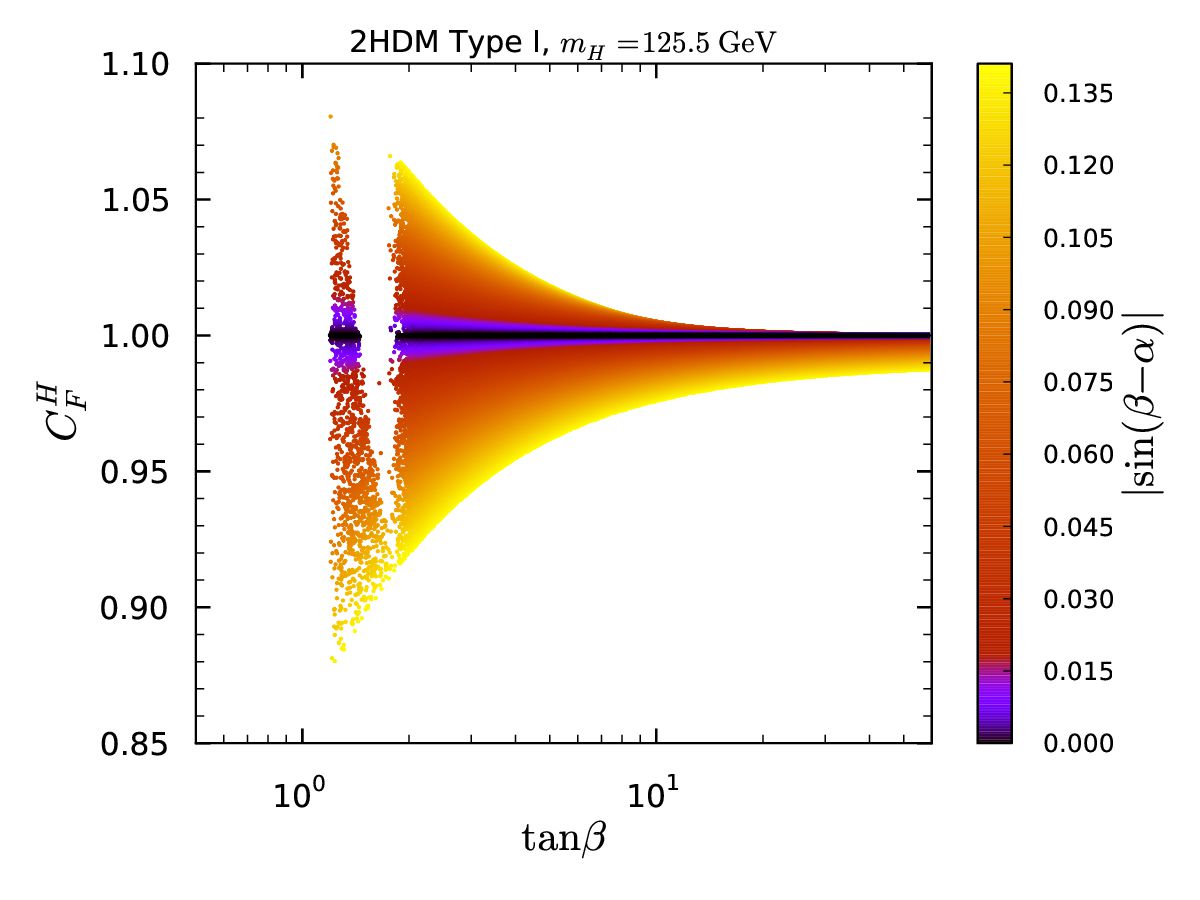}\\
\includegraphics[width=0.5\textwidth]{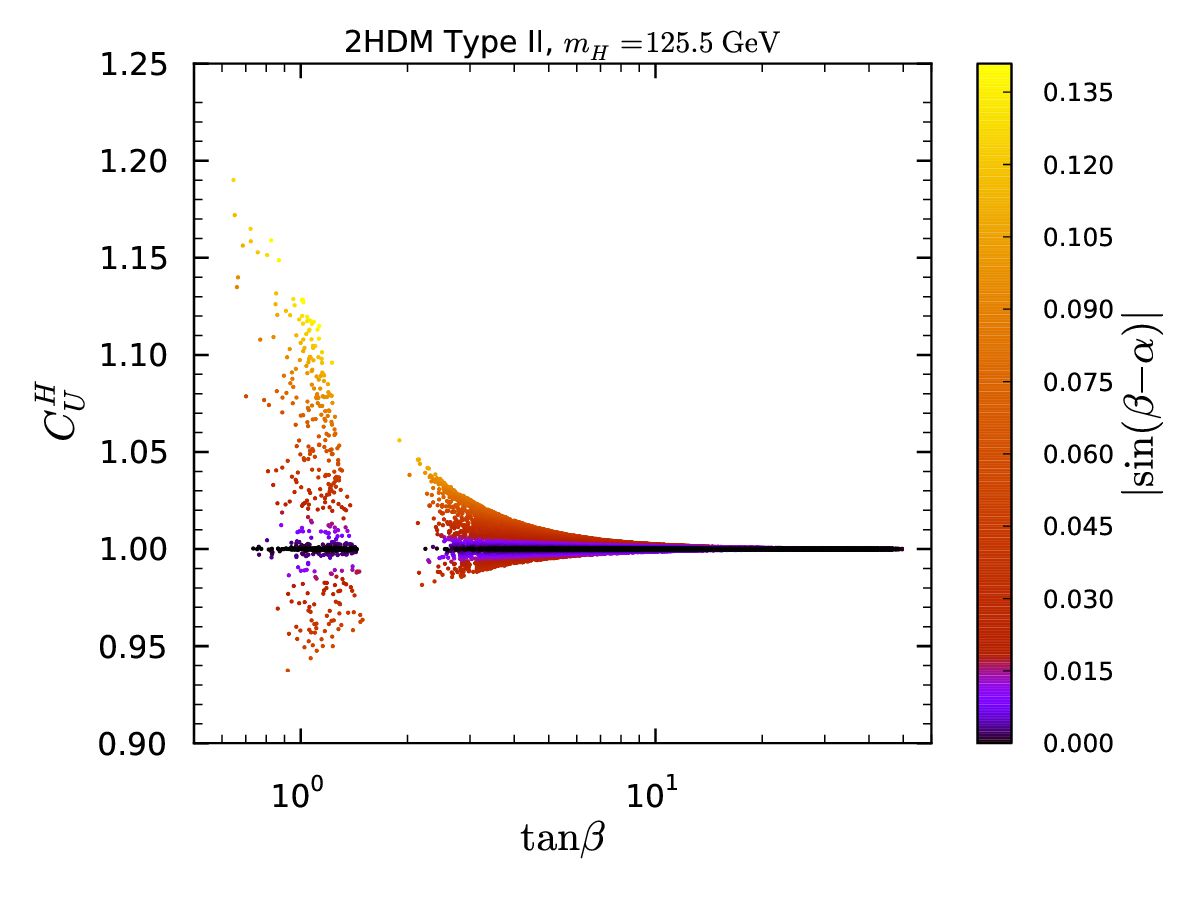}\includegraphics[width=0.5\textwidth]{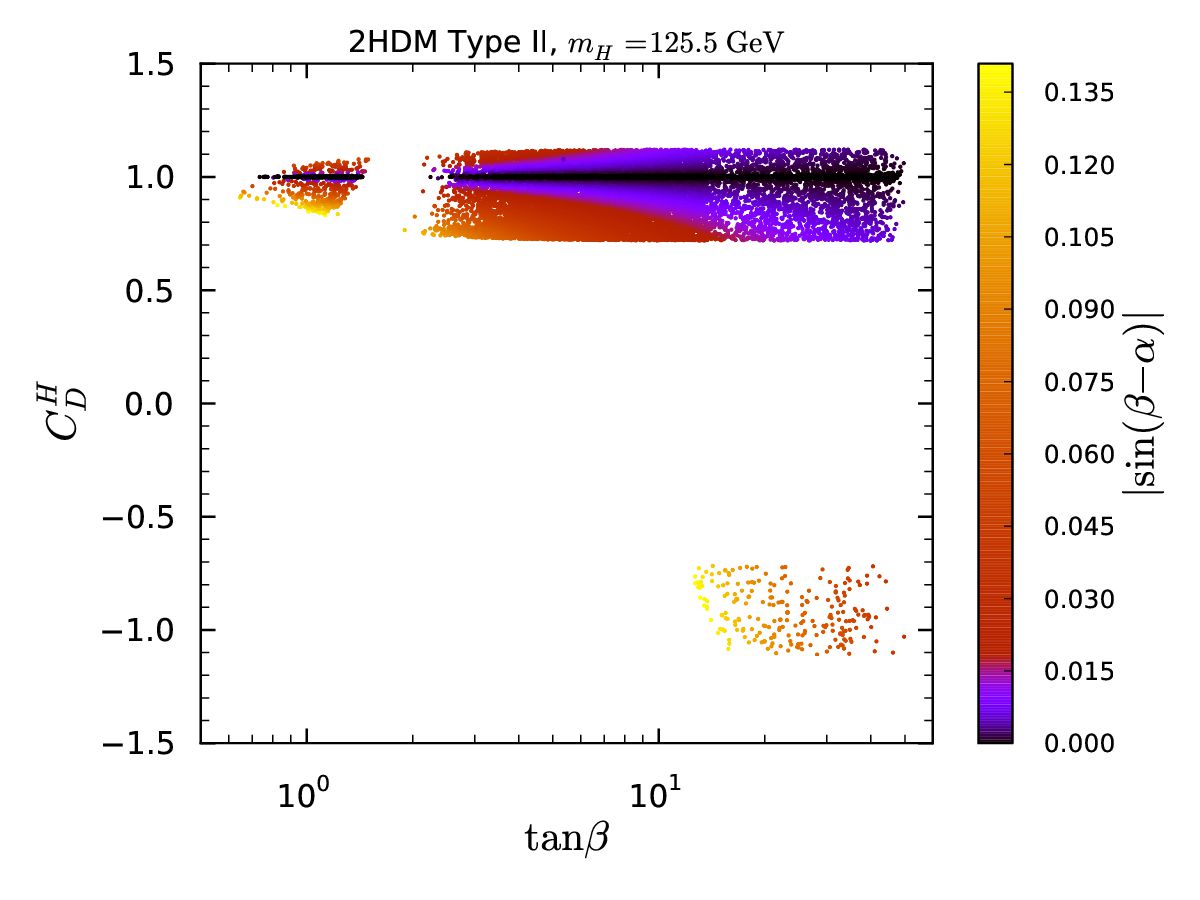}
  \caption{Fermionic couplings versus $\tan\beta$ in Type~I (upper panel) and Type~II (lower panels) with $|\sba|$ color code. Points are ordered from high to low $|\sba|$.}
  \label{tb_CF_H125}
\end{figure}

The $\tan\beta$ dependence of the fermionic couplings, shown in Fig.~\ref{tb_CF_H125}, is also noteworthy.  
In both Type~I and Type~II, sizable deviations from $C_U^H=\cba-\sba/\tb=1$ are possible only for small $\tan\beta$. 
In Type~II, $C_U^H$ very quickly converges  to 1 once  $\tan\beta\gtrsim 7$--$8$ because
 the allowed range of $\sba$ decreases with increasing $\tan\beta$, as can be seen in the right panel of  Fig.~\ref{CV_CD_H125}.
In Type~I, the convergence of the fermionic couplings to their SM values is less pronounced due to the fact that, even for $\tan\beta = 60$, the full $|\sba|$ range considered is allowed. 
For $C_D^H$ in Type~II, the situation is quite different, as this coupling is given by $\cos\alpha/\cos\beta$ instead of $\sin\alpha/\sin\beta$.  For the normal-sign region, as soon as $\tan\beta$ is at least moderate in size ($\tan\beta\approx 10$), $C_D^H$ saturates the full range allowed by the measured signal strengths, even for small values of $|\sba|$ of a few times $10^{-3}$.
In contrast, as discussed in the previous paragraph, the opposite-sign solution is only possible for large enough negative $\sbma$, concretely $\sbma\lesssim -0.04$, cf.\ the right panel of Fig.~\ref{CV_CD_H125}. 
Overall, in these plots, the impact of the CMS limit on $A\to Zh$ is even more striking than in Fig.~\ref{CV_CF_H125}, as it excludes most points with  $\tanb\approx 1.2$--$1.8$ in \typei\ and the entire range of $\tanb\approx 1.5$--$2$ in \typeii.

%
%

Turning to the loop-induced Higgs couplings to gluons and to photons, we first note that
the $H$ coupling to gluons, $C_g^H$, is dominated by the top-quark loop, and 
its behavior is thus practically the same as that of $C_U^H$ in Figs.~\ref{CV_CF_H125} and \ref{tb_CF_H125}. 
We therefore do not show separate plots for $C_g^H$. However, an exception occurs for the opposite-sign $C_D^H$ solution,  for which the $b$-loop contribution interferes constructively with the $t$-loop contribution, resulting in $C_g^H\approx 1.06$. (The same happens in the $m_h=125\gev$ case, see 
\cite{Bernon:2015qea,Ferreira:2014naa}.) 

\begin{figure}[t!]\centering
\includegraphics[width=0.5\textwidth]{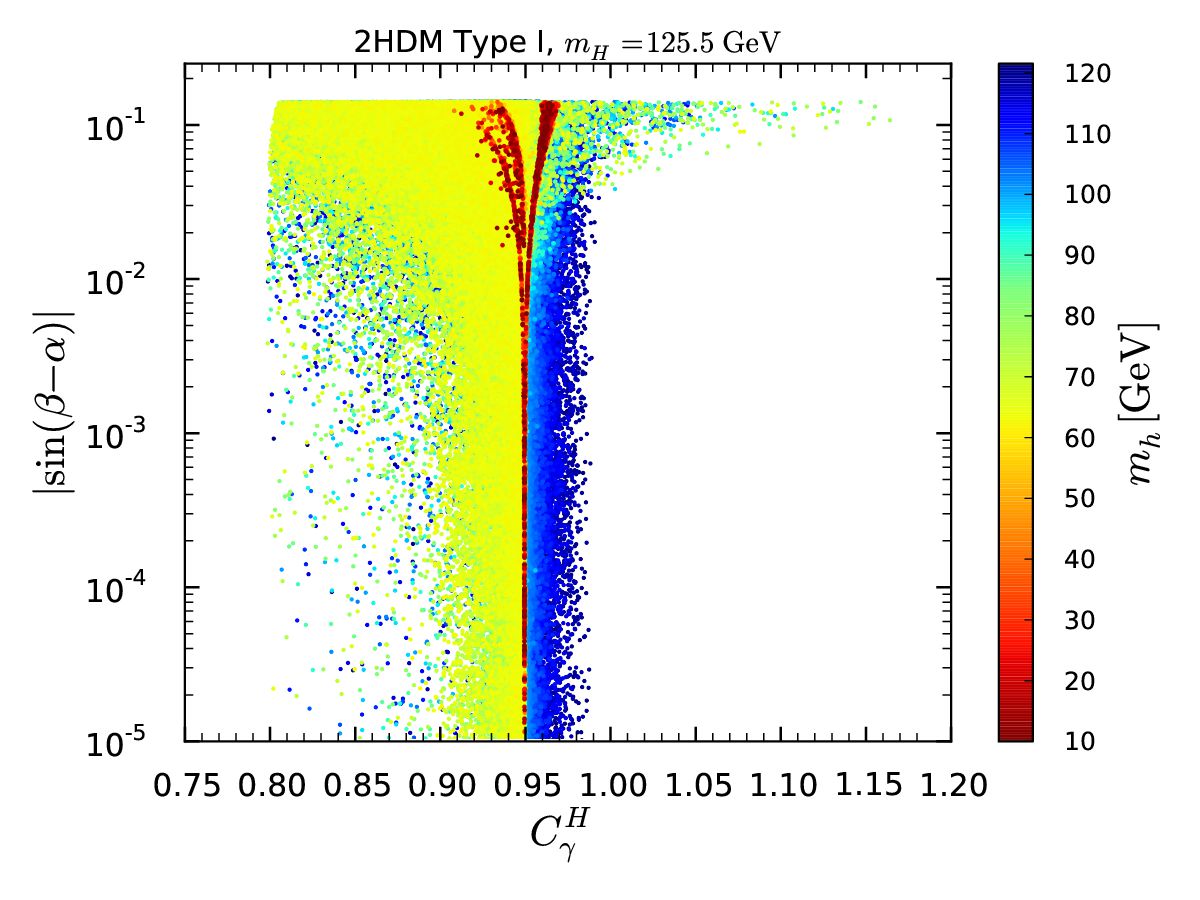}\includegraphics[width=0.5\textwidth]{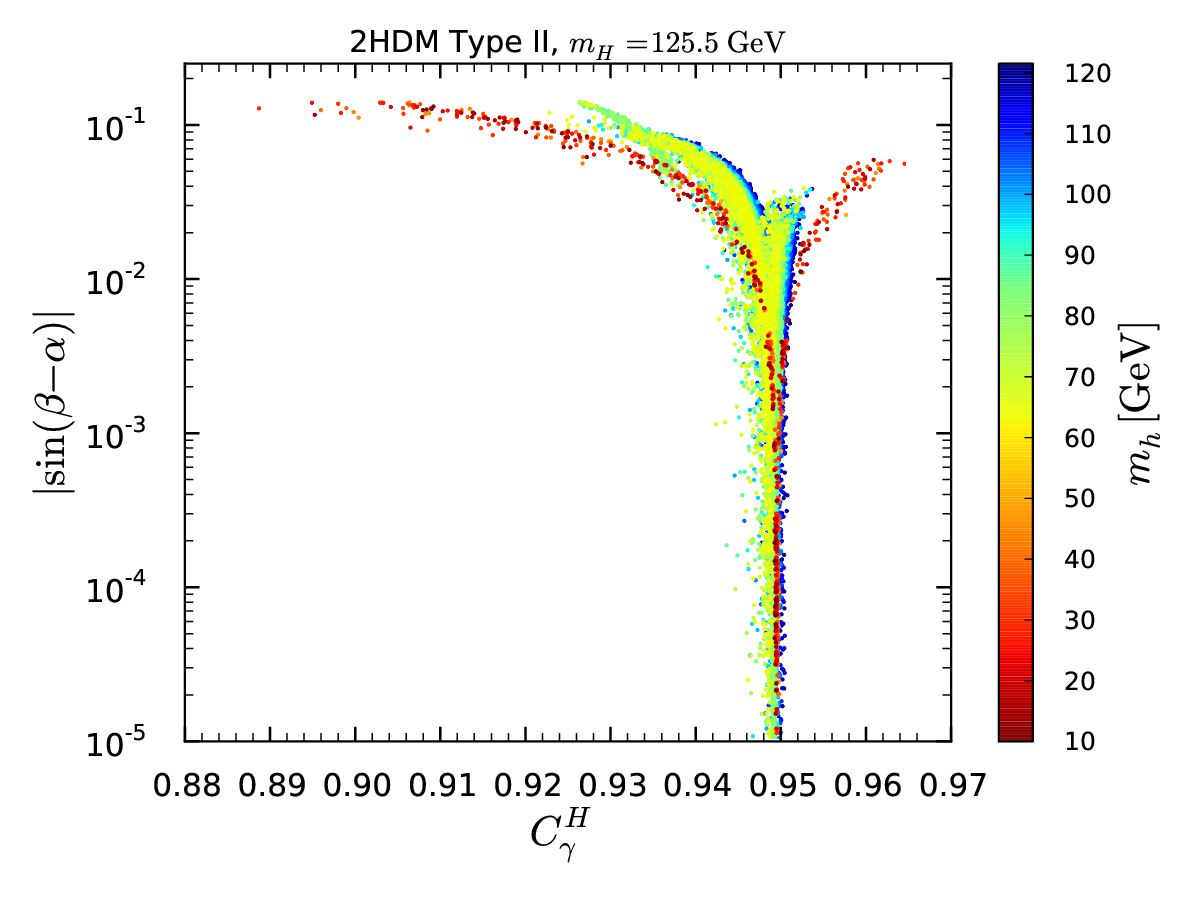}\\
\includegraphics[width=0.5\textwidth]{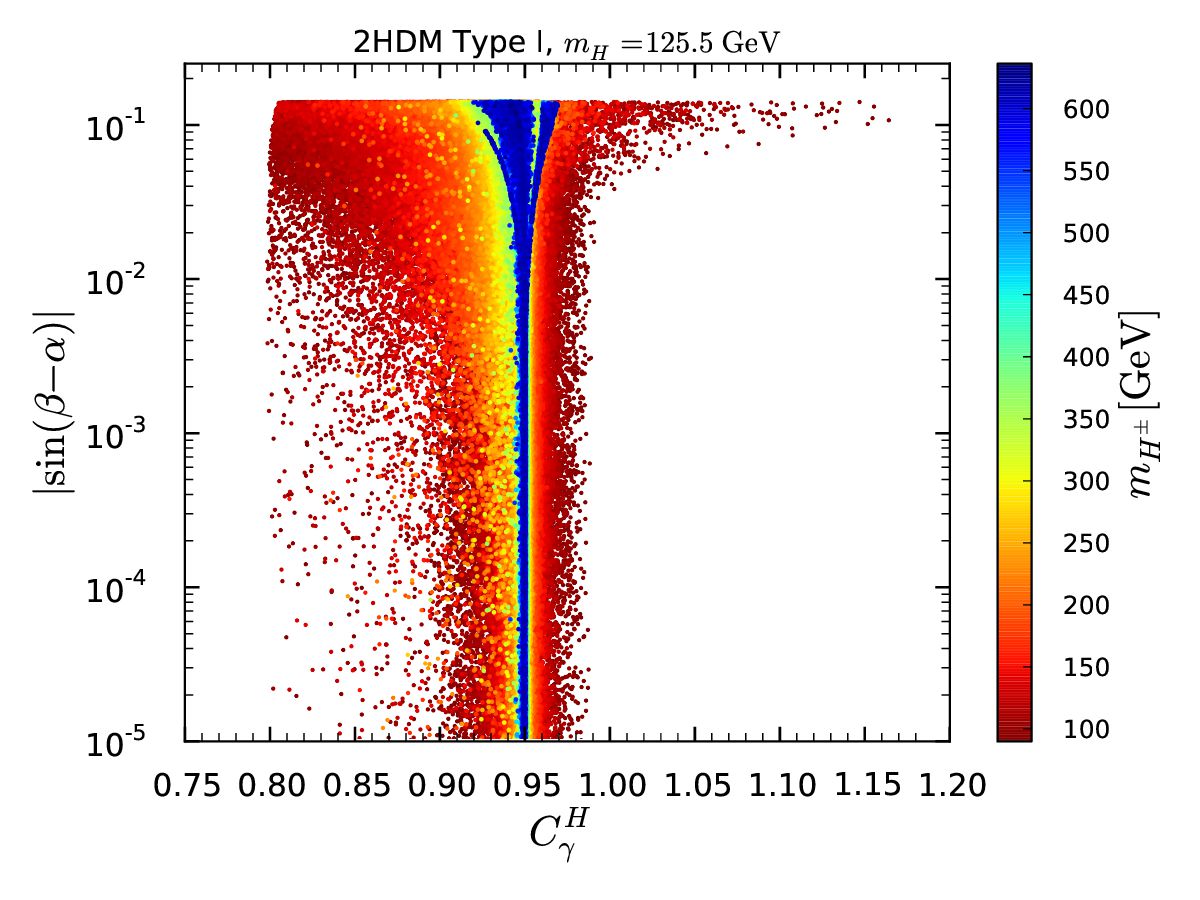}\includegraphics[width=0.5\textwidth]{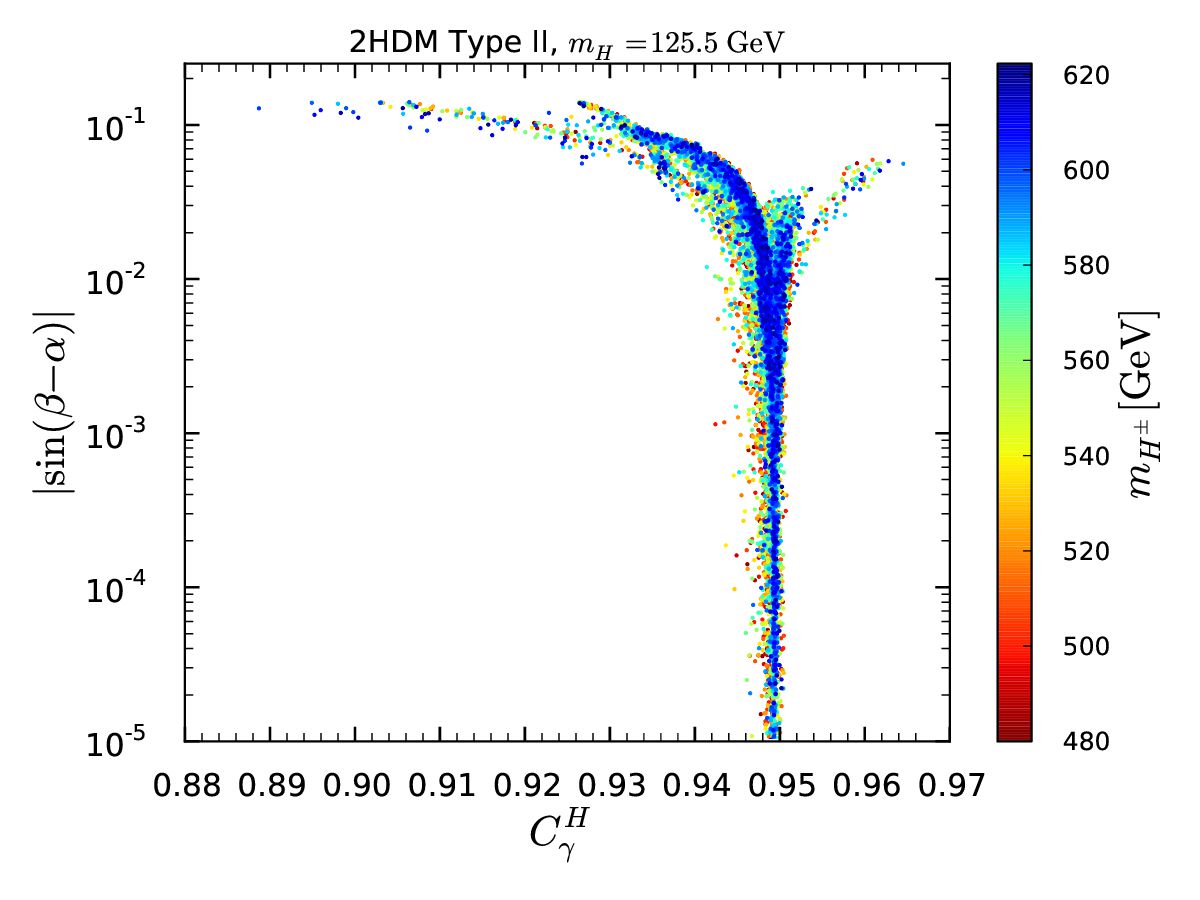}\\
  \caption{$|\sba|$ versus $C_\gamma^H$ in Type~I (left) and Type~II (right) with $m_h$ color code (upper panels) and with $m_{H^\pm}$ color code (lower panels). Points are ordered from high to low $m_h$ in the upper panels and from low to high  $m_{H^\pm}$ in the lower panels.} 
\label{Cgamma_H125}
\end{figure}

\begin{figure}[t!]\centering
\includegraphics[width=0.5\textwidth]{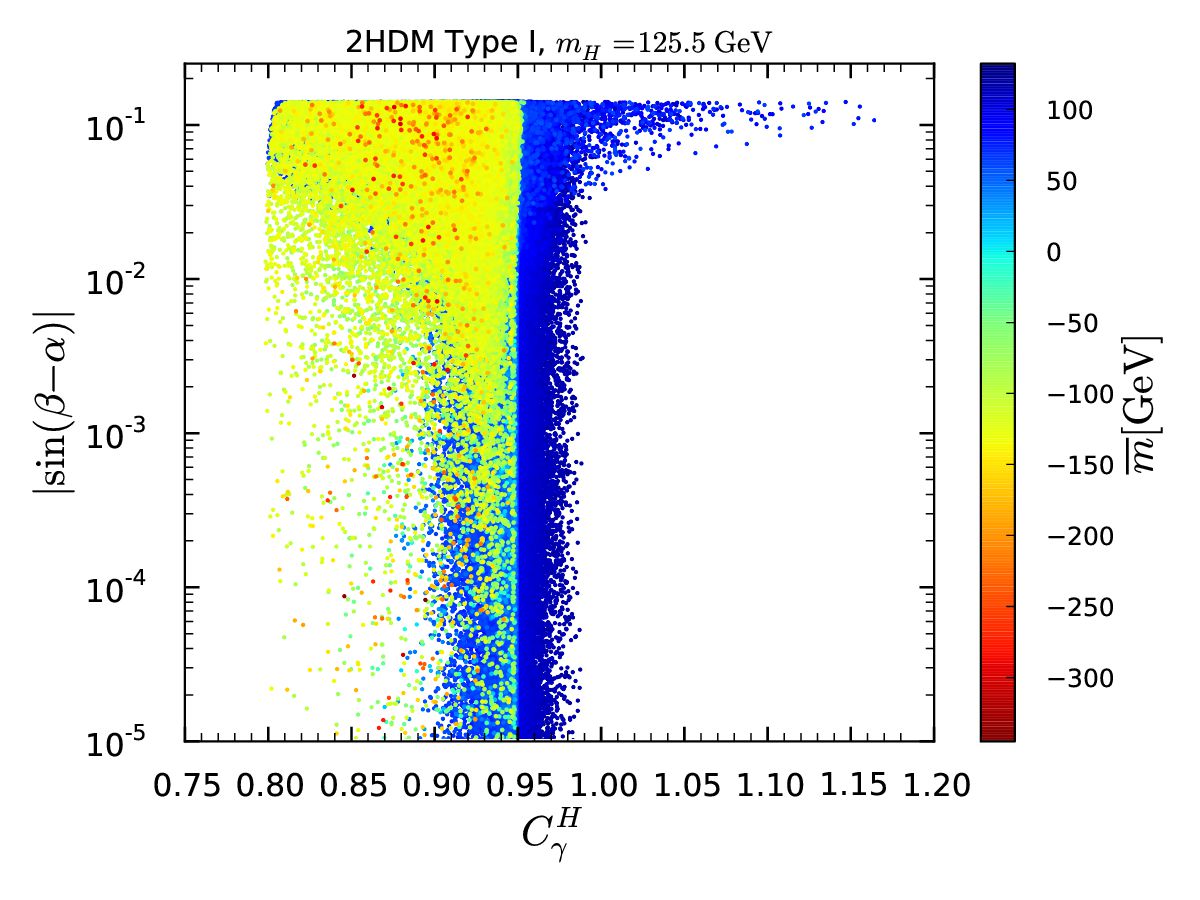}\includegraphics[width=0.5\textwidth]{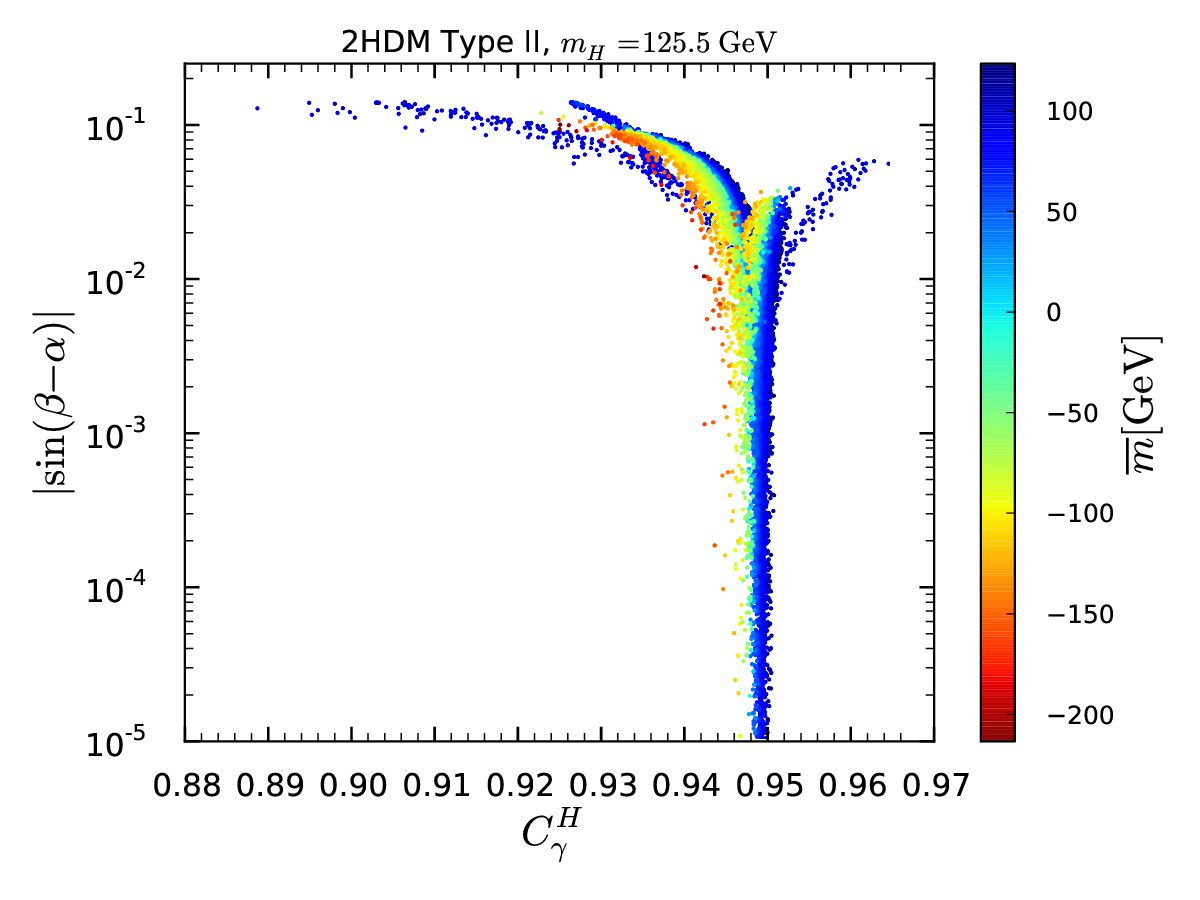}\\
  \caption{As for Fig.~\ref{Cgamma_H125} but showing the dependence of $C_\gamma^H$ on $\overline{m}\equiv \sgn{\mhat^2}\sqrt{|\mhat^2|}$. Points are ordered from high to low $\overline{m}$. }
\label{Cgamma_H125_mhat}
\end{figure}

The coupling to photons, $C_\gamma^H$, is more complicated. 
Here, the main contributions come from $W$ and top-quark loops as in the SM, as well as from loops with charged Higgs bosons. The $W$ and top-quark loops contribute with opposite signs, and thus the values of
$C^H_U >1$ $(C^H_U<1)$ seen in Fig.~\ref{CV_CF_H125} will lead to smaller (larger) $C^H_\gamma$, respectively.
The $H^\pm$ loop typically also has the opposite sign relative to the $W^\pm$ loop and can thus substantially suppress $C_\gamma^H$ even at very small $\sba$.   (However, positive interference of the $W^\pm$ and $\hpm$ loops is possible for low $m_\hpm$, as noted at the end of Section~\ref{modelreview}.)
The net effect on $C_\gamma^H$ is shown in Fig.~\ref{Cgamma_H125}. 
In particular, we observe a large variation in $C_\gamma^H$ in Type~I, where the charged Higgs boson can be light.
For $|\sbma|\lesssim 10^{-2}$ and  $\mhpm \gtrsim 500$~GeV, we find $C_\gamma^H\approx 0.95$ in both Type~I and Type~II, in agreement with the expected 5--6\% reduction of $C_\gamma^H$ relative to the SM in the limit $|\sbma|\to 0$ with heavy $\mhpm$ [cf.~\Eq{ghhhphmform}].
For $|\sbma|\gtrsim 10^{-2}$ (but still assuming that $\mhpm$ is large), this reduction can be more or less than 5\% depending on the sign of $\mhat^2\equiv 2m_{12}^2/\sin 2\beta$. 
Note however, that while $\mhat^2<0$ can reach values as large as $-(350\gev)^2$ $\bigl[-(200\gev)^2\bigr]$ in Type~I [Type~II], respectively, $\mhat^2>0$ does not exceed $\sim (150\gev)^2$. Therefore, in Type~II where $\mhpm > 480$~GeV,  $C_\gamma^H$ is always below 1 (although one will need linear collider precision to pin this down with sufficient accuracy~\cite{Peskin:2013xra}). 
In contrast, in Type~I, 
for $\mhpm \lesssim 160$~GeV a value of $\mhat^2$ between about $(60\gev)^2$ and $(120\gev)^2$
can lead to a switch in sign of $g_{HH^+H^-}$, giving $C_\gamma^H>1$. 
The dependence on $\mhat^2$ is illustrated explicitly in Fig.~\ref{Cgamma_H125_mhat}. 
Of course such a light charged Higgs boson can also (and in fact more easily) suppress $C_\gamma^H$, down to $C_\gamma^H\approx 0.8$, irrespective of the value of $|\sbma|$. 

\begin{figure}[h!]\centering
\includegraphics[width=0.5\textwidth]{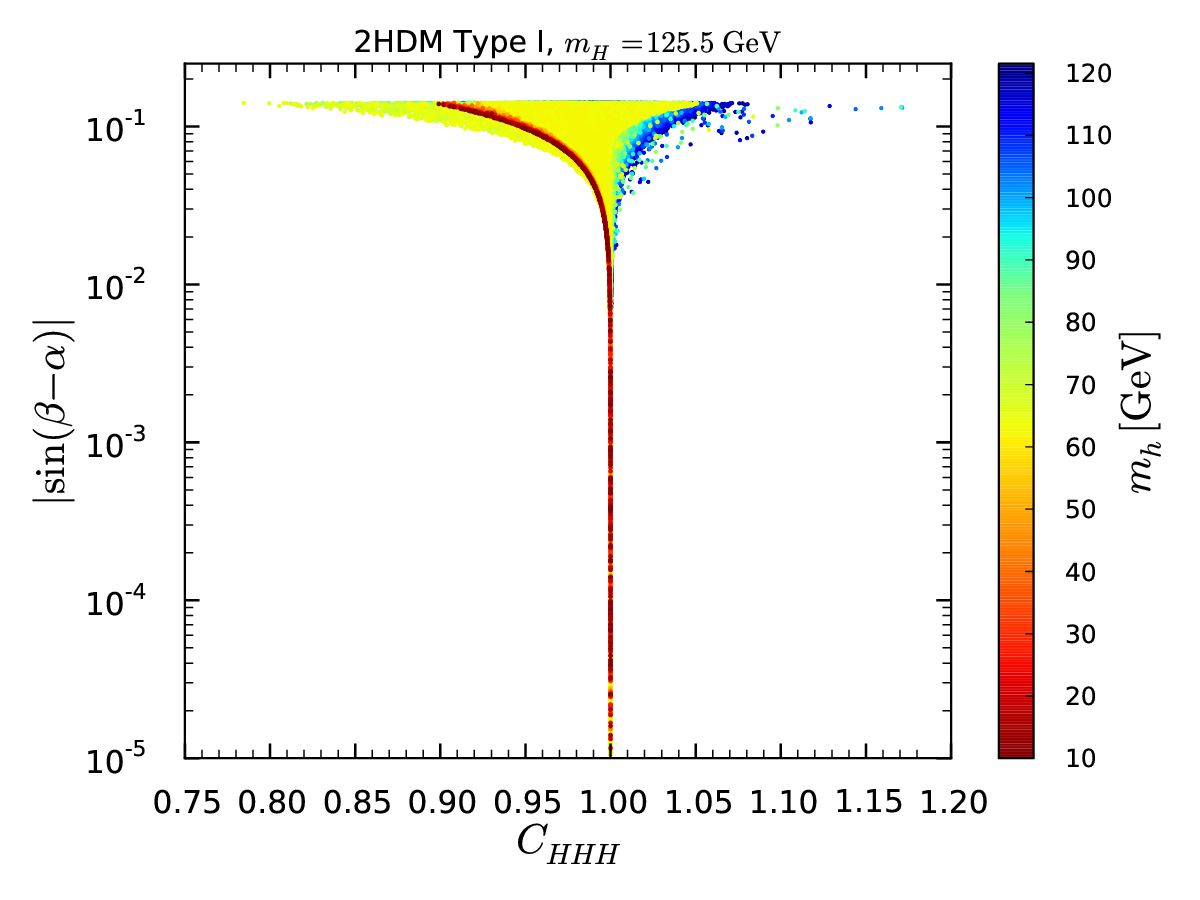}\includegraphics[width=0.5\textwidth]{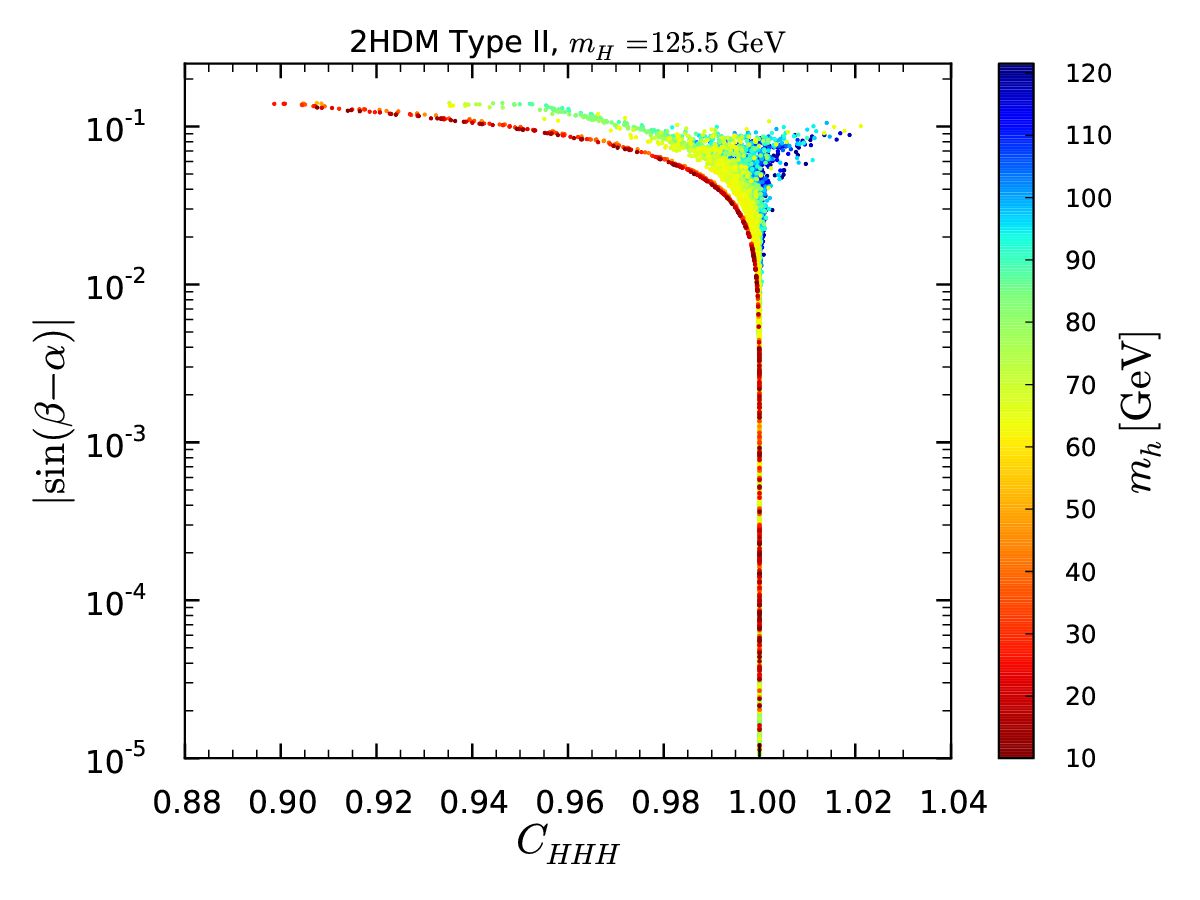}
\includegraphics[width=0.5\textwidth]{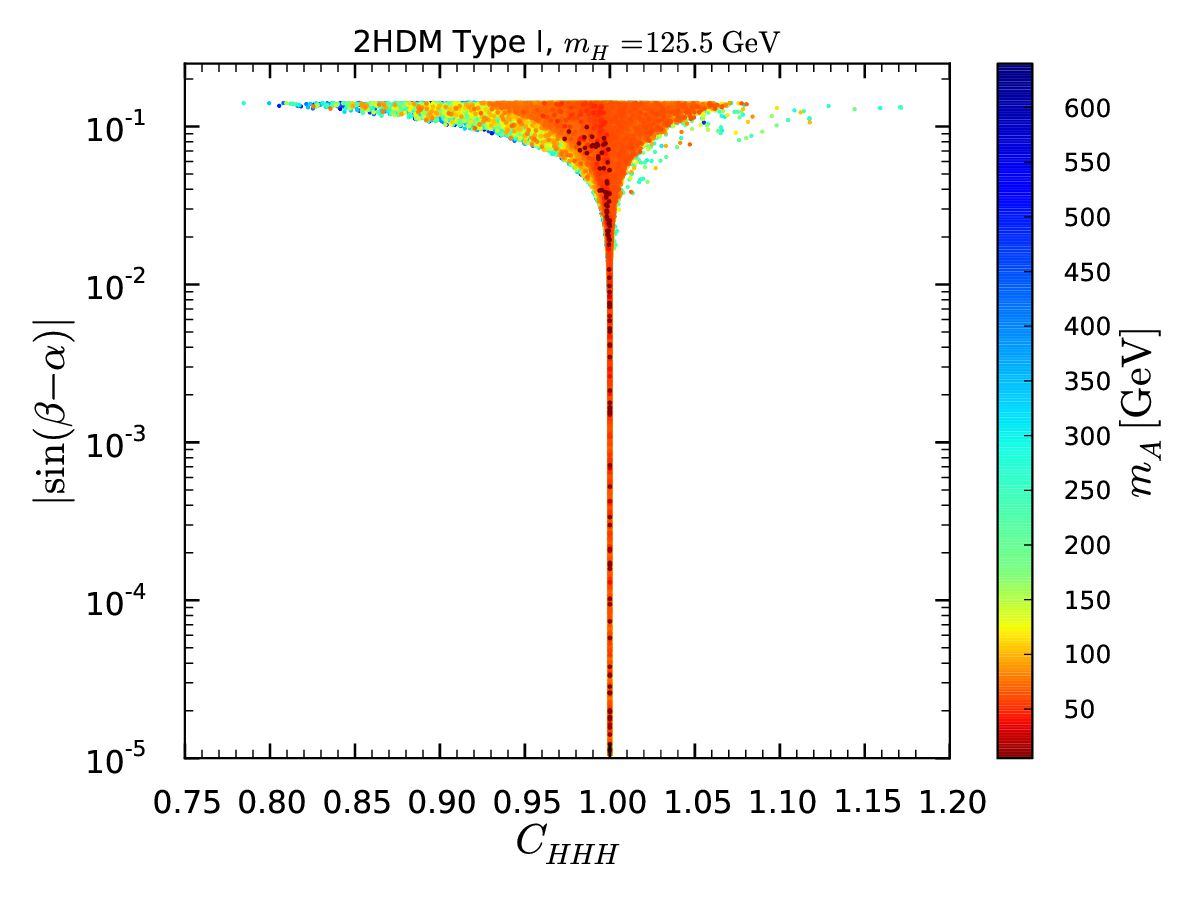}\includegraphics[width=0.5\textwidth]{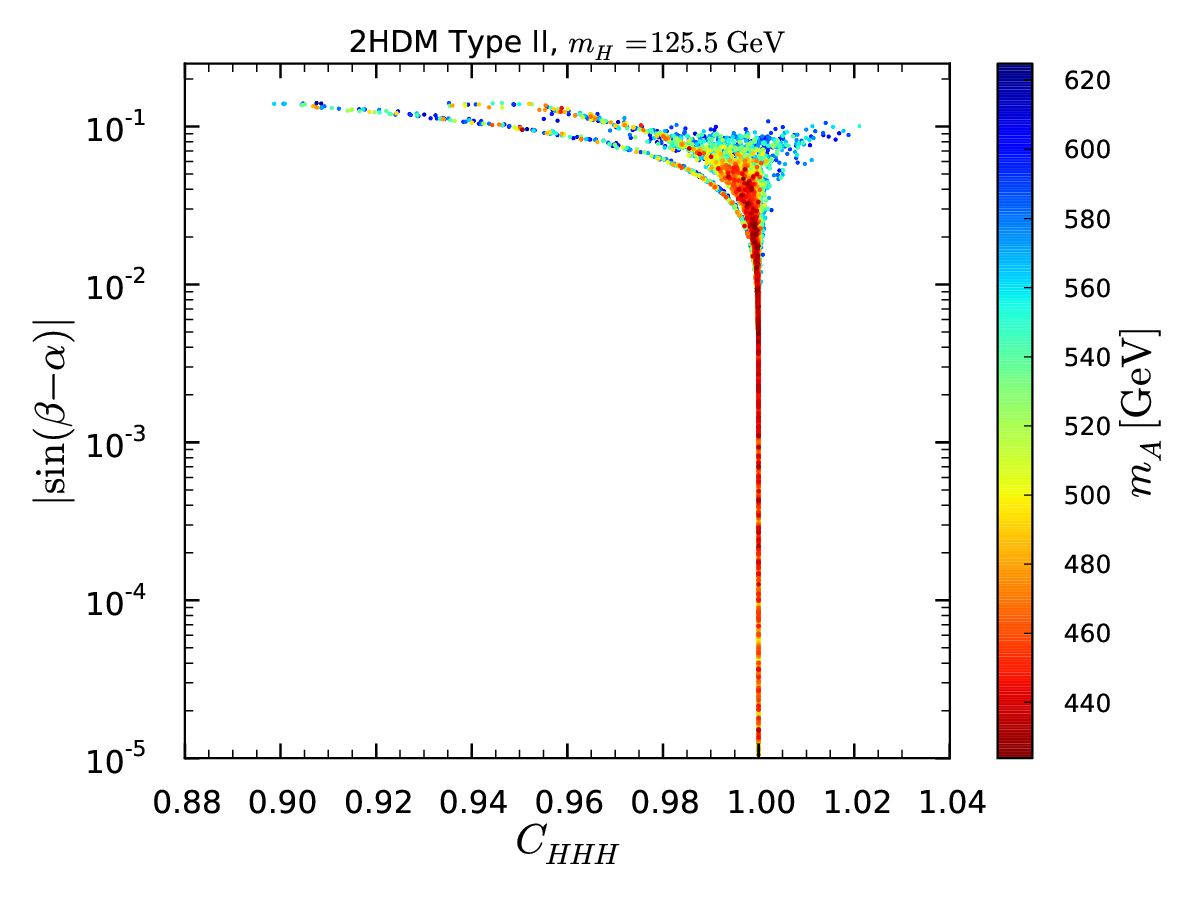}
\caption{Reduced triple Higgs coupling $C_{HHH}$ in Type~I (left) and Type~II (right), in the top panels with 
$m_h$ and in the bottom panels with $m_A$ color coding. Points are ordered from high to low $\overline{m}$, $m_h$ or $m_A$ values.\\[-40pt]}
  \label{mH_CHHH_H125}
\end{figure}
%
%
\begin{figure}[t!]\centering
\includegraphics[width=0.5\textwidth]{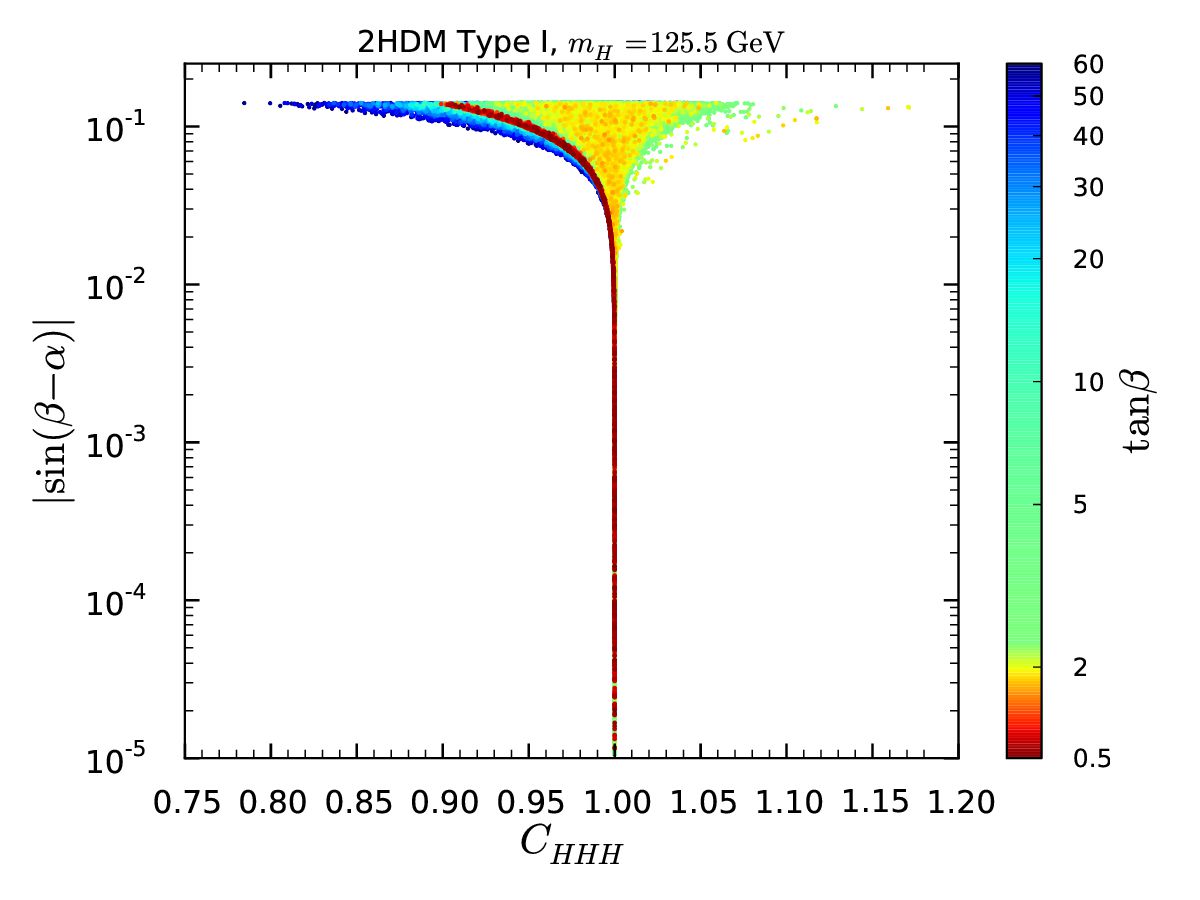}\includegraphics[width=0.5\textwidth]{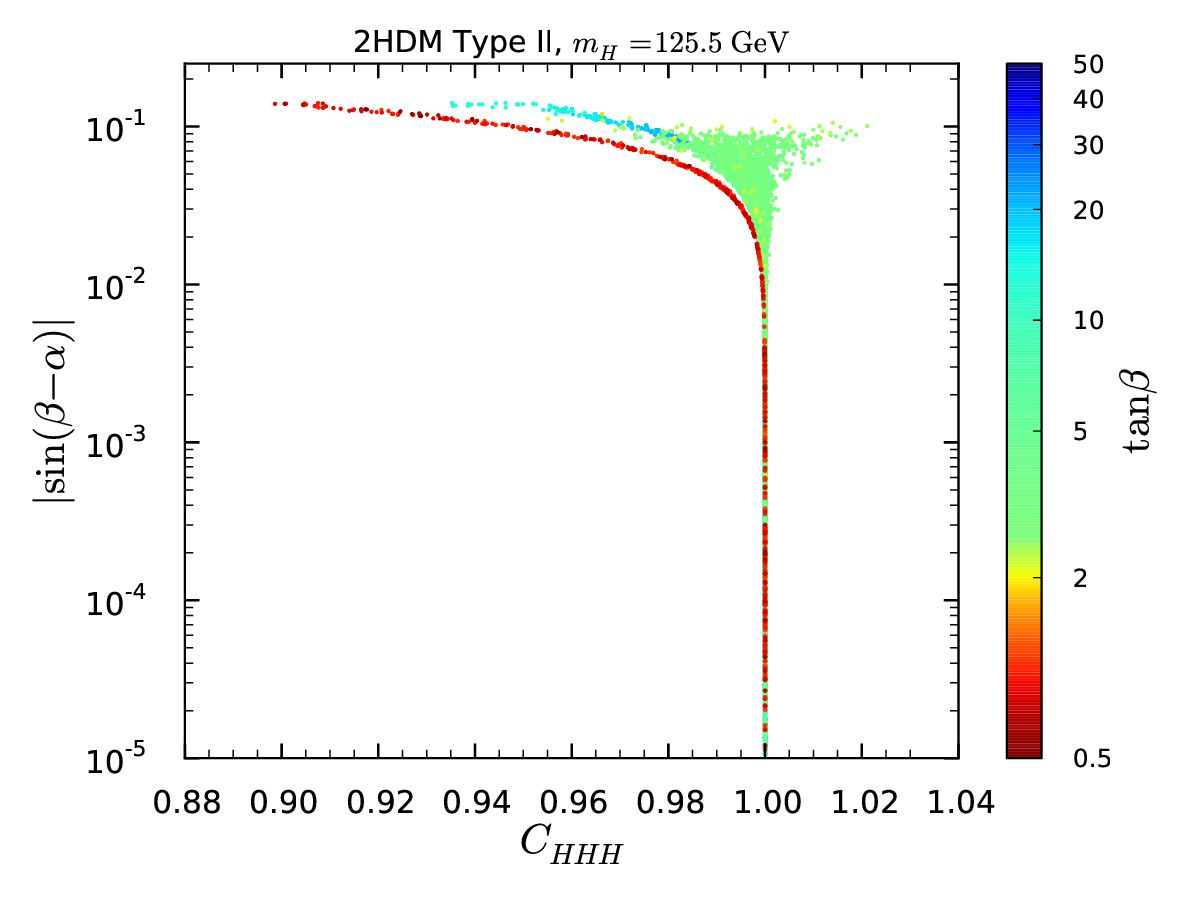}
\caption{As in Fig.~\ref{mH_CHHH_H125} but with $\tan\beta$ color coding. 
Points are ordered from high to low $\tanb$ values.
}
  \label{mH_CHHH_H125_tb}
\end{figure}

Finally, we consider the trilinear $HHH$ coupling, which is useful for consistency checks of the model, provided it can be measured precisely enough. The dependence of $C_{HHH}\equiv g_{HHH}/g_{HHH}^{\rm SM}$ on $|\sba|$ and $m_h$ (top panels) as well as $m_A$ (bottom panels) is shown in Fig.~\ref{mH_CHHH_H125}. 
Similar to $C_{hhh}$ in the non-decoupling regime of the $m_h=125\gev$ scenario \cite{Bernon:2015qea}, 
large values of the triple Higgs coupling beyond 1 can be achieved for $|\sba|$ values of the order of $0.1$ in the alignment regime of the $m_H=125\gev$ scenario. 
But, there is no direct analogue to the decoupling regime of the $m_h=125\gev$ scenario,  where the triple Higgs coupling was always suppressed as compared to its SM prediction. Instead, in the $m_H=125\gev$ scenario,  
$C_{HHH}$ can be enhanced or suppressed for any value of $m_A$. However, most of the points which might have had $C_{HHH}\gg 1$ are associated with $\tanb\approx 1$--$2$, which is precisely the range eliminated by the CMS limits on $gg\to A\to Zh$. In the end we are left with $C_{HHH}\approx 0.8$--$1.2$ in Type~I and $C_{HHH}\approx 0.9$--$1.02$ in Type~II.\footnote{Without the $gg\to A\to Zh$ limits,  we find values up to $C_{HHH}\approx 1.4$,  as discussed in Appendix~A.} 
In Type~I, the possible variation is less important for smaller $m_A$, in particular for $m_A$ below about $100\gev$. 
Moreover,  we note that  $C_{HHH}\le 1$ for $m_h\lesssim 60\gev$ in both Type~I and Type~II. 
Finally, the smallest values of $C_{HHH}<1$ in Type~I are found for large $\tan\beta$, while in Type~II $C_{HHH}$ converges to 1 with increasing $\tan\beta$, as shown in Fig.~\ref{mH_CHHH_H125_tb}.

\subsection{Signal strengths}\label{sec:signalstrengths}

\begin{figure}[t!]\centering
\includegraphics[width=0.5\textwidth]{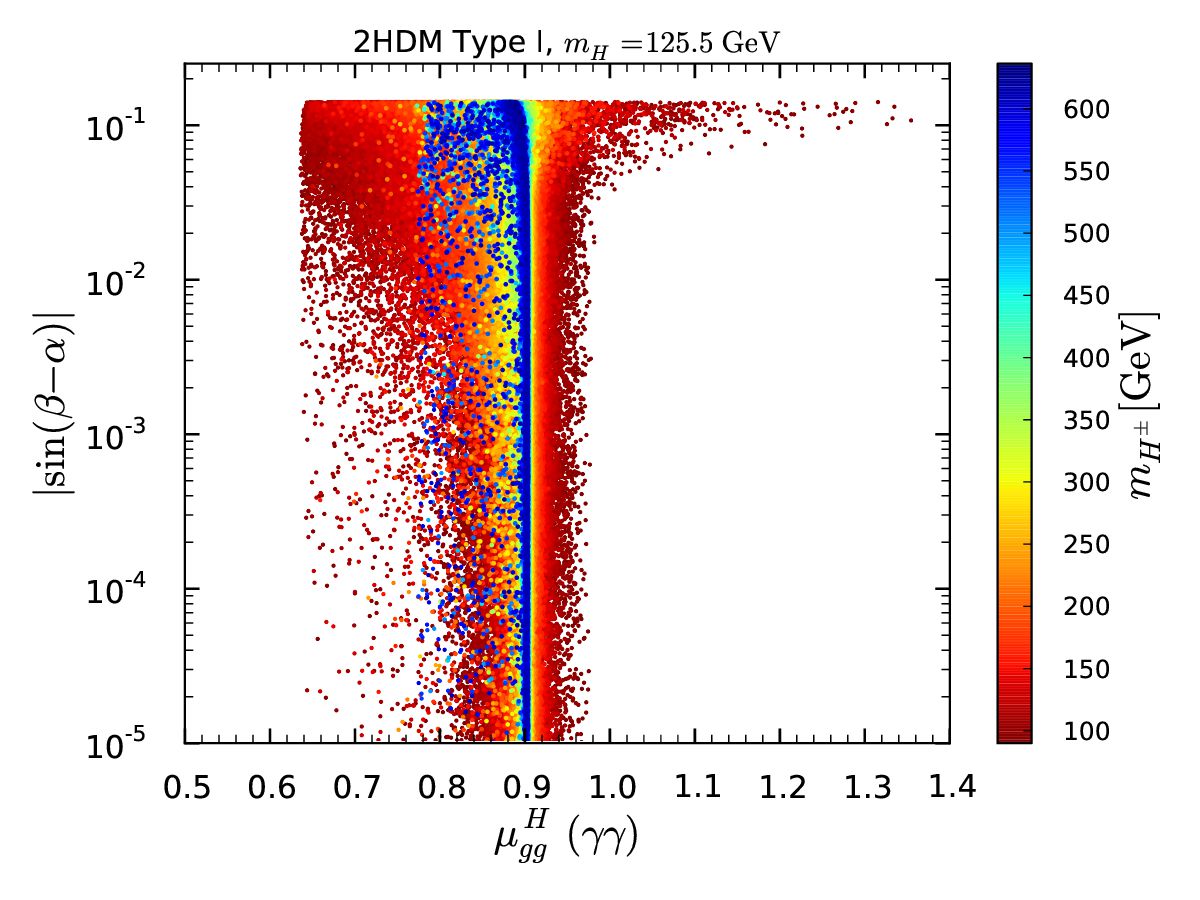}\includegraphics[width=0.5\textwidth]{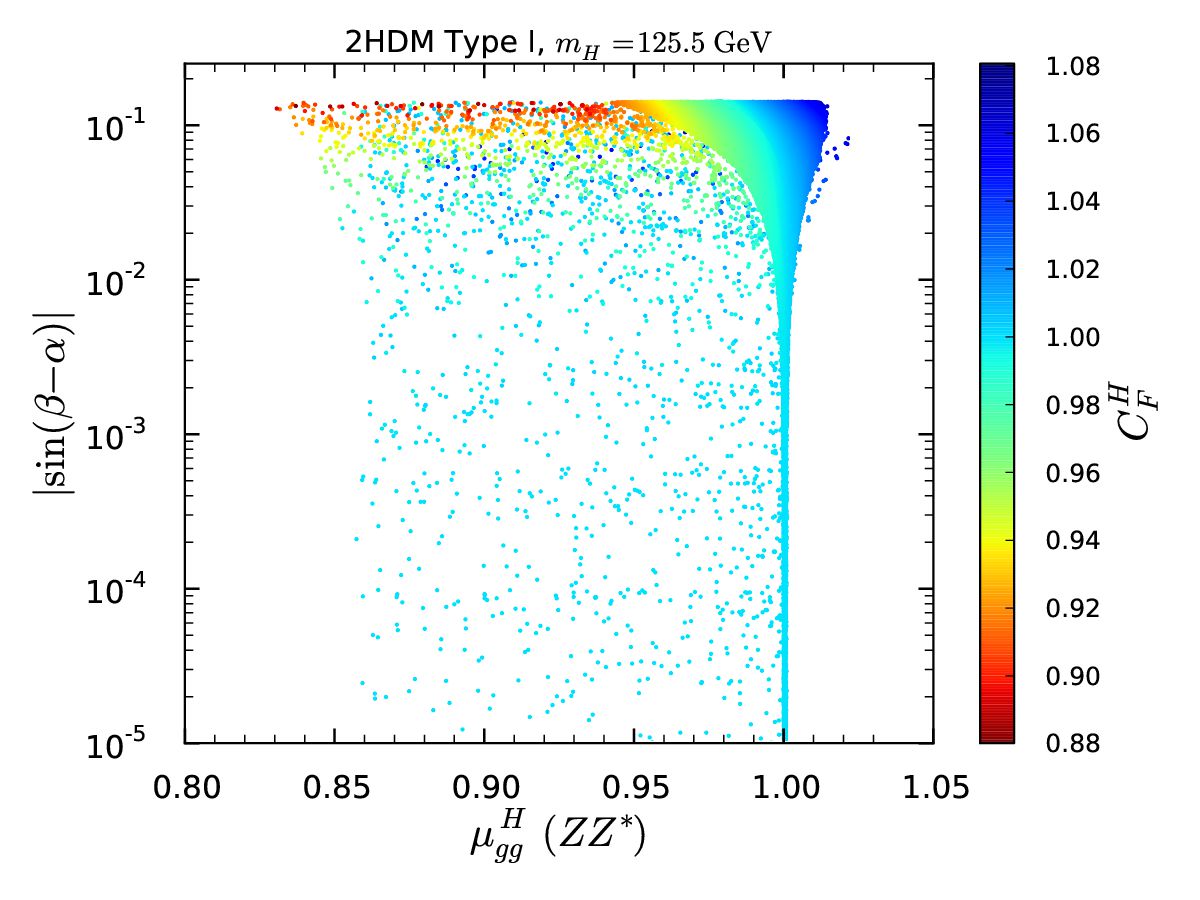}\\
\includegraphics[width=0.5\textwidth]{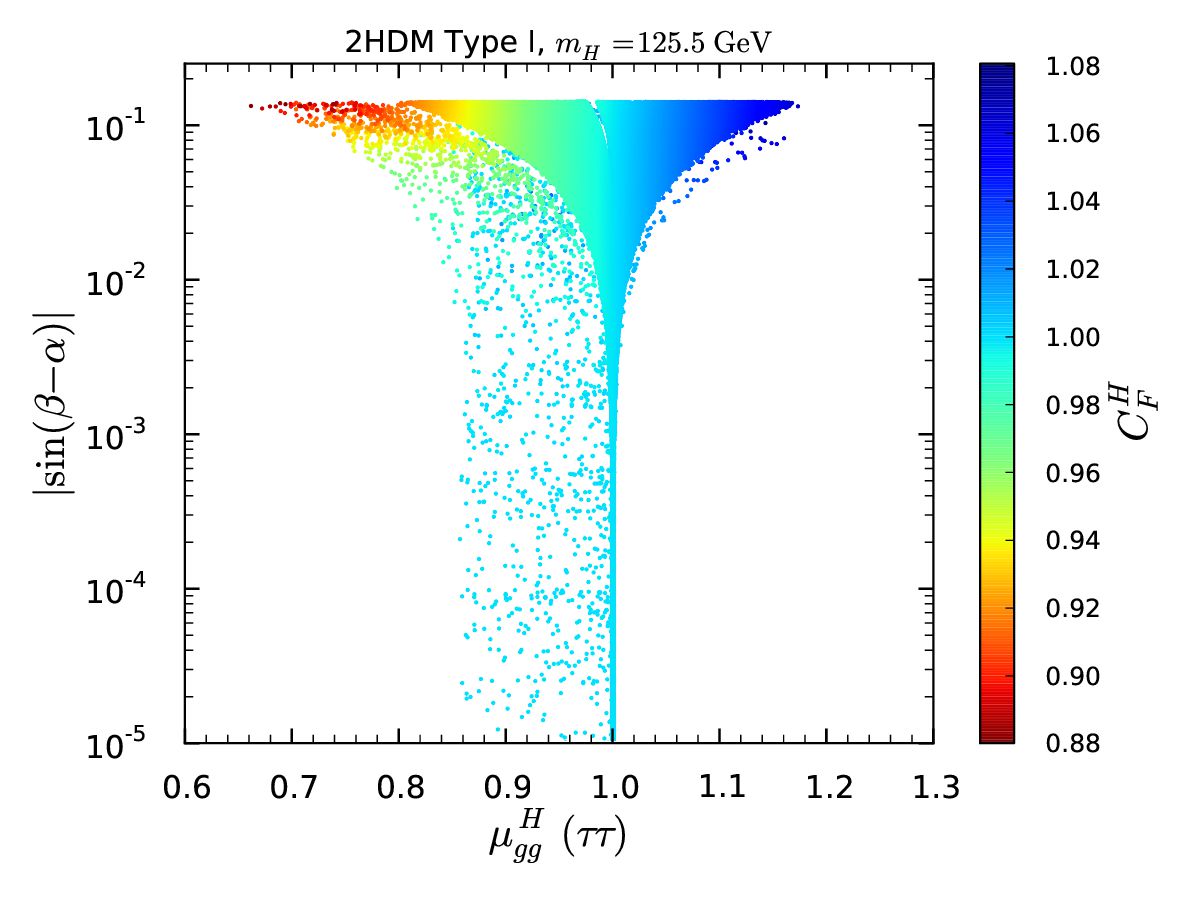}\includegraphics[width=0.5\textwidth]{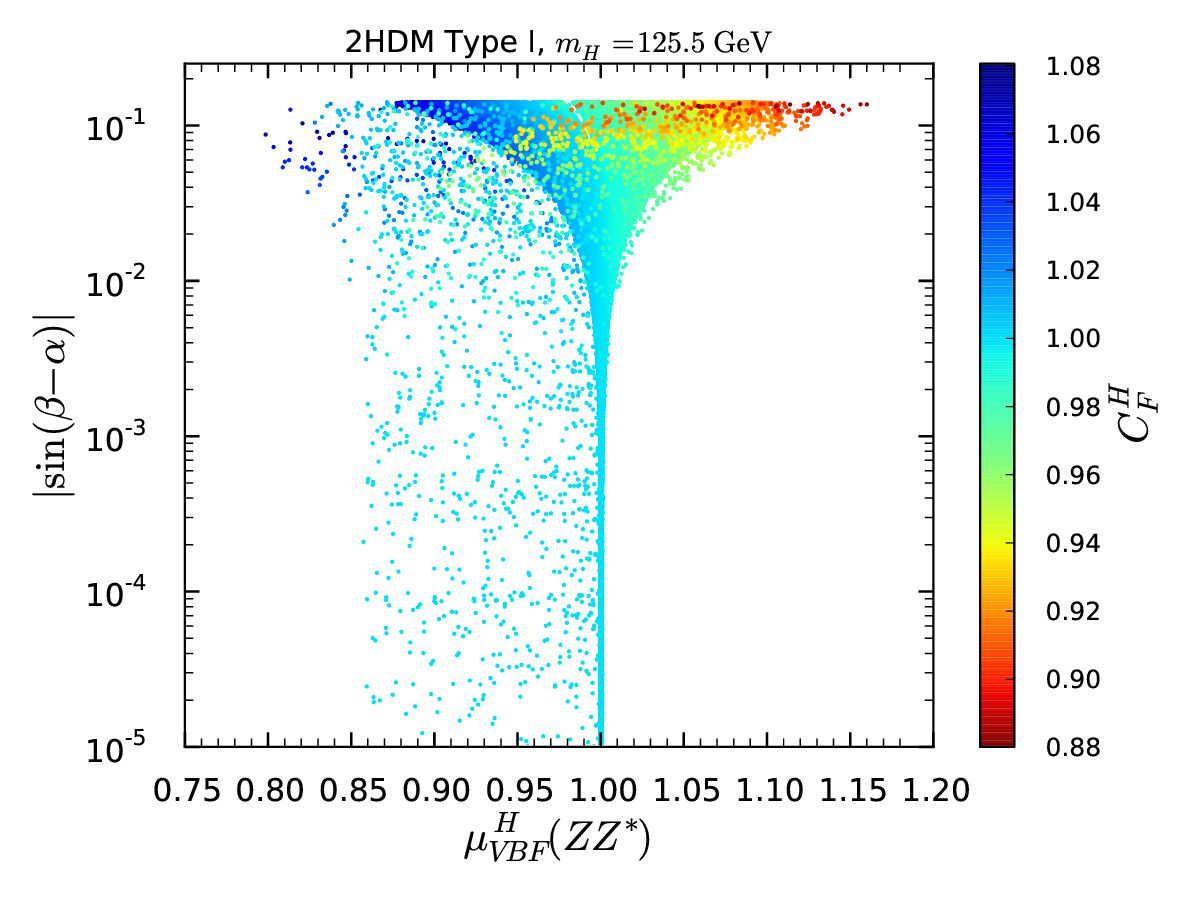}
\vskip -0.1in
\caption{Signal strengths in Type~I.}
  \label{mu_1d_type1}
\end{figure}

\begin{figure}[t!]\centering
\includegraphics[width=0.5\textwidth]{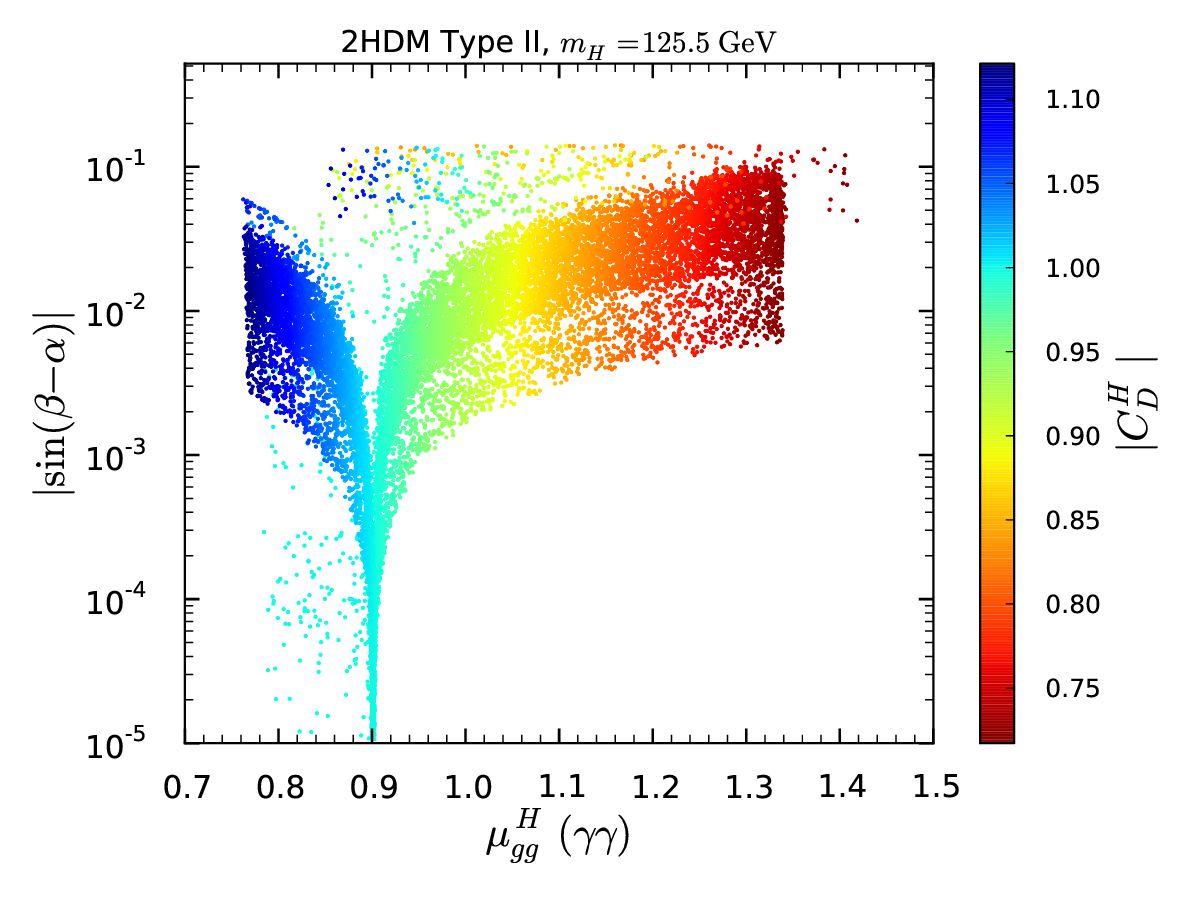}\includegraphics[width=0.5\textwidth]{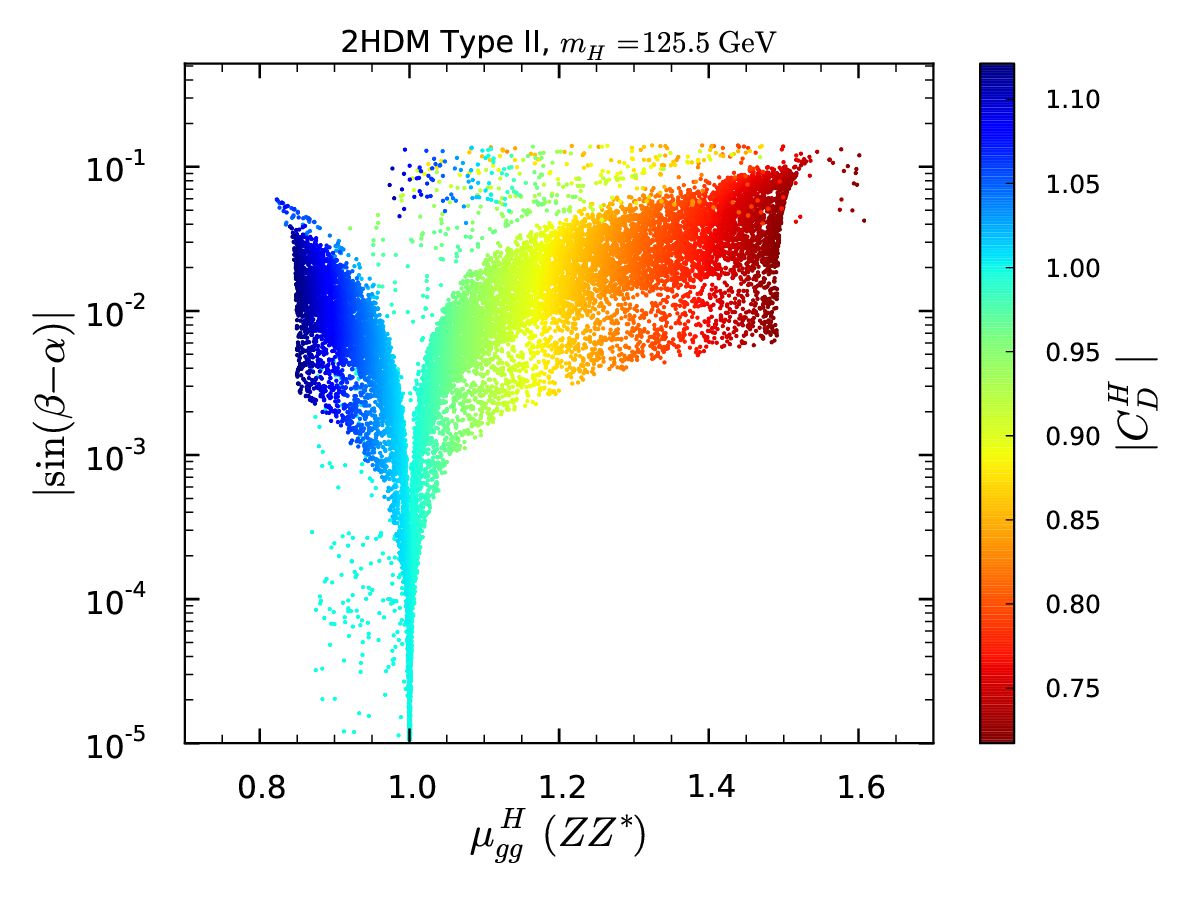}\\
\includegraphics[width=0.5\textwidth]{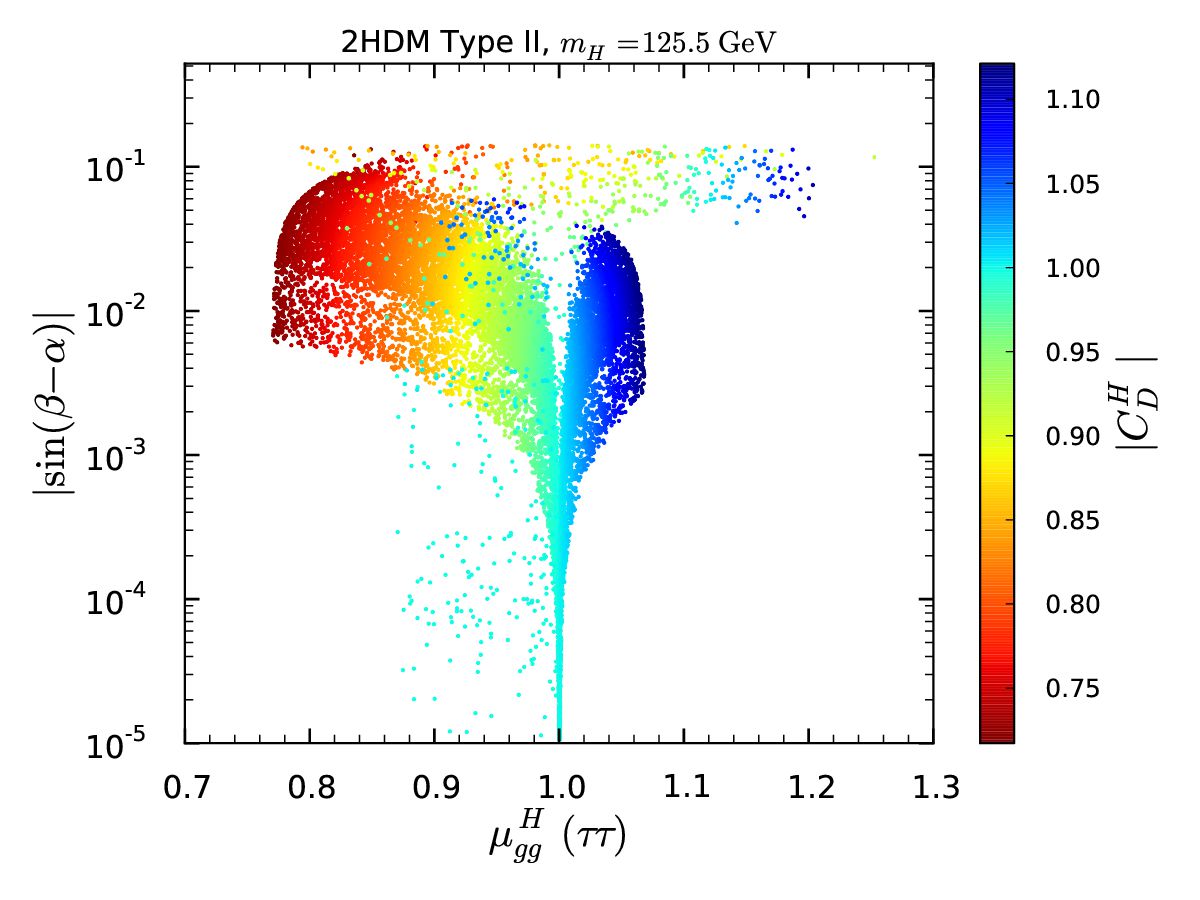}\includegraphics[width=0.5\textwidth]{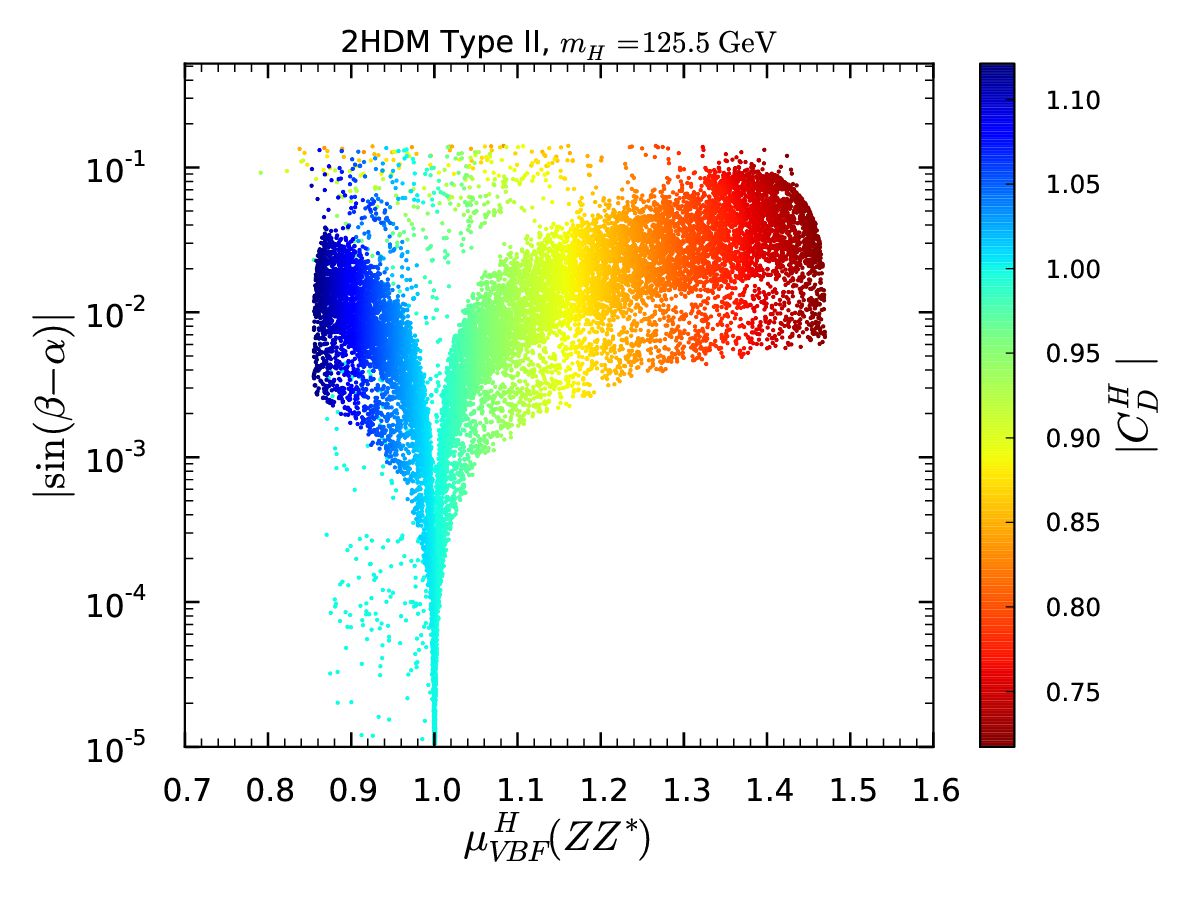}\caption{Signal strengths in Type~II. The horizontal bars near $|\sba|\approx 0.1$ arise from the opposite-sign $C_D^H$ solution.}
  \label{mu_1d_type2}
\end{figure}

The variations in the couplings to fermions discussed above have direct consequences for the signal strengths of the SM-like Higgs boson.
In Type~I, the signal strengths in the $H\to\gamma\gamma$ decay mode are driven by the value of $\mhpm$, while for the 
$H\to VV^*$ ($VV^*=ZZ^*,WW^*$) and $H\to\tau\tau$ decay modes they depend mostly on $C_F^H$, as illustrated in Fig.~\ref{mu_1d_type1}.\footnote{We employ the notation $\mu_{X}^H(Y)$ for the signal strength in the production mode $X$ and decay mode $H\to Y$.} Notice that the $C_F^H$ dependence in the $ZZ^*$ mode is opposite for $gg$ fusion and vector boson fusion (VBF) production. In the case of VBF, a smaller value of $C_F^H$ implies a  smaller $b\bar{b}$ partial width and therefore a larger $ZZ^*$ branching ratio, whereas in $gg$ production  $C_F^H$ determines the size of the top-quark loop contribution which is enhanced for a larger  value of $C_F^H$. 
In contrast, in Type~II, the signal strengths are always dominantly driven by $C_D^H$, as this determines the $H\to b\bar b$ partial width (and $\mhpm$ is constrained to be heavy). This is illustrated in Fig.~\ref{mu_1d_type2}. In this case, the dependence is always the same for $gg$ and VBF production.  

\begin{figure}[t!]\centering
\includegraphics[width=0.33\textwidth]{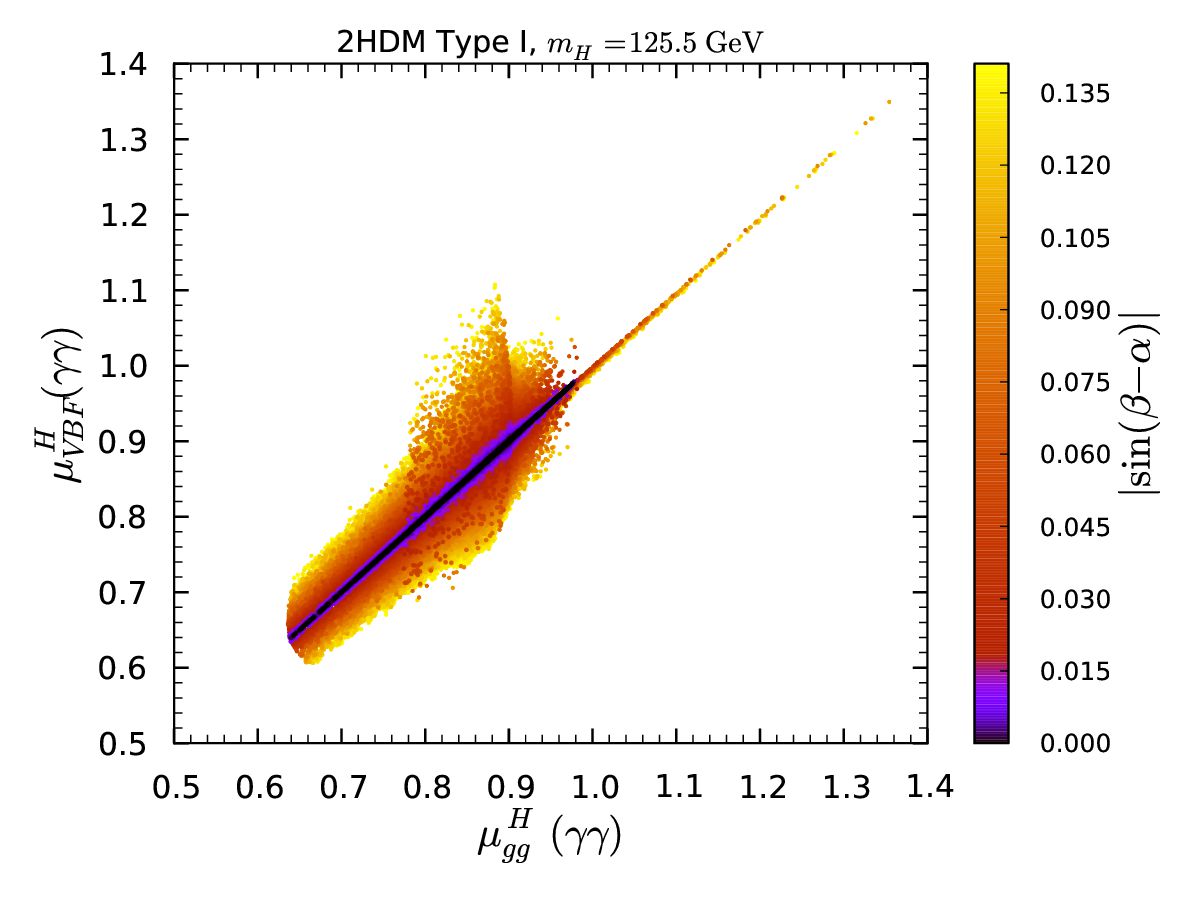}\includegraphics[width=0.33\textwidth]{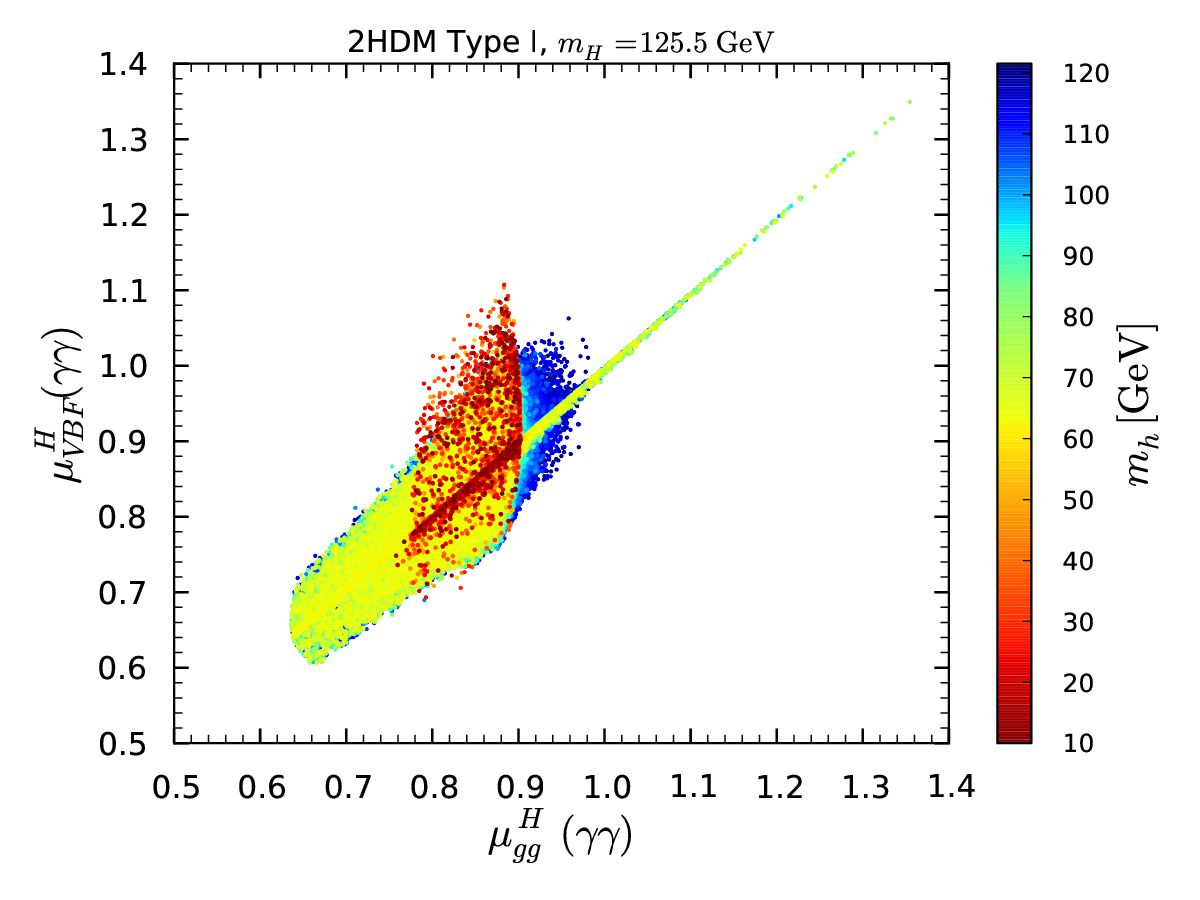}\includegraphics[width=0.33\textwidth]{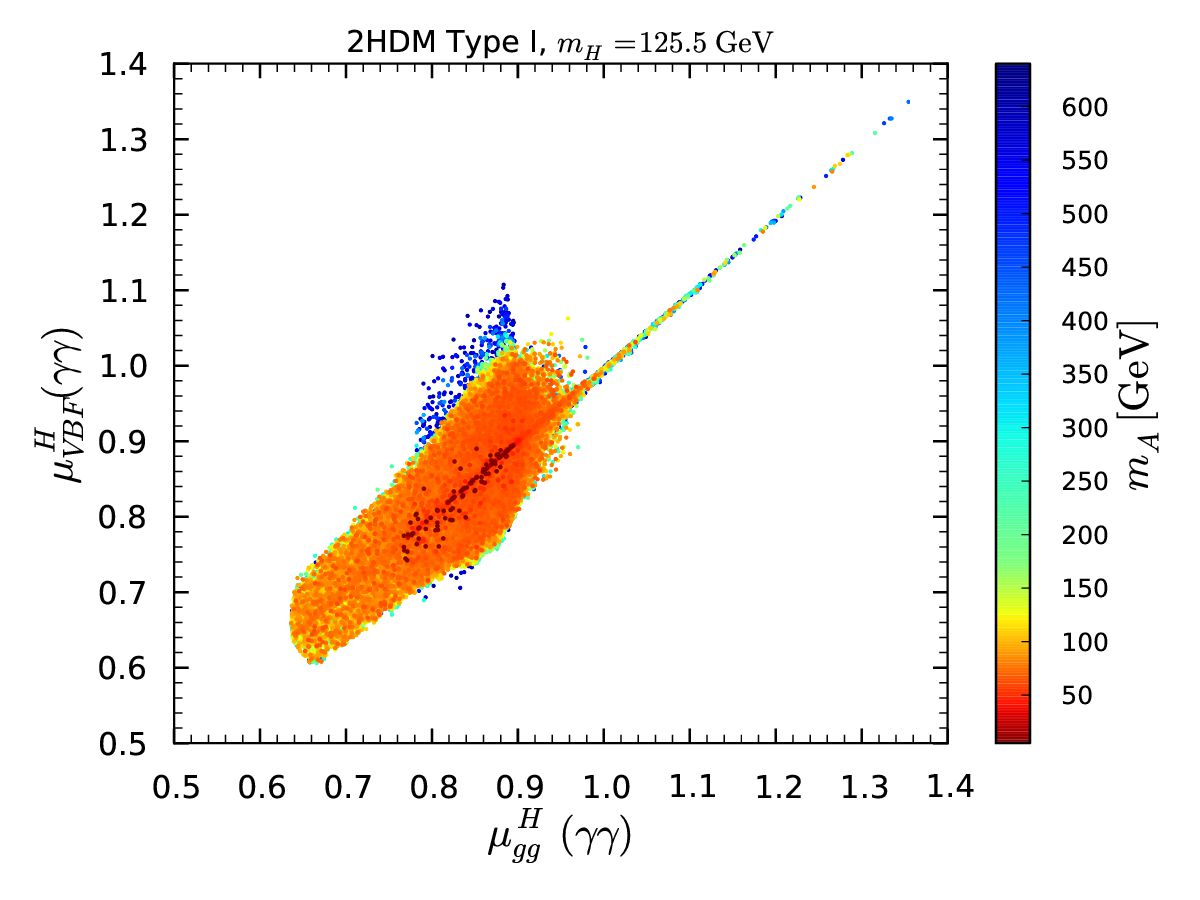}\\\includegraphics[width=0.33\textwidth]{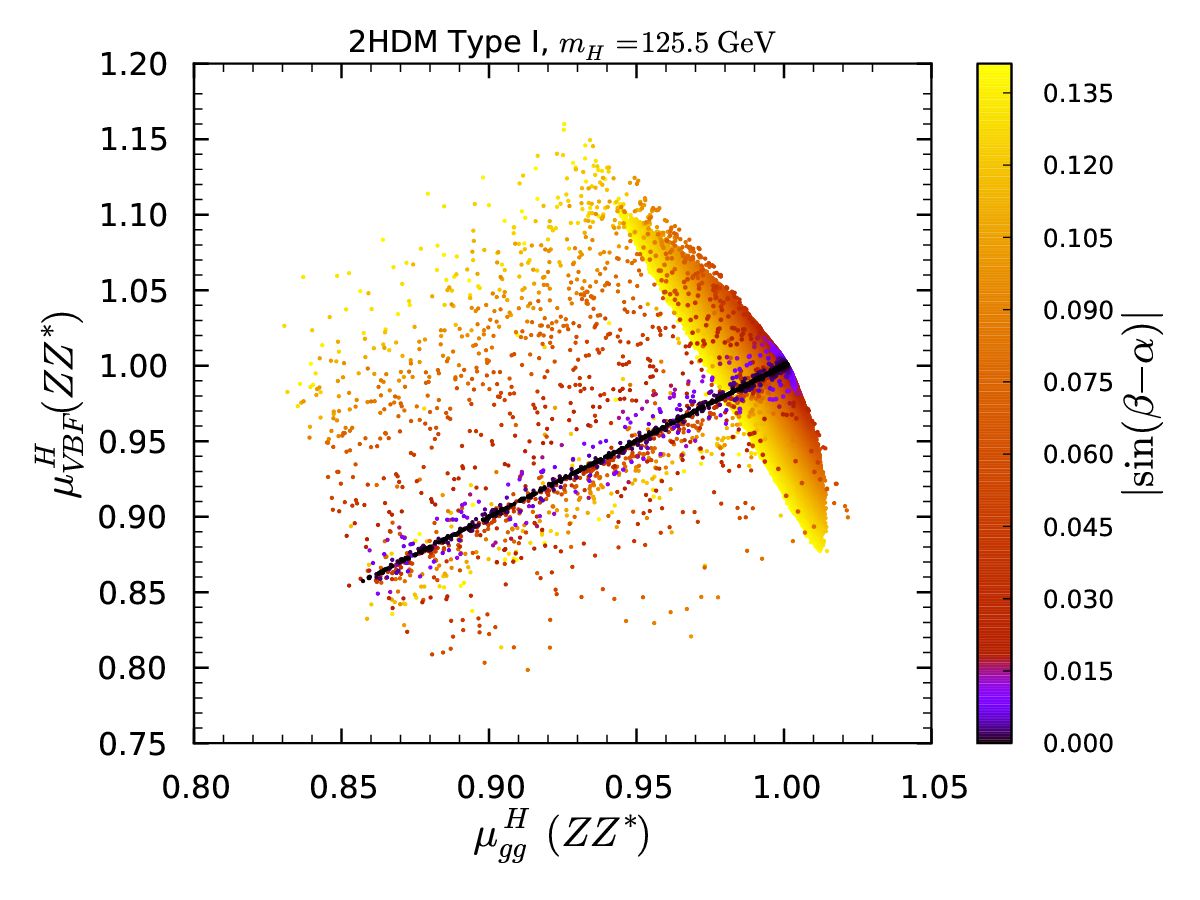}\includegraphics[width=0.33\textwidth]{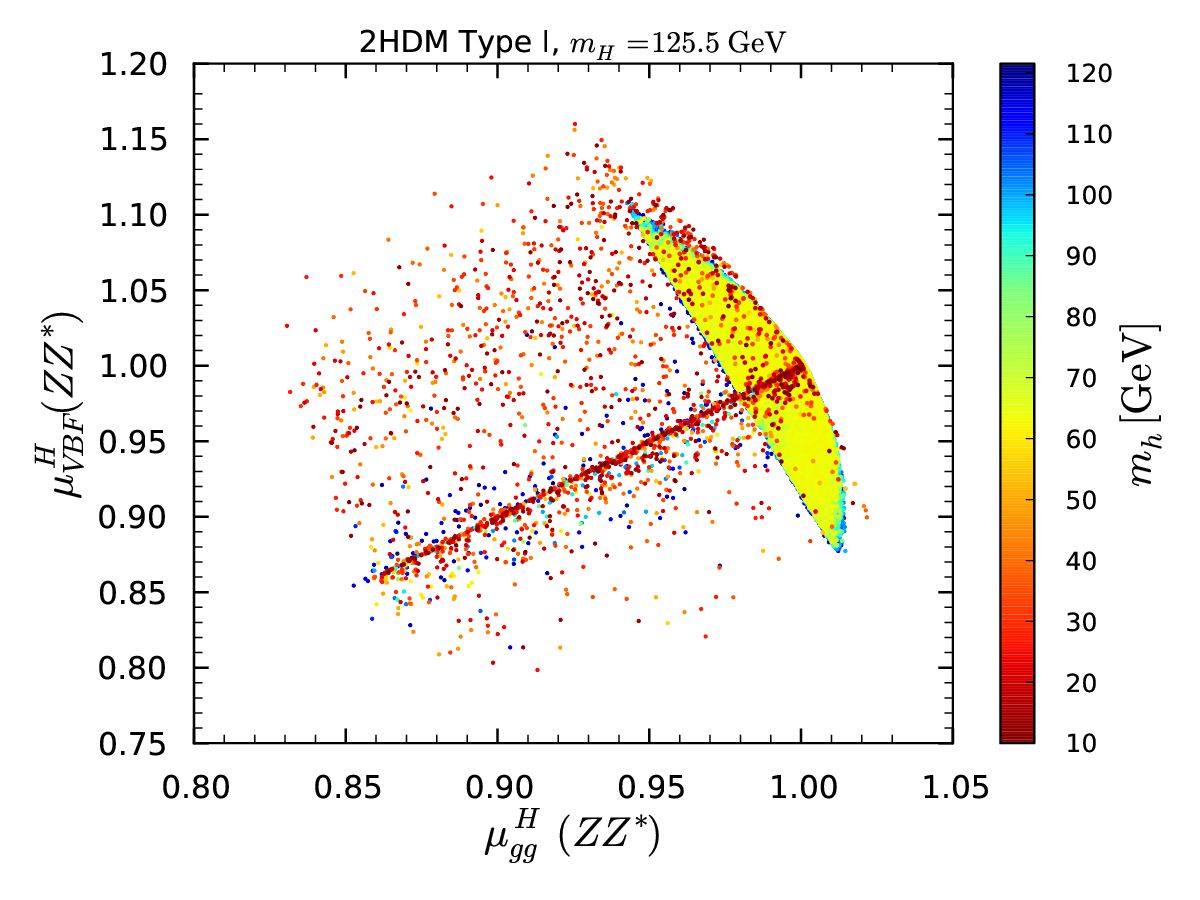}\includegraphics[width=0.33\textwidth]{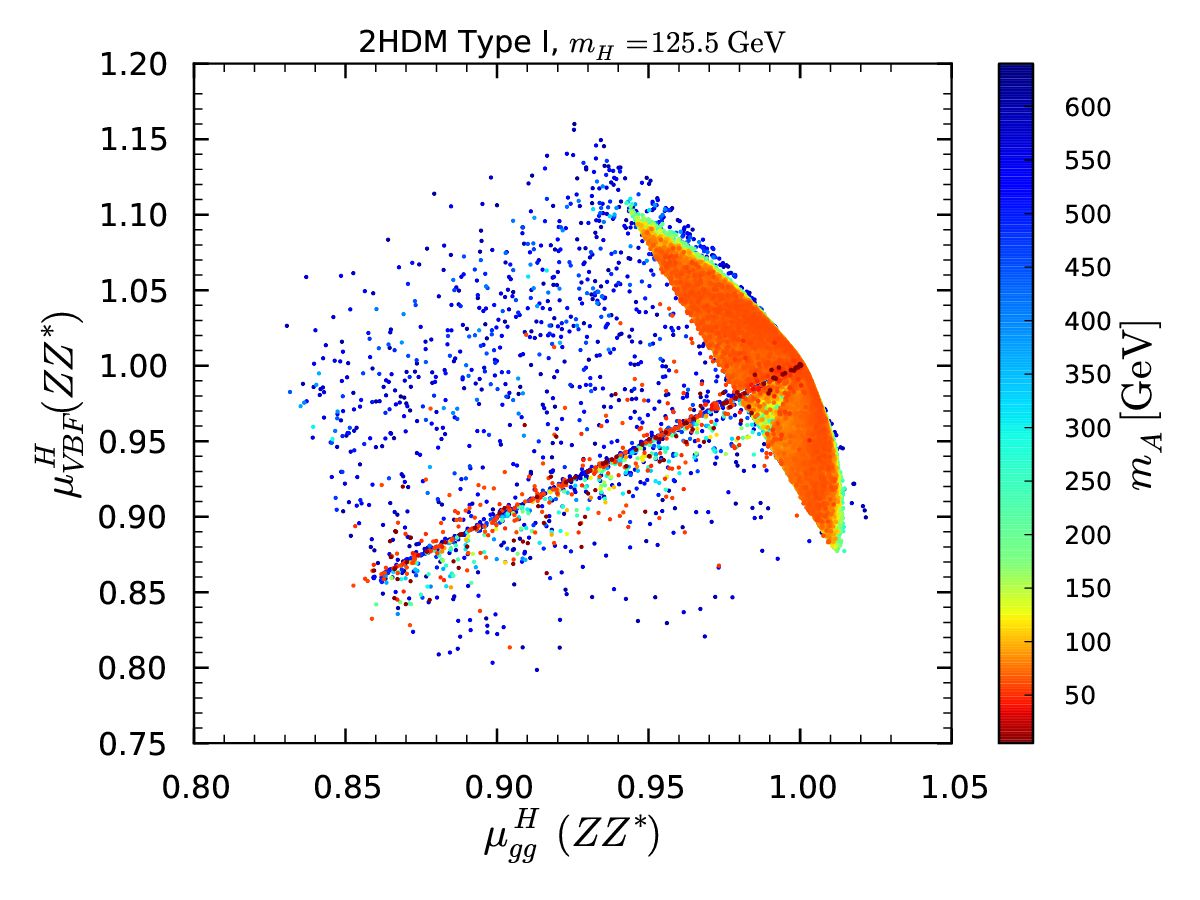}\\\includegraphics[width=0.33\textwidth]{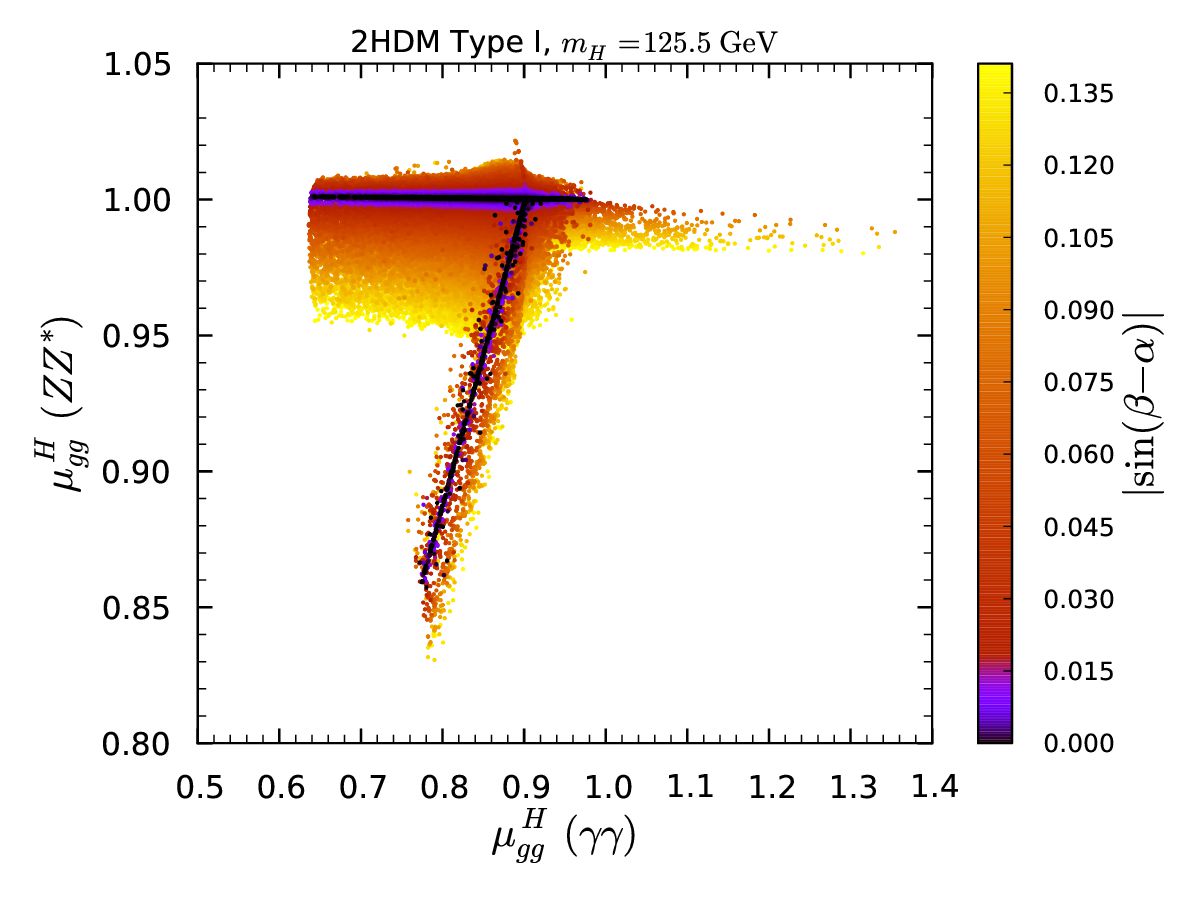}\includegraphics[width=0.33\textwidth]{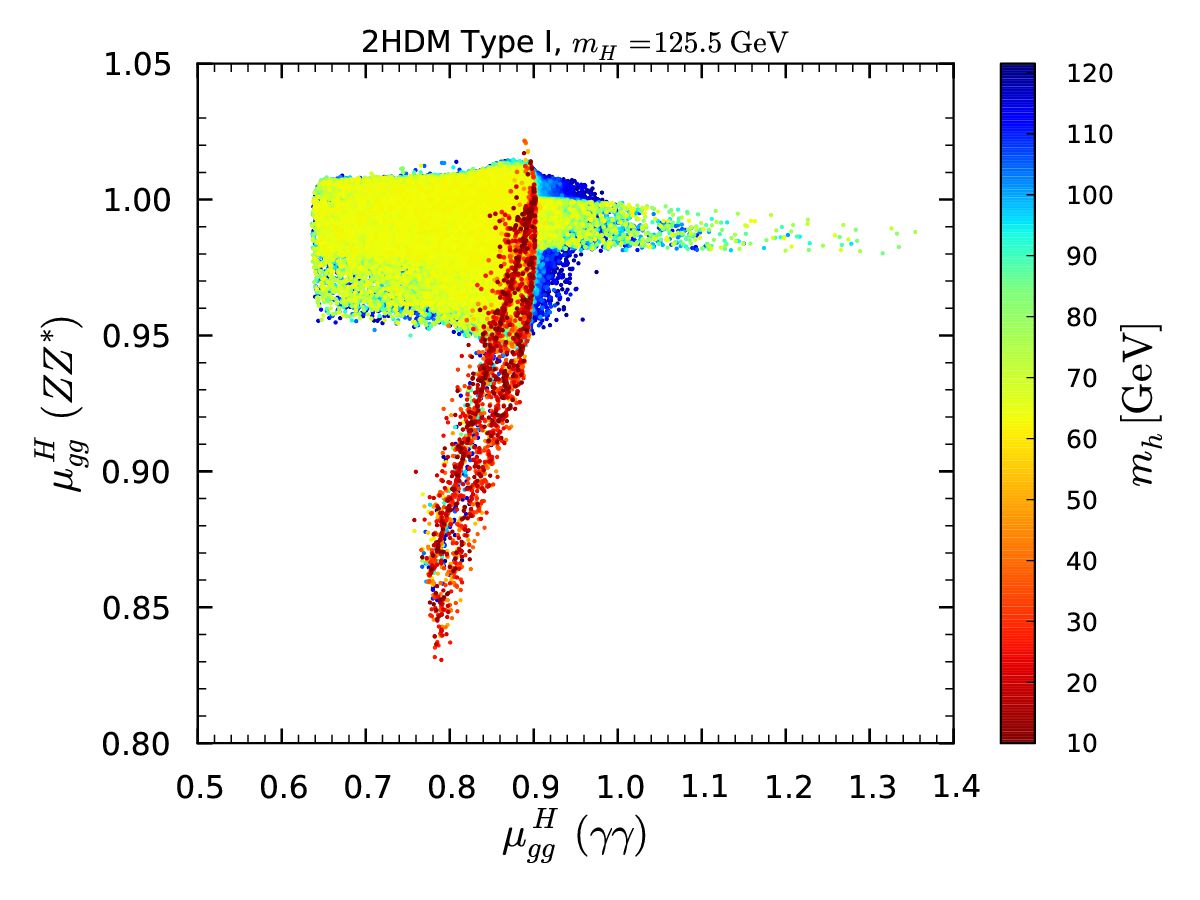}\includegraphics[width=0.33\textwidth]{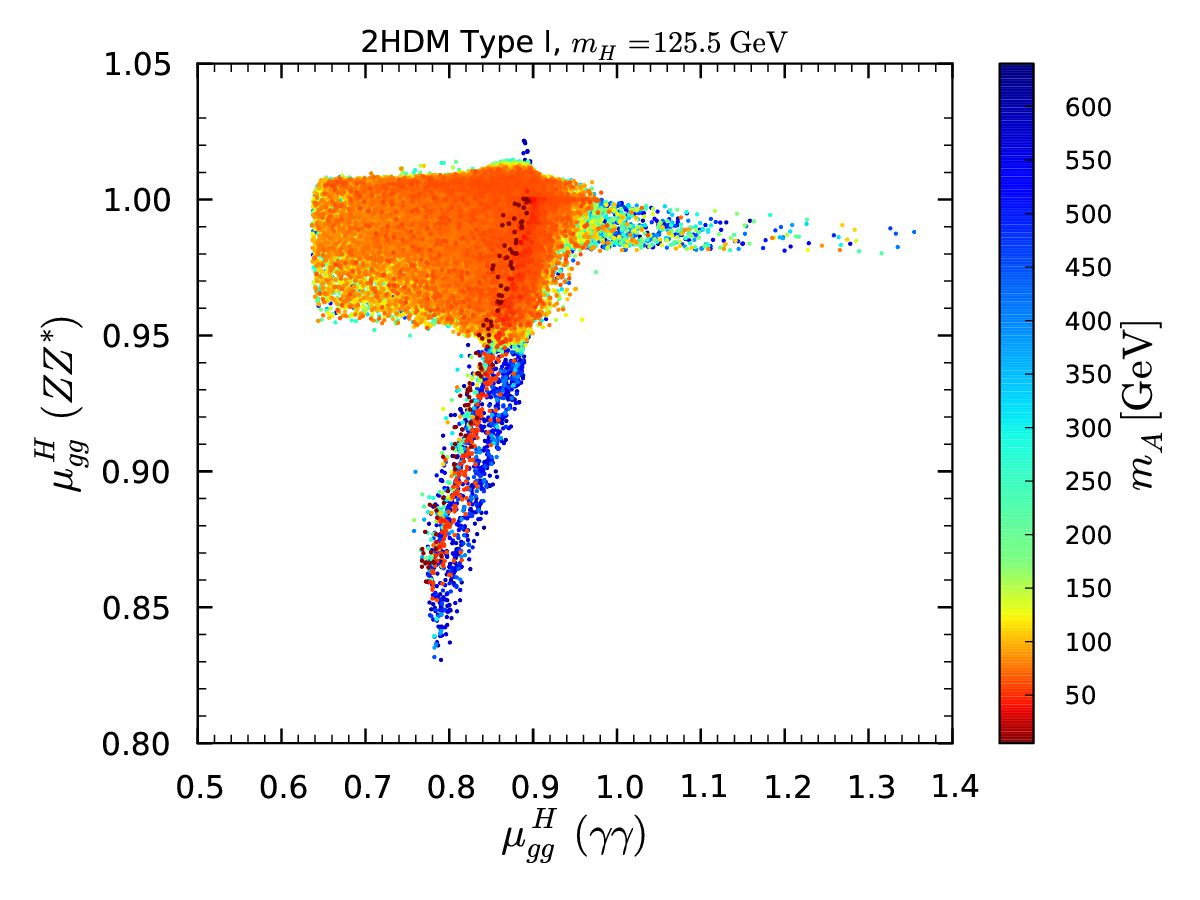}\\\includegraphics[width=0.33\textwidth]{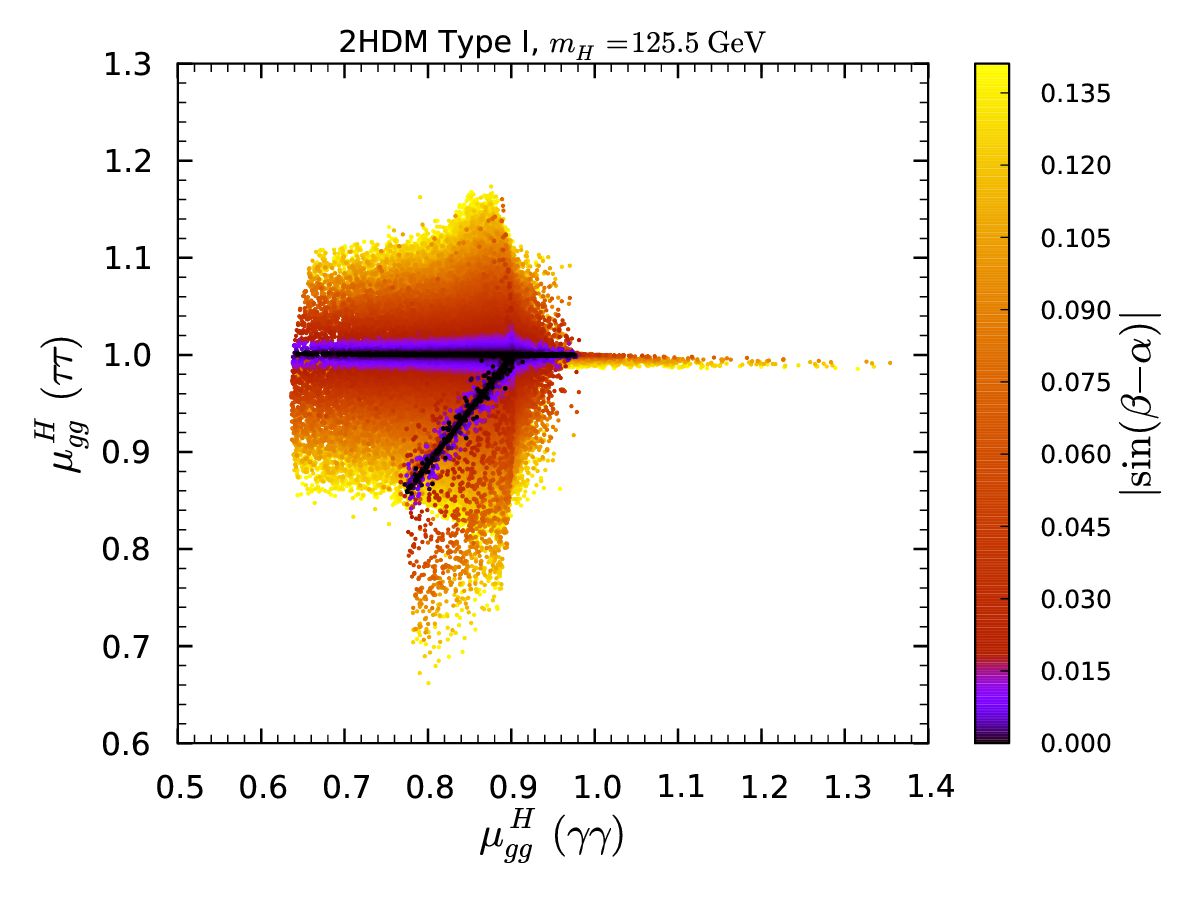}\includegraphics[width=0.33\textwidth]{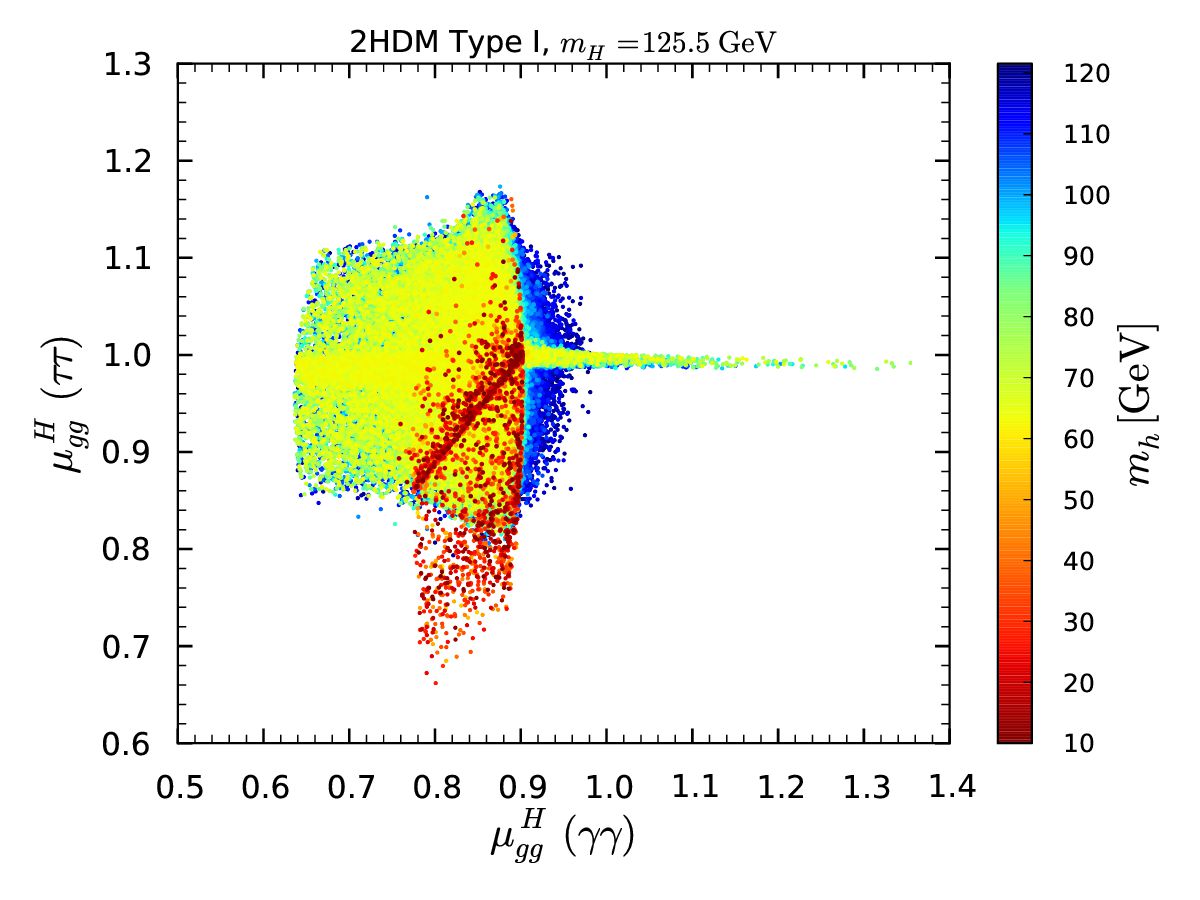}\includegraphics[width=0.33\textwidth]{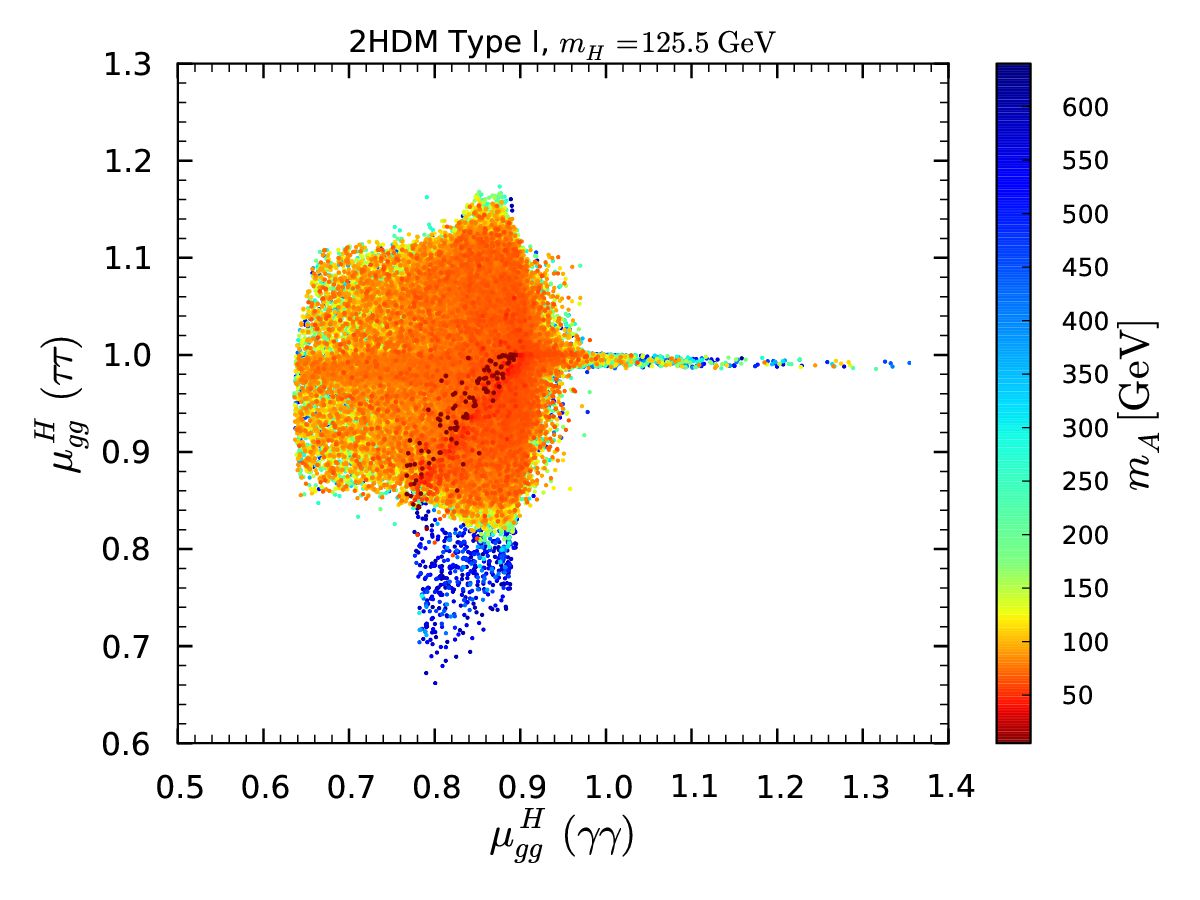}\\
\caption{Examples of correlations between signal strengths in Type~I. The top panels show $\mu_{\rm VBF}^H(\gamma\gamma)$ vs.\ $\mu_{gg}^H(\gamma\gamma)$, the upper middle panels show $\mu_{\rm VBF}^H(ZZ^*)$ vs.\ $\mu_{gg}^H(ZZ^*)$, 
the lower middle panels show  $\mu_{gg}^H(ZZ^*)$ vs.\ $\mu_{gg}^H(\gamma\gamma)$ and the bottom panels show $\mu_{gg}^H(\tau\tau)$ vs.\ $\mu_{gg}^H(\gamma\gamma)$. The color code indicates, from left to right, the dependence on $|\sbma|$, $m_h$ and $m_A$.  Points are ordered from high to low $|\sbma|$, $m_h$ and $m_A$ values.}
\label{correlations1}
\end{figure}

\begin{figure}[t!]\centering
\includegraphics[width=0.33\textwidth]{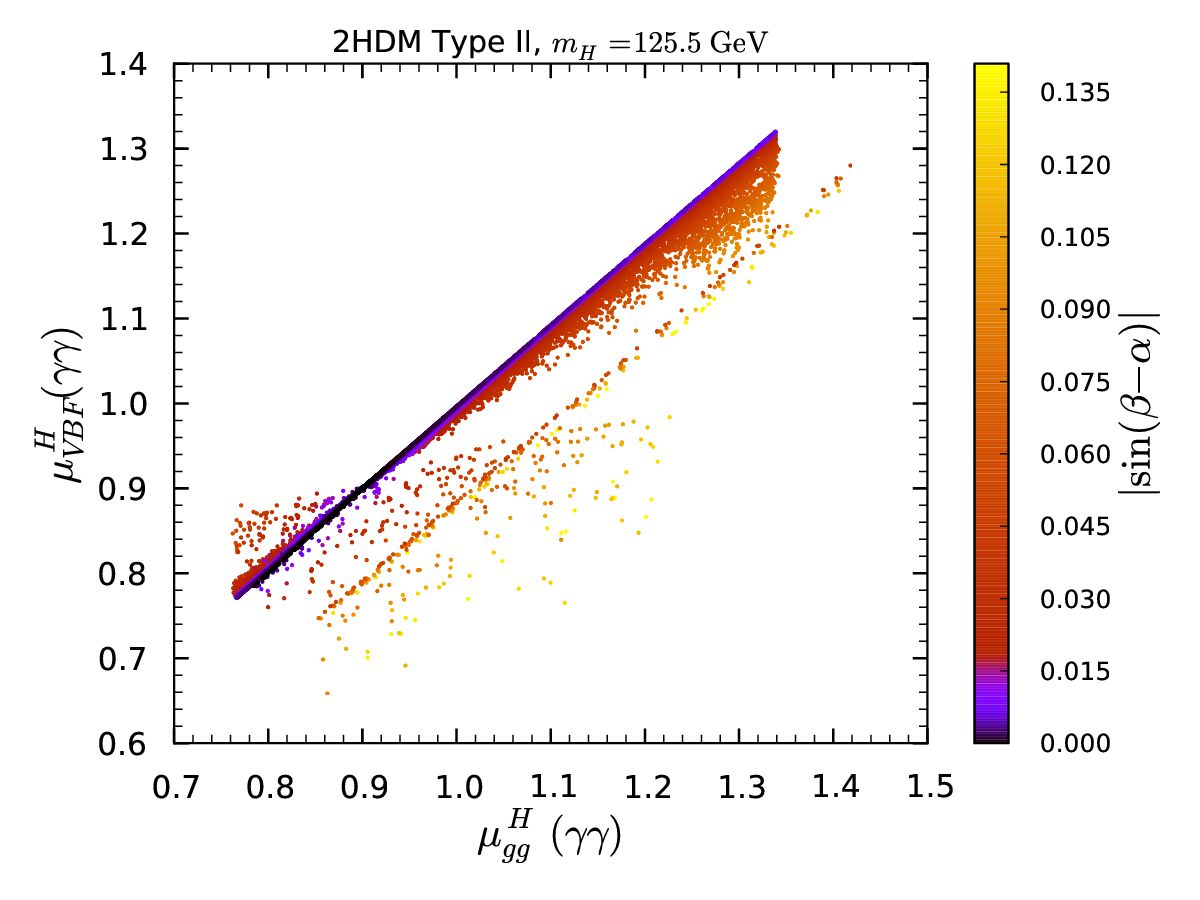}\includegraphics[width=0.33\textwidth]{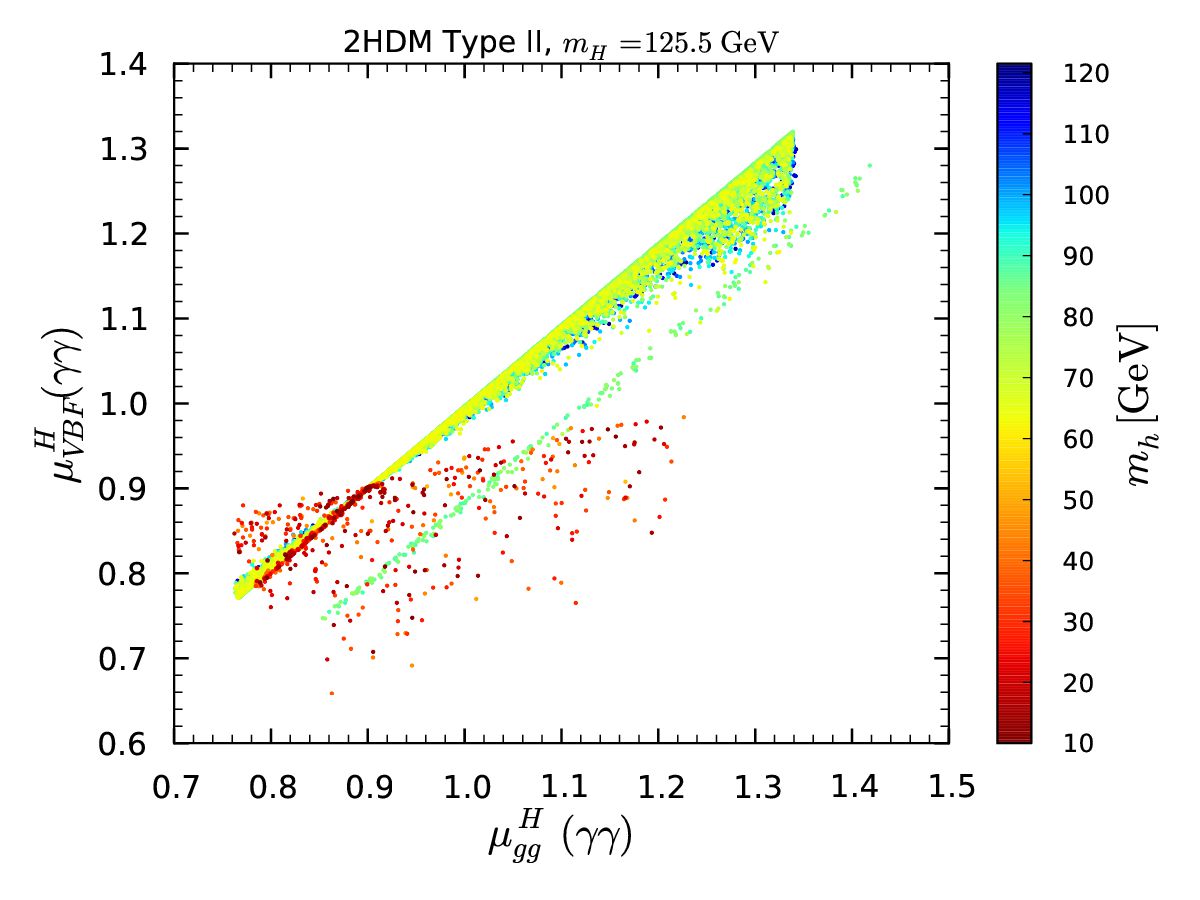}\includegraphics[width=0.33\textwidth]{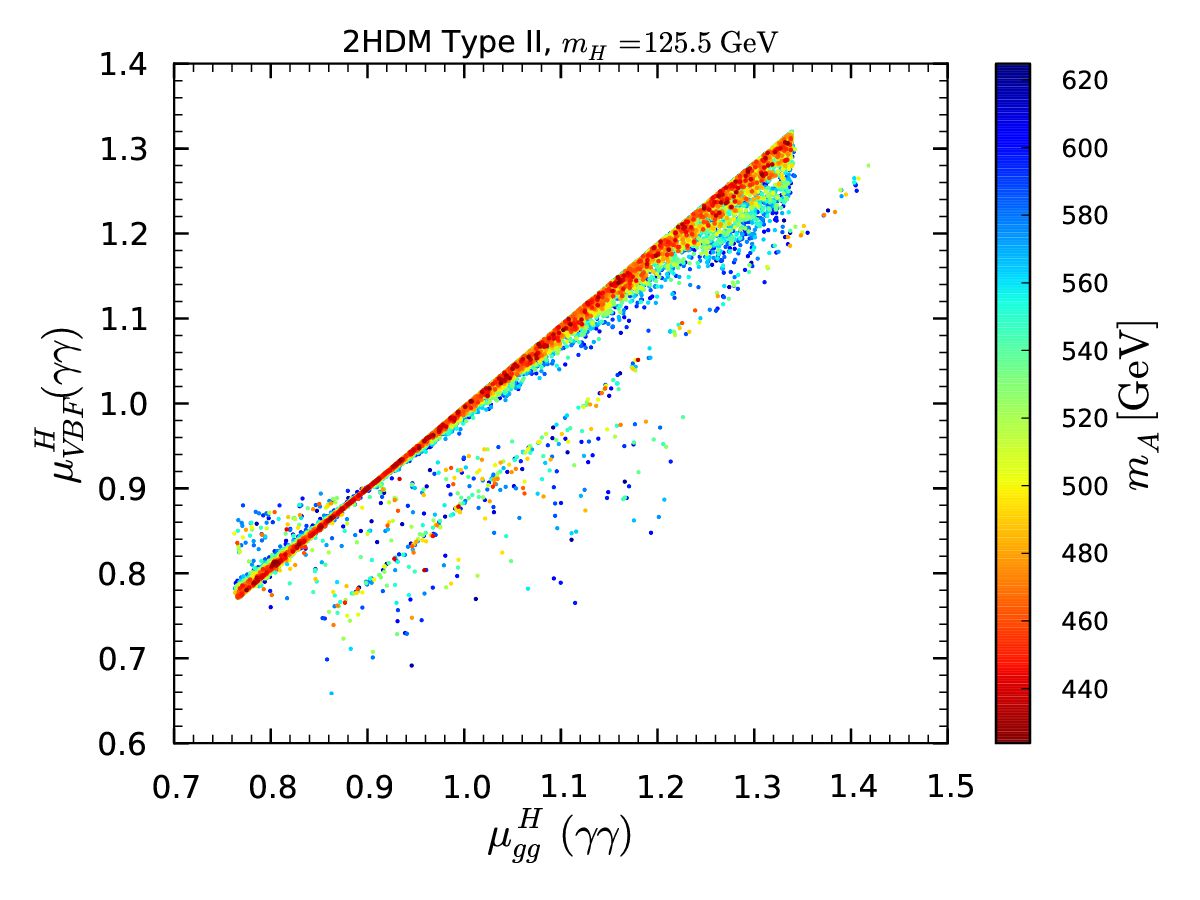}\\
\includegraphics[width=0.33\textwidth]{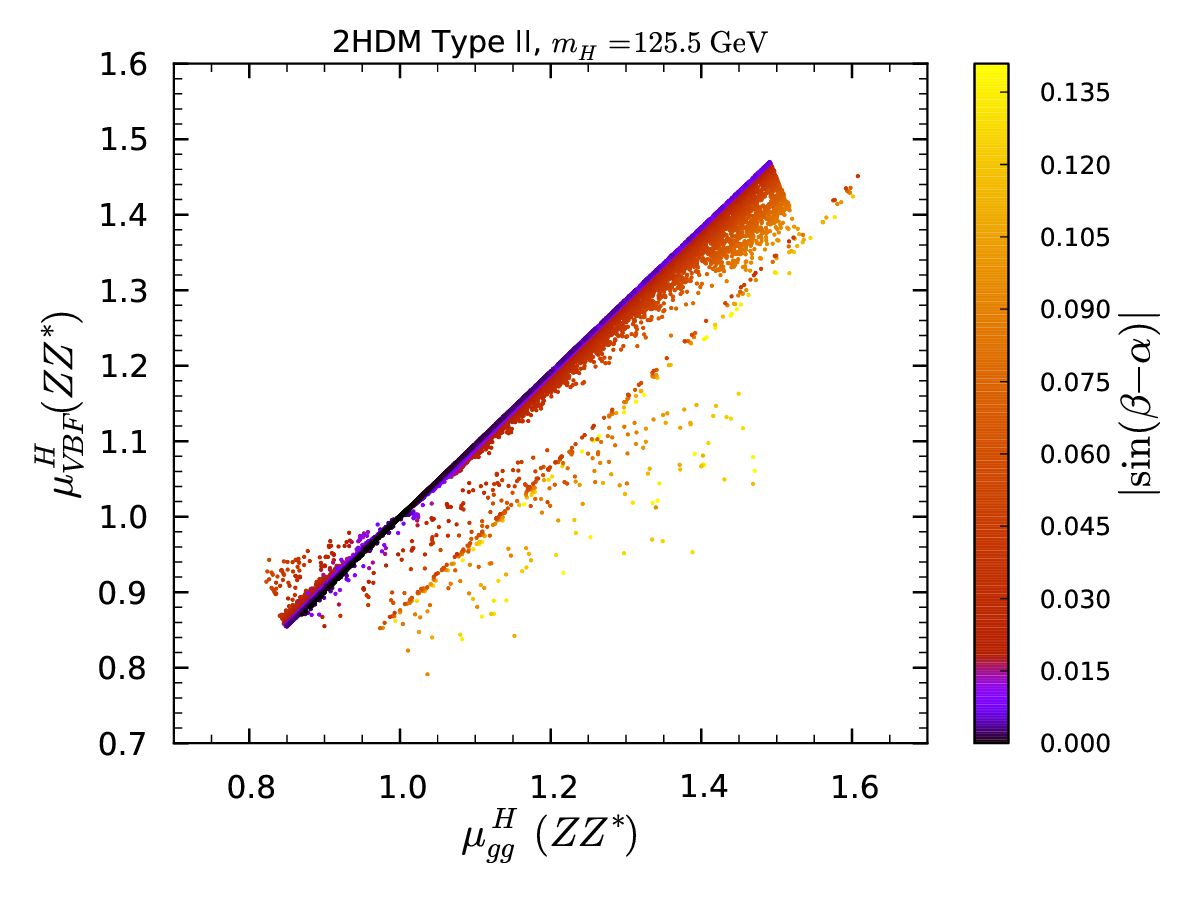}\includegraphics[width=0.33\textwidth]{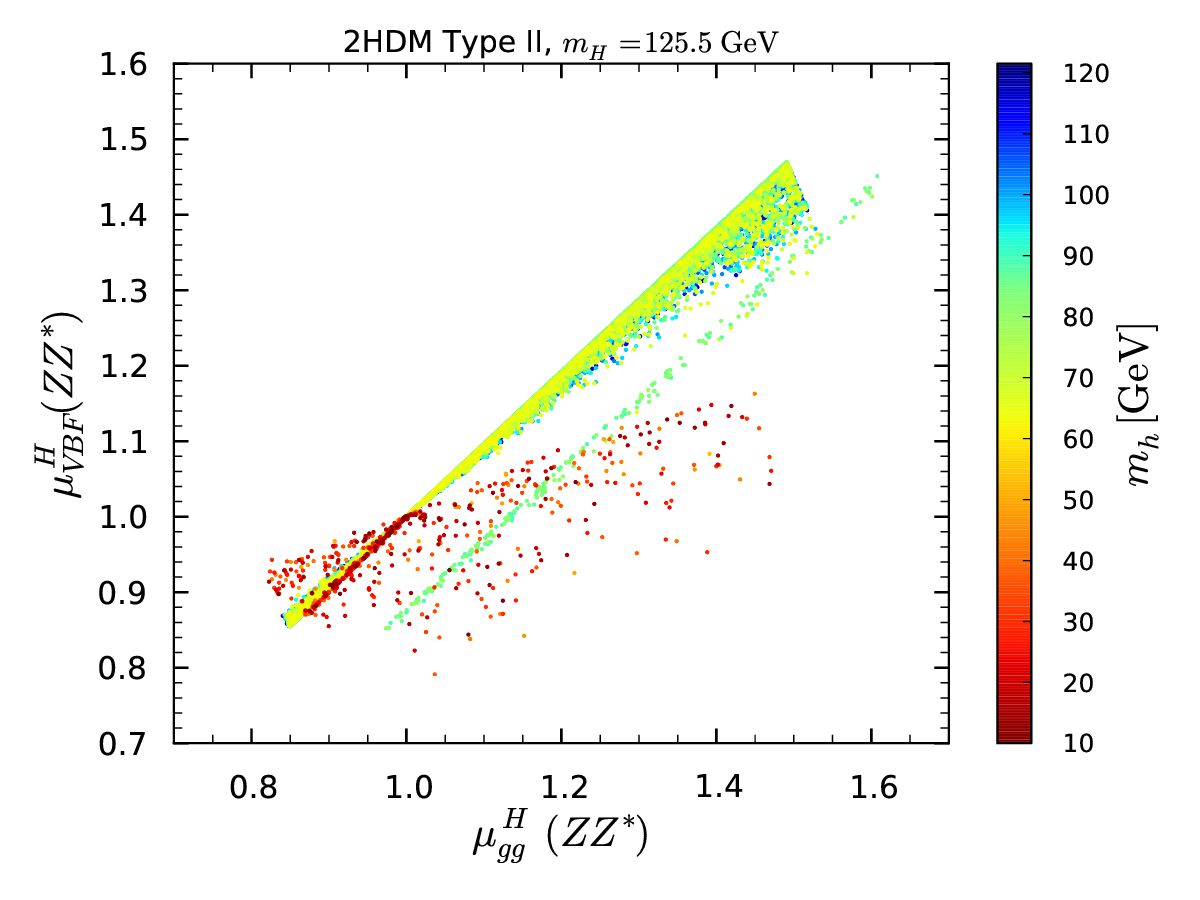}\includegraphics[width=0.33\textwidth]{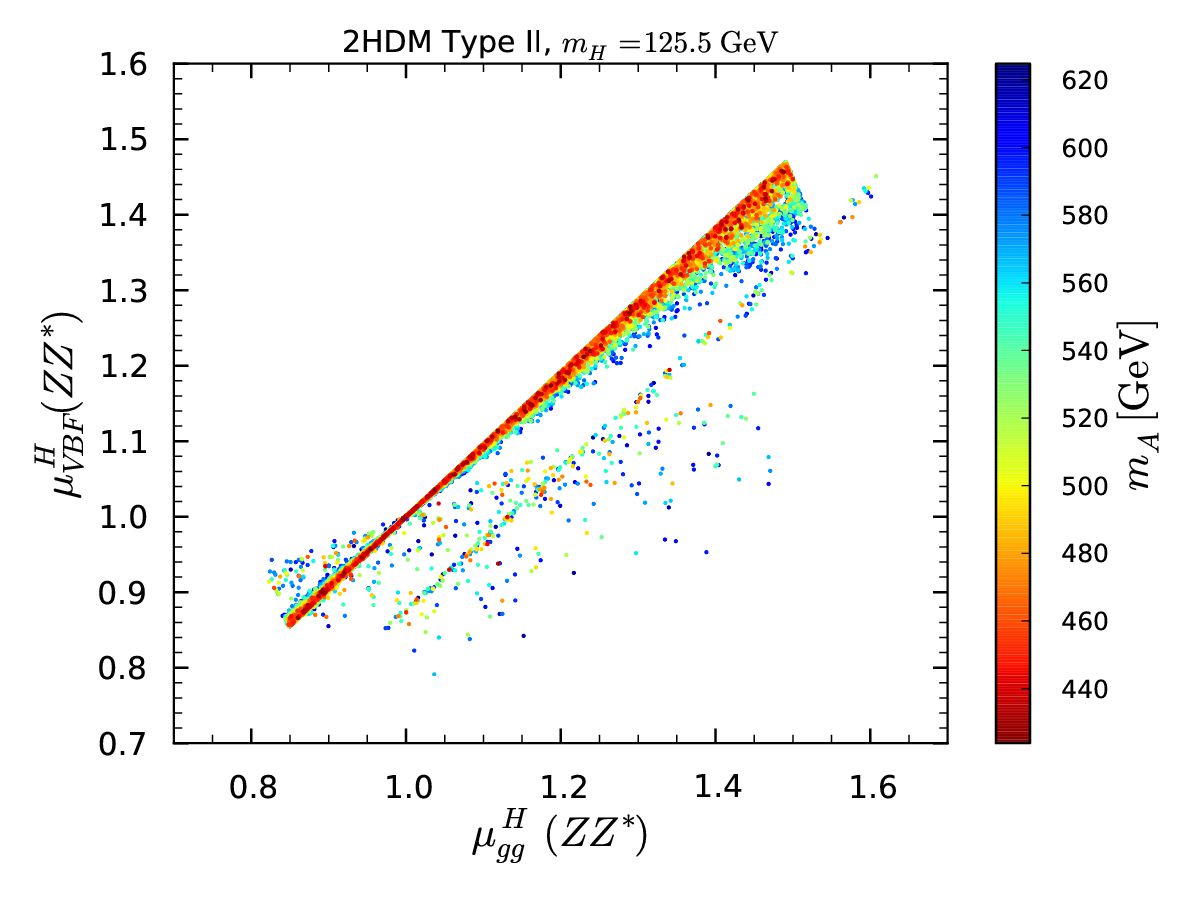}\\
\includegraphics[width=0.33\textwidth]{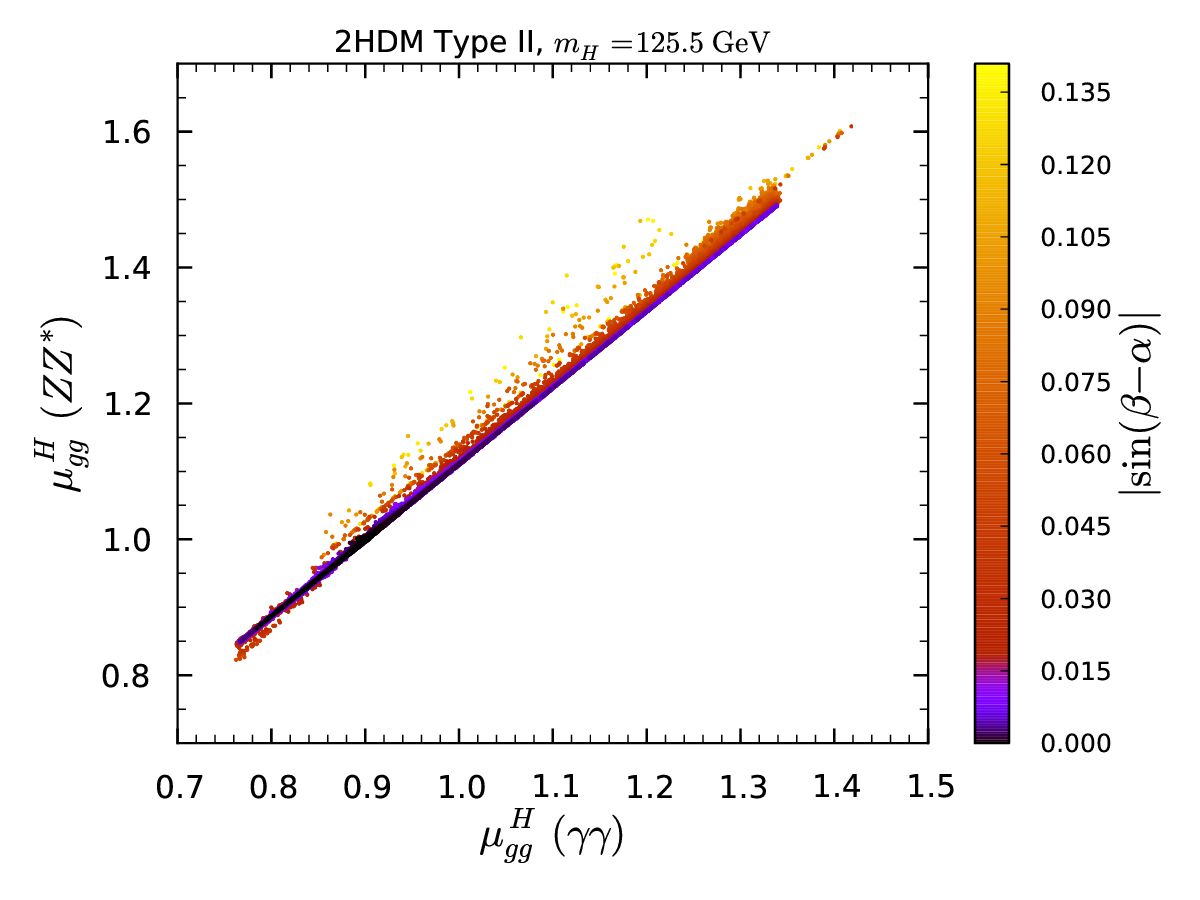}\includegraphics[width=0.33\textwidth]{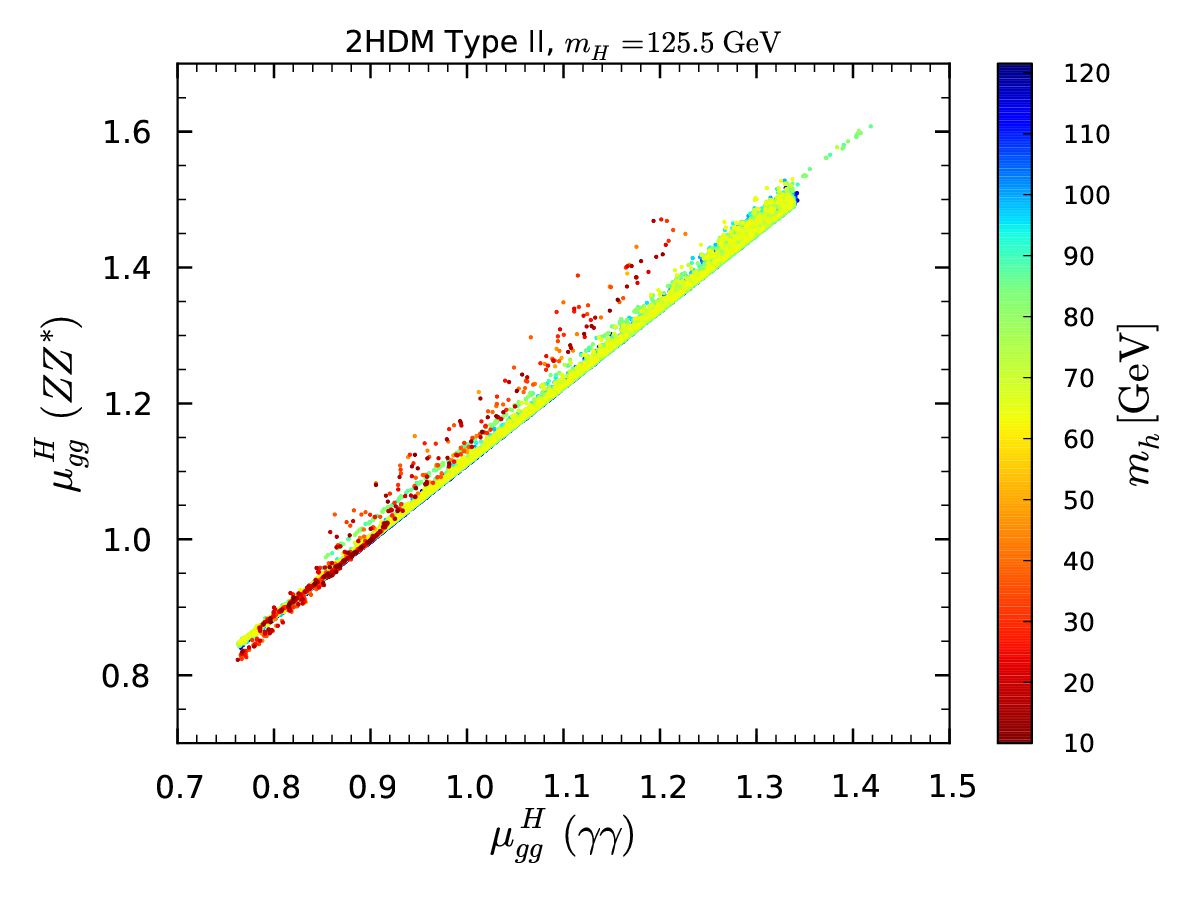}\includegraphics[width=0.33\textwidth]{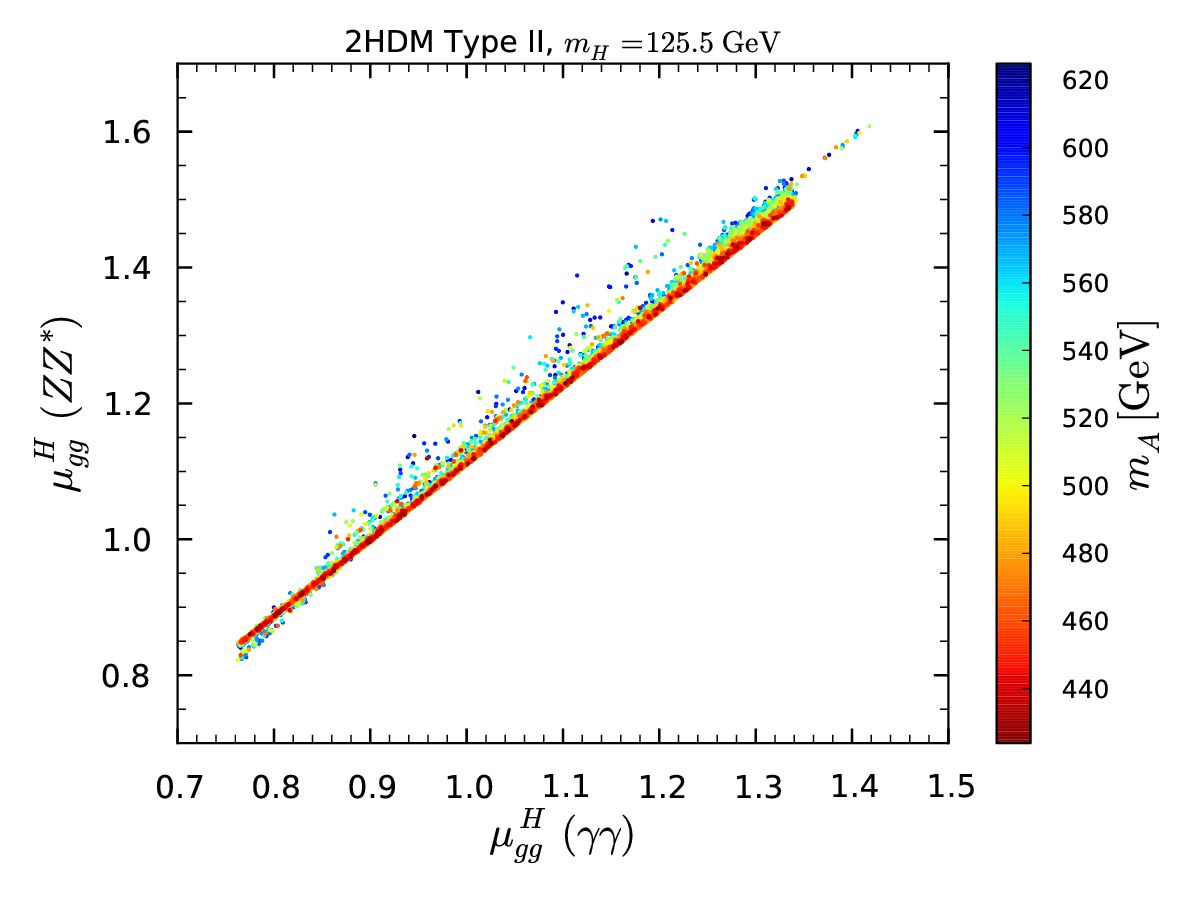}\\\includegraphics[width=0.33\textwidth]{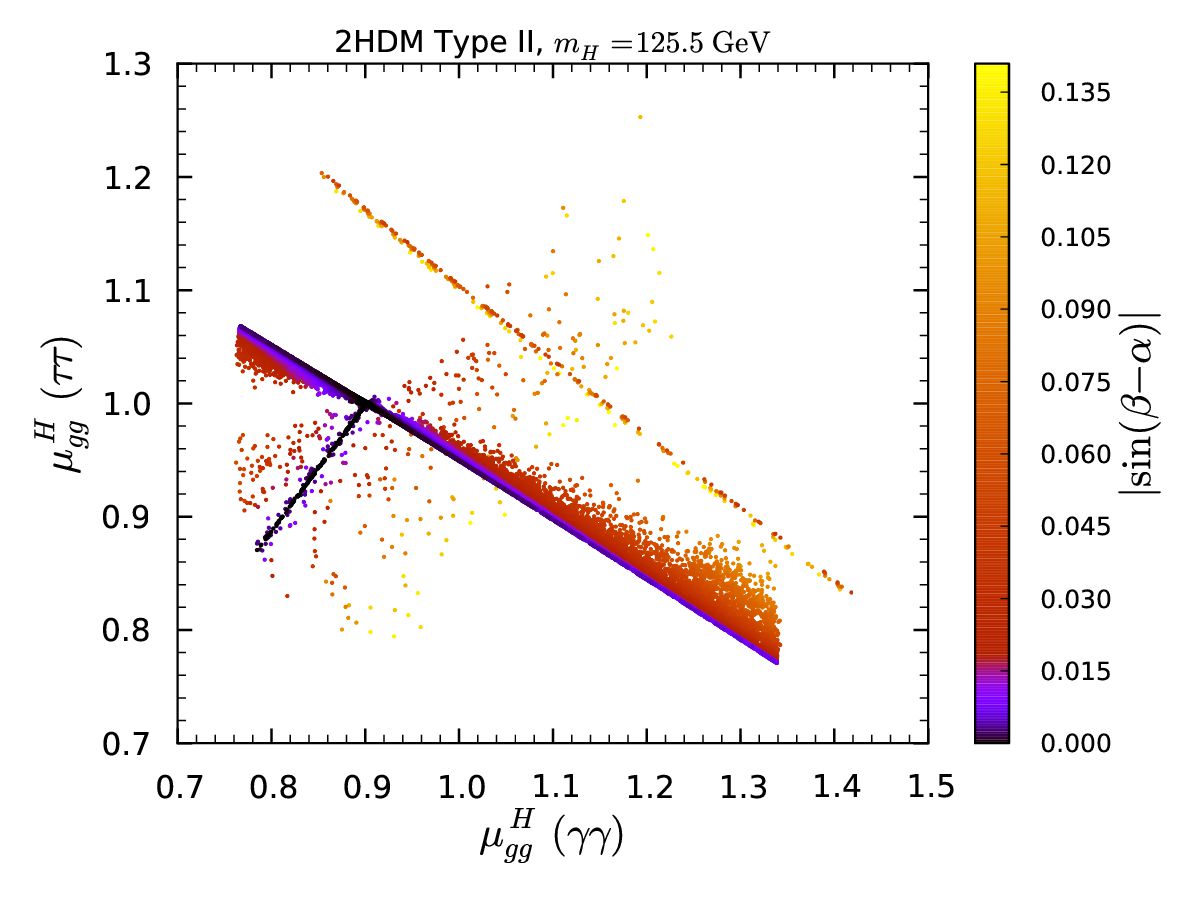}\includegraphics[width=0.33\textwidth]{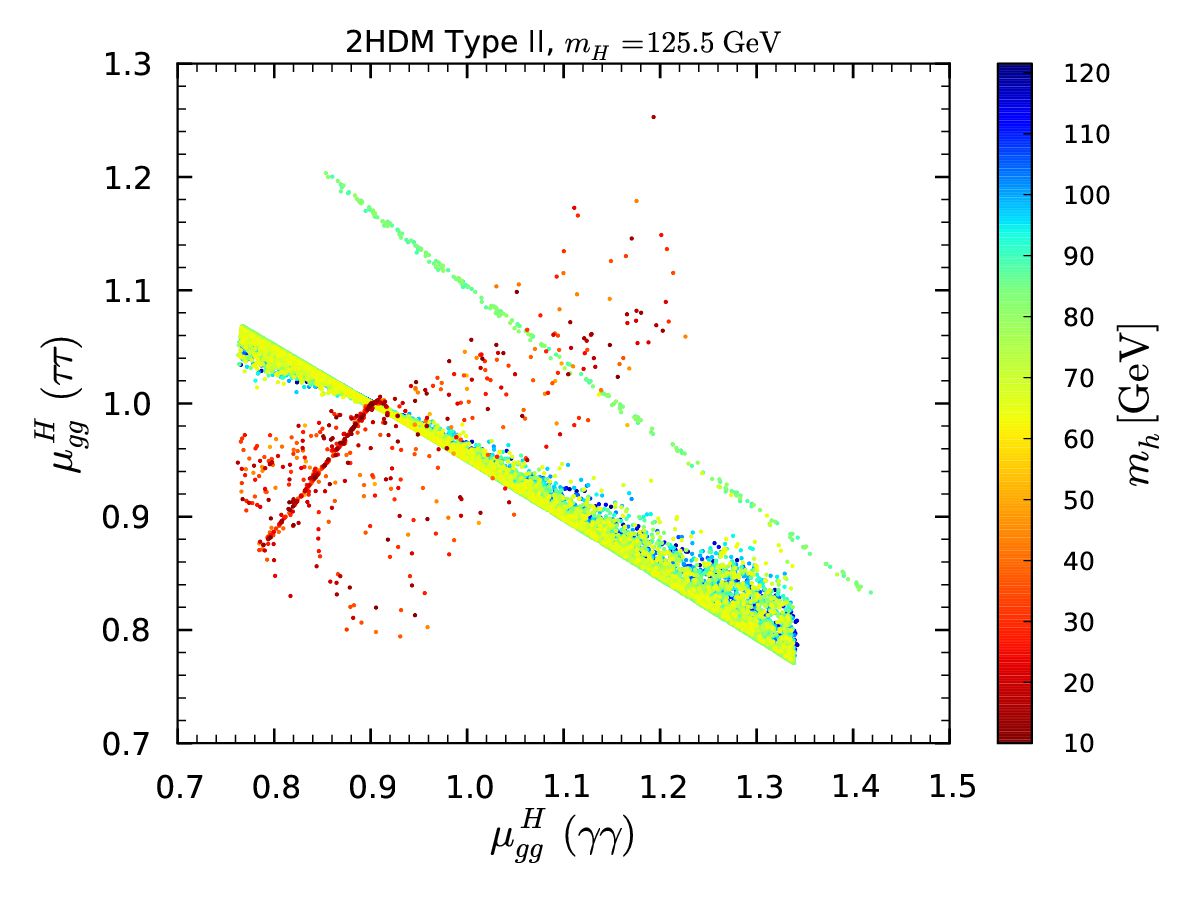}\includegraphics[width=0.33\textwidth]{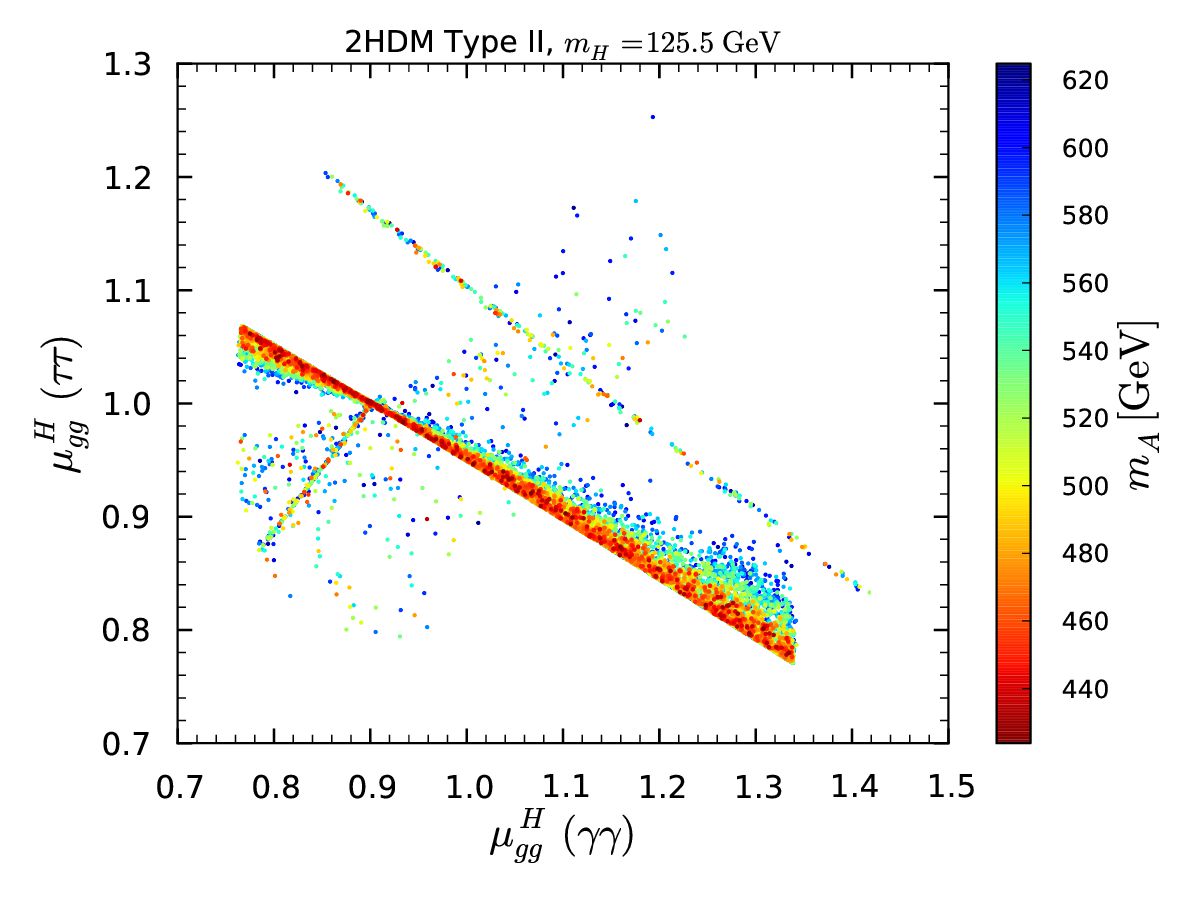}\\
\caption{Examples of correlations between signal strengths in Type~II. The top panels shows $\mu_{\rm VBF}^H(\gamma\gamma)$ vs.\ $\mu_{gg}^H(\gamma\gamma)$, the upper middle panels show $\mu_{\rm VBF}^H(ZZ^*)$ vs.\ $\mu_{gg}^H(ZZ^*)$, 
the lower middle panels show  $\mu_{gg}^H(ZZ^*)$ vs.\ $\mu_{gg}^H(\gamma\gamma)$ and the bottom panels show $\mu_{gg}^H(\tau\tau)$ vs.\ $\mu_{gg}^H(\gamma\gamma)$. The color code indicates, from left to right, the dependence on $|\sbma|$, $m_h$ and $m_A$.  Points are ordered from high to low $|\sbma|$, $m_h$ and $m_A$ values.
Note that the correlations look the same in the first two rows of plots (\ie\ for VBF vs.\ $gg$ production in $\gamma\gamma$ or $ZZ^*$ final state) but the actual $\mu$ values are different. The opposite-sign $C_D^H$ solution is visible as a separate narrow line with: $|\sba|\approx 0.1$ (left-hand panels), $m_h\gsim 65\gev$ (middle panels) and $\mha\in[420,630]\gev$ (right-hand panels).}
\label{correlations2}
\end{figure}

Putting everything together we find quite distinct correlations of signal strengths as shown in Fig.~\ref{correlations1} for Type~I and in Fig.~\ref{correlations2} for Type~II. It is especially noteworthy that even in the deep alignment limit, the signal strengths can significantly differ from the corresponding SM predictions, but the ratio of $gg$ and VBF production is very close to 1. Moreover, certain combinations can only be reached for specific ranges of $m_h$ and/or $m_A$ values. For example, $\mu^H_{gg}(\gamma\gamma)\simeq\mu^H_{\rm VBF}(\gamma\gamma)\simeq0.7$ requires $m_h\gtrsim 60\gev$ in Type~I while it is not reached at all in Type~II. Likewise, a suppression of $\mu^H_{gg}(\gamma\gamma)$ while $\mu^H_{\rm VBF}(\gamma\gamma)\gtrsim 1$ would point towards a somewhat heavy $A$ in Type~I with a slight departure from strict alignment, while again this combination is not possible in Type~II. Another example is the relation between $\mu^H_{gg}(\gamma\gamma)$ and $\mu^H_{gg}(ZZ^*)$.  In the alignment limit in Type~II we expect $\mu^H_{gg}(\gamma\gamma)/\mu^H_{gg}(ZZ^*)\simeq 0.9$, with both enhancement or suppression of the individual $\mu^H_{gg}(\gamma\gamma),\mu^H_{gg}(ZZ^*)$ with respect to the SM being possible. In Type~I, there is a band in which this ratio also applies (for all $\mhl$) and the signals are always suppressed. For values of $m_h\gsim 60$~GeV in \typei,  in the deep (near) alignment limit we find $\mu^H_{gg}(ZZ^*)\sim 1$ ($\in[0.95,1.02]$), while  $\mu^H_{gg}(\gamma\gamma)$ can range from $0.64$ to $0.98$ (1.4). An analogous discussion is possible for $\mu^H_{gg}(\tau\tau)$ versus $\mu^H_{gg}(\gamma\gamma)$.

In general, when $H\to hh$ or $H\to AA$ decays are kinematically allowed, these (so far) unobserved decay modes suppress the $H$ branching ratios into SM final states, thus leading to a simultaneous suppression of all the $\mu_{X}^H(Y)$ even in the deep alignment regime. This is apparent in all the $\mu$ correlations shown in Figs.~\ref{correlations1} and \ref{correlations2}, but is most notably visible as the upward-sloping diagonal lines of points in the $\mu_{gg}^H(ZZ^*)$ vs.\ $\mu_{gg}^H(\gamma\gamma)$ and $\mu_{gg}^H(\tau\tau)$ vs.\ $\mu_{gg}^H(\gamma\gamma)$ plots for Type~I. 
(However, note that in Type~II due to the non-universal nature of the Yukawa couplings the signal strengths can also  be larger than 1 when the $H \to hh$ decay mode is open.) 

Comparing these results with the corresponding results of \cite{Bernon:2015qea}, it seems very difficult to
distinguish $m_h\simeq125\gev$ from $m_H\simeq125\gev$ with signal strength measurements and coupling fits alone. One possibility for such a distinction might be that the measured values point towards Type~I or Type~II but are excluded by $A\to Zh$ in the case of $m_H\simeq125\gev$ for a particular model type.
Such a result would obviously favor the $m_h\simeq125\gev$ scenario.

\clearpage
\subsection{Cross sections for $h$ and $A$ production}\label{cross-sections}

\begin{figure}[b!]\centering
\includegraphics[width=0.51\textwidth]{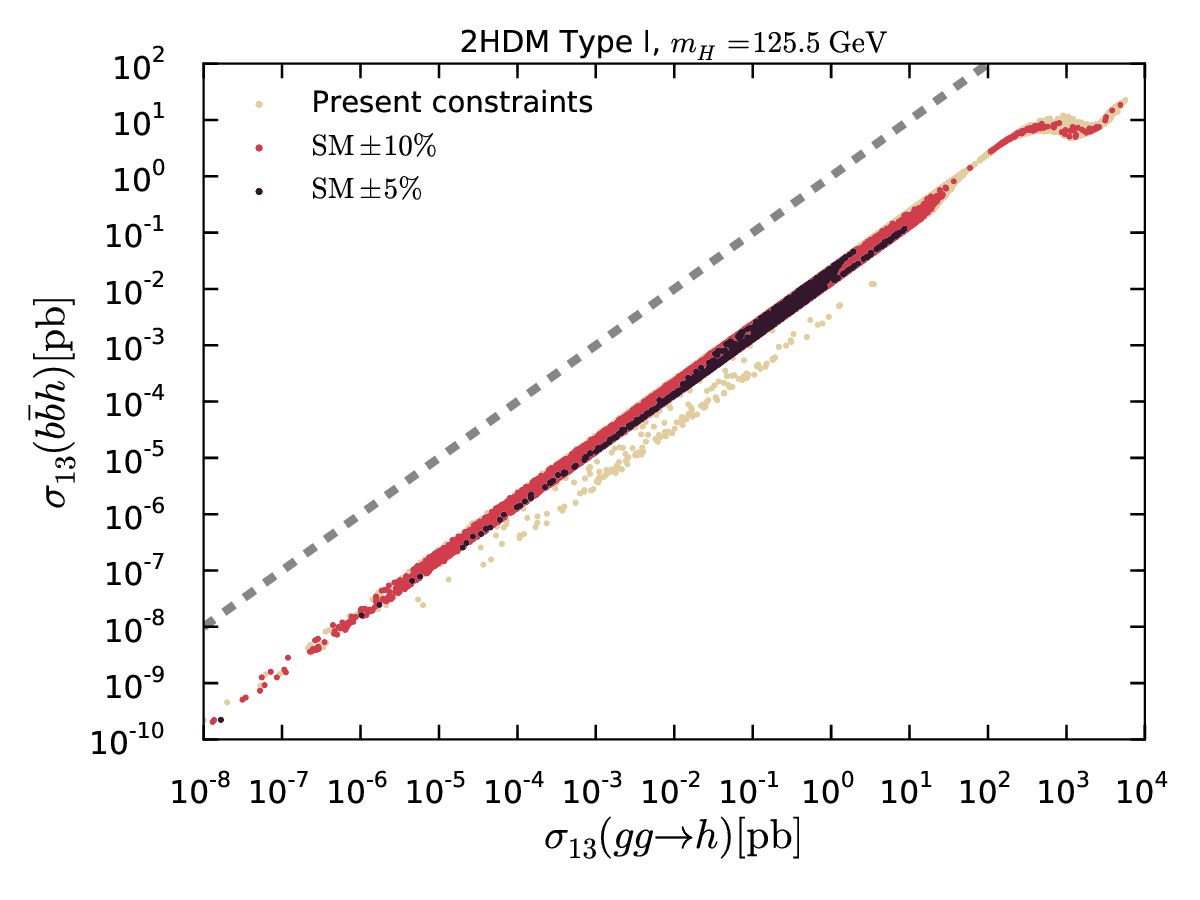}\includegraphics[width=0.51\textwidth]{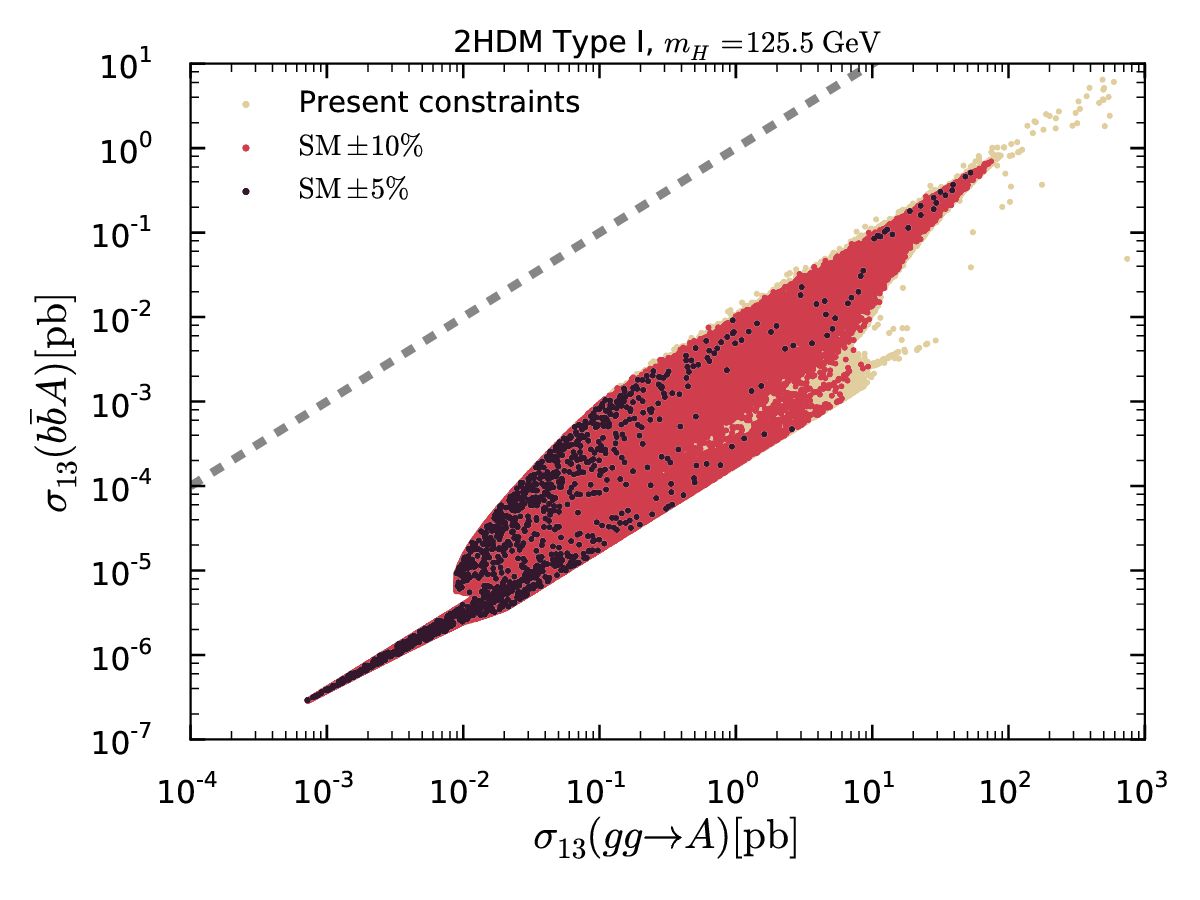}
  \caption{$\sigma(b\bar{b}X)$ versus $\sigma(gg\to X)$ for $X=h$ (left) and $X=A$ (right) in Type~I at the 13~TeV LHC for points satisfying all present constraints (in beige) as well as  points for which the signals strengths from Eq.~\eqref{mu5} are within $5\%$ and $2\%$ of the SM predictions (in red and dark red, respectively). The dashed lines indicate $\sigma_{13}(b\bar{b}X)=\sigma_{13}(gg\to X)$.}
  \label{xsec_correlation_I_13}
\end{figure}

\begin{figure}[t!]\centering
\includegraphics[width=0.51\textwidth]{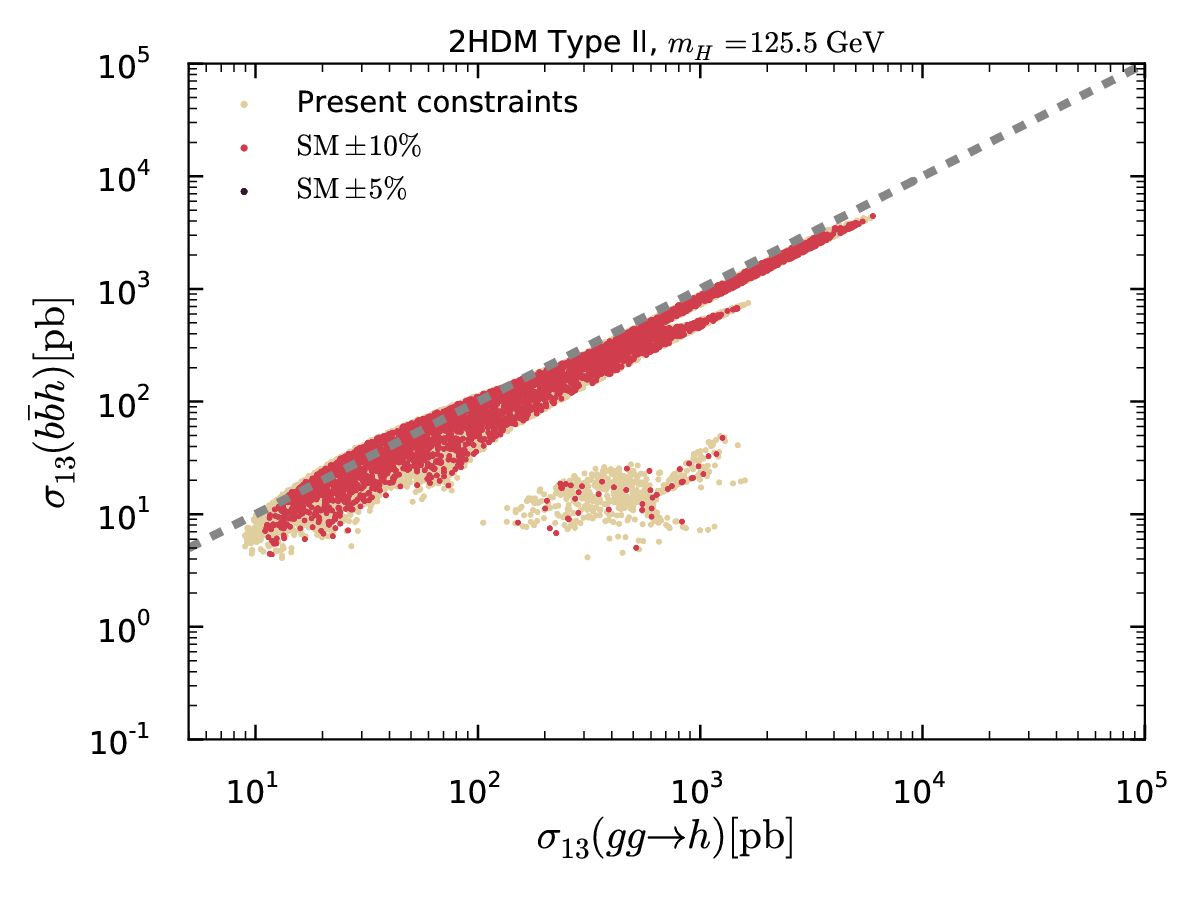}\includegraphics[width=0.51\textwidth]{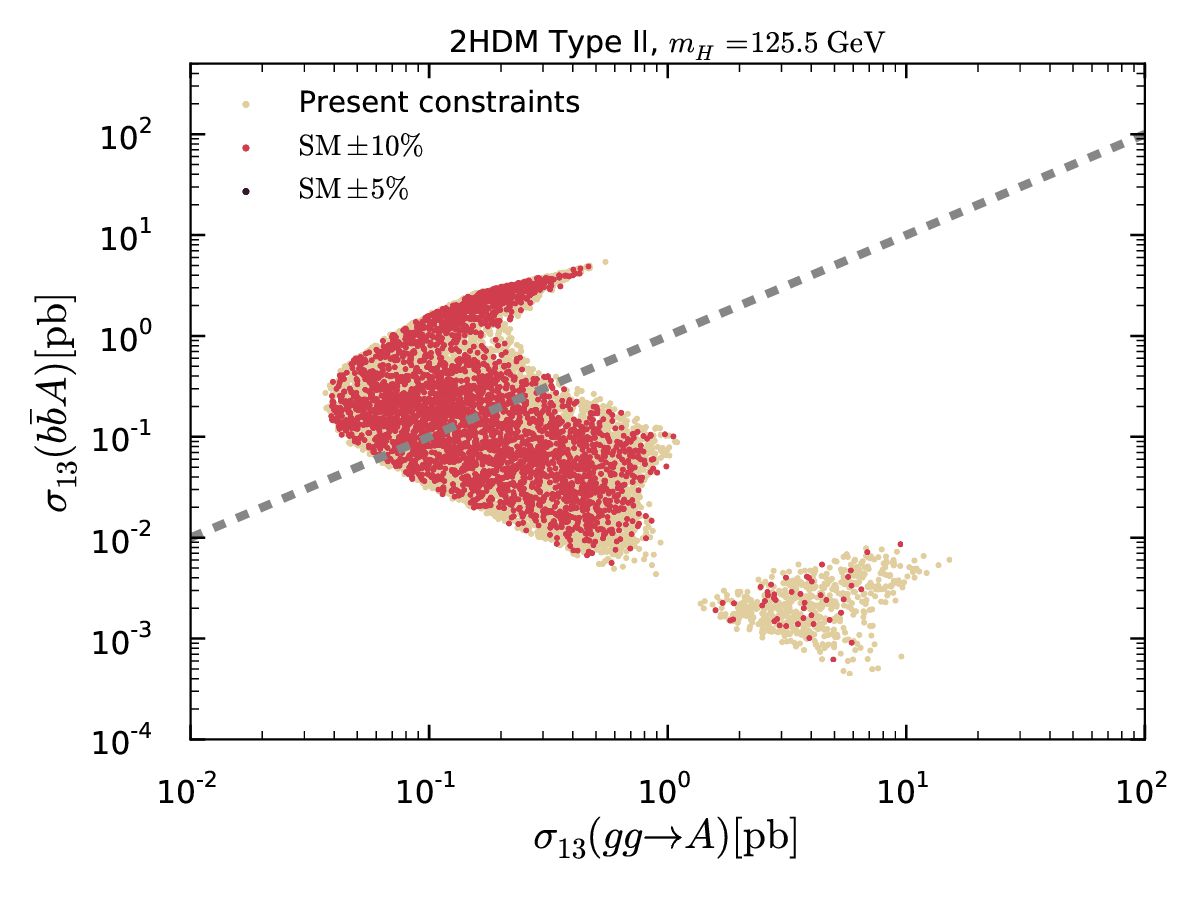}
\caption{As in Fig.~\ref{xsec_correlation_I_13} but for Type~II. }
  \label{xsec_correlation_II_13}
\end{figure}

Let us now turn to the prospects for discovering the additional neutral scalar states. 
The two largest production modes at the LHC are gluon fusion, $gg\to X$, and the associated production with a pair of $b$-quarks, $b\bar{b}X$, with $X=h,A$. 
The correlations of the $gg\to X$ and $b\bar{b}X$ cross sections at the 13 TeV LHC are shown in Fig.~\ref{xsec_correlation_I_13} for the Type~I model and in Fig.~\ref{xsec_correlation_II_13} for the Type~II model. 
We show the points that pass all present constraints (in beige) and highlight those that have a very SM-like 125~GeV Higgs state by constraining all the following signal strengths to be within $5\%$ or $2\%$ of their SM values, respectively, denoted as SM$\pm 5\%$ (red) and SM$\pm 2\%$ (dark red): 
\beq
\label{mu5}
   \mu_{gg}^H(\gamma\gamma),\ \mu_{gg}^H(ZZ^*),\ \mu_{gg}^H(\tau\tau),\ \mu_{\rm VBF}^H(\gamma\gamma),\ \mu_{\rm VBF}^H(ZZ^*),\ \mu_{\rm VBF}^H(\tau\tau),\ \mu_{VH}^H(b\bar{b}),\ \mu_{t\bar{t}}^H(b\bar{b}) \,.
\eeq


Regarding the production of $h$ and $A$ in Type~I, shown in Fig.~\ref{xsec_correlation_I_13},
there is a strong correlation between the two production modes, gluon fusion and $b\bar{b}$ associated production, which stems from the fact that the relevant couplings are the same (up to a sign in the case of the $A$): $C_U^h=C_D^h=\cos\alpha/\sin\beta$ and $C_U^A=-C_D^A=\cot\beta$, respectively. 
The larger spread in $\sigma(b\bar{b}A)$ observed for $\sigma(gg\to A)>10^{-2}$~pb comes from the fact that for $m_A\lesssim 400\gev$ the $b\bar{b}A$ cross section grows faster with decreasing $m_A$ than that of  $gg\to A$. Therefore, along a line of fixed $\sigma(gg\to A)$ in the plot, a point with higher $\sigma(b\bar{b}A)$ has a smaller $m_A$. Note also that there is an interference of the top and bottom loop diagrams in $gg\to A$ which changes sign depending on the value of $m_A$. 
However, $\sigma(gg\to A)$ is always at least two orders of magnitude larger than $\sigma(b\bar{b}A)$.

Turning to Type~II scenarios,  we observe that in the case of the $A$ either the $b\bar b A$ or the $gg\to A$ cross section can be dominant due to the fact that the $Ab\bar b$ and $At\bar t$ couplings have different $\tanb$ dependence, the former being proportional to $\tanb$ and the latter to $\cotb$. Also, as already noted in \cite{Dumont:2014wha}, the $\mhh\simeq 125\gev$  scenarios can be either eliminated or confirmed when the LHC measurements reach a precision such that the rates in the various initial$\times$final state channels can be measured to 5\% accuracy.  This is because for this scenario the charged Higgs boson loop contribution to the $H \gam\gam$ coupling results in at least a 5\% reduction in the $H\to\gam\gam$ coupling.

\begin{figure}[t!]\centering
\includegraphics[width=0.51\textwidth]{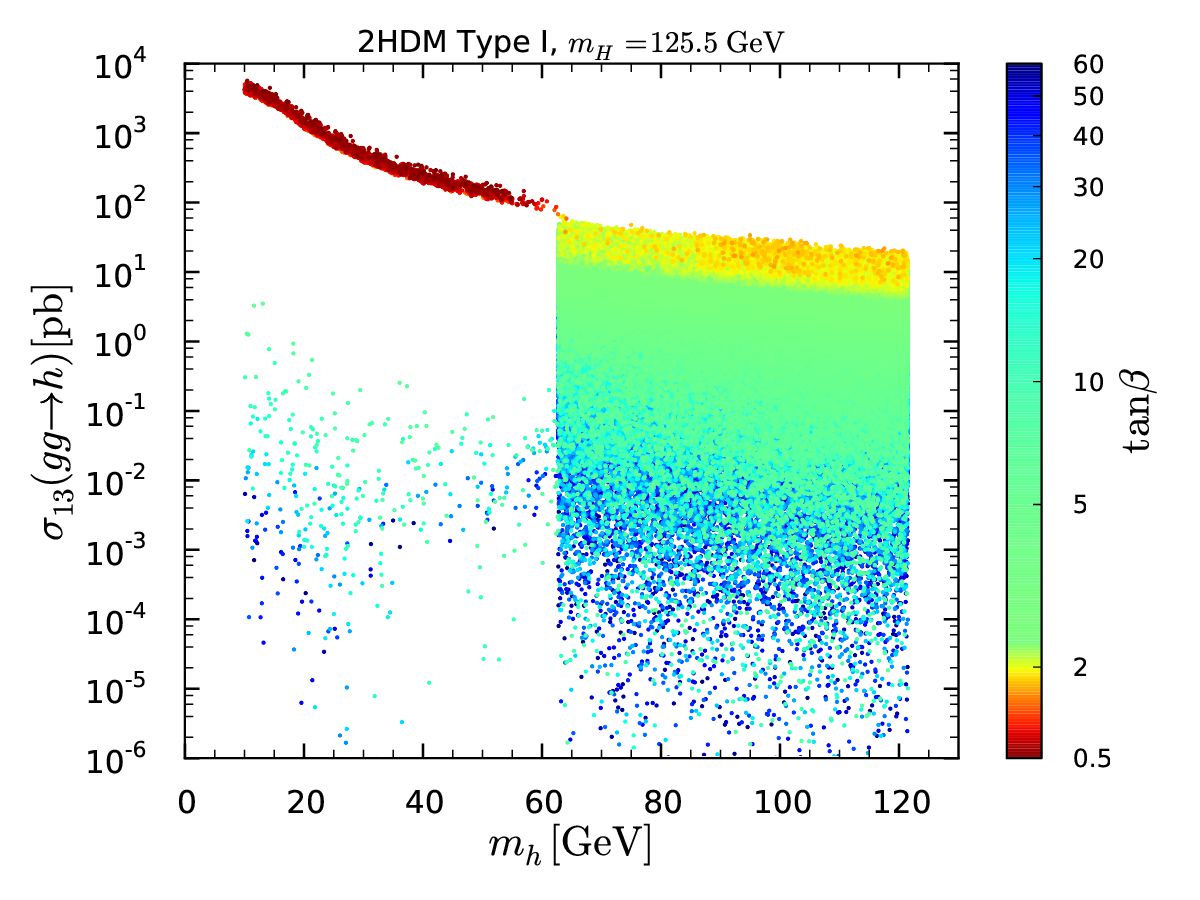}\includegraphics[width=0.51\textwidth]{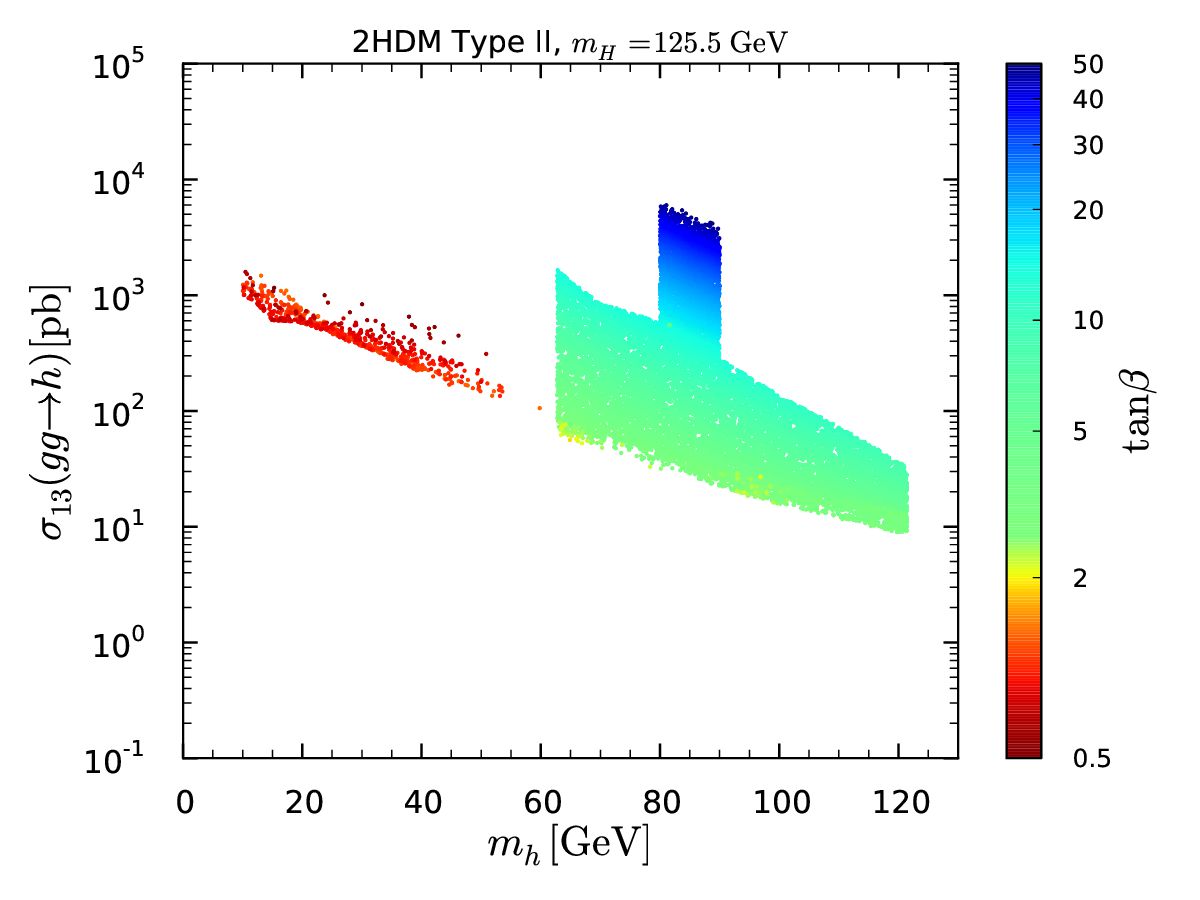}
\includegraphics[width=0.51\textwidth]{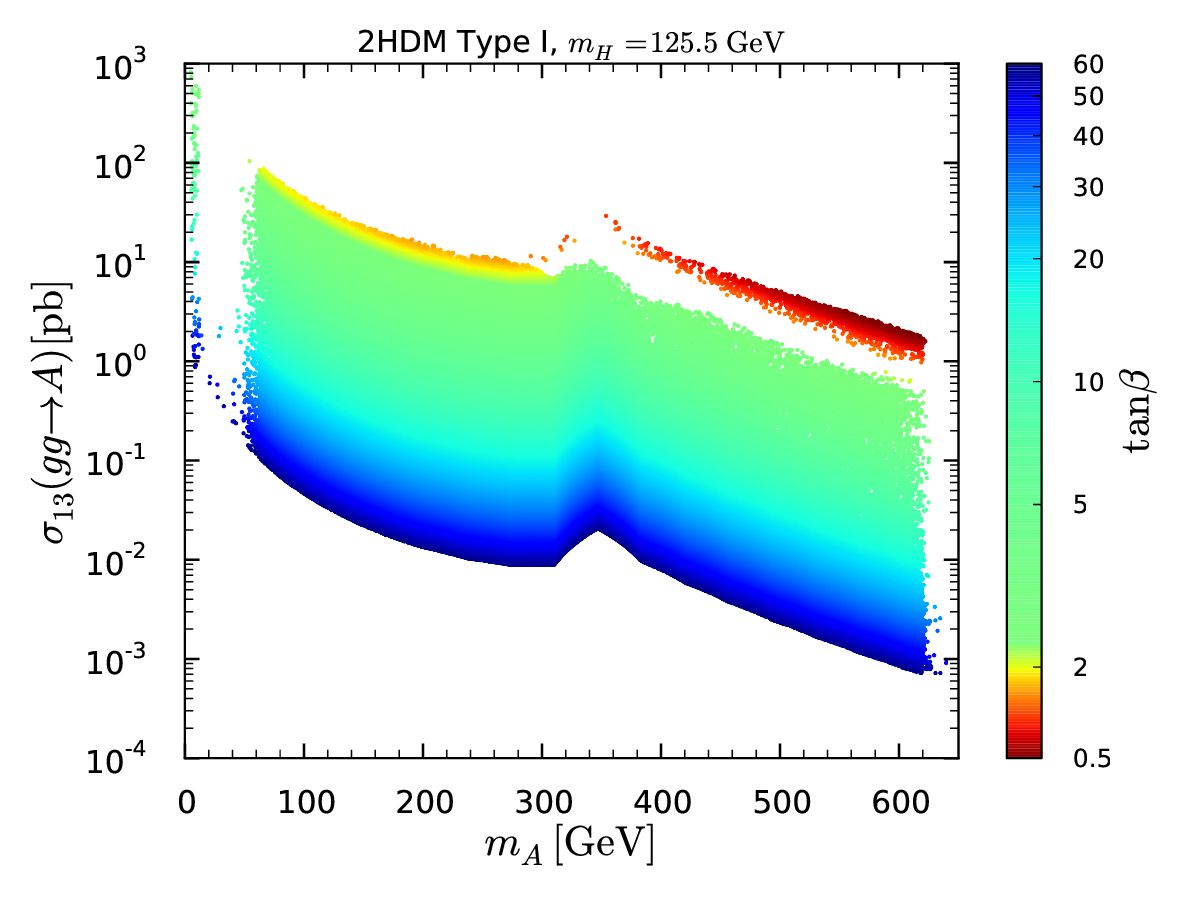}\includegraphics[width=0.51\textwidth]{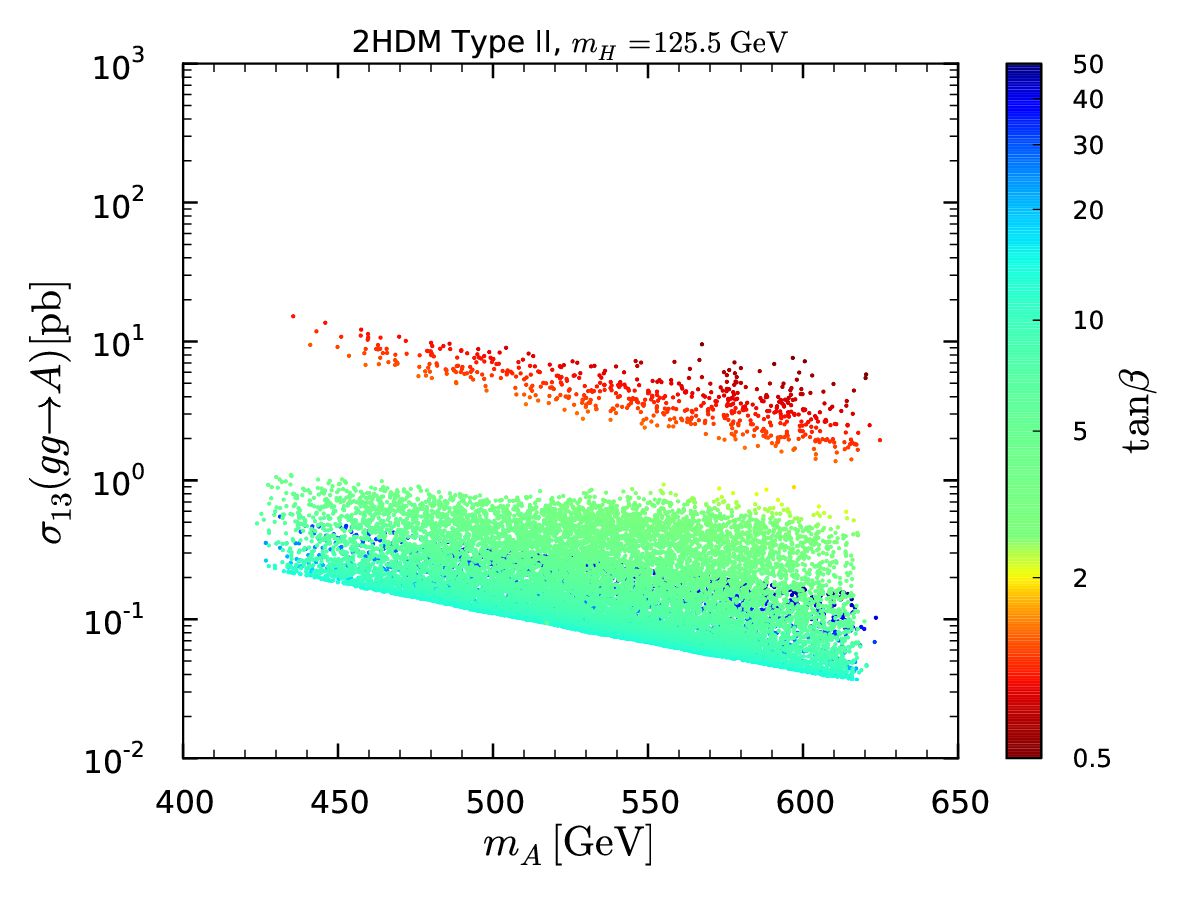}
\includegraphics[width=0.51\textwidth]{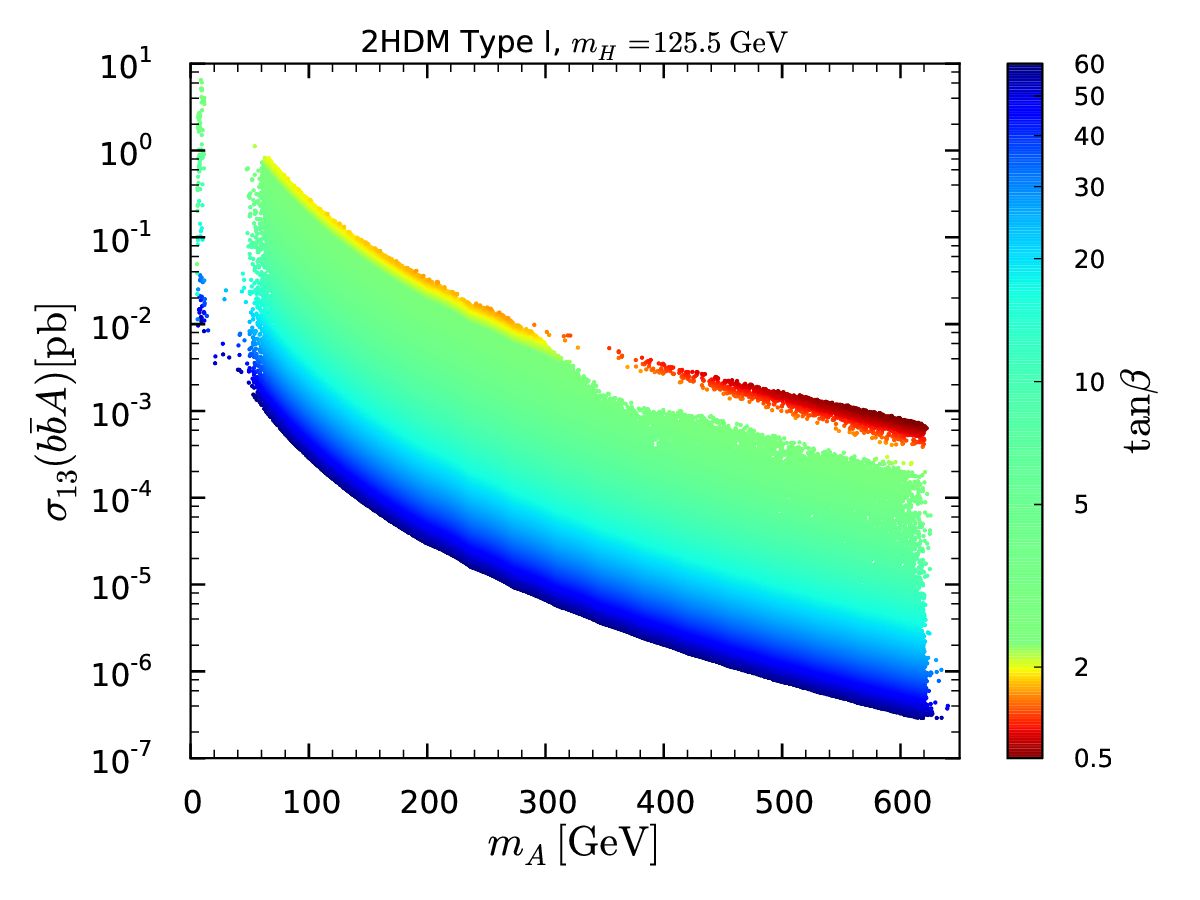}\includegraphics[width=0.51\textwidth]{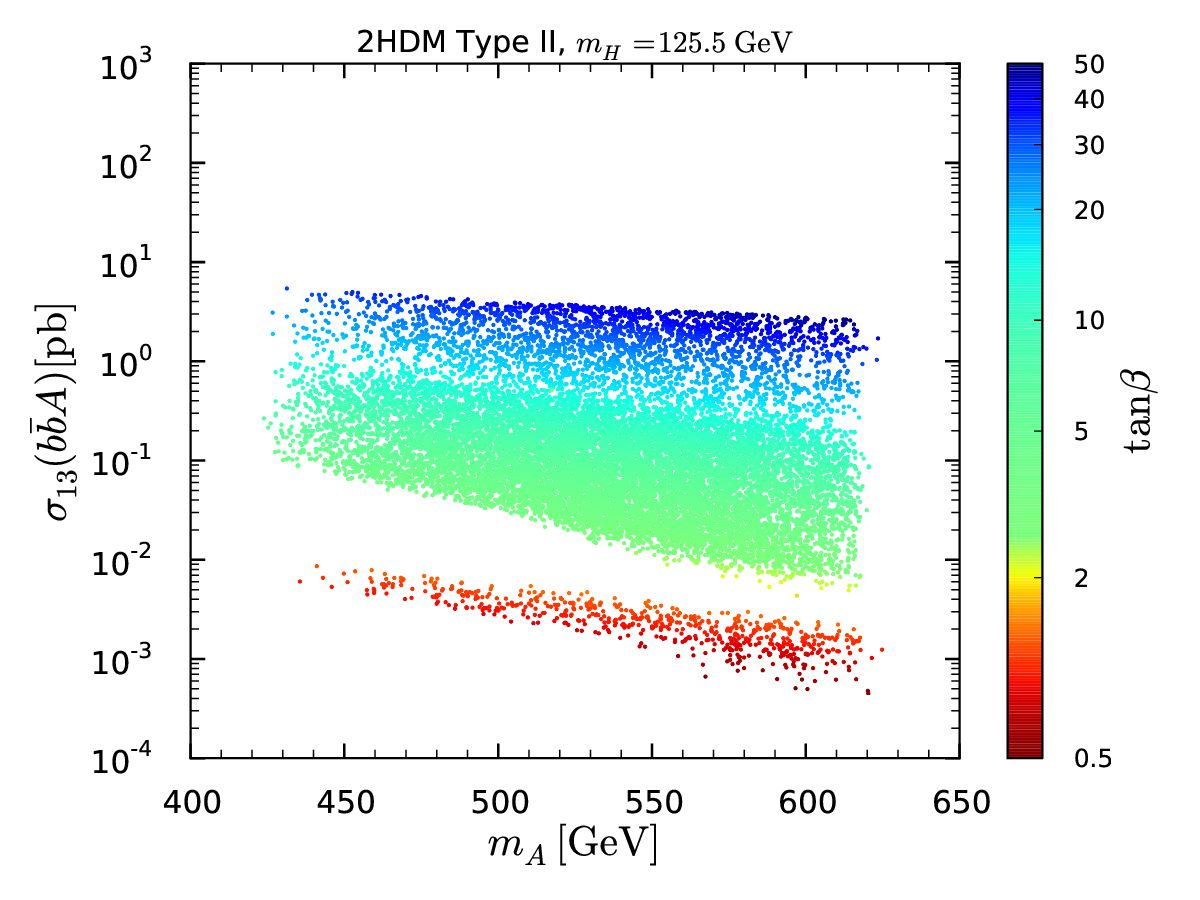}
\caption{Cross sections in Type~I (left) and Type~II (right) for $gg\to X$ as functions of $m_X$ for $X=h$ (upper panels) and $X=A$ (middle panels) with $\tan\beta$ color code. The bottom panels depict the  cross sections for $b\bar bA$ production as function of $m_A$ with $\tan\beta$ color code.  In all six plots, points are ordered from high to low $\tan\beta$.}
  \label{xsec13}
\end{figure}

Before considering specific decay channels of $h$ and $A$, we present in Fig.~\ref{xsec13} the gluon-fusion cross sections, as well as the $b\bar b A$ cross section, as functions of $m_h$ and $m_A$ in Type~I and Type~II at the 13 TeV LHC. Here, the color code shows the dependence on $\tan\beta$.
In Type~I, the $gg\to A$ cross section is proportional to $\cot^2\beta$; this explains why it is larger (smaller) at lower (higher) $\tan\beta$. A cross section of 1~fb is guaranteed for $m_A$ up to the maximum possible mass of $\sim 600$~GeV. At very small $m_A$ and low $\tanb$ it can reach 100~pb.   On the other hand, the
$gg\to h$ cross section in Type~I is proportional to $(C_F^h)^2$ and can take on extremely small values. The reason is that $\sba$ can take either sign and the values allowed in our scan are high enough such that a cancellation between the two terms of $C_F^h=\sba+\cbma\cotb$ can occur and leads to an almost vanishing coupling.  
For $m_h<\half m_H$, this cancellation is also possible for fine-tuned points at large $\tanb$, but for most points 
$\tanb$ is small and the cross section can be as large as $5\times 10^3$~pb at $m_h\sim 10$ GeV.
In Type~II, any phenomenologically viable mass in the range $420$--$630$~GeV gives a $gg\to A$ cross section larger than 30~fb with values as large as 10~pb possible for $\tanb\lesssim 1$.  Moreover, note that the $\tanb$ dependence is opposite for $b\bar bA$ compared to $gg\to A$ so that one or the other cross section is always large, as illustrated in the right panel of Fig.~\ref{xsec_correlation_II_13}. 
As for $h$ in Type~II, the smallest $gg\to h$ cross section is of order 8~pb at $m_h\simeq120$~GeV, with values as large as $4-6\times 10^3$~pb for  $m_h\sim 80-90$~GeV at very large $\tanb$. As previously mentioned, large values of $\tanb$ are excluded for $m_h<80\gev$ and $m_h>90\gev$ because of the severe constraints from $h\to\tau\tau$ decays. 
For values of $m_h<\half\mhh$, for which $\tanb$ must be small, the $gg\to h$ cross section takes a minimum value of $\sim 100$~pb reaching $2\times 10^3$~pb at $m_h\sim 10$~GeV.  To summarize, the prospects for observing the $h$ are good in Type~II, while the prospects for discovering the $A$ look promising in both Type~I and Type~II.  

\begin{figure}[t!]\centering
\includegraphics[width=0.51\textwidth]{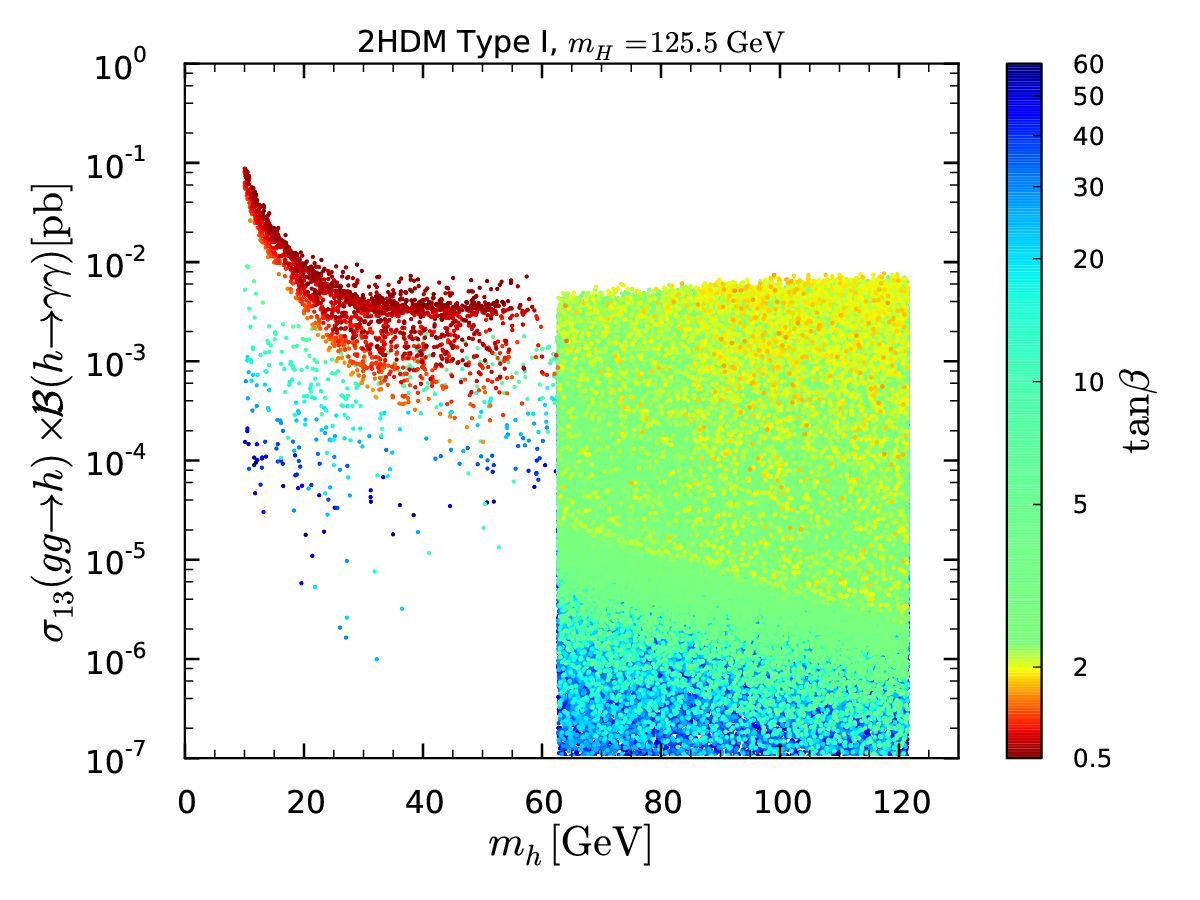}\includegraphics[width=0.51\textwidth]{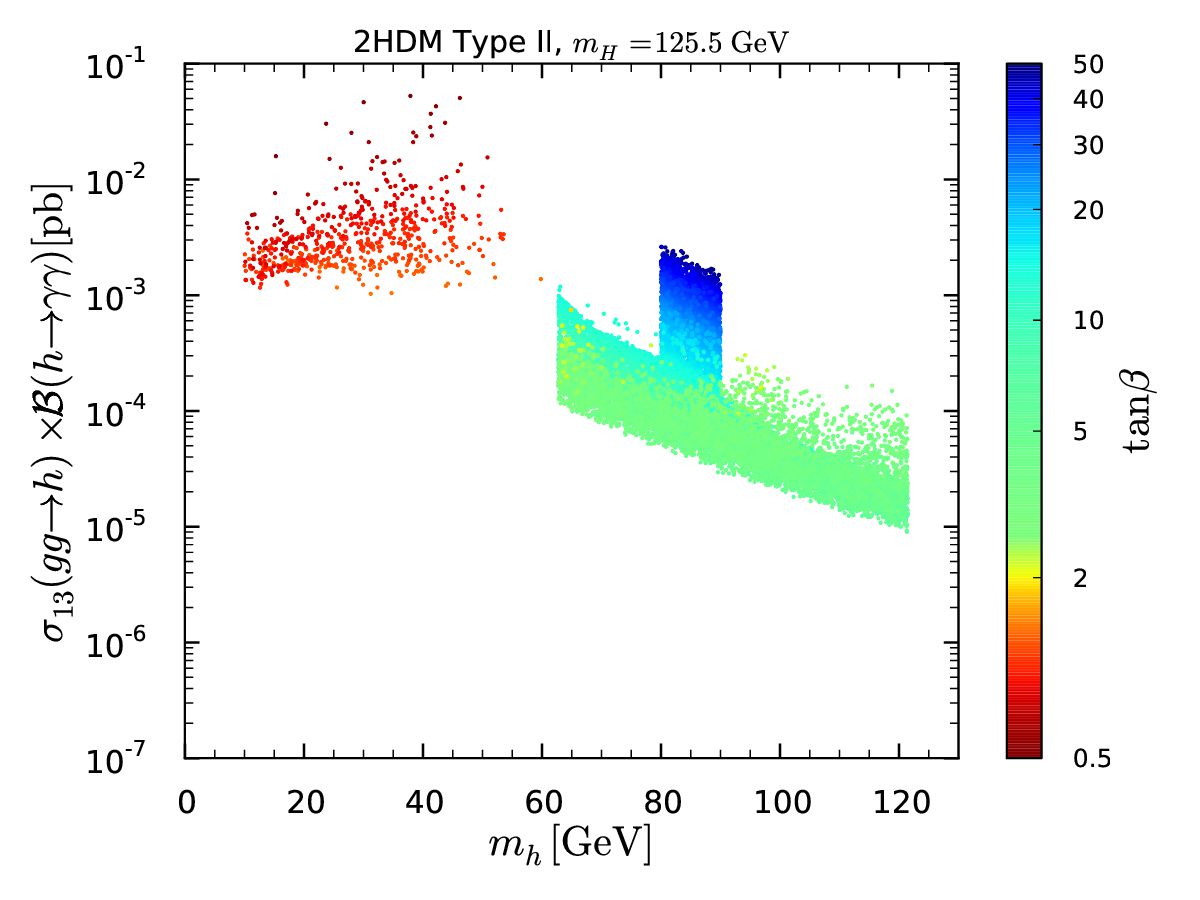}\\
\includegraphics[width=0.51\textwidth]{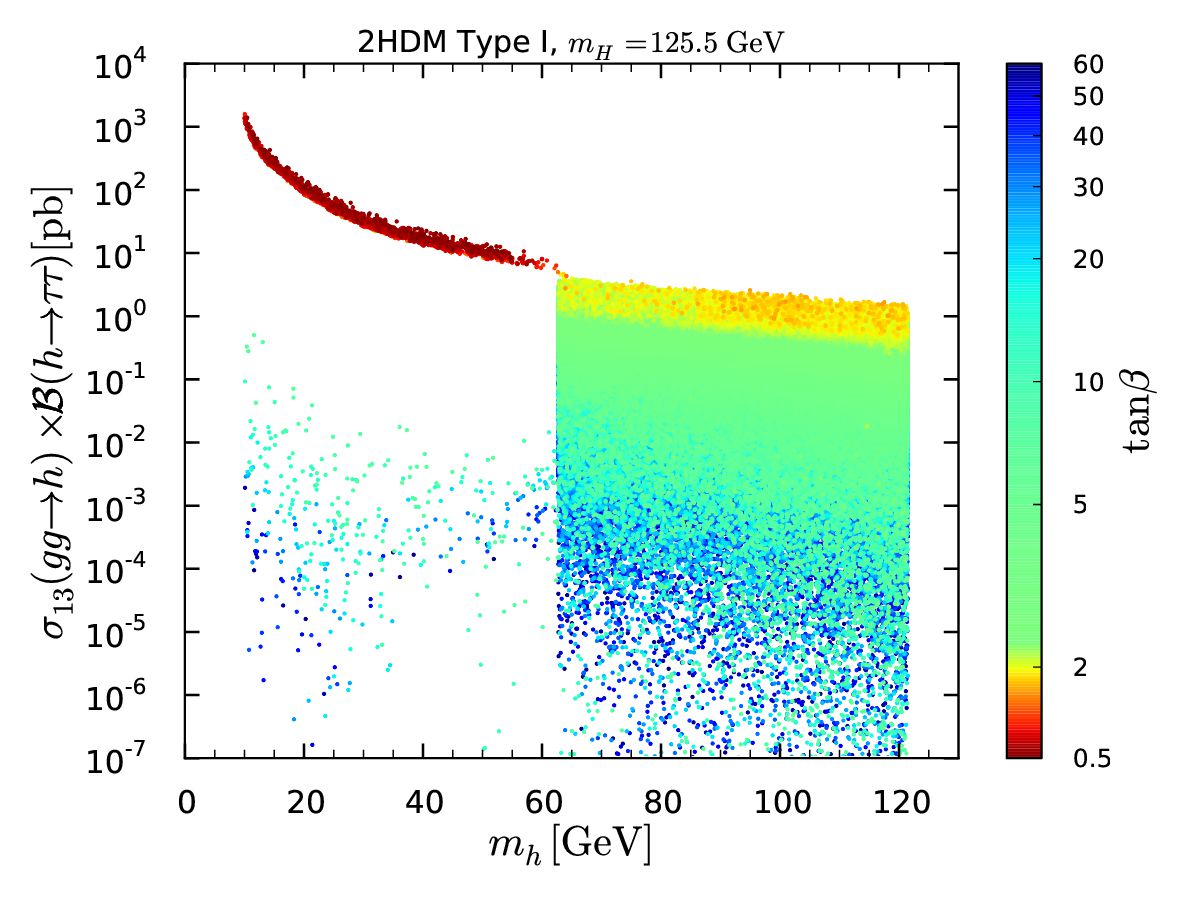}\includegraphics[width=0.51\textwidth]{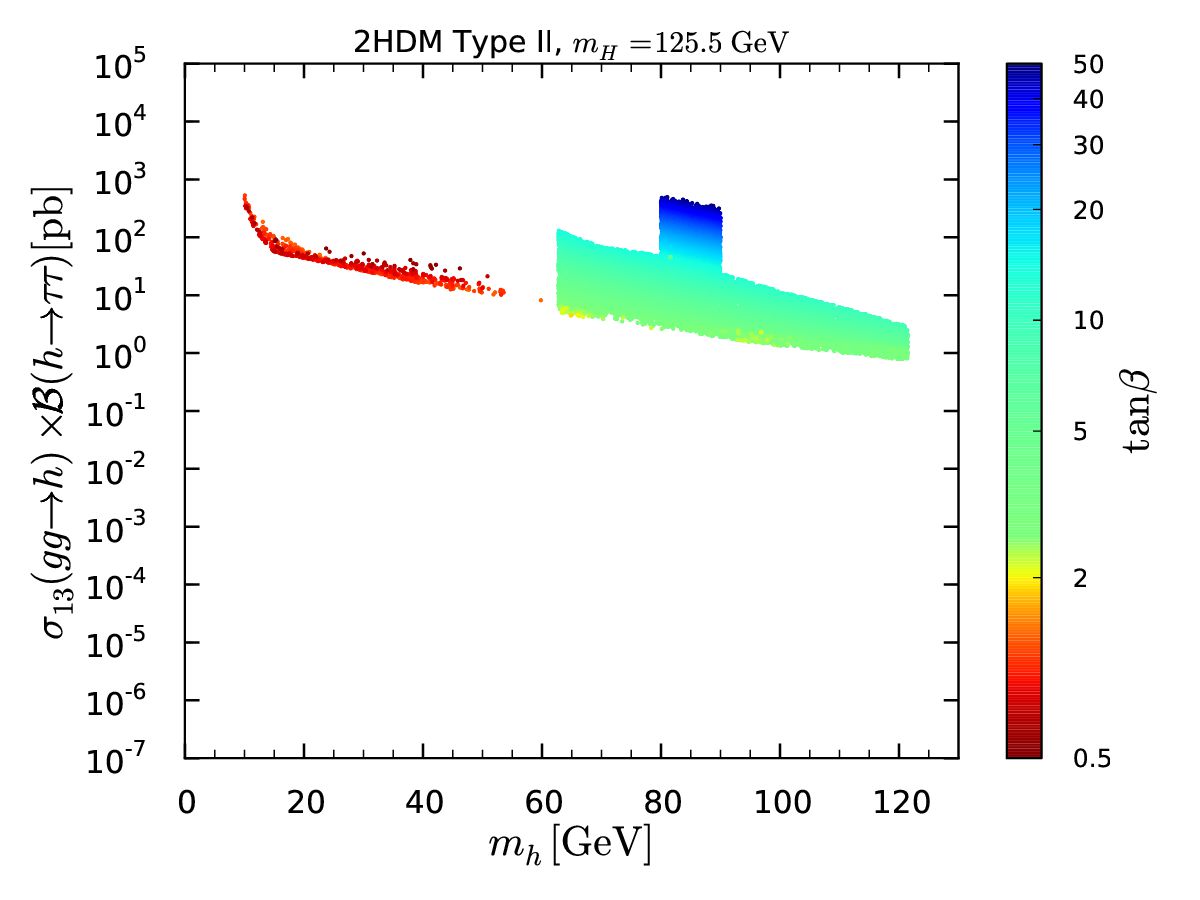}
\caption{Cross sections times branching ratios in Type~I (left) and in Type~II (right) for $gg\to h\to Y$ at the 13~TeV LHC as functions of $m_h$ for $Y=\gamma\gamma$ (upper panels) and $Y=\tau\tau$ (lower panels) with $\tan\beta$ color code. Points are ordered from high to low $\tan\beta$.}
\label{xsecBRh13} 
\end{figure}

Let us now turn to specific signatures.\footnote{To avoid a proliferation of plots, we focus here primarily on the results for gluon fusion; all corresponding results for the $b\bar{b}$ cross section can be provided upon request.}
The cross sections for $gg\to h\to Y$ with $Y=\gamma\gamma$ and $\tau\tau$ in Types~I and II are exhibited in Figure~\ref{xsecBRh13}. Note that the $y$-axis is cut off at $10^{-7}$~pb.  Although much lower values of the cross section are possible in some cases, we do not show these lower values since they will certainly not be observable at the LHC.
In Type~II, $gg\to h \to \gam\gam$ cross sections of  at least 1~fb are guaranteed  if $\mhl<\half \mhh$. In this same region one finds that $\sigma(gg\to h\to \gam\gam)>0.1$~fb in Type~I if $\tanb<2$. For $\mhl\gsim\half\mhh$, cross sections in the $\gam\gam$ final state can reach 10~fb (3~fb) in Type~I (Type~II),  though they can also be much lower, especially in Type~I.  
The behavior of $\sigma(gg\to h \to \tau\tau)$ is similar with cross sections above 1~pb over the full $m_h$ range in Type~II, and also in Type~I if $\tanb$ is small enough. 
Existing limits on $\sigma(gg\to \hl\to \gam\gam)$ at 8~TeV from CMS  for $m_h=80$--$110\gev$ are roughly $0.05$--$0.1\pb$ at 68\%~CL ~\cite{CMS-PAS-HIG-14-037}. 
Thus, we expect that future Run~2 data will eventually provide a sensitive probe in this channel, which
will be particularly interesting if the analyses can be extended to cover the whole 
$m_h$ range down to about 10~GeV.

\begin{figure}[t!]\centering
\includegraphics[width=0.51\textwidth]{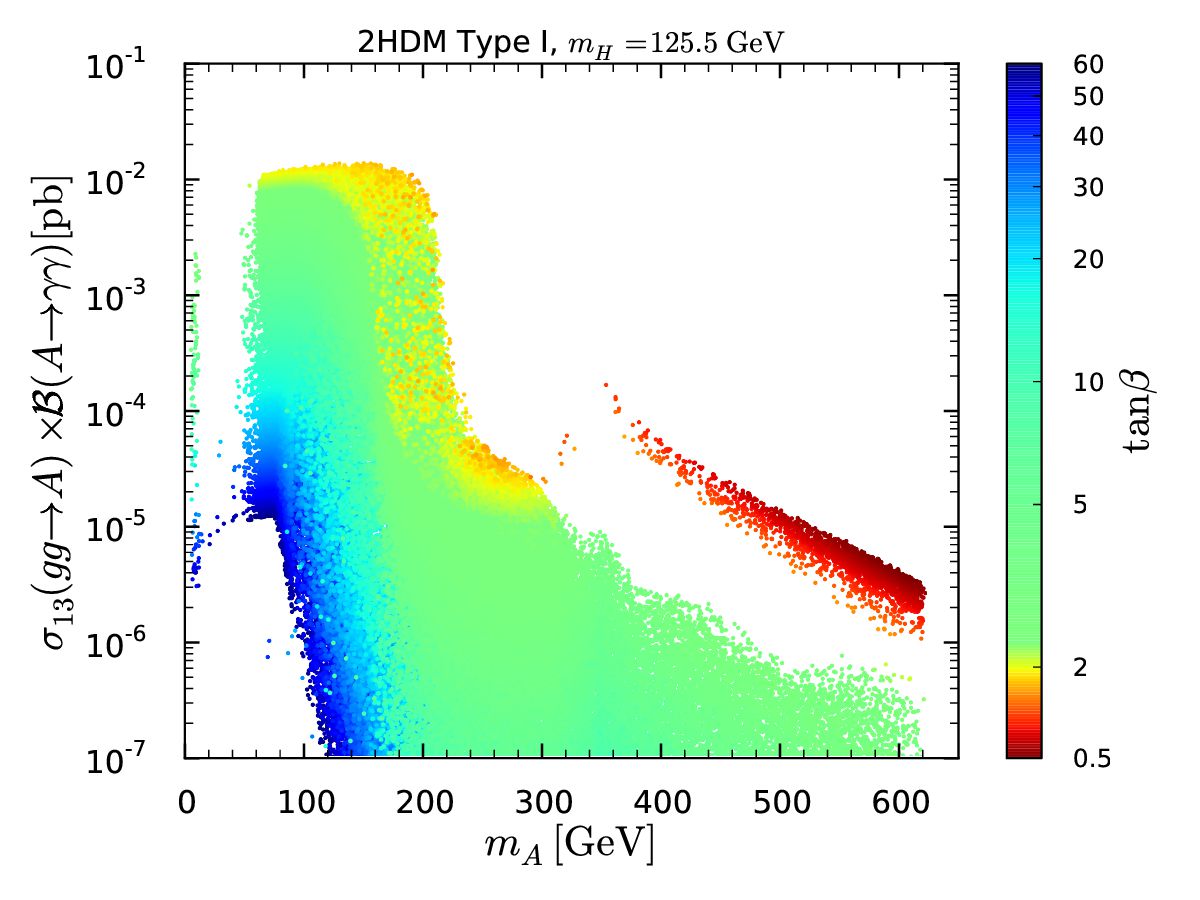}\includegraphics[width=0.51\textwidth]{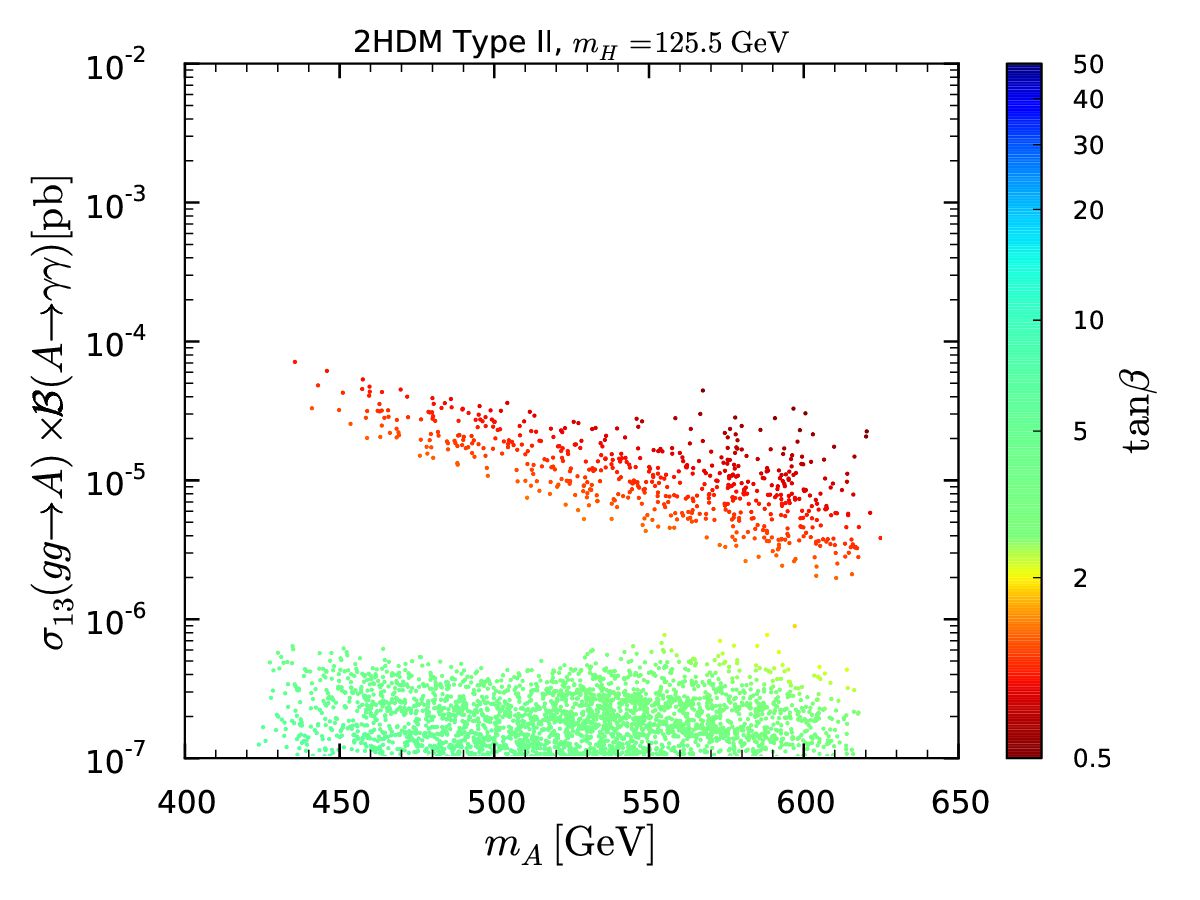}\\
\includegraphics[width=0.51\textwidth]{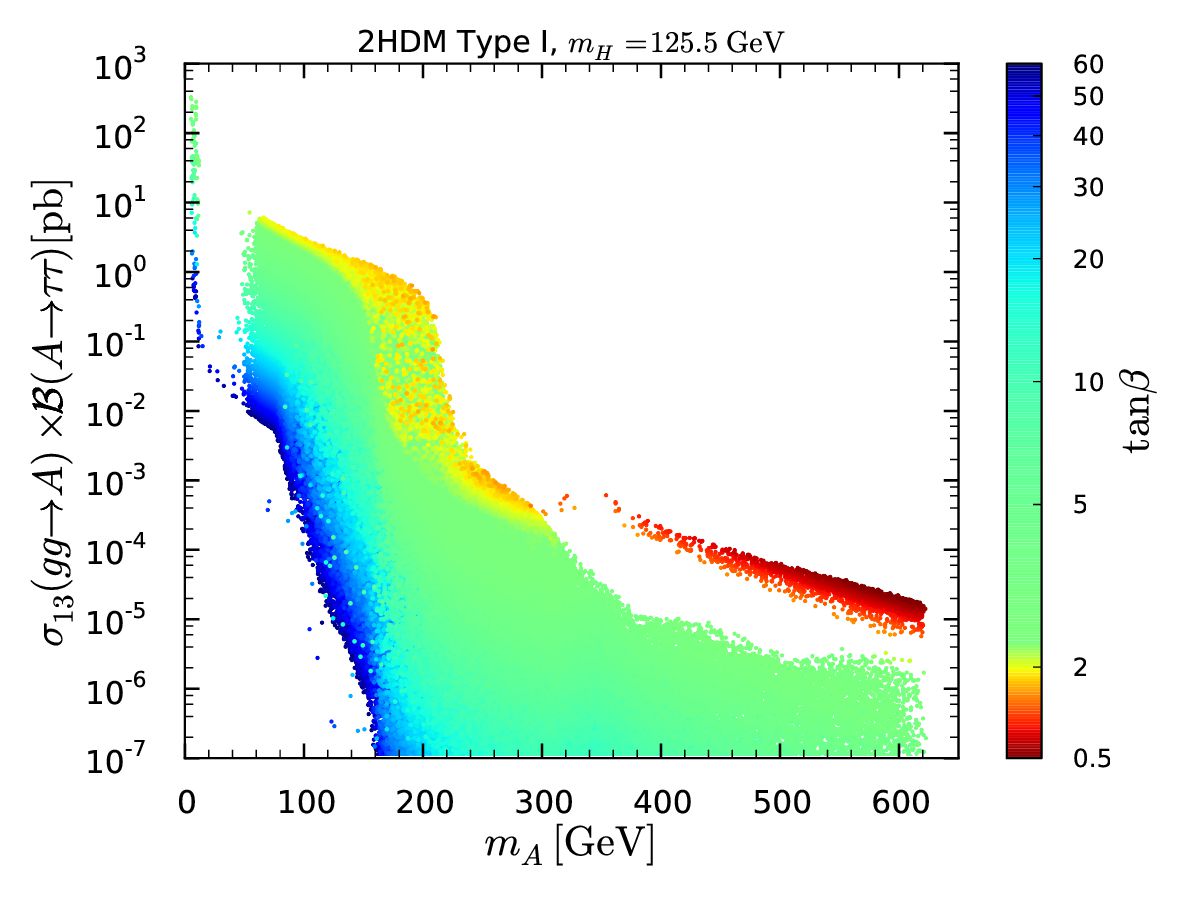}\includegraphics[width=0.51\textwidth]{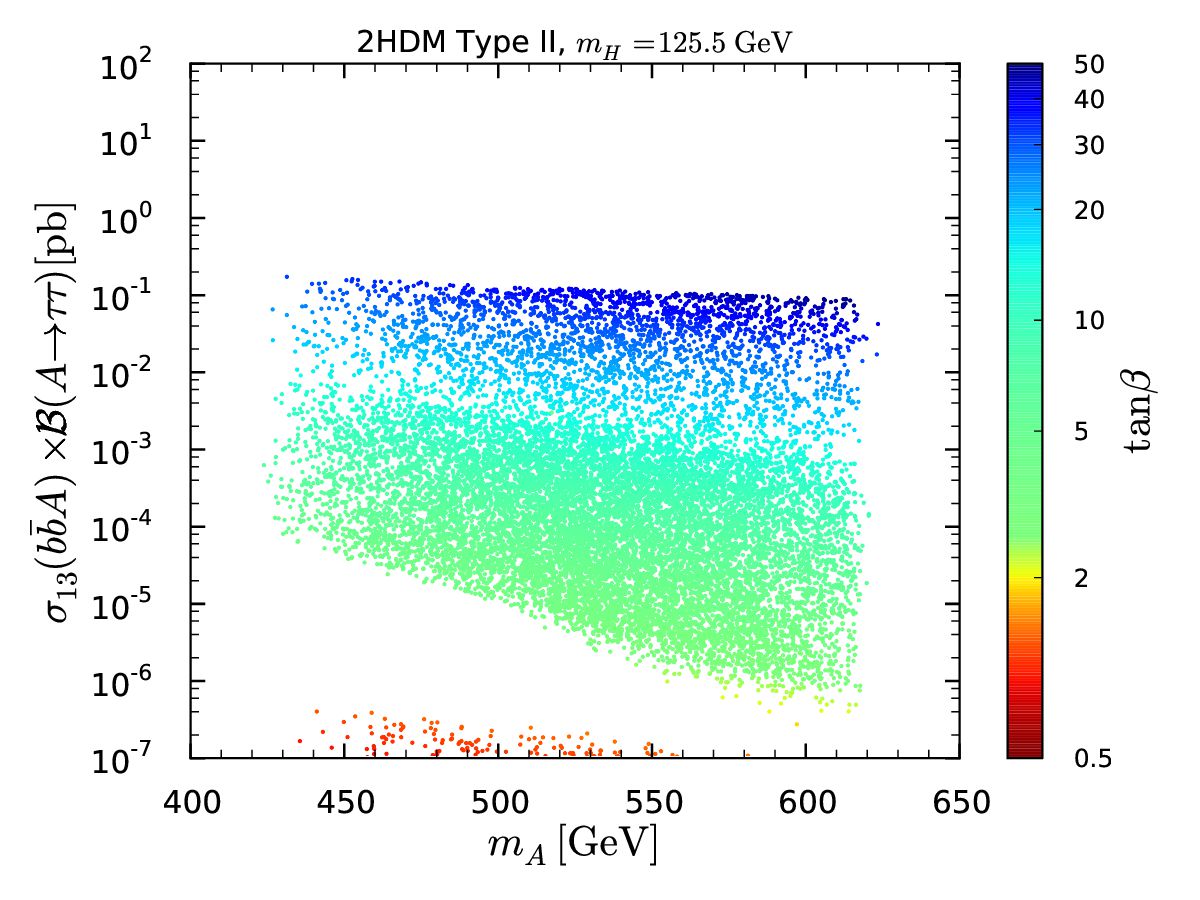}
\includegraphics[width=0.51\textwidth]{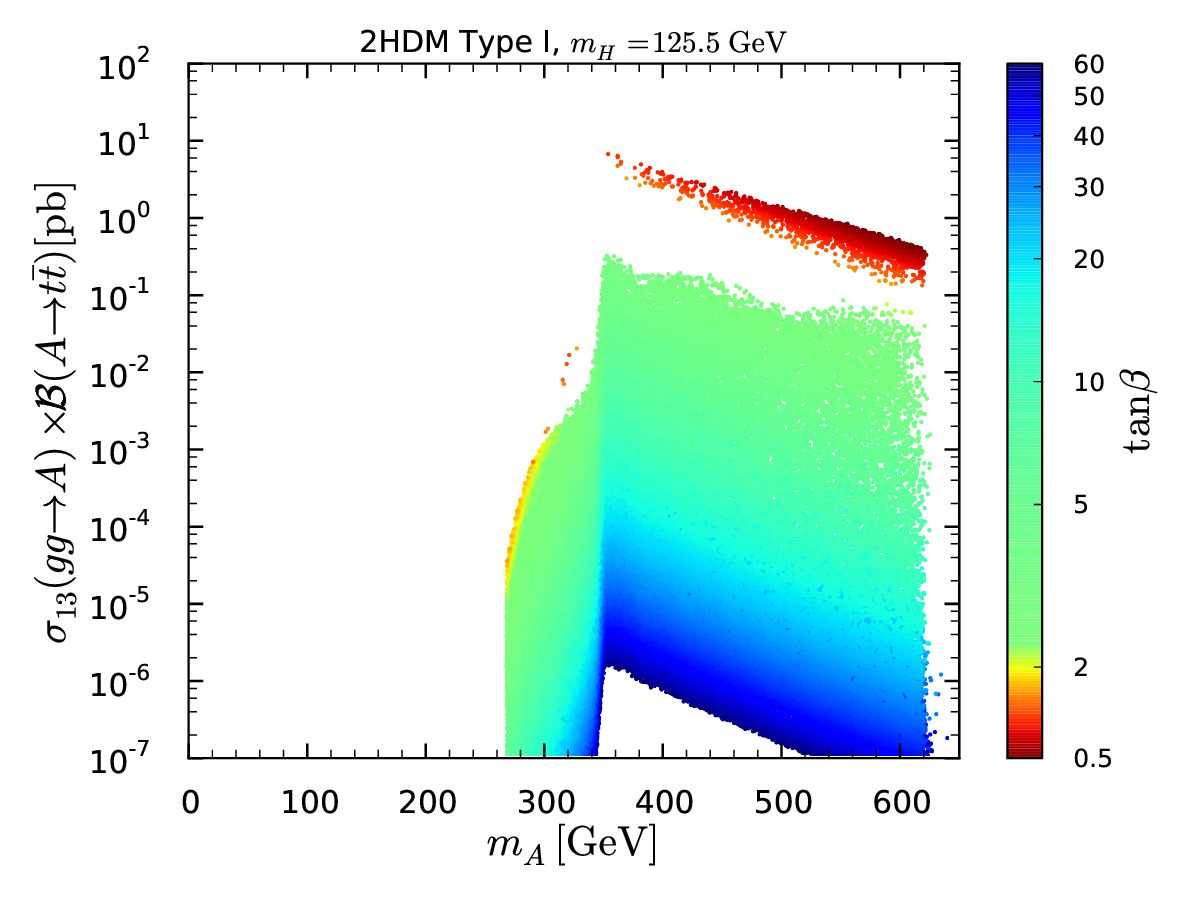}\includegraphics[width=0.51\textwidth]{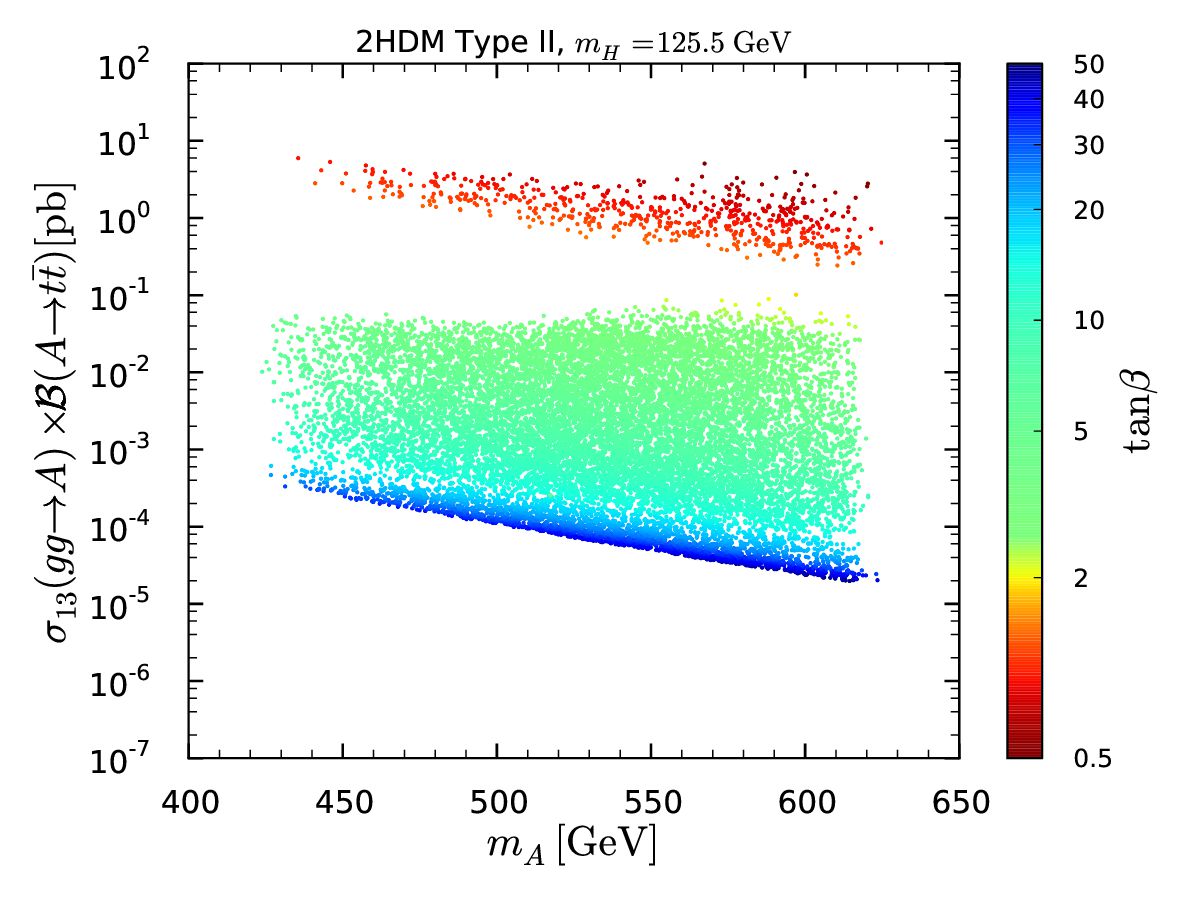}
\caption{Cross sections times branching ratios in Type~I (left) and in Type~II (right) for $A\to Y$ signatures at the 13~TeV LHC as functions of $m_A$ with $Y=\gamma\gamma$ (upper panels), $Y=\tau\tau$ (middle panels) and $Y=t\bar t$ (lower panels) with $\tan\beta$ color code. Points are ordered from high to low $\tan\beta$.}
\label{xsecBRA13} 
\end{figure}

The cross sections for $A$ production with decays into SM channels, $A\to \gam\gam,\, \tau\tau,\, t\bar t$, are presented in Fig.~\ref{xsecBRA13}.   In Type~I, the $gg\to A\to \gam\gam$ ($\tau\tau$) cross section can be as large as roughly 15~fb  (10 to 1~pb), respectively, for $m_A\in[60,200]\gev$, with minimum values that would still be potentially observable for $m_A\lsim 100\gev$ for a very large integrated luminosity.  Of course, the maximal (minimal) values arise for small (large) $\tanb$, implying an indirect determination of $\tanb$ would be possible by measuring these cross sections. We also note a narrow band of non-excluded points with $m_A$ between 10 and 60~GeV (cf.~\cite{Bernon:2014nxa}),  
with very large $gg\to A\to \tau\tau$ cross sections for $m_A\sim 10\gev$. For $\mha\in[200,2m_t]$ the $A$ cross section is small in both the $\gam\gam$ and $\tau\tau$ channels, and the $t\bar t$ cross section is either zero or very tiny.  Once $m_A>2m_t$, a very substantial $gg\to A\to t\bar t$ cross section (up to about 0.3~pb for $\tan\beta>2$ and roughly 0.1--6~pb for $\tan\beta<2$)
is possible for small $\tanb$, but as $\tanb$ increases this cross section declines rapidly. 
Turning to \typeii, we see that observation of the $A$ in the $\gam\gam$ final state will be, at best, extremely difficult.  In contrast, observation of $b\bar b A$ production with $A\to\tau\tau$ may be possible at large $\tanb$.  Moreover,  $b\bar b A$ production is useful for observing the $A\to t\bar t$ decay in \typeii. The cross section (not shown) ranges from  $60$--$0.2$~fb for $m_A\simeq 420$--$630$~GeV, with only little dependence on $\tan\beta$.   
The cross section for $gg\to A\to t\bar t$ is sizeable (up to 8~pb) for very small $\tan\beta$, but below 0.1~pb for $\tanb\gtrsim 2$. 

In evaluating the potential for the discovery of $A$ via the $t\bar{t}$ final state, it
is noteworthy that $gg\to A\to t\bar t$  strongly interferes with the $pp\to t\bar t$ SM background, which 
yields a peak-dip structure in the $t\bar t$ invariant mass distribution~\cite{Dicus:1994bm,Frederix:2007gi,Jung:2015gta}. 
One should also consider the
set of complementary modes, $t\bar t A$ associated production in Types I and II, and $b\bar b A$ associated production in Type II, followed (in both cases) by $A\to t\bar t$, as recently explored in \cite{Craig:2015jba,Hajer:2015gka,Chen:2015fca}.\footnote{We thank Ning Chen and Tao Liu for bringing these studies to our attention.}

\begin{figure}[t!]\centering
\includegraphics[width=0.51\textwidth]{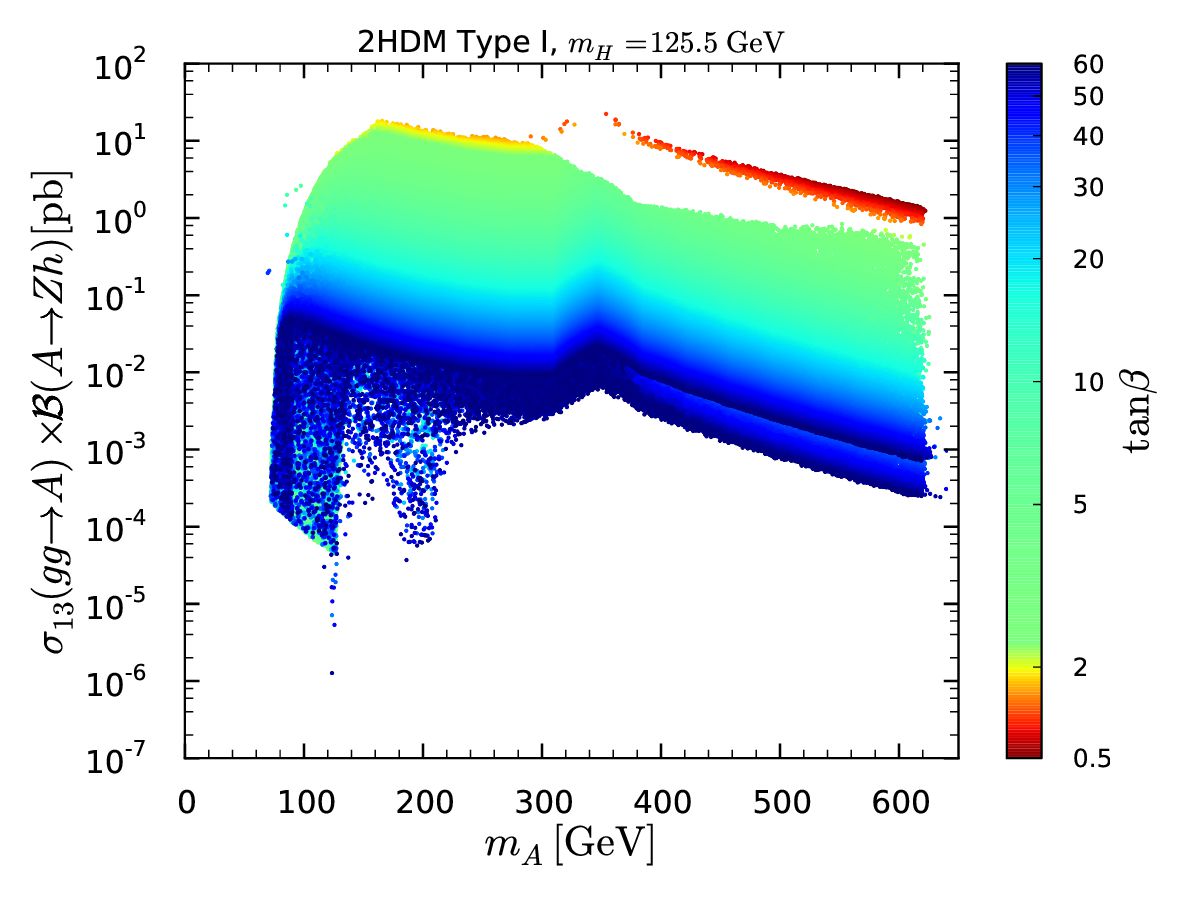}\includegraphics[width=0.51\textwidth]{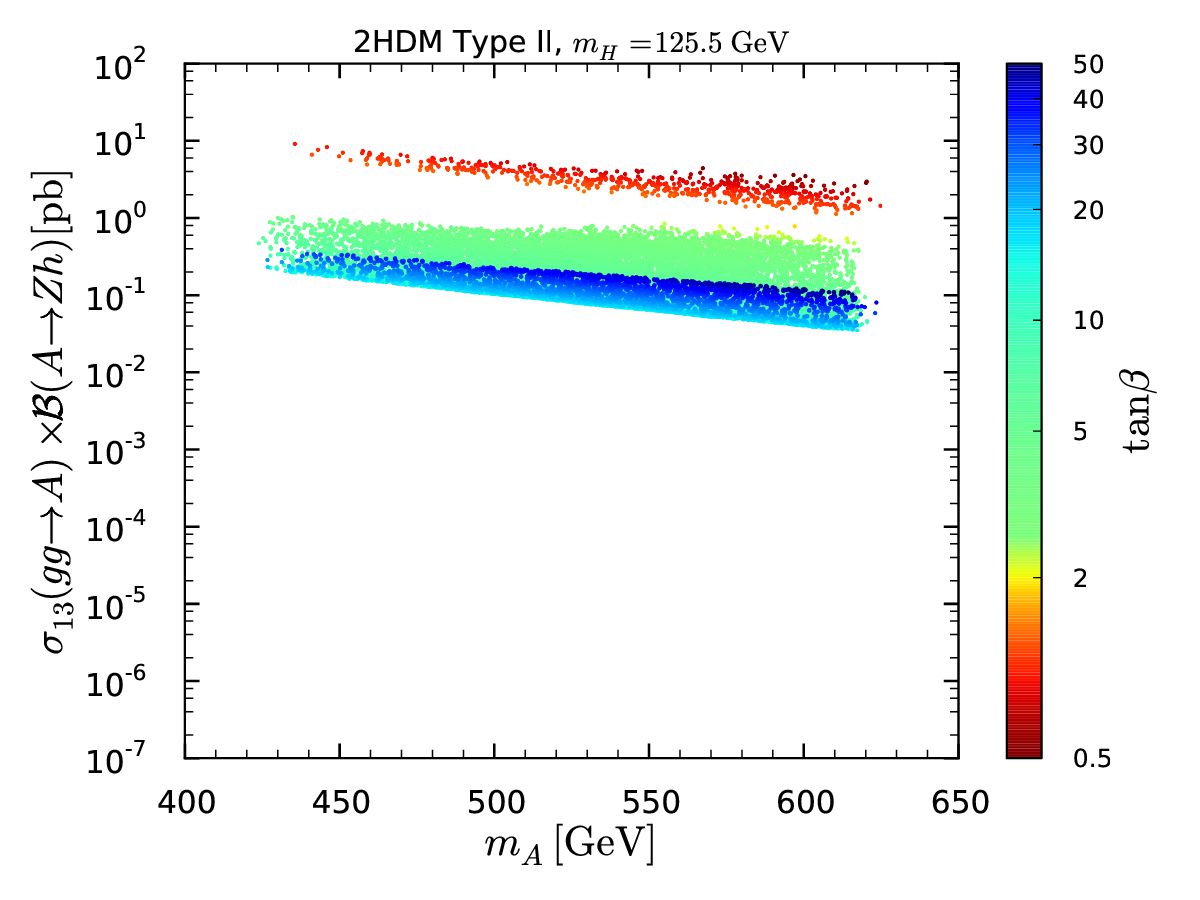}\\
\includegraphics[width=0.51\textwidth]{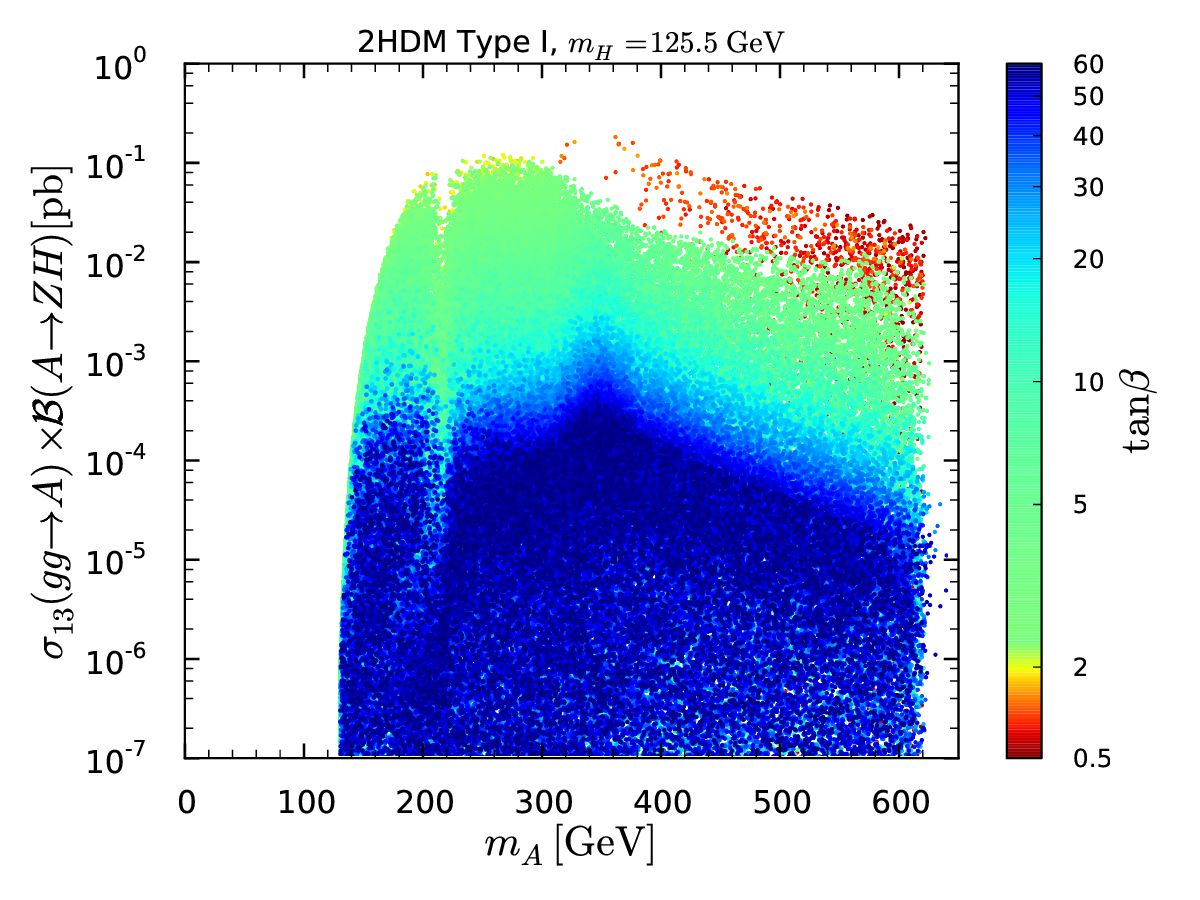}\includegraphics[width=0.51\textwidth]{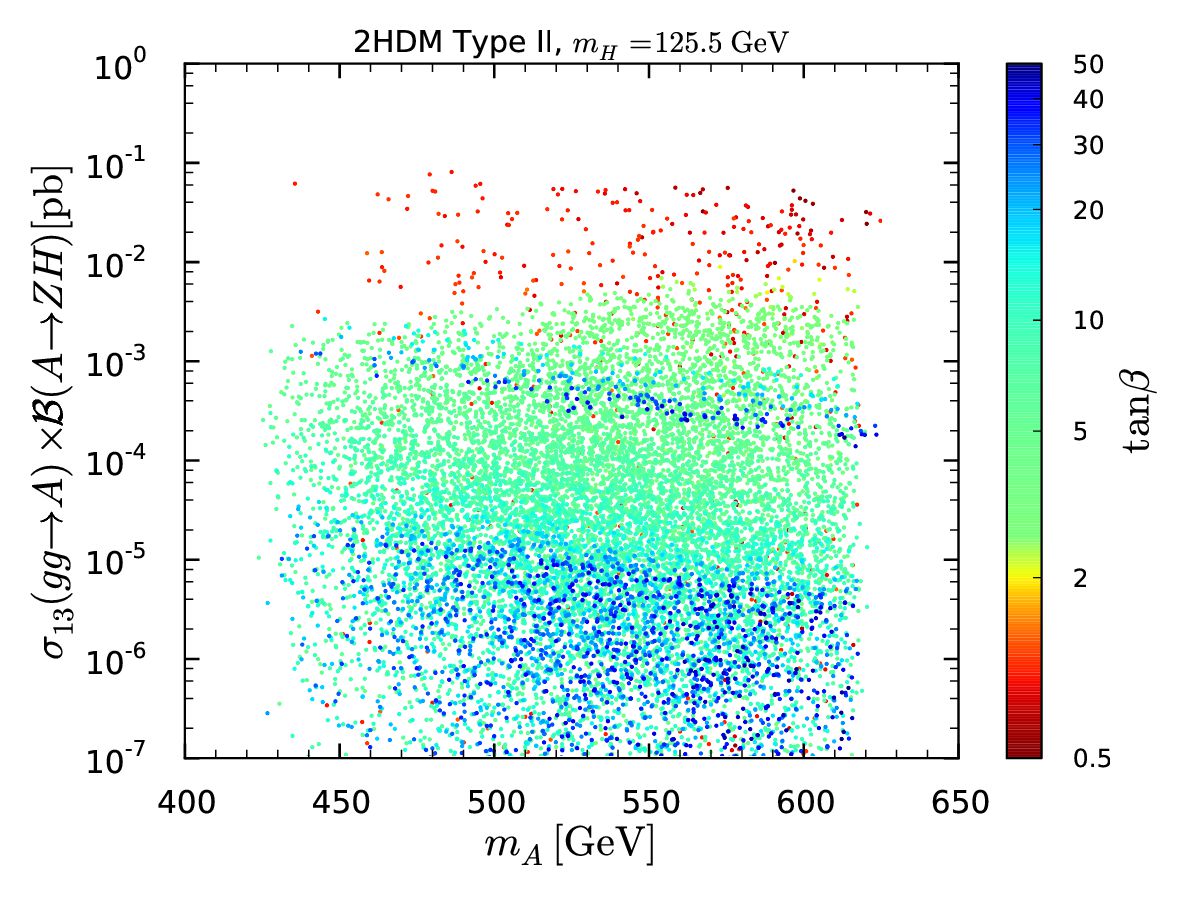}
\caption{Cross sections times branching ratios in Type~I (left) and in Type~II (right)  at the 13~TeV LHC as functions of $m_A$ for $gg\to A\to Zh$ (upper panels) and for $gg\to A\to ZH$ (lower panels) with $\tan\beta$ color code. Points are ordered from low to high $\tan\beta$. 
\label{xsecBRA13exotic} }
\end{figure}

While the sizable cross sections discussed above provide interesting probes of the extended Higgs sector in the alignment limit,  the non-standard signatures of $A\to Zh$ and/or $A\to ZH$ shown in Fig.~\ref{xsecBRA13exotic} appear to be even more 
promising.\footnote{The LHC reach for $gg\to A\to ZH$ and $gg\to A\to Zh$ at $\sqrt{s}=14$~TeV was previously investigated in \cite{Coleppa:2014hxa}.}
In \typeii, there is a strict lower bound on the $gg\to A\to Z\hl$ cross section, with values above 1~pb at small $\tanb$ and at least of order 25~fb at large $\tanb$ even at the maximal value of $\mha=630\gev$. In \typei, at low $\tanb$ the $gg\to A\to Z\hl$ cross sections fall in the range $\sim 1\pb$ to $20\pb$ while at large $\tanb$ this cross section could be as small as $\sim 0.1\fb$ (or even smaller for $m_A<220\gev$). 
Given that the Run~1 searches in this channel remove a significant portion of the low $\tanb$ 2HDM \typei\ and \typeii\ points, it seems certain that Run~2 results would either be substantially more constraining or reveal a signal.

The cross sections for $gg\to A\to ZH$ are typically at least a factor of 100 smaller than those for the $Zh$ final state. Nevertheless, the $A\to ZH$ decay could provide an additional probe of the small-to-intermediate $\tanb$ regime, but will likely be unobservable at large $\tanb$ in both Type~I and Type~II.

\clearpage
\section{Conclusions}\label{conclusions}

In this paper, we studied the approach to the alignment limit of the 2HDM with Type~I and Type~II Yukawa couplings under the assumption that the observed Higgs boson with mass 125~GeV is the heavier CP-even scalar, $H$. 
With the $H$ mass eigenstate being approximately aligned with the direction of the scalar field vacuum expectation value in field space, its coupling to the $W$ and $Z$ bosons tends towards the SM value, $C^H_V\to 1$. Allowing for at most a 1\% deviation from unity in $C_V^H$, we found that deviations  in the couplings to fermions of 10--20\% are possible while maintaining consistency at 95\%~CL with the LHC Run~1 Higgs measurements. While $C_F^H$ in Type~I and $C_U^H$ in Type~II rather quickly approach unity as $|\sbma|\to 0$, the approach of the bottom Yukawa coupling to its SM value in the alignment limit is delayed in Type~II, with $C_D^H\approx 0.72$--$1.12$ even for values of $|\sbma|\sim 10^{-2}$.
Moreover, there can be significant deviations from 1 in the loop-induced coupling to photons: $C_\gamma \approx 0.80$--$1.17$ ($0.88$--$0.97$) in Type~I~(II). 
In the case of \typei, the reason for the larger range of $C_\gam$ and for its extending also to values above 1 is that the charged Higgs can be light. 
All these variations in the couplings feed into distinctive behaviors of the signal strengths. 
Thus, even in the deep alignment regime, where one might naively expect everything to be very SM-like, precise measurements of the signal strengths at 125~GeV can help determine the existence of the extended Higgs sector. Furthermore, correlations between signal strengths are characteristic for the model and can point towards a 2HDM of Type~I or Type~II. 

Comparing with the results from \cite{Bernon:2015qea}, distinguishing $m_h\simeq 125$~GeV from $m_H\simeq 125$~GeV with signal strength measurements and coupling fits alone seems very difficult, unless one finds
values that are excluded by $A\to Zh$ in the $m_H\simeq 125$~GeV case.
Preferably, and certainly more definitively, one would wish to observe the second CP-even scalar, $h$. 
Indeed, in the $m_H\simeq125\gev$ case studied in this paper, the $h$ must lie below $125\gev$ by definition (which also implies that there is no decoupling limit for this type of model).  Moreover, for $\cba\geq 0.99$, precision electroweak observables impose an upper limit on the masses of the CP-odd and the charged Higgs of $m_{A,H^\pm}\lesssim 630\gev$, suggesting that all the extra Higgs states of the 2HDM are at least kinematically accessible at the LHC in this setup. While we did not study the potential for observing the $H^\pm$, direct detection of the $h$ and/or $A$ might be possible in a variety of production $\times$ decay channels. Most exciting and enticing is the channel $gg\to A\to Zh$ which would reveal the presence of both the $h$ and $A$ simultaneously. The associated cross section at $\sqrt{s}=13$~TeV is at least 20~fb (and can be as large as 10~pb) in Type~II.  In Type~I, $\sigma(gg\to A \to Zh)$ is also large, 10~fb to 30~pb, over most of the parameter space, although for very large $\tanb$ it can drop below 1~fb in the ranges $m_A\simeq 90$--$250\gev$ and $m_A\gtrsim 500\gev$.
The searches for $A\to Zh$ with $Z\to \ell\ell$ and $h\to b\bar b$ or $\tau\tau$ are therefore excellent probes for discovering or excluding the 2HDM scenarios with a SM-like $H$, provided that they are performed {\em without} requiring a SM-like $h$ with $m_h=125\gev$. In fact, CMS has already performed such a search for $A\to Zh$ at $\sqrt{s}=8$~TeV for general $m_A$ and $m_h$ values (down to 40~GeV), and the limits they obtained are among the most severe constraints for the scenario studied in this paper. 

Other channels of high interest include $gg\to h \to \gam\gam$ (for $m_h\lesssim 90\gev$)  as well as $gg\to h \to \tau\tau$ (or $\mu\mu$) in Type~II. The $\gam\gam$ channel may also reveal a light $A$ in Type~I if $\tanb$ is small. Moreover, $gg\to A\to \tau\tau$ (or $\mu\mu$) can be used to search for a light $A$ in the 10--250~GeV mass range in Type~I, while in Type~II it would be preferable to exploit the $b\bar b A$ production mode to search for the same $A$ decays (over the relevant mass range of $m_A\simeq 420$--$630\gev$). Finally the $t\bar t$ final state can be relevant  for $m_A$ above 350~GeV in both Type~I and Type~II if $\tanb$ is small.

In short, it is possible that the observed 125~GeV Higgs boson appears SM-like due to
the alignment limit of a multi-doublet Higgs sector. However, the alignment limit does not 
necessarily imply that the additional Higgs states of the model are heavy. 
Indeed, it is possible that the observed Higgs boson at 125~GeV is the heavier CP-even $H$,  
in which case ${m_h\in[10,121.5]\gev}$ (since we intentionally avoided the case of $h$--$H$ mass degeneracy) 
and $m_{A,H^\pm} < 630\gev$. 
Such a scenario, if realized in nature, would lead to exciting new effects to be probed at Run~2 of the LHC.

\section*{Acknowledgements}

This work was supported in part by the ``Investissements d'avenir, Labex ENIGMASS'', the ``Theory-LHC-France Initiative'' of CNRS (INP/IN2P3), the ANR project DMASTROLHC, ANR-12-BS05-0006, and the Research Executive Agency (REA) of the European Union under the Grant Agreement PITN-GA2012-316704 (HiggsTools). 
H.E.H. is supported in part by U.S. Department of Energy grant DE-FG02-04ER41286. 
J.F.G.\ and Y.J.\ are supported in part by the US DOE grant DE-SC-000999. 
Y.J.\  also acknowledges generous support by the LHC-TI fellowship US NSF grant PHY-0969510 
and the Villum Foundation. In addition, he thanks the LPSC Grenoble for its hospitality.
J.F.G., H.E.H. and S.K. are grateful for the hospitality and the inspiring working atmosphere  
of the Aspen Center for Physics, supported by the National Science Foundation Grant No.\ PHY-1066293, 
where this project was initiated.

\bigskip

\appendix

\numberwithin{equation}{section}  

\section*{APPENDIX A: Impact of the CMS $\boldsymbol{A\to Zh}$ exclusion}\label{CMS-AZh}

Both ATLAS and CMS have performed searches at $\sqrt{s}=8$~TeV for a new heavy resonance decaying to a $Z$ boson and a light resonance, with the $Z$ decaying to $\ell\ell=ee,\,\mu\mu$ and the light resonance decaying to $b\bar b$ or $\tau\tau$. 
While the ATLAS analysis~\cite{Aad:2015wra} required that the light resonance be consistent with the observed 125~GeV Higgs boson, the CMS analysis~\cite{CMS-PAS-HIG-15-001} treated the masses of the two resonances as free parameters and published limits on cross section times branching ratios as functions of the two masses. 
We can therefore use this CMS result as a constraint on $A\to Zh$ in our study. 

\begin{figure}[t!]\centering
\includegraphics[width=0.51\textwidth]{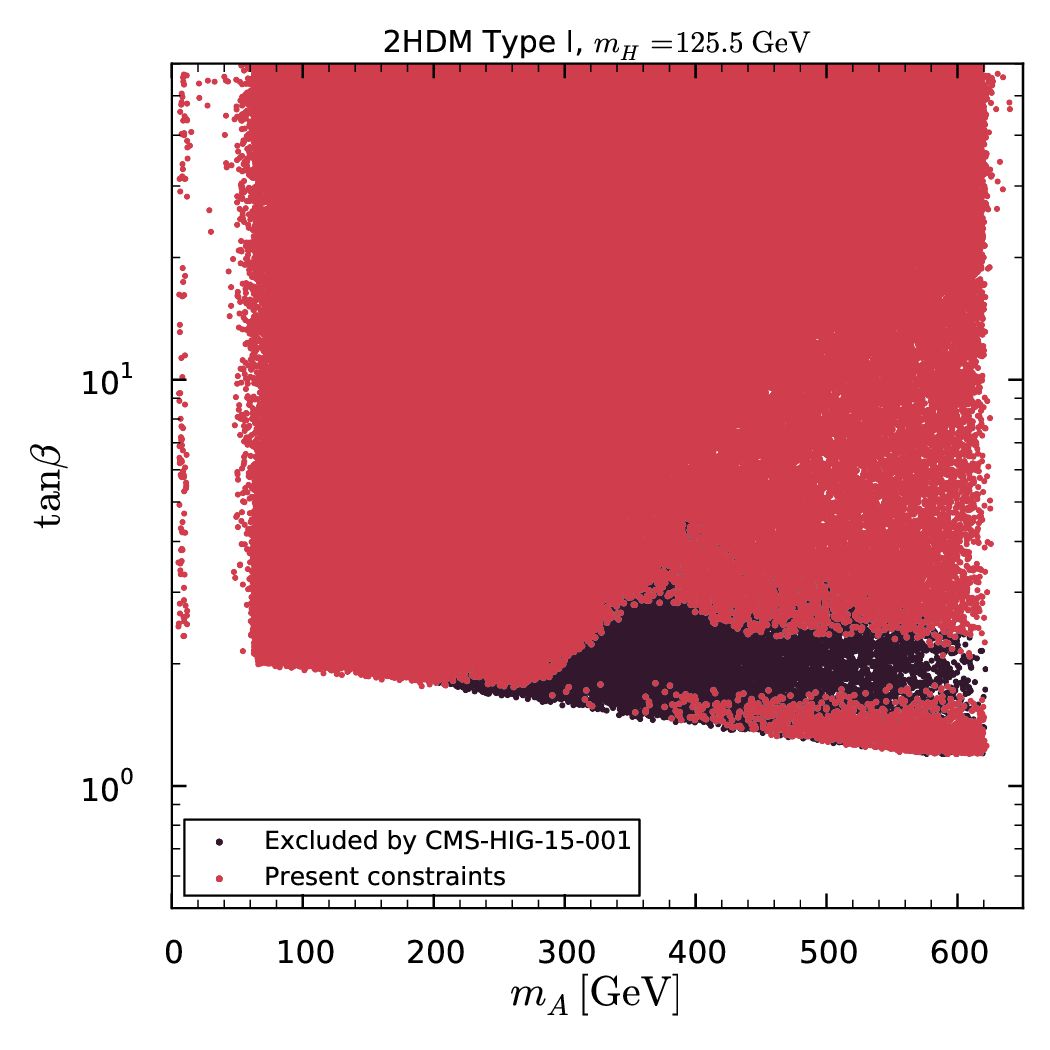}\includegraphics[width=0.51\textwidth]{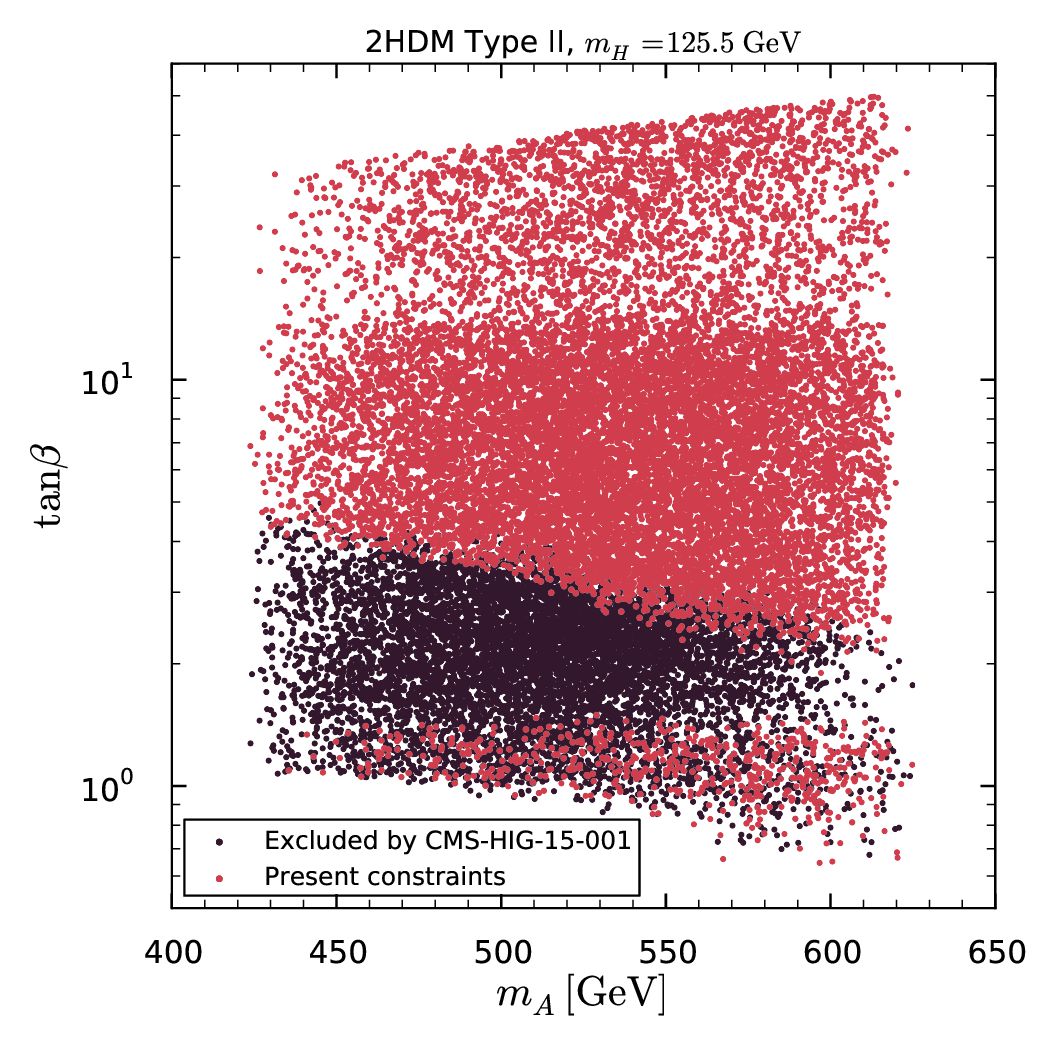}
\caption{Projection of the scan points in the plane $\tanb$ vs.\ $m_A$, on the left for Type~I, on the right for Type~II. The red points are consistent with all constraints used in this paper, while the underlying black points are those which are excluded by the CMS $A\to Zh$ ($Z\to \ell\ell$, $h\to b\bar b,\tau\tau$) limit~\cite{CMS-PAS-HIG-15-001} after all other constraints have been applied.}
\label{mA-tb_AZh_comparison} 
\end{figure}

\begin{figure}[h!]\centering
\includegraphics[width=0.51\textwidth]{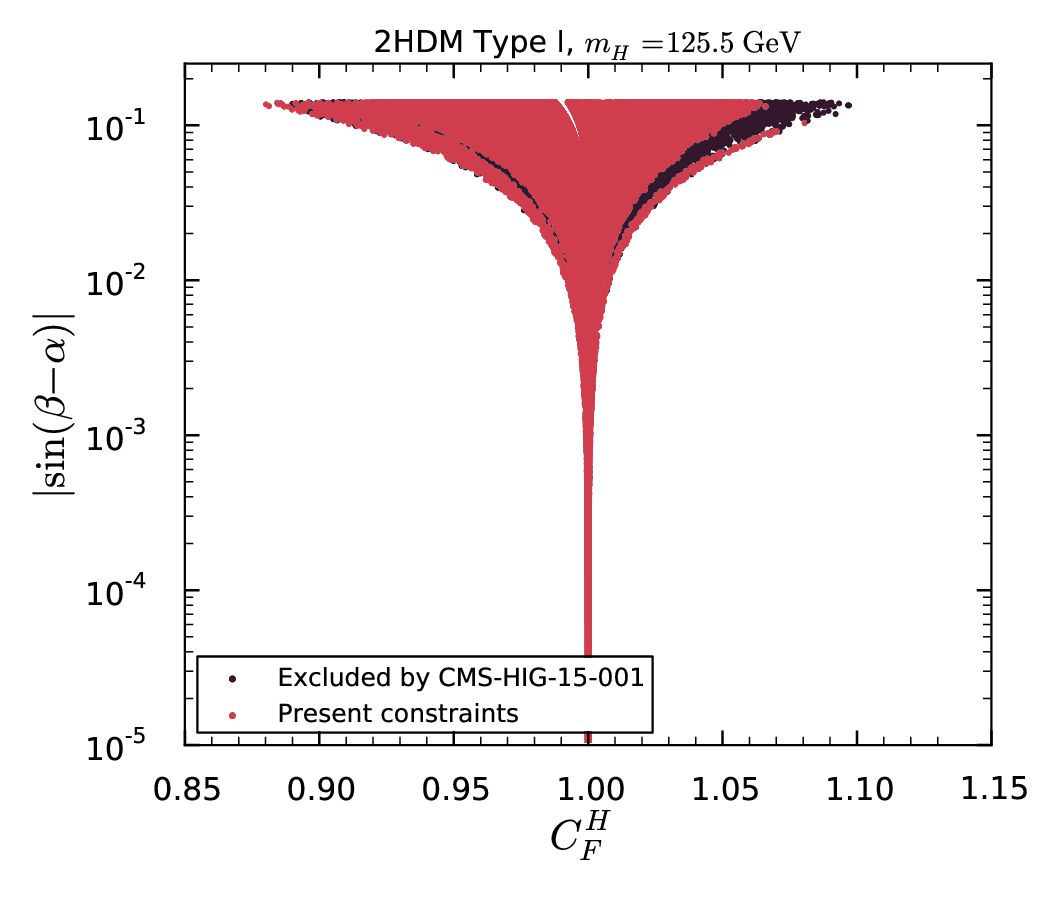}\includegraphics[width=0.51\textwidth]{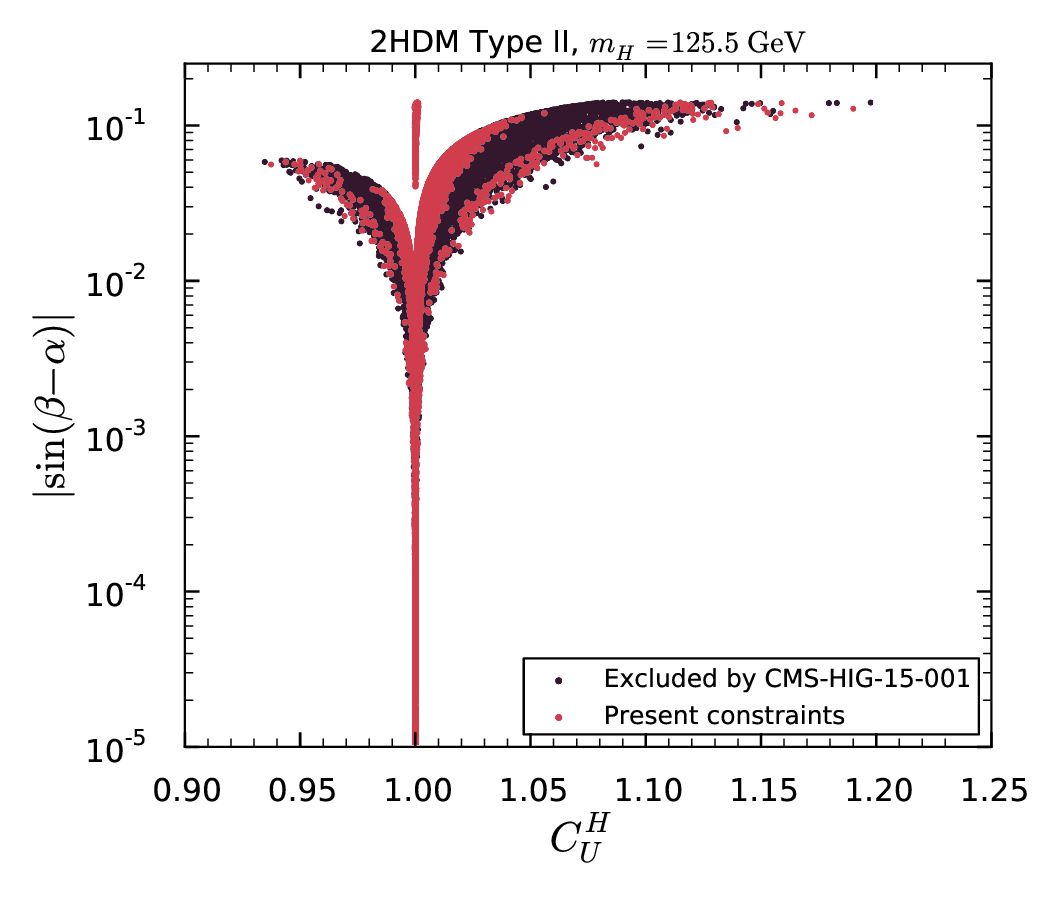}\\
\caption{As in Fig.~\ref{mA-tb_AZh_comparison} but in the plane $|\sba|$ vs.\ $C_F^H$ for Type~I (left) and $|\sba|$ vs.\ $C_U^H$ for Type~II (right). To be compared to Fig.~\ref{CV_CF_H125}.} 
\label{couplings1_AZh_comparison} 
\end{figure}

\begin{figure}[h!]\centering
\includegraphics[width=0.51\textwidth]{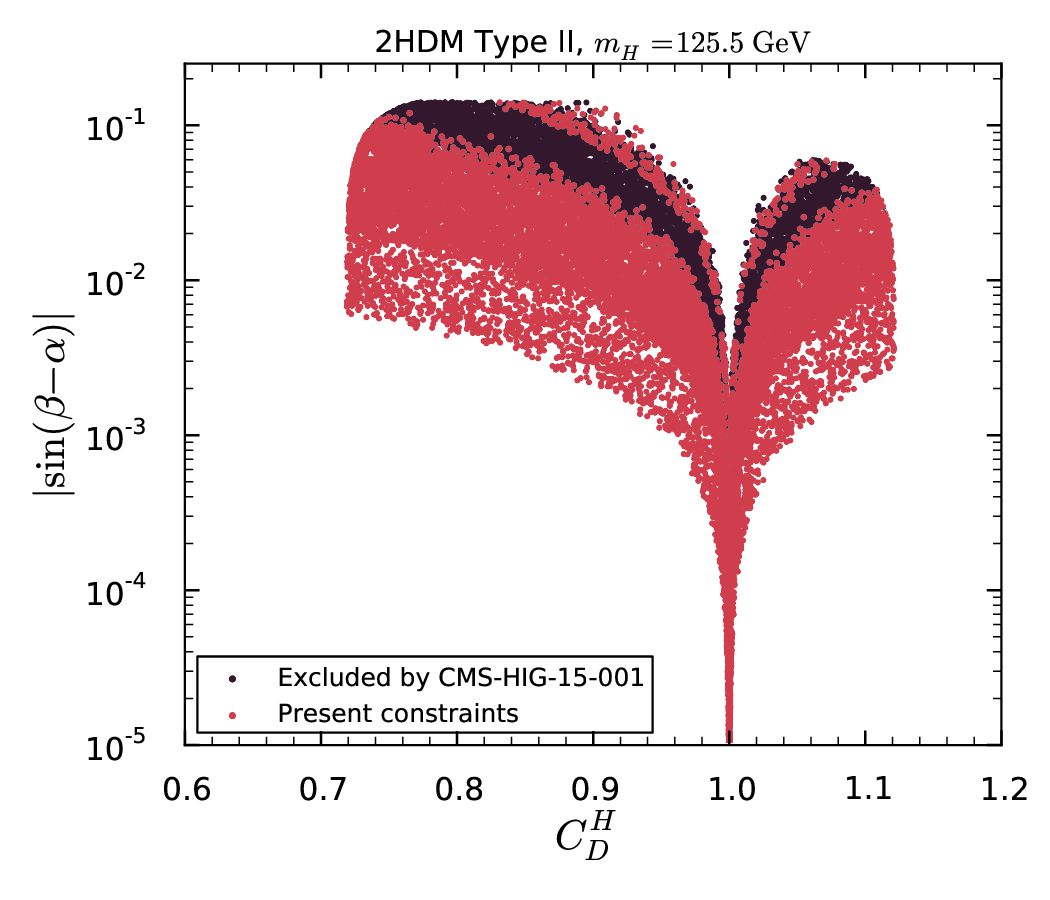}
\caption{As in Fig.~\ref{mA-tb_AZh_comparison} but in the plane $|\sba|$ vs.\ $C_D^H$ for Type~II. To be compared to the left panel of Fig.~\ref{CV_CD_H125}. Note that the figure shows only the positive $C_D^H$ region, the opposite-sign solution not being affected by the CMS $A\to Zh$ constraint.}
\label{couplings1b_AZh_comparison} 
\end{figure}

\begin{figure}[h!]\centering
\includegraphics[width=0.51\textwidth]{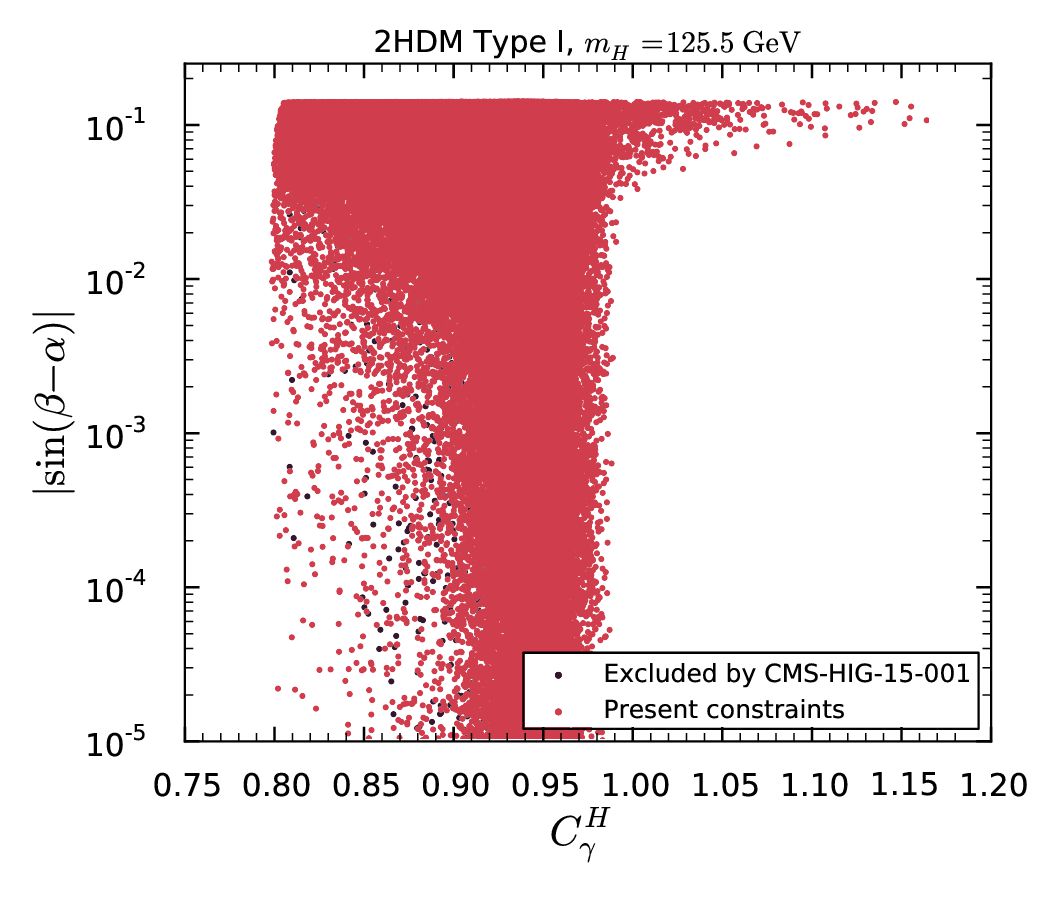}\includegraphics[width=0.51\textwidth]{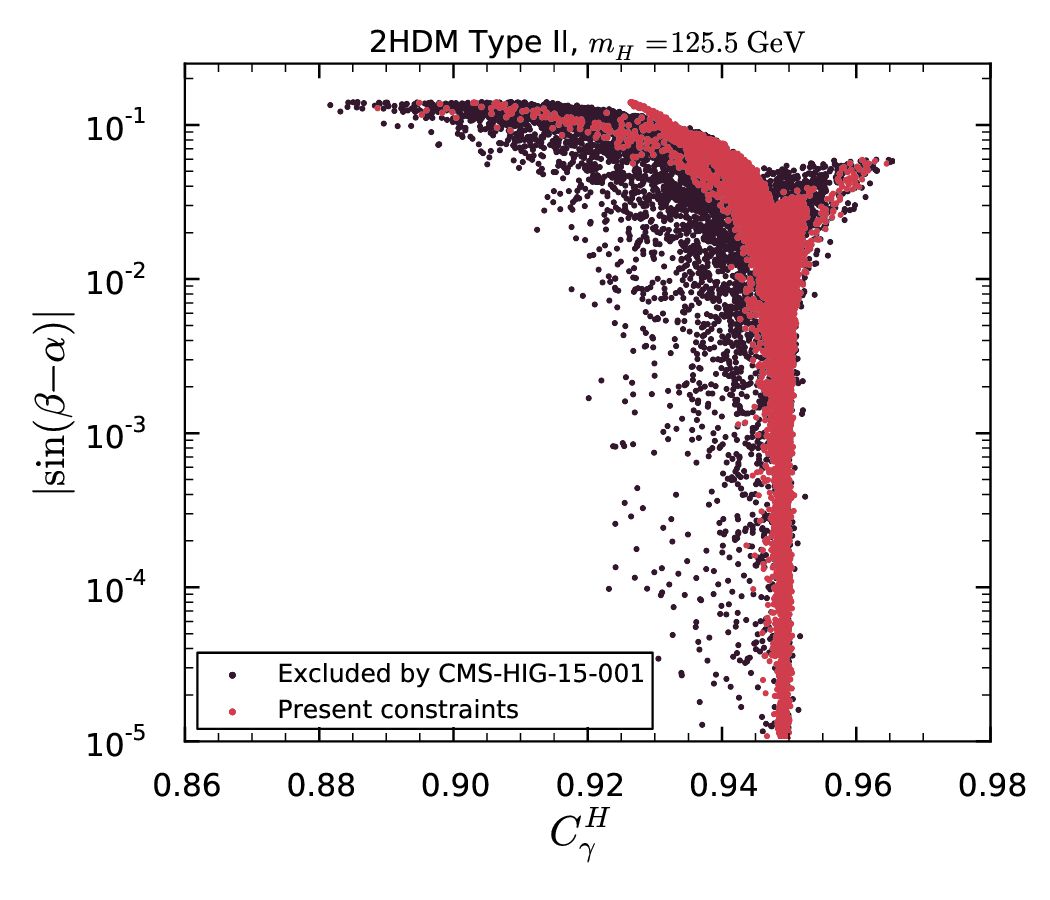}\\
\caption{As in Fig.~\ref{mA-tb_AZh_comparison} but in the plane $|\sba|$ vs.\ $C_\gamma^H$. To be compared to Fig.~\ref{Cgamma_H125}.}
\label{couplings2_AZh_comparison} 
\end{figure}

\begin{figure}[h!]\centering
\includegraphics[width=0.51\textwidth]{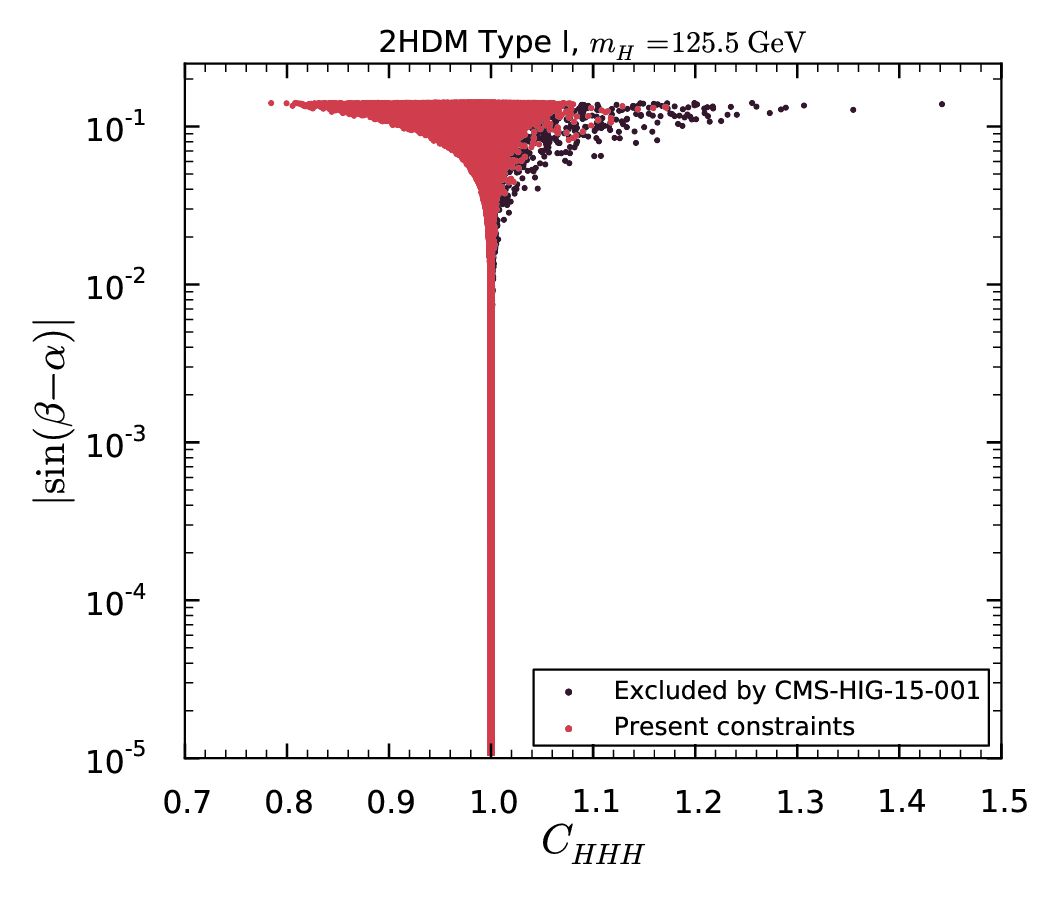}\includegraphics[width=0.51\textwidth]{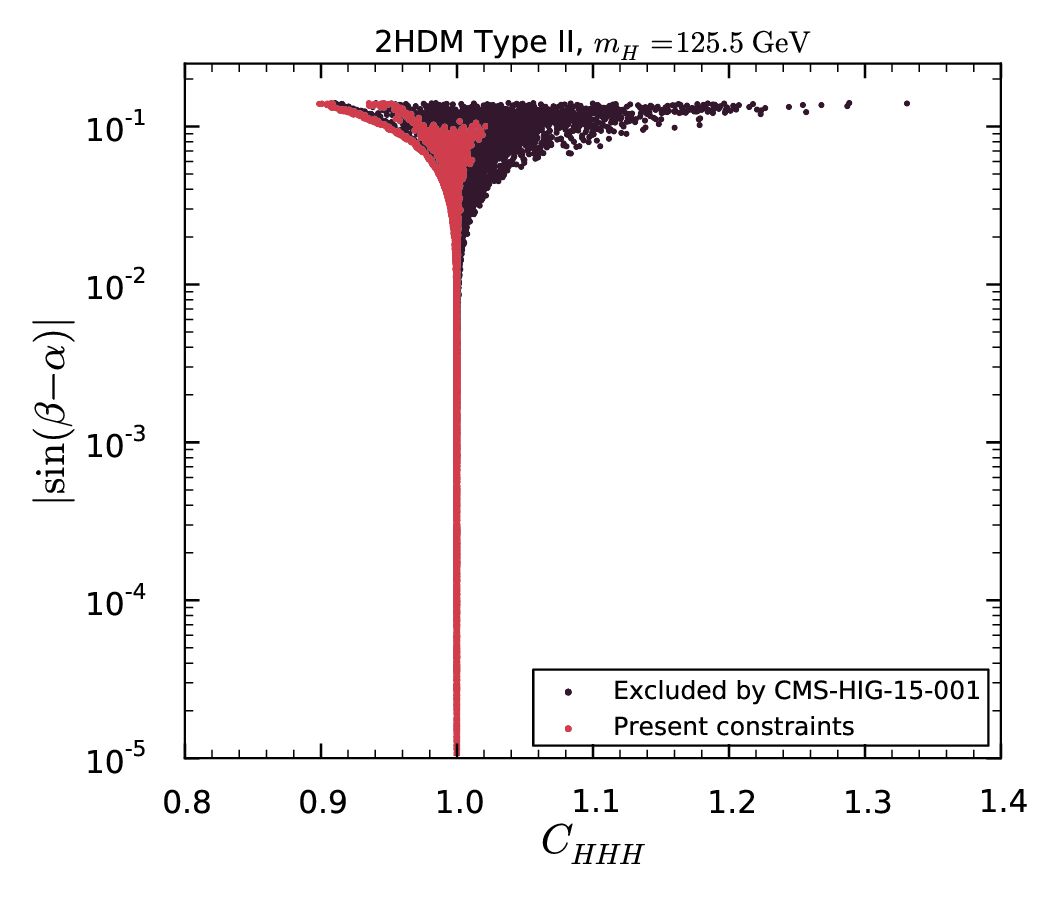}
\caption{As in Fig.~\ref{mA-tb_AZh_comparison} but in the plane $|\sba|$ vs.\ $C_{HHH}$. To be compared to Fig.~\ref{mH_CHHH_H125}.}
\label{couplings2b_AZh_comparison} 
\end{figure}

\begin{figure}[h!]\centering
\includegraphics[width=0.51\textwidth]{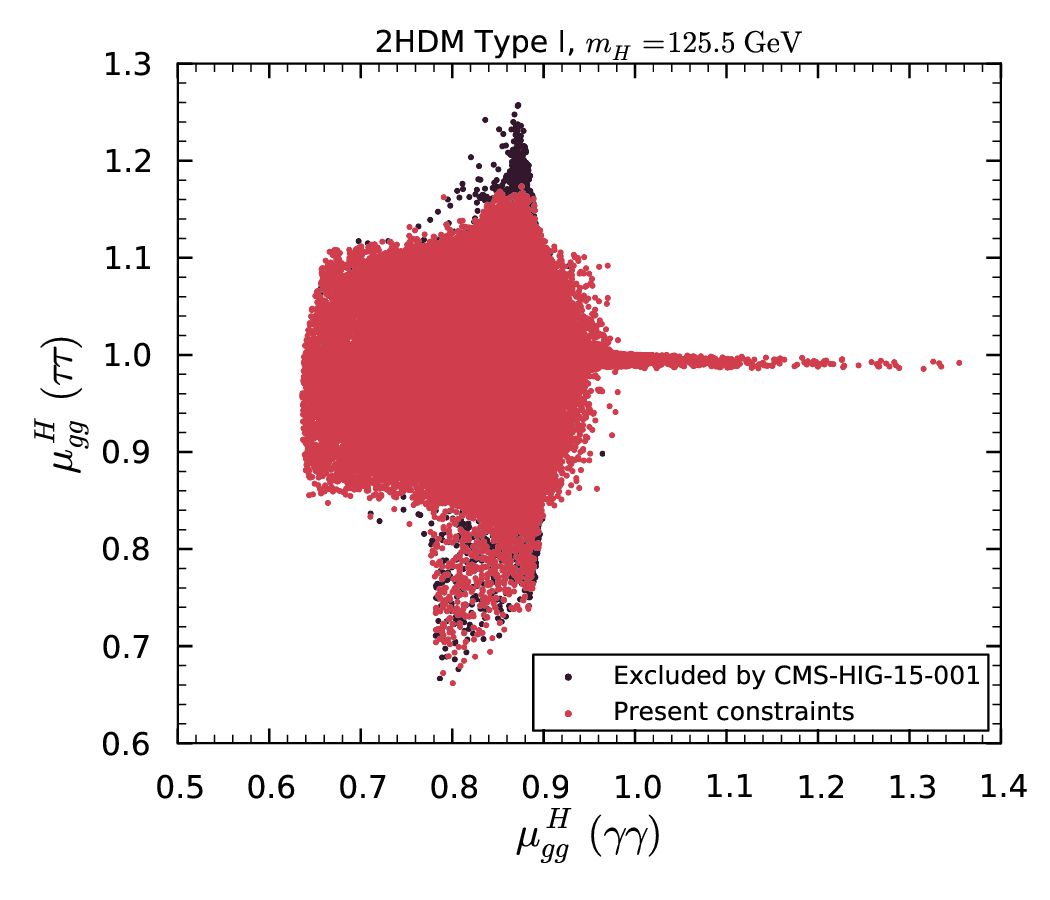}\includegraphics[width=0.51\textwidth]{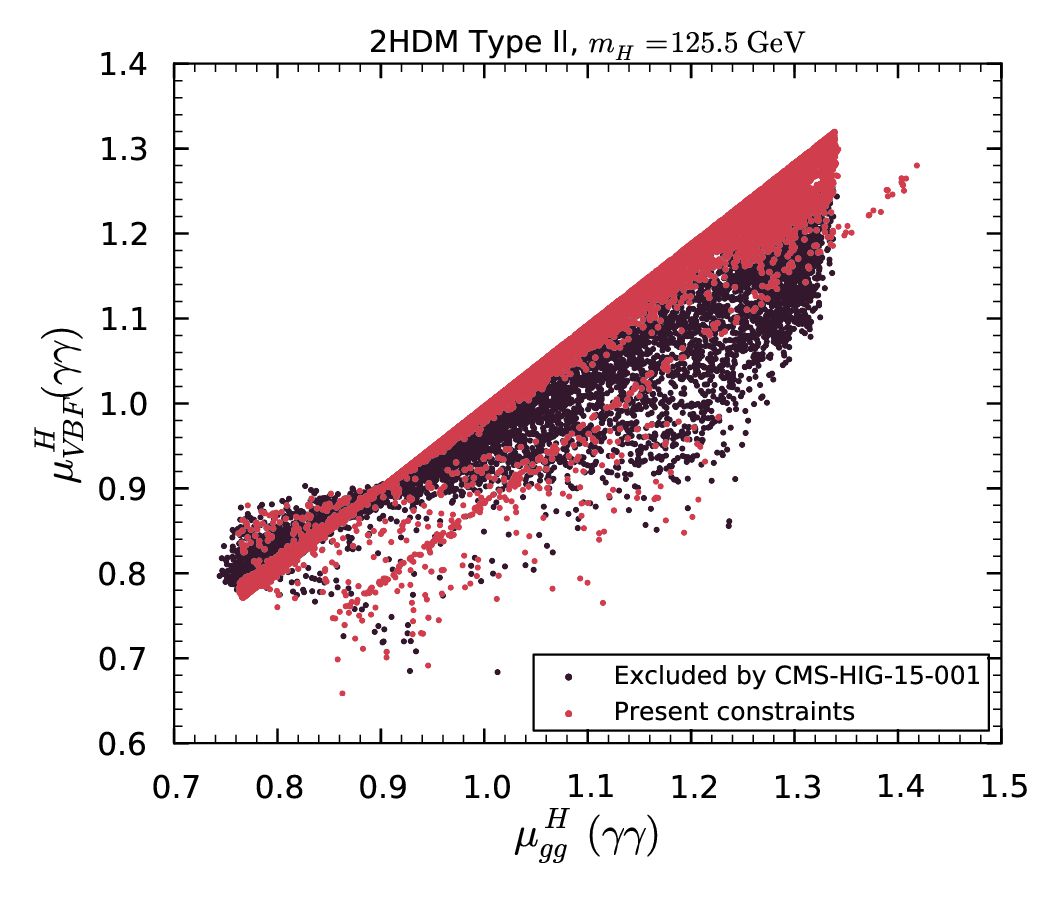}
\caption{As in Fig.~\ref{mA-tb_AZh_comparison} but in the $\mu_{gg}^H(\tau\tau)$ vs.\ $\mu_{gg}^H(\gamma\gamma)$ in Type~I (left panel) and  $\mu_{VBF}^H(\gamma\gamma)$ vs.\ $\mu_{gg}^H(\gamma\gamma)$ in Type~II (right panel) planes. To be compared to the fourth and first rows of Fig.~\ref{correlations1} and Fig.~\ref{correlations2} respectively.}
\label{mu_mu_AZh_comparison} 
\end{figure}

\begin{figure}[h!]\centering
\includegraphics[width=0.51\textwidth]{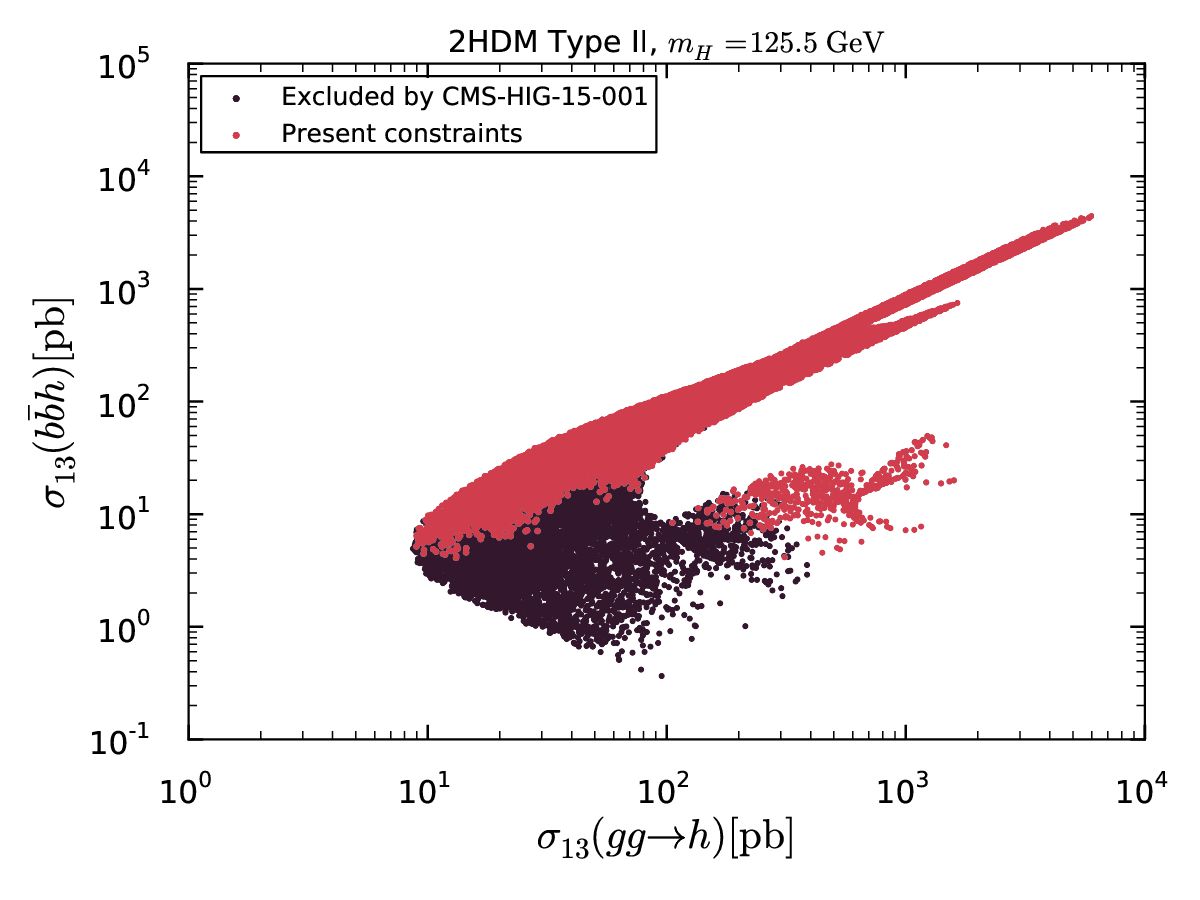}\includegraphics[width=0.51\textwidth]{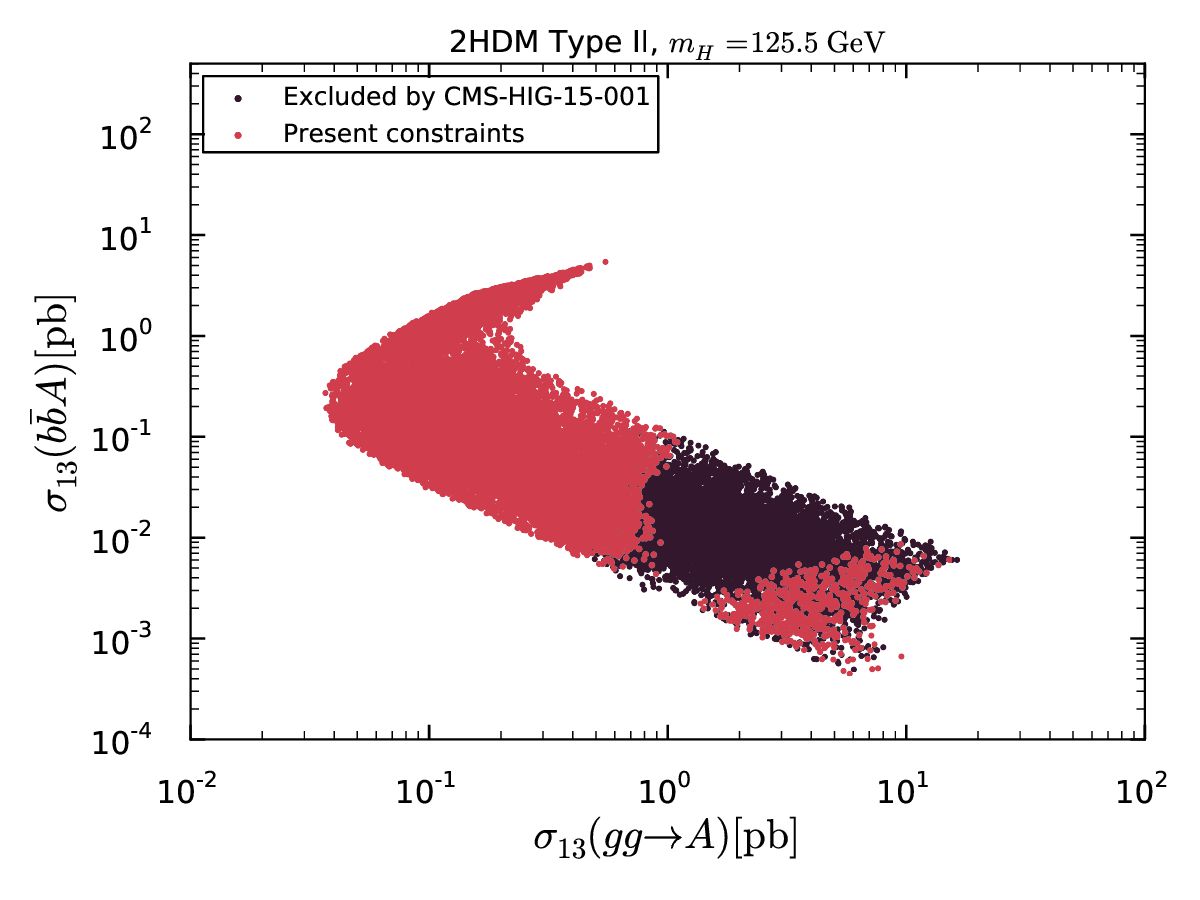}
\caption{As in Fig.~\ref{mA-tb_AZh_comparison}, but showing cross sections $\sigma(b\bar{b}X)$ versus $\sigma(gg\to X)$ for $X=h$ (left) and $X=A$ (right) in Type~II at the 13~TeV LHC. To be compared to Fig.~\ref{xsec_correlation_II_13}. The effect on the analogous Type~I results is very small.}
\label{xsecTypeII_AZh_comparison} 
\end{figure}

\begin{figure}[t!]\centering
\includegraphics[width=0.51\textwidth]{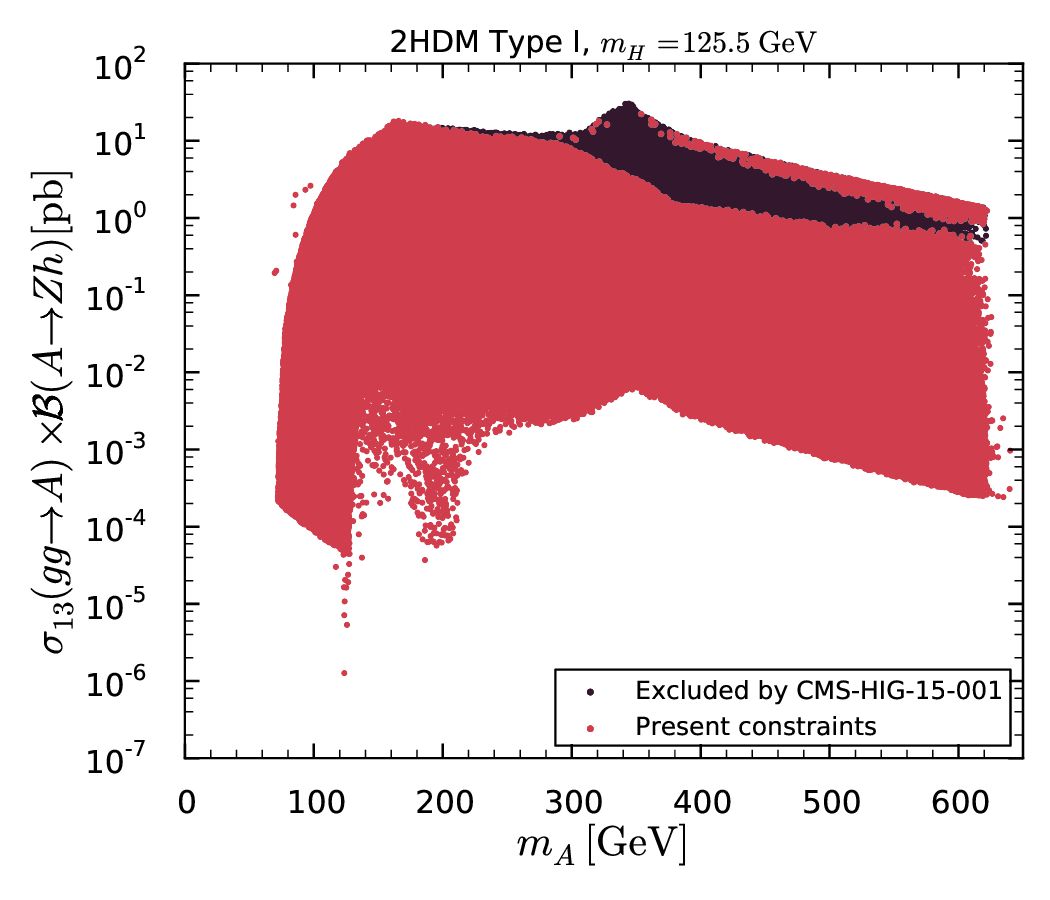}\includegraphics[width=0.51\textwidth]{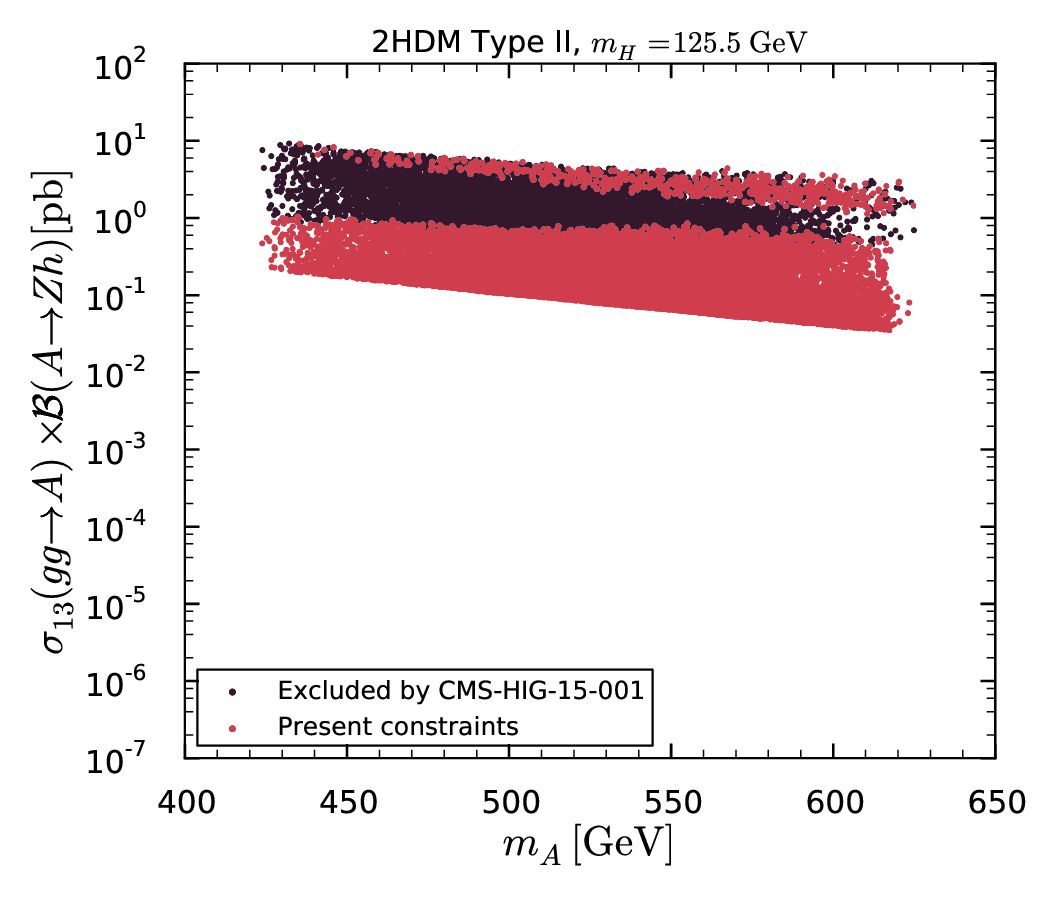}
\caption{As in Fig.~\ref{mA-tb_AZh_comparison}, but showing cross section times branching ratio as function of $m_A$ in Type~I (left) and in Type~II (right) for $gg\to A\to Zh$ at the 13~TeV LHC. To be compared to the upper row of plots of Fig.~\ref{xsecBRA13exotic}. 
The strip of red points with $m_A>400\gev$ and high cross section corresponds to the red points in the bottom-right corners of Fig.~\ref{mA-tb_AZh_comparison}; the CMS $A\to Zh$ limit is evaded because of a light $h$, $m_h\lesssim 40\gev$.
}
\label{xsecAZh13TeV_AZh_comparison} 
\end{figure}

For values of $m_A\approx 200$--$600$~GeV, which is the mass range of particular interest for our analysis, the 95\%~CL limit on 
$\sigma(gg\to A\to Zh)\times \br(Z\to \ell\ell)\times \br(h\to b\bar b)$ obtained by CMS is about 100~fb for $m_h\approx 40$--$45$~GeV, corresponding to the lowest $m_h$ considered in \cite{CMS-PAS-HIG-15-001}. For heavier $m_h$, the limit is about 100~fb at $m_A\approx 200$~GeV going down to about 5~fb at $m_A\approx 600$~GeV. As previously mentioned, this is a very severe constraint for the 2HDM in the alignment limit with $m_H\simeq 125\gev$, cutting out whole slices of parameter space, in particular at low $\tanb$. The limit on $\sigma(gg\to A\to Zh)\times \br(Z\to \ell\ell)\times \br(h\to \tau\tau)$ has a weaker impact. Indeed, most of the points excluded by the $\ell\ell\tau\tau$ search channel are also excluded by the $\ell\ell b\bar b$ channel.  Nonetheless, in Type~I there is a small corner of parameter space at $\tanb\approx 2$ and $m_A\lesssim 400\gev$ that is mostly constrained by the $A\to Zh\to \ell\ell\tau\tau$ CMS limit. The impact of the CMS $A\to Zh$ exclusion is illustrated explicitly in Figs.~\ref{mA-tb_AZh_comparison}--\ref{xsecAZh13TeV_AZh_comparison}. 

In Fig.~\ref{mA-tb_AZh_comparison} we show the projections of the scan points onto the plane $\tanb$ versus  $m_A$ for both Type~I and Type~II. The red points are consistent with all constraints used in this paper, while the underlying black points are those which are excluded by the CMS $A\to Zh$ ($Z\to \ell\ell$, $h\to b\bar b,\tau\tau$) limits~\cite{CMS-PAS-HIG-15-001} after all other constraints have been applied. We see that the $A\to Zh$ limit from Run~1 excludes a whole slice of parameter space at low $\tanb$ and $m_A$ above about 300~GeV.  The surviving red points with $m_A>400\gev$ and $\tanb<2$ have $m_h\lesssim 40\gev$. Had the CMS analysis been sensitive to light resonance masses below 40 GeV, the entire parameter space with $m_A>400\gev$ and $\tanb\lesssim 2$ would have been ruled out.

The effect on the allowed ranges of the reduced couplings is shown in Figs.~\ref{couplings1_AZh_comparison}--\ref{couplings2b_AZh_comparison}. These plots can be directly compared to the ones in Section~3.2. Particularly striking is the impact on $C_\gamma^H$ in Type~II, where the previously possible deviations below the canonical value of  $C_\gamma^H\approx 0.95$ are now largely ruled out. Likewise, large deviations of $C_{HHH}>1$ are very much restricted  by the $A\to Zh$ limit in both Type~I and Type~II, as shown in Fig.~\ref{couplings2b_AZh_comparison}. 
Examples for the impact on the signal strength correlations in Type~I and Type~II are shown in Fig.~\ref{mu_mu_AZh_comparison}.

Finally, we examine the impact of the CMS data for $A\to Zh$ on the cross sections for the production of $A$ and $h$. 
Significant parts of the parameter space are eliminated, as exemplified in Figs.~\ref{xsecTypeII_AZh_comparison} and \ref{xsecAZh13TeV_AZh_comparison}. 

\clearpage

\bigskip\bigskip

\bibliography{references}

\end{document}